\documentclass[a4paper,10pt]{book}
\usepackage[english]{babel}
\usepackage[utf8]{inputenc}
\usepackage{multicol}
\usepackage{multirow}
\usepackage[hypcap]{caption}
\usepackage{amssymb}
\usepackage{enumitem}
\usepackage{eqnarray,amsmath}
\usepackage{color}
\usepackage{footnote}
\usepackage{arydshln}
\usepackage{tikz}
\usepackage{graphicx}
\usepackage{hyperref}
\usepackage{appendix}
\usepackage{geometry}
\usepackage{afterpage}

\usepackage [authoryear]{natbib}
\newcommand\blankpage{%
	\null
	\thispagestyle{empty}%
	\newpage}

\begin{document}

%_____PORTADA___________________________________________________________________________________
\begin{titlepage}
	\begin{center}
		\vspace*{2cm}
		\textbf{\LARGE{\textsf{Evolution of Line-Force Multiplier Parameters in Radiation Driven Winds of Massive Stars}}}\\
		\vspace*{2.5cm}
		\large{\textsc{Alex Camilo Gormaz Matamala}}\\
		\vspace*{2.5cm}
		Instituto de Física y Astronomía\\
		Facultad de Ciencias\\
		\vspace*{2.5cm}
		\begin{figure}[htb]
			\begin{center}
				\includegraphics[width=2.9cm]{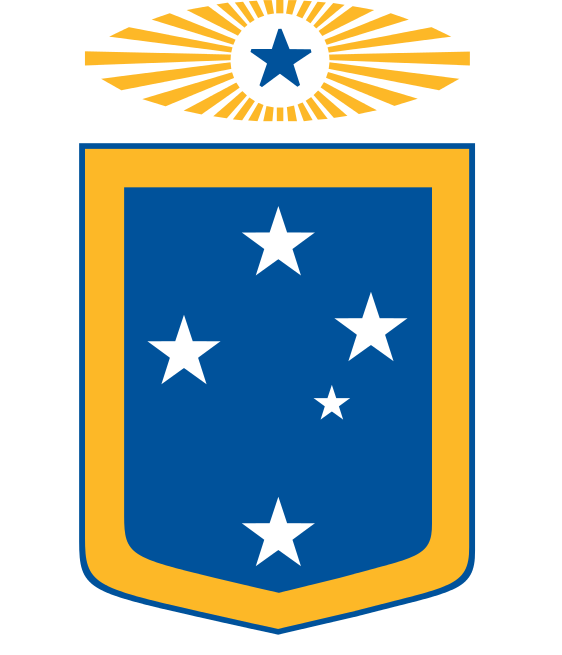}
			\end{center}
		\vspace*{3cm}
		\end{figure}
		Universidad de Valparaíso\\
		Programa de Doctorado en Astrofísica\\
		\vspace*{3cm}
	\end{center}
%	\begin{flushright}
%		Tutor Professor: Dr. Michel Curé Ojeda\\
%		Instituto de Astrofísica, Facultad de Ciencias\\
%		Universidad de Valparaíso\\
%	\end{flushright}
\end{titlepage}

%_____Dedicatory________________________________________________________________________________________
\newpage
\begin{flushright}
	\vspace*{5cm}
	\textit{A la memoria de mi abuela,\\Brígida del Rosario Vega Palma.\\}
	\vspace*{5cm}
	\textit{To the memory of my granny,\\Brígida del Rosario Vega Palma.}
\end{flushright}
\afterpage{\blankpage}

%_____Declaration of not plagiarism__________________________________________________________________________
\newpage
\begin{center}
	\vspace*{8cm}
	This thesis is solely my own composition,\\
	except where specifically indicated in the text.\\
	\vspace*{1cm}
	Total or partial reproduction, for scientific or academic purposes,\\
	is authorised including a bibliographic reference to this document.\\
	\vspace*{8cm}
	Alex Camilo Gormaz Matamala\\
	November 2019\\
	Valparaíso, Chile
\end{center}
\newpage
%_____AGRADECIMIENTOS__________________________________________________________________________
\chapter*{Aknowledgements}\addcontentsline{toc}{chapter}{Aknowledgements}	
	\textit{“The most terrifying fact about the Universe is not that it is hostile but that it is indifferent; but if we can come to terms with this indifference and accept the challenges of life within the boundaries of death – however mutable [hu]man may be able to make them – our existence as a species can have genuine meaning and fulfilment.
	However vast the darkness, we must supply our own light.”}
	\begin{flushright}
		Stanley Kubrick.
	\end{flushright}

	\textit{"All we have to decide is what to do with the time that is given to us."}
	\begin{flushright}
		Gandalf.
	\end{flushright}
	
	The present doctoral work would not have been possible to exist without the great help and support from many people.
	
	Firstly, I would like to express my sincere thanks to my advisor Prof. Michel Curé for the continuous support of my Ph.D study and related research, for his patience, motivation, and immense knowledge.
	His guidance helped me throughout these years of research.
	I also say thanks to the members of my former commission of my PhD Thesis Project defence, Prof. Radostin Kurtev and Prof. Francisco Najarro, who trusted in the success of this work.
	To Prof. Lydia Cidale and Roberto Venero, for their helpful support and feedback from beyond the Andes.
	I sincerely thank J. Puls for helpful discussions that improved this work and for having put to our disposal his code FASTWIND.
	My sincere thanks also goes to Prof. D.~John Hillier and Prof. Jose Groh, who provided me the opportunity to work with them as visitor at University of Pittsburgh (USA) and Trinity College Dublin (Ireland) respectively.
	And to Prof. Alex Lobel, who accepted me as an intern at the Royal Observatory of Belgium.
	
	Besides, I would like to say thanks to each LOC and SOC of the International Workshop on Wolf-Rayet Stars 2015, the Potsdam Astrophysical Summer School 2016, the 10th IAU Symposia 329 \textit{Lives and Death-throes of Massive Stars} 2016, the XXIX Winter School \textit{Application of radiative transfer to stellar and planetary atmospheres} 2017 and the SOCHIAS General Meetings of 2015 and 2016, for letting me participate in each one of these events.
	All these meetings, conferences and schools have been a great experience for me to know colleagues and make my work known for other researchers.
	
	Special mention to my fellows of the Stellar Winds Group at UV, Ignacio, Catalina and Maxi, for their infinite help several times.
	To my office mates Miguel Ángel, Daniela, Clément and Mohsen, for tolerating me these last days.
	To my mates of the PhD in Astrophysics program at Valparaíso: Ana Carolina, Alejandra, Nicolás, Nicolás, Daniela, Alejandro, Javier, Felipe, Stephania, Irma, Rosa, Alexandra, Aurora, Murat, Elena.
	To all the people belonging to the Instituto de Física y Astronomía UV.
	To the people who were my fellows during my abroad internships: Lluís, Kevin, Brian, Dritan, Azarin, Christian, Kara, Amro, who gave me a warm welcome in Pittsburgh.
	Also to the people at Dublin: Eoin, Ioana, Robert, Stephen and specially Laura for helping me a lot to implement the Geneva evolutive code.
	
	Last but not the least, I would like to thank my family: my parents Ana and Carlos and to my sisters Aurora, María José, Margarita and my cousin Arielle for supporting me spiritually throughout writing this thesis and in my life in general.
	Also to all my friends here in Chile and around the world, this would not be the same without all your support.
	
	The present doctoral work has been financially supported by the PhD Scholarship folio Nº 2116 1426 from National Commission for Scientific and Technological Research of Chile (CONICYT), besides receiving funding from the European Union’s Framework Programme for Research and Innovation Horizon 2020 (2014-2020) under the Marie Sk\l{}odowska-Curie grant Agreement No. 823734, from FONDECYT project 1190485 and
from the Chilean Astronomical Society (SOCHIAS) and from Centro de Astrofísica de Valparaíso.

\tableofcontents
%\listoffigures
%\listoftables
%_____PREFACIO__________________________________________________________________________________
%\newpage
\chapter*{Abstract}\addcontentsline{toc}{chapter}{Abstract}
	Massive stars expell strong stellar winds which are described by the theory of radiation-driven wind.
	Accurate mass-loss rates are needed to properly describe the stellar evolution across the Hertzsprung-Russel Diagram.
	
	We present two self-consistent procedures that couple the hydrodynamics with calculations of the line-force in the frame of radiation wind theory.
	These procedures give us the line-force parameters, the velocity field, and the mass-loss rate.
	The first one is based on the so-called m-CAK theory.
	Such computations contemplate the contribution to the line-force multiplier from more than $\sim 900,000$ atomic transitions, an NLTE radiation flux from the photosphere and a quasi-LTE approximation for the occupational numbers.
	A full set of line-force parameters for $T_\text{eff}\ge 32,000$ K and surface gravities higher than 3.4 dex for two different metallicities are presented, along with their corresponding wind parameters (terminal velocities and mass-loss rates).
	Here, we find that the already known dependence of line-force parameters on effective temperature is enhanced by the dependence on $\log g$.
	Terminal velocities present a steeper scaling relation with respect to the escape velocity, this might explain the scatter values observed in the hot side of the bistability jump.
	For the case of homogeneous winds (i.e., without clumping) comparison of self-consistent mass-loss rates shows a good agreement with empirical values.
	We also consider self-consistent wind solutions that are used as input in FASTWIND to calculate synthetic spectra.
	By comparison with the observed spectra for three stars with clumped winds, we found that varying the clumping factor the synthetic spectra rapidly converge into the neighbourhood region of the solution.
	Therefore, this self-consistent m-CAK procedure significantly reduces the number of free parameters needed to obtain a synthetic spectrum.
	
	The second procedure (called \textit{Lambert-procedure}) provides a self-consistent solution beyond m-CAK theory and its approximations, and line-acceleration is calculated by the full NLTE radiative transfer code CMFGEN.
	Both the mass-loss rate and the clumping factor are set as free parameters, hence their values are obtained by spectral fitting after the respective self-consistent hydrodynamics is calculated.
	Since performing the Lambert-procedure requires significant computational power, the analysis is made only for the star $\zeta$-Puppis.
	It is found that fitted wind-parameters are close to those predicted by the m-CAK prescription.
	This suggests that both methodologies providing a lower clumping effect on the wind that those suggested by previous authors.
	
	We illustrate the future potential of the self-consistent m-CAK prescription, showing the first results of two ongoing works: the spectral fitting for a set of high resolution spectra observed by \textsc{Hermes} and the development of new evolutionary tracks with the Geneva evolutive code using self-consistent mass-loss rates.
	The promising results gives a positive balance about the future applications for the self-consistent solutions presented on this thesis.

%_____INTRODUCTION_____________________________________________________________________________
\chapter{Introduction to Massive Stars}\label{generalintro}
	The study of massive stars (i.e., stars with  $M_* >10\,M_\odot$) is a relevant topic in the framework of stellar astrophysics, because these stars exhibit some of the most extreme physical conditions, such as the hottest temperatures, the highest outflows of matter and a complex nucleosynthesis.

	Strong outflowing stellar winds of massive stars eject high amounts of matter that contribute to the chemical enrichment of the interstellar medium in a relatively short timescale.
	Moreover, it has been found that differences on a factor of two in the mass-loss rate affects considerably the final fate of a massive star \citep{meynet94,smith14}.
	Therefore, a better understanding about massive stars and their evolution strongly requires accurate determination of their fundamental parameters, with the amount of matter released being the most relevant \citep[][]{kudritzki00,puls08}.
	Subsequently, it is necessary to understand more in detail the mechanism responsible for driving the wind on massive stars in order to predict more accurately their mass-loss rates.
	The motivation for the present thesis is then, to have a better understanding about the physics involved in the generation of the strong stellar winds on massive stars, in order to perform new prescriptions capable to quantify their mass-loss rates for future issues such as the already mentioned stellar evolution and chemical enrichment.

\section{On the fundamental physics of stars}
	\textit{Stars} are giant spheres of gas at high temperatures emitting energy (as electromagnetic radiation mainly, although, as we will see later, this is not the only way) to the interstellar medium.
	According with \citet[][Section 1.1]{prialnik10}, a star can be defined as a body satisfying the following two conditions:
	\begin{itemize}
		\item It is bound by self-gravity. From this condition it is undergone that stars must have a spherical shape because of gravity, or spheroidal in the case of the existence of axisymmetric forces such as rotation.
		\item It radiates energy supplied by an internal source.
		This source is normally thermonuclear energy, although sometimes gravitational potential energy may play a role due to contractions and collapse.
	\end{itemize}
	For the purposes of the present thesis, we will focus on the thermonuclear energy produced in the interior of stars by thermonuclear fusion: reactions where atomic nuclei are transformed into another species releasing the excess of mass as energy according with the Einstein's equation $E=mc^2$.
	The most important thermonuclear reactions that happen in a star are those where hydrogen is burnt in: the \textit{proton-proton process}\footnote{\textit{Proton-proton process} consists on four hydrogen atoms fusing themselves to generate one atom of helium, together with energy as gamma radiation. This way of hydrogen-burning is predominant is low-mass stars. On the other hand, \textit{CNO cycle} consists in a cyclic chain of thermonuclear reactions that uses \textit{CNO elements}, carbon, nitrogen and oxygen as catalysers. This is the predominant process in more massive stars.} (pp) and the \textit{CNO cycle} (carbon-nitrogen-oxygen).
	Energy produced by nuclear fusion, in form of photons (electromagnetic radiation), passes through all the stellar structure from the nucleus until the surface to be finally released from there to the space.

	The stars are formed by the gravitational collapse of an interstellar gaseous cloud (being these nebulae, supernova remnants or molecular complexes).
	Because of the perturbations produced by shock waves from nearby supernova explosions or collisions with other clouds \citep[][Section 12.2]{prialnik10} or simply because random matter movements, the gas cloud (originally thought as homogeneous) starts to form different regions with over-densities.
	If the density is high enough, it will collapse on these regions in a process called \textit{fragmentation}.
	Compression due to the disruption causes temperature to increase: it increases until conditions for nuclear reactions are reached.
	The \textit{radiation pressure}, produced by the triggered energy, counteracts the gravitational collapse and each point of concentred matter reaches the equilibrium again.
	So, it is generated a sphere in hydrostatic equilibrium whose photons produced in its inner parts because of nuclear fusion will be liberated into the interstellar medium: a star is born.
	
	Since our initial cloud was not homogeneous on density, the different points where the cloud collapses do not concentrate the same quantity of matter.
	Some agglomerations will be bigger and the others smaller.
	This leads to, the mass of the incipient stars be varied: some of them with a mass of one tenth of Sun mass only (hereafter solar-mass, or $M_\odot$), and others fifty or up to hundred times more massive than the Sun.
	It has been found that the distribution of these masses is not homogeneous but it follows a distribution known as \textit{initial mass function} $\xi(m)\propto m^{-\alpha}$, being $\alpha$ a factor around $\sim2.35$ for stars with masses greater than $m=1\,M_\odot$ \citep{salpeter55,kroupa01}.
	In other words, whereas the more massive a star is, the less abundant in the Universe is.
	
	More mass for a star implies a greater compression in its core; a greater compression in the core implies a higher temperature; and a higher temperature implies in turn more collisions among the particles, therefore the rate of thermonuclear reactions increases.
	Given this, whether more mass has a star when it is born, it will have a higher temperature and a higher luminosity (it releases more photons to the space per time unit).
	There is one last consequence: the higher rate of nuclear reactions makes the fuel hydrogen to exhaust faster, so the star lives less time if it is more massive.
	
	Therefore, from all the previously remarked, it results easy to understand why the fate of a star is strongly linked to the mass with which it is born.
	Massive stars (stars whose initial mass is ten or more times the mass of the Sun) are hotter, brighter and have a shorter lifetime than their smaller siblings.
	Besides, the big mass leads into other consequences not included in the previous description.
	The extreme high temperature reached in the core of massive stars (in the order of $\sim10^8$ Kelvin) does not make the rate of nuclear reactions to increase only, but also makes new kind of nuclear reactions appear whose existence would not be possible in a not so extremely hot environment (such as $\sim7\times10^6$ K).
	Moreover, there is one more element playing a role: not only photons are emitted from the star, but also an outflow of particles called \textit{stellar wind}.
	Because massive stars release more energy to the interstellar medium (hereafter ISM), even stellar wind will be stronger in these stars.
	Hence, a star with a great mass will experiment an evolutionary path completely different from a solar like star.
	
	We will go deeper into the evolution followed by a massive star in the next sections, but we will bring more theory about massive stars and stellar winds at first.
	\begin{figure}[t!]
		\begin{center}
			\includegraphics[width=0.85\textwidth]{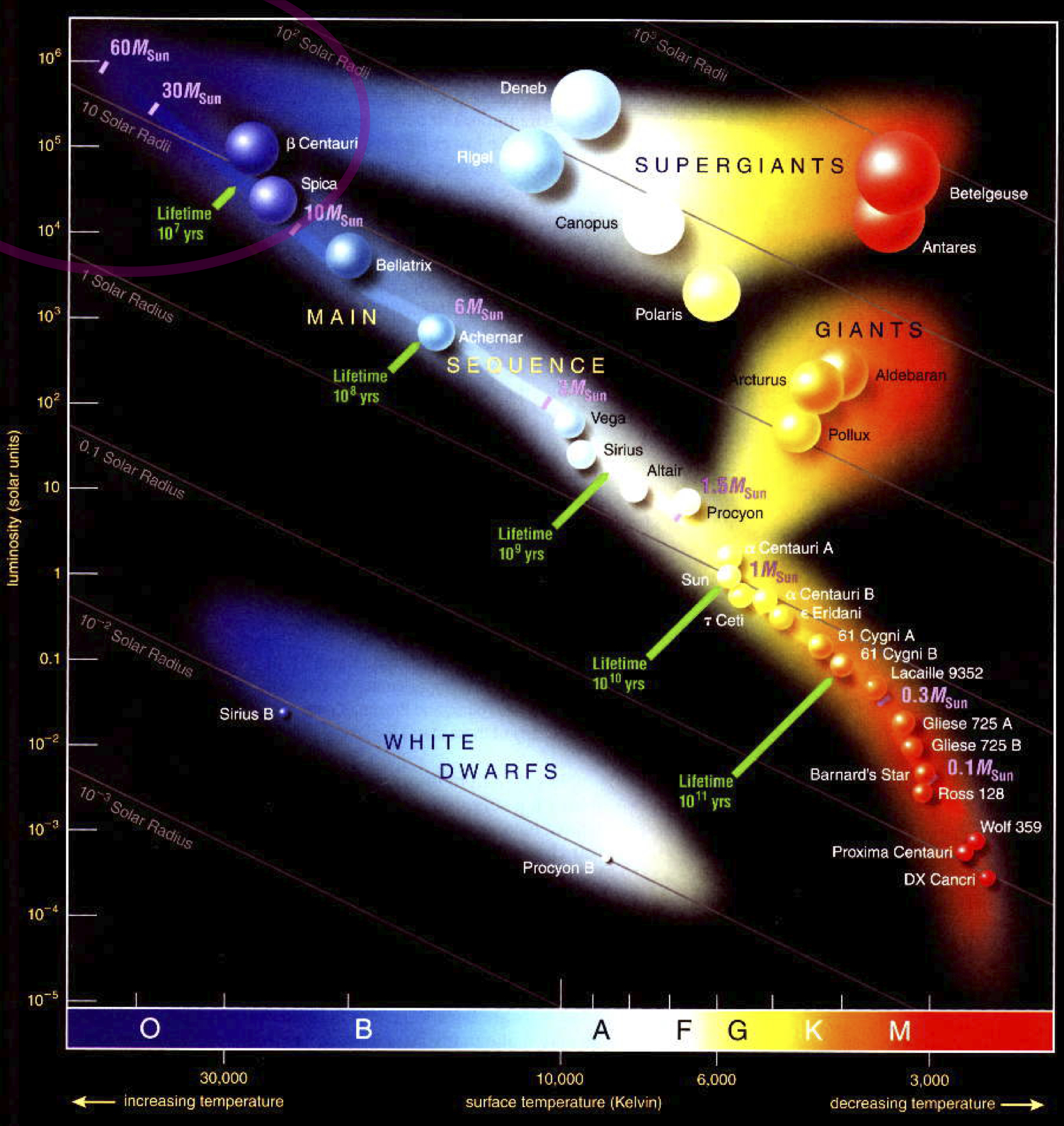}
			\caption[Hertzsprung-Russell diagram]{\small{Hertzsprung-Russell diagram. Location of massive stars appears delimited by the magenta ellipse.}}\label{HRdiag}
		\end{center}
	\end{figure}

%_____Propiedades de las Estrellas Masivas______________________________________________________________
\section{Massive stars}
	As we previously mentioned, massive stars are those with a stellar mass ten or more times the solar mass $M_\odot$.
	
	In accordance with their spectral classification they are stars type O and type B (typically referred then as just \textit{OB stars}), which corresponds to the hottest spectral types, together with being the most luminous stars \citep[][Section 3.1]{spectralclassification}.
	In their spectra lines of ionised helium, neutral helium and hydrogen are mainly observed, together with the so-called ''metals''.
	%\footnote{Do not confuse with the significance of metal in Chemistry. In Astronomy, \textit{metals} are named all elements different from hydrogen and helium.}: carbon, nitrogen, oxygen, etc.
	Their properties are summarised in the Table~\ref{datosmasivas}.
	\begin{table}[h]
		\begin{center}
			\begin{tabular}{cl}
				\hline\hline
				Initial Mass & $\gtrsim10\,M_\odot$\\
				Temperature & $\gtrsim10$ kK\\
				Luminosity & $\gtrsim10^4\,L_\odot$\\
				Mass-loss rate & $\gtrsim10^{-9}\,M_\odot$ yr$^{-1}$\\
				Lifespan & $\sim10^7$ yrs\\
				\hline
			\end{tabular}
			\caption[Properties of massive stars]{\small{Properties of massive stars.}}\label{datosmasivas}
		\end{center}
	\end{table}
	
	Given their high brightness and temperature, massive stars are located in the top-left region of the Hertzsprung-Russell diagram (diagram that organises stars as function on their luminosity and temperature, see Figure~\ref{HRdiag}).
	Their peaks of emission (range of their spectrum where radiation reaches it maximum value) are in the ultraviolet region, the reason why we see these stars as blue-coloured.
	
	As it has been previously mentioned, massive stars are also characterised by presenting a strong stellar wind which makes them lose a big amount of matter during their lifetimes (i.e., a high mass-loss rate). 
	This feature is crucial for their future evolutionary stages, reason why we need to have a better understanding about what stellar wind consists to later go deeper into evolution of stars with high mass.

%_____Evolución de las Estrellas Masivas________________________________________________________________
\subsection{Evolution of massive stars}
	Previously, we explained why the massive stars have reserved a different future compared to their smaller siblings.
	The main consequences of the big amount of mass that will determine later the evolutionary track are, as we know: the greater mass-loss rate due to the stronger stellar wind and a more complex nucleosynthesis in the hotter core.
	Concerning this last point, it will be relevant only at the moment of the final fate of the star, when it explodes as a supernova and eventually becomes a black hole.
	The evolutionary track is then, mostly affected by the high value of mass-loss rate.
	
	Hereunder, we will describe the evolutionary track for a $60\,M_\odot$ star presented by \citet{maeder87} and available in the book \textit{Introduction to Stellar Winds} \citep[][Section 13.2]{stellarwind}.

	The most critical effect produced by the mass-loss rate upon a star is the "dismantling" of this one, i.e., the stellar wind destroys the outer layers of the star letting the inner layers exposed.
	It also produces instability: a very massive star will never become a red supergiant (as low and intermediate-mass stars) because its great mass-loss rate prevents to reach equilibrium when core starts to burn helium and it must expand.
	Instead of that, there will be as a result an unstable star, variables and capable to sent shocks of matter towards the space: so-called Luminous Blue Variable stars (or simply LBV).

	Dismantling will produce later that, once the star have consumed all the hydrogen in its core and begins to burn helium, the remnants of the hydrogen-burning processes will appear in the surface of the star (helium due to the proton-proton process and nitrogen due to the CNO-cycle process mainly).
	As consequence, these elements (initially hidden in the inner layers of the star) turn to be observable in the stellar spectrum.
	Given that the star will exhibit an \textit{extended atmosphere}\footnote{The \textit{atmosphere} of a star is the boundary between the stellar interior and the ISM. Photosphere, the surface of the star, is then the most inner layer of the atmosphere where photons can finally escape from the interior, then, spectral lines are formed in these zones \citep{lanz00}.}, it will be seen in the spectrum broad emission lines of helium and nitrogen: we will observe a Wolf-Rayet star (WR star).
	These WRs are considered the final stage in the life sequence of a massive star, previous to the final explosion as supernova and subsequent stage as black hole.
	
	It is important to remark, however, that the previous description is a general screenshot, and it neglects many details that makes evolutive scenario more complex.
	Some of these issues are:
	\begin{itemize}
		\item The accurate constraint in the initial mass, to delimitate whether the massive star will reach the LBV and WR stages.
		Current acceptable values are given by \citep{crowther07} for solar-metallicity and non-rotating stars\footnote{Nomenclature are: LBV for \textit{Large Blue Variables}, WN for \textit{Nitrogen Wolf-Rayet stars}, WC for \textit{Carbon Wolf-Rayet stars} and SN for \textit{supernovae}.}.
		\begin{center}
			$M_*\gtrsim75M_\odot$: O V $\rightarrow$ WNh $\rightarrow$ LBV $\rightarrow$ WN $\rightarrow$ WC $\rightarrow$ SN Ic\\
			\vspace{2mm}
			$M_*\sim40-75M_\odot$: O V ($\rightarrow$ LBV) $\rightarrow$ WN $\rightarrow$ WC $\rightarrow$ SN Ic\\
			\vspace{2mm}
			$M_*\sim25-40M_\odot$: O V $\rightarrow$ (LBV)/RSG $\rightarrow$ WN ($\rightarrow$ WC) $\rightarrow$ SN Ib\\
			\vspace{2mm}
			$M_*\sim20-25M_\odot$: O V $\rightarrow$ RSG $\rightarrow$ WN $\rightarrow$ SN II/Ib\\
			\vspace{2mm}
			$M_*\sim10-20M_\odot$: O V $\rightarrow$ RSG $\rightarrow$ BSG $\rightarrow$ SN II
		\end{center}
		\item Related with the first point: what is the impact of metallicity and rotational effects upon the evolution of a star?
	\end{itemize}
	
	Over the last decade, different studies have performed more detailed analysis about evolutionary tracks for a wide range of masses, rotational speeds and metallicities \citep{ekstrom12,georgy12,georgy13,groh19}, all of them developed using the Geneva evolutive code (\textsc{Genec}, see Chapter~\ref{tcd} for details).
	All these studies have been a great contribution in order to understand the whole picture about evolution of massive stars.
	Yet mass-loss rates employed by them comes from a prescription for the stellar wind which is not self-consistent.
	On the following chapter of this thesis we present a new prescription that provides new theoretical values for the mass-loss rate, undergone from a self-consistent calculation for the stellar wind.
	However, before giving the details, it is necessary to give a brief general picture of stellar winds.

%_____Viento Estelar________________________________________________________________________________
\section{Stellar wind on massive stars}\label{stellarwind}
	We call \textit{stellar wind} to the outflow of particles which, as same as the photons, are released from the photosphere of the star towards the interstellar medium.
	
	The main mechanism that explains the existence of stellar winds is the fact that in the photosphere of the star the forces making hydrostatic equilibrium, total pressure from the inner part (generated by the radiation and by the gas of the star) and gravity are not longer in equilibrium at all.
	Pressure force coming from the interior wins over gravity and, due to this imbalance, an outflow of matter is produced \citep{stellarwind}.
	This explains partially the fact that in hotter stars, where the radiation pressure is higher, the stellar wind is stronger.
	
	The two main parameters of the stellar wind, which can be determined by spectral analysis, are:
	\begin{itemize}
		\item\textit{Terminal velocity:} ($v_\infty$), understood as the asymptotic velocity reached by the particles of the wind at large distances, measures in km~s$^{-1}$.
		\item\textit{Mass-loss rate:} ($\dot M$), corresponding to the amount of matter released by the star per unit time, measured in $M_\odot$~yr$^{-1}$.
	\end{itemize}
	
	Both terms allow us, for instance, to know the amount of energy and momentum released into the interstellar medium.
	
\begin{figure}[h]
		\begin{center}
			\includegraphics[width=0.85\textwidth]{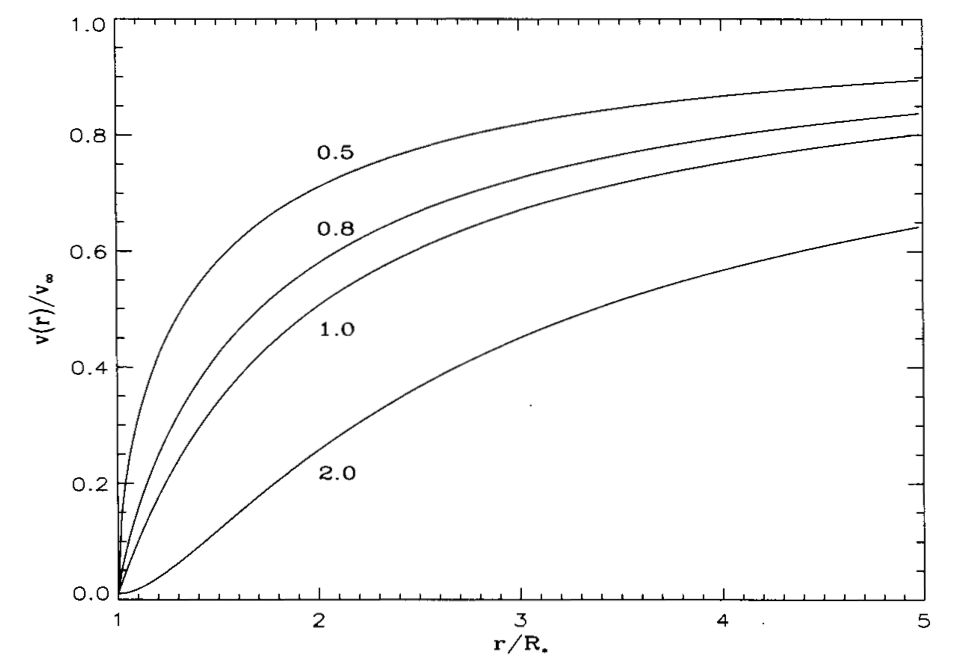}
			\caption[Velocity field with different values of $\beta$]{\small{Velocity field with different values of $\beta$ \citep{stellarwind}.}}\label{betas}
		\end{center}
	\end{figure}

	An approximate function that describes the velocity field of the wind is the so-called $\beta$-law, and it is expressed as:
	\begin{equation}
		v(r)\simeq v_\infty\left(1-\frac{r_0}{r}\right)^\beta\;,
	\end{equation}
	with $r$ the radial coordinate and:
	$$r_0=R_*\left[1-\left(\frac{v_0}{v_\infty}\right)^{1/\beta}\right]\;,$$
	being $v_0$ the wind velocity in the photosphere of the star, i.e., $v(R_*)=v_0$.
	Here, $\beta$ is a factor indicating how steep the increment in velocity along the path is: the higher the $\beta$ value, the less pronounced the increase in speed will be (Fig.~\ref{betas}).
	
	Nevertheless, it is important to remark that $\beta$ value is just an empirical value which describes an approximate behaviour of the wind and agrees with observed spectra with quite acceptance.
	\citet{kudritzki00} argued that the use of the parameter $\beta$ to describe the velocity field is only justified \textit{a posteriori} once the fit is achieved.
	However, it is also possible the existence of velocity fields that can be determined (always with $v_\infty$ as asymptotic limit) without a specific value of $\beta$.
	We will discuss this point with more details in the following chapters, where we will perform our own self-consistent velocity fields.
	
	For the Sun-like stars, mass-loss rates are in the order of $\dot M\sim10^{-13}\;M_\odot$ yr$^{-1}$, which is ten thousand times less intense than the minimum that massive stars exhibit (see Table~\ref{datosmasivas}).
	Hence, it is possible to see that the previously mentioned dismantling due to the stellar wind is not significant for the ordinary stars.
	However, for massive stars it will play a key role that will condition the future evolutionary stages.
	In order to figure out how these evolutionary stages are determined by the features of the stellar wind, it is required to understand how this wind is produced, how wind accelerates and what values can we derive for the stellar wind parameters.

%_____General objectives____________________________________________________________________________________________
\section{General Objectives}
	In order to accurately characterise stellar winds and theoretically predict their parameters, it is necessary to analyse the physical processes behind.
	This is the motivation behind this thesis study.
	For that reason, our first main objective is the obtention of self-consistent solutions (i.e., acceleration of the wind and hydrodynamics must be in agreement) for stellar wind parameters (mass-loss rate and terminal velocity) given different initial set of stellar parameters (effective temperature, mass, radius, metallicity and abundances mainly).
	The self-consistent characteristic of the wind properties to be determined implies that they correspond to a unique solution for a given set of stellar parameters, and then they do not depend of \textit{a priori} assumptions such as a $\beta$-law for velocity profile.
	Besides, we have to evaluate the influence on the final self-consistent solution of different approximations such as treatment for atomic populations and radiation field.
	And finally, we proceed to explore the potential future works derived from the results for the self-consistent solutions.
	
	The process responsible for driving the wind for hot massive stars outwards is called \textit{line-driven}, because it is produced by absorption and reemission of photons by the matter of the wind, and it will be extensively explained in Chapter~\ref{linedriven}.
	In Chapter~\ref{alfakdelta27}, we will employ the m-CAK theory for line-driven winds to calculate self-consistent solutions for the acceleration of the wind and their velocity and density profiles (i.e., the hydrodynamics of the wind) for a set of hot massive stars.
	Results, comparison with observed wind parameters and the new synthetic model spectra obtained from the prescription are also presented.
	In Chapter~\ref{lambert}, we calculate self-consistent hydrodynamically solutions for the stellar wind beyond m-CAK theory under a full non-LTE scenario, the so-called \textit{Lambert-procedure}.
	A complete analysis on the differences with m-CAK prescription and their consequences are also included.
	In Chapter~\ref{rob}, we perform several synthetic spectra for a set of massive stars from the self-consistent solutions calculated under the m-CAK prescription.
	In Chapter~\ref{tcd}, we use the mass-loss rates derived from the self-consistent hydrodynamics to perform new evolutionary tracks for standard non-rotating massive stars.
	Finally, summary and conclusions of our work is presented on Chapter~\ref{conclusions}.

%_____LINE-DRIVING WINDS_________________________________________________________________________
\chapter{Line-driven Winds}\label{linedriven}
%	Through this chapter, we will discuss in detail the line-driven mechanism described by \citet[][hereafter CAK]{cak} and the improvements done by \citet{abbott82} and \citet{ppk}, which lead us into the so-called m-CAK theory.
	Through this chapter, we will discuss in detail the mechanism that allows the wind on massive stars accelerate outwards.
	Most part of the content presented here can be also seen in the Chapter 8 of the book \textit{Introduction to Stellar Winds} \citep{stellarwind}, together with the study made by \citet{puls00} and the reviews from \citet{kudritzki00} and \citet{puls08}.

	\begin{figure}[htbp]
		\begin{center}
			\includegraphics[width=0.6\textwidth]{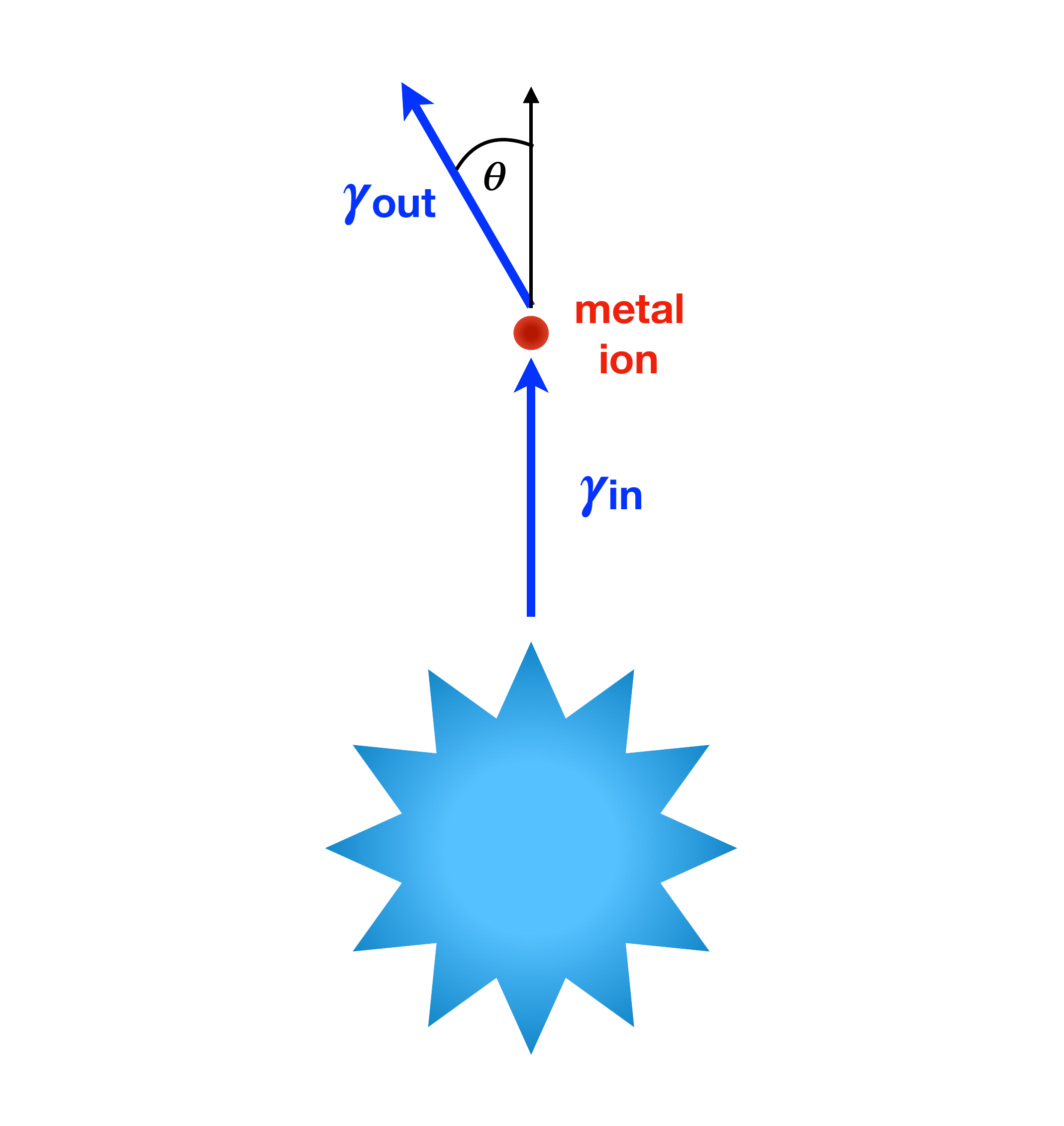}
			\caption[Absorption and reemission in the wind]{\small{Schema showing the absorption of a photon coming from the stellar photosphere, and their further reemission in a random direction.}}\label{absandreemit}
		\end{center}
	\end{figure}
	
	The key to explain both the big amount of matter released to the space by a massive star by means of its stellar winds and the high acceleration reached by this outflow lies in the capacity of the atoms and ions along the wind of absorbing and reemiting photons coming from the photosphere of the star.
	 Because photons are coming from a specific region whereas the reemission is released in any arbitrary direction (see Figure~\ref{absandreemit}), in average the ions gain momentum generating then a force: we called to this process \textit{line-driving}.
	 Due to Doppler effect, line-driving process is not limited only to a specific rest frequency $\nu_0$ where the transitions take place but also occurs in a wide range of radial velocities where the relative frequency matches with the rest frequency by means of:
	 \begin{equation}\label{dopplerfreq}
	 	\nu=\nu_0\left(1+\frac{v(r)}{c}\right)\;.
	\end{equation}
	 As consequence, the effect of the absorption and further reemission is enhanced along different parts of the wind, resulting in a force larger than the gravitational one.
	 
	Notice that the gain of momentum due to the previously described line-driving processes is applied over the individual ions instead the entire fluid.
	This leads to the fact that the ions having more lines (i.e., metal ions) accelerate more than hydrogen and helium.
	However, the higher momentum gained by metal ions is shared with the more abundant and with lighter elements hydrogen and helium through Coulomb collisions.
	This scenario is fulfilled when the timescale necessary to transfer momentum due to collisions is small enough to decelerate the metal ions before letting them escape.
	For stellar winds with high densities, this condition is easily reached and then the acceleration from the line-driving process is transferred to all the plasma, whereas for atmospheres with low mass-loss rates and large terminal velocities the \textit{ionic runaway effect} (i.e., ions that escape without sharing all their momenta) becomes more relevant \citep{springmann92}.
	
	The force generated by means of the line-driving mechanism will play a key role in the calculation of the mass-loss rate: the \textit{line-acceleration} $g_\text{line}$.
	Because the line-driving processes involves a large number of features from the wind, it is necessary to take them into account under different levels of relaxation in order to properly calculate the line-acceleration.
	But before analysing how to determine $g_\text{line}$, it is important to examine how wind parameters are obtained from a specific value for line-acceleration.
	
%_____Hydrodynamics of the Wind____________________________________________________________________
\section{Hydrodynamics of the wind}\label{generalhydrodynamics}
	We call \textit{wind hydrodynamics} to the coupled density $\rho(r)$ and velocity $v(r)$ profiles characterising the wind of a star.
	Both fields are related each other by means of the isothermal and non-rotating stationary equation of momentum on spherical coordinates\footnote{Hereafter, every spatial equation is considered in spherical coordinates because of the geometry of the star. Besides, all of them are solved assuming spherical symmetry, reason why only relevant spatial coordinate is the radius $r$. This assumption is done in order to have consistency with codes such as FASTWIND and CMFGEN, which solves their equations in 1D.}:
	\begin{equation}\label{eqmomentum}
		 v\frac{d v}{dr}=-\frac{1}{\rho}\frac{dP}{dr}-\frac{GM_*(1-\Gamma_e)}{r^2}+g_\text{line}\;,
	\end{equation}
	and the equation of continuity:
	\begin{equation}\label{eqcontinuity}
		\dot M=4\pi\rho(r)r^2 v(r)=\text{constant}\;,
	\end{equation}
	with $dP/dr$ being the pressure gradient, $M_*$ the total mass of the star and $\Gamma_\text{e}$ the Eddington factor.
	\begin{equation}\label{eddington}
		\Gamma_e=\frac{\sigma_\text{e}L_*}{4\pi cGM_*}\;.
	\end{equation}
	
	Equation~\ref{eqmomentum} shows that there will be a positive acceleration when left-hand side is greater than zero, i.e., the radiative components of the acceleration must be greater than the gas pressure and gravitational components\footnote{It is important to remark that, sometimes in the literature the equation of momentum is written using the term $g_\text{rad}$ instead $g_\text{line}$. In that case, the radiative acceleration due to the continuum (i.e., not produced by the line-driving but by the photons doing Thomson scattering with the electrons of the wind) is not included in the gravitational term and then the equation of momentum reads as: $$v\frac{d v}{dr}=-\frac{1}{\rho}\frac{dP}{dr}-\frac{GM_*}{r^2}+g_\text{rad}\;.$$} above the photosphere: 
	
	\begin{equation}\label{grad}
		g_\text{rad}=\frac{GM_*\Gamma_\text{e}}{r^2}+g_\text{line}>\frac{1}{\rho}\frac{dP}{dr}+\frac{GM_*}{r^2}\;.
	\end{equation}
	Besides, from Eq.~\ref{eqcontinuity} is clearly seen that both fields determine the wind parameters: mass-loss rate $\dot M$ and terminal velocity $v_\infty$.
	Therefore, calculation of wind hydrodynamics means calculation of these wind parameters which are later constrained by observations in the stellar spectra.
	
	In order to solve equation of momentum, we can assume isothermal conditions in order to introduce the equation of state for an ideal gas:
	\begin{equation}\label{idealgas}
		P=a^2\rho(r)\;,
	\end{equation}
	with $a$ being the isothermal sound speed:
	\begin{equation}\label{souldspeed}
		a:=\sqrt{\frac{k_B T_\text{eff}}{\mu m_\text{H}}}
	\end{equation}
	and with $k_B$ being the Boltzmann's constant, $\mu$ the mean particle mass and $m_\text{H}$ the hydrogen atom mass.
	In this case, equation of momentum becomes the \textit{equation of motion}\footnote{Unfortunately there is not a consensus about the strict name of the equations of motion and momentum. On this Thesis we are using the names assigned by \citet{puls08}, but \citet{stellarwind} calls Eq.~\ref{eqmomentum} as equation of motion and Eq.~\ref{eqmotion} as equation of momentum.}:
	\begin{equation}\label{eqmotion}
		\left(1-\frac{a^2}{v^2}\right)v\frac{dv}{dr}=\frac{2a^2}{r}-\frac{GM_*(1-\Gamma_e)}{r^2}+g_\text{line}\;.
	\end{equation}
	As consequence of the fact that we are using a constant temperature equivalent to the effective temperature (isothermal wind), sound speed is considered as a constant.
	Besides, notice the fact that Eq.~\ref{eqmotion} does not longer depend explicitly on density $\rho(r)$.
	Actually, dependence on density is implicitly included inside term for $g_\text{line}$, one of the reasons why it is necessary provide a hydrodynamics to solve line-acceleration self-consistently.
	However, as we will examine later in Chapter~\ref{lambert}, for the case where $g_\text{line}$ is directly obtained from the solution of the radiative transfer equation and its dependence on density is not directly known, we can consider $\rho$ (and therefore mass-loss rate $\dot M$) as a free parameter.

	Treatments employed to solve Eq.~\ref{eqmotion} depends on the formulation used to calculate the line-acceleration and what variables were considered for its calculation; therefore, we will focus the discussion through this chapter into knowing how $g_\text{line}$ is determined under the 'classical' line-driving theory performed in the decade of the 70s: the CAK (and later m-CAK) theory.

%_____The m-CAK Theory___________________________________________________________________________
\section{The m-CAK theory}\label{mcaktheory}
	\citet{lucy70} described the mechanism that drives the strong stellar winds observed in hot stars: the so-called radiation driven winds.
	The process of absorption and further re-emission of photons and Coulomb interactions previously described at the beginning of this chapter is the mechanism responsible to give momentum to the wind of hot stars, then producing an outwards line-force.
	According to these authors, the effectivity of line-driving mechanism lies in the fact that the most part of the atomic transitions involved come from the ultraviolet resonance lines, which in turn is where the peak of radiation field for hot stars is located.
	The foundation of the theory of radiation driven winds was later developed by \citet[][hereafter CAK theory]{cak}, who, based on the Sobolev and the point-star approximations, modelled the line-acceleration analytically in terms of the acceleration produced by electron scattering times a force multiplier factor.
	This factor represents the contribution of absorption and re-emission processes depending on the optical depth only, and it was parametrised by two constant parameters through the wind, namely $k$ and $\alpha$.

\subsection{Theoretical background}
	In order to understand the theoretical bases of line-driven winds, let us analyse first the case of gaining momentum from a single line.
	Assuming the wind is optically thick for this transition, implies that all the photons coming from the photosphere which could be absorbed by the atomic transition will do.
	Therefore, due to Doppler effect there is not a unique frequency $\nu_0$ to be absorbed by the wind, but a range going from the rest frequency (in the region of the wind where $v=0$) to the external parts of the wind where $v=v_\infty$ (with $\nu$ given by Eq.~\ref{dopplerfreq}).
	As consequence, the total photospheric radiation being absorbed by the wind (per unit of time) is given by:
	\begin{equation}\label{lumabssingleline}
		L_{*,\text{abs}}=4\pi R_*^2\int_{\nu_0}^{\nu_0(1+v_\infty/c)}\mathcal F_\nu\,d\nu\;,
	\end{equation}
	with $\mathcal F_\nu$ being the flux at the line frequency:
	\begin{equation}\label{fluxnu}
		\mathcal F_\nu=\frac{L_\nu}{4\pi r^2}\;,\text{ satistying: }L_*=\int_0^\infty L_\nu\,d\nu\;.
	\end{equation}
	
	The associated momentum is then given by:
	\begin{equation}\label{momentumsingleline}
		p_{\text{line},i}=\frac{L_{*,\text{abs}}}{c}=\frac{4\pi R_*^2}{c}\int_{\nu_0}^{\nu_0(1+v_\infty/c)}\mathcal F_\nu\,d\nu\;.
	\end{equation}
	
	The total momentum gained by the wind will be then equivalent to the sum of all the lines where the wind is fully optically thick:
	\begin{equation}\label{momentumoptthick}
		p_\text{line}=\frac{4\pi R_*^2}{c}\sum_{i=1}^N\int_{\nu_0}^{\nu_0(1+v_\infty/c)}\mathcal F_\nu\,d\nu\;.
	\end{equation}
	
	Hence, line-acceleration could be obtained once an accurate calculation of all the possible optically thick lines that take part in the line-driving process.
	However, full optically thick lines are an idealisation for illustrative purposes.
	Instead, each line will present a different value for its opacity depending on the atomic properties of its associated transition and the zone of the wind where the absorption takes place.
	Then, it is necessary to define the \textit{mass absorption coefficient} $\kappa_l$ for a single line:
	\begin{equation}\label{kappal}
		\kappa_l(r)=\frac{\pi e^2}{m_ec}f_l\frac{N_l}{\rho(r)}\left(1-\frac{N_u}{N_l}\frac{g_l}{g_u}\right)\;,
	\end{equation}
	with $N_l$ and $N_u$ being the number density of the ion for the lower and upper excitation levels respectively, $g_l$ and $g_u$ being the respective statistical weights and $f_l$ being the oscillator strength of the atomic transition\footnote{The usage of the subindex $l$ seems to be misleading, meaning "lower" for the atomic densities and the statistical weights, and meaning "line" for the oscillator strengths. However, since physically a spectral line is produced due to an atomic transition from a specific lower level of excitation to an upper one (or viceversa), it is possible to use the subindex $l$ without leading into errors.}.
	Considering that the photon generated by this atomic transition has an energy of $h\nu_0$, absorption coefficient can be written in terms of frequency:
	\begin{equation}\label{kappanu}
		\kappa_\nu=\kappa_l\phi(\Delta\nu)\;,%\text{ with: }\Delta\nu=\nu-\nu_0\;,
	\end{equation}
	with $\phi(\Delta\nu)$ being a normalised profile function describing the range on the frequency domain where the transition occurs:
	\begin{equation}\label{profiledeltanu}
		\phi(\Delta\nu)=\frac{1}{\sqrt{\pi}\Delta\nu_G}e^{-(\Delta\nu/\Delta\nu_G)^2}\;,
	\end{equation}
	with $\Delta\nu_G$ being the Gaussian width of the profile, determined by the thermal (see Eq.~\ref{thermalvelocity}) and turbulent motions of the wind.
	\begin{equation}\label{gaussianwidth}
		\Delta\nu_G=\frac{\nu_0}{c}\sqrt{\frac{2}{3}(\langle v_\text{th}\rangle^2+\langle v_\text{turb}\rangle^2)}\;.
	\end{equation}
		
\begin{figure}[t!]
		\begin{center}
			\includegraphics[width=0.75\textwidth]{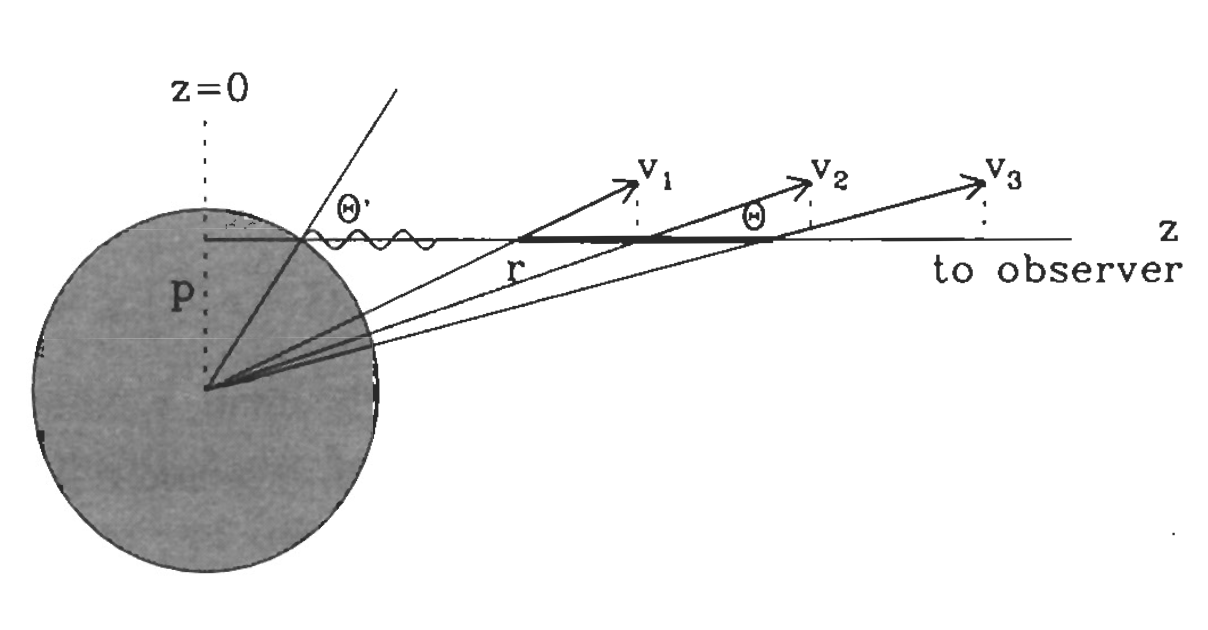}
			\caption[Schema of line of sight]{\small{Schema showing the velocity field $v_z=v(r)\cos\theta$ along the line of sight. A photon emitted by the photosphere at frequency $\nu_p$ will be absorbed by the wind at a specific frequency $\nu_0$ determined by both the velocity field $v(r)$ and the angle $\theta$. Image taken from \citet{stellarwind}.}}\label{impactrange}
		\end{center}
	\end{figure}

	Because of the dependence on density and the influence of Doppler effect, absorption coefficient depends then not only on the frequency of the photon and the atomic information for the involved ion, but also depends implicitly on the point of the wind there the absorption will take place.
	For the case of considering the wind for a star with finite disk (i.e., the star has a specific radius $R_*$ and it is not assumed as a point source), we can parametrise the location on the wind there the transition occurs in terms of the line of sight $z$:
	\begin{equation}\label{lineofsight}
		z=r\cos\theta=\sqrt{r^2-p^2}\;,
	\end{equation}
	with $\theta$ being the angle between the radial direction and the line of sight, and $p$ being the \textit{impact parameter}, perpendicular to the line of sight (see Fig.~\ref{impactrange}).
	Notice that it is always satisfied that $p\le R_*$.
		
	Location of this $z$ point represents the place where the peak of the profile function $\phi(\Delta\nu)$ is located.
	Since the neighbourhood around $z$ represents the region where the radiation coming from the photosphere is absorbed, the atmosphere beyond this zone becomes opaque for photons at that range of frequencies (although, as we explained before, this opacity is not infinite).
	In order to represent mathematically this situation, we introduce the \textit{optical depth} defined as:
	\begin{equation}\label{opticaldepth1}
		\tau_{\nu_p}(z_1)=\int_{z_1}^\infty \kappa_\nu(z)\rho(z)dz\;,
	\end{equation}
	and using Eq.~\ref{kappanu} and Eq.~\ref{kappal}:
	\begin{equation}\label{opticaldepth2}
		\tau_{\nu_p}(z_1)=\frac{\pi e^2}{m_ec}f_l\int_{z_1}^\infty N_l(z)\left(1-\frac{N_u(z)}{N_l(z)}\frac{g_l}{g_u}\right)\phi(\Delta\nu)\,dz\;.
	\end{equation}
		
	The interval covered by $\Delta\nu$ depends on the wind velocity at the point $z_1$, which is given by Doppler effect:
	\begin{equation}\label{deltanu}
		\Delta\nu(z)=\nu_p\left(1-\frac{v_z(r)}{c}\right)-\nu_0=\nu_p\left(1-\frac{z}{r}\frac{v(r)}{c}\right)-\nu_0
	\end{equation}
		
	Calculation of optical depth implies to introduce Eq.~\ref{profiledeltanu} inside Eq.~\ref{opticaldepth2} and later integrate all over $z$, which implies moreover to know the behaviour of our atomic populations in function of the radius.
	However, as we have pointed out previously, the absorption takes place inside a region determined by the velocity of the wind.
	If we consider a steep velocity gradient $dv/dr$, the $\Delta v$ where the transition occurs leads into a even narrower $\Delta r$; and moreover, a narrow Gaussian profile (see Eq.~\ref{gaussianwidth}), will also lead into a narrow region on radius.
	Given these scenarios, it is possible to take this assumptions in order to simplify the calculation of the optical depth.
	
\subsection{Sobolev approximation}\label{sobolevapprox}
	The consideration of the absorption region as small enough in length (and therefore $z$) in order to reduce the zone to a single point, is called \textit{Sobolev approximation} \citep{sobolev60}.
	In this limit, the profile $\phi(\Delta\nu)$ becomes a delta-function and so:
	\begin{eqnarray}\label{opticaldepthso}
		\tau_{\nu_p}(z_1)&=&\frac{\pi e^2}{m_ec}f_l\int_{\Delta\nu(z_1)}^{\Delta\nu(z\rightarrow\infty)} N_l(z)\left(1-\frac{N_u(z)}{N_l(z)}\frac{g_l}{g_u}\right)\phi(\Delta\nu)\left(\frac{dz}{d\,\Delta\nu}\right)d\Delta\nu\nonumber\;,\\
		&=&\frac{\pi e^2}{m_ec}f_l N_l(r_s)\left(1-\frac{N_u(r_s)}{N_l(r_s)}\frac{g_l}{g_u}\right)\left(\frac{dz}{d\,\Delta\nu}\right)_{r_s}\nonumber\;,\text{ with }z_1\le r_s\;,\\
		&=&\kappa_l(r_s)\rho(r_s)\left(\frac{dz}{d\,\Delta\nu}\right)_{r_s}\;.
	\end{eqnarray}
	Thanks to Sobolev approximation, optical depth can be expressed as a step function where the absorption coefficient (see Eq.~\ref{kappal}) is evaluated in a single point (called \textit{Sobolev point}).
	In order to obtain the final expression for $\tau_s$ (where $s$ also means Sobolev), we solve the derivative $(dz/d\Delta\nu)_{r_s}$ using Eq.~\ref{deltanu} and Eq.~\ref{lineofsight}:
%		\begin{eqnarray}\label{dzddeltanu}
%			\left(\frac{dz}{d\,\Delta\nu}\right)_{r_s}&=&\frac{c}{\nu_p}\frac{d}{d\Delta\nu}\left(\frac{r}{v(r)}(\nu_p-\nu_0-\Delta\nu)\right)\nonumber\;,\\
%			&=&\frac{c}{\nu_p}\left[(\nu_p-\nu_0-\Delta\nu)\frac{dz}{d\Delta\nu}\frac{dr}{dz}\frac{d}{dr}\left(\frac{r}{v(r)}\right)-\frac{r}{v(r)}\right]\nonumber\;,\\
%			&=&\frac{c}{\nu_p}\left[\frac{\nu_p}{c}\frac{z}{r}v(r)\frac{dz}{d\Delta\nu}\frac{dr}{dz}\frac{d}{dr}\left(\frac{r}{v}\right)-\frac{r}{v}\right]
%		\end{eqnarray}
	\begin{equation}\label{dzdeltanu}
		\left(\frac{dz}{d\,\Delta\nu}\right)_{r_s}=\frac{c}{\nu_p}\left[\sin^2\theta\frac{v(r)}{r}+\cos^2\theta\frac{dv}{dr}\right]_{r_s}^{-1}\;.
	\end{equation}
		
	For simplicity, we can define $\mu=\cos\theta=z/r$.
	Then, Sobolev optical depth in terms of the rest frequency where the absorption takes place is written as:
	\begin{eqnarray}\label{sobolevopticaldepth}
		\tau_{s,\nu_0}&=&\kappa_l(r_s)\rho(r_s)\frac{c}{\nu_p}\left[(1-\mu^2)\frac{v(r)}{r}+\mu^2\frac{dv}{dr}\right]_{r_s}^{-1}\nonumber\;,\\
		&=&\kappa_l(r_s)\rho(r_s)\frac{c}{\nu_0}\left[\frac{v(r)}{r}+\mu^2\left(\frac{dv}{dr}-\frac{v}{r}\right)\right]_{r_s}^{-1}\nonumber\;,\\
		&=&\kappa_l(r_s)\rho(r_s)\frac{c}{\nu_0}\left(\frac{v}{r}\right)^{-1}\left[1+\mu^2\left(\frac{r}{v}\frac{dv}{dr}-1\right)\right]_{r_s}^{-1}\nonumber\;,\\
		&=&\kappa_l(r_s)\rho(r_s)\lambda_0\left(\frac{v}{r}\right)^{-1}\left[1+\mu^2\sigma\right]_{r_s}^{-1}\;,
	\end{eqnarray}
	where we have introduced the variable $\sigma$:
	\begin{equation}\label{sigmadvdr}
		\sigma=\frac{d\ln v}{d\ln r}-1=\frac{r}{v}\frac{dv}{dr}-1\;.
	\end{equation}
	
	Besides, we have used the approximation $\nu_p\simeq\nu_0$ valid for not relativistic wind velocities.
	Decomposing the absorption coefficient the Sobolev optical depth gives:
	\begin{equation}\label{finsobolevopticaldepth}
		\tau_{s,\nu_0}=\frac{\pi e^2}{m_ec}\lambda_0f_l N_l(r_s)\left(1-\frac{N_u(r_s)}{N_l(r_s)}\frac{g_l}{g_u}\right)\left(\frac{r/v}{1+\mu^2\sigma}\right)_{r_s}\;.
	\end{equation}
		
	Hence, optical depth finally depends on the atomic and wind conditions on the point where it is being evaluated.
	This point comes from the fact that, under Sobolev approximation, we are considering the interaction region where the atomic transition takes place as being infinitely narrow.
	However, this assumption is not real at all because of random motion (thermal and turbulence velocities), expressed in the term $\Delta\nu_G$ inside Eq.~\ref{profiledeltanu}.
	In spite of that, Sobolev approximation works well as far as the velocity field $v(r)$ is larger enough compared with $v_\text{th}$ and $v_\text{turb}$ (both usually in the order of $\sim10-30$ km s$^{-1}$) and whether the velocity $dv/dr$ is large enough to keep the absorption region as narrow as possible (in order to keep the particles density $N_l$ and $N_u$ almost constant).
	This means, the region should have a width of:
	\begin{equation}\label{sobolevlength}
		\Delta r\simeq \frac{v_\text{th}}{(dv/dr)}\equiv L_s \;,
	\end{equation}
	with $L_s$ being the \textit{Sobolev length}.
	Previous conditions are easily satisfied downstram from sonic point, with sound speed in the order of $\sim25$ km s$^{-1}$ and a high acceleration on the wind, outwards.
	Thus, we establish the range of validity for Sobolev approximation (and therefore m-CAK line-acceleration) to be from the sonic point to infinite.
	This discussion will be retaken in Chapter~\ref{alfakdelta27}.
		
\subsection{The CAK line-acceleration}
	Once we have obtained an analytical expression for optical depth, we proceed to derive an expression for the acceleration due to line-driving.
		
	Previously, we had derived a temptative expression for the momentum gained by the wind given the ideal case of being absorbing all the radiation at that range of frequencies.
	In reality, the amount of momentum to be gained is proportional to the opacity of the wind for that line at that point of the wind, which implies a complex problem of radiative transfer since we need to know how much flux of energy from the photosphere reaches to the point $r$.
	However, thanks to the Sobolev approximation this problem is easily reduced to consider that the photon emitted by the photosphere will interact with the wind at that transition only in the Sobolev point $r_s$.
	Thus, the amount of radiation accelerating the wind at point $r$ will depend only on the intensity coming from the stellar photosphere and the local conditions around $r_s$ and then determined by the optical depth $\tau_\nu(r_s)$.
		
	If we assume that radiation coming from the photosphere is homogeneous, the intensity of radiation reaching the Sobolev point is given by:
	\begin{equation}\label{radiationsobolev}
		I_{\nu_p}(\mu)=I_{\nu_p}^*e^{-\tau_{\nu_p}(\mu)}\;,
	\end{equation}
	where for $\tau_{\nu_p}$ we adapted the expression given by Eq.~\ref{opticaldepth1} to run from the photosphere to the radius $r$ where intensity is being evaluated\footnote{We use the variable $d\mathfrak{l}$ to determine an infinitesimal segment instead $dl$, in order to not confuse with $l$ for 'line'.}:
	\begin{eqnarray}\label{intensityopticaldepth1}
		\tau_{\nu_p}(\mu)&=&\int_\text{phot}^r \kappa_{\nu_p}\rho d\mathfrak{l}\nonumber\;,\\
		&=&\int_\text{phot}^r \kappa_l \rho\phi(\Delta\nu) d\mathfrak{l} \nonumber\;,\\
		&=&\kappa_l(r)\rho(r)\left(\frac{d\mathfrak{l}}{d\Delta\nu}\right)\int_{\Delta\nu(\text{phot})}^{\Delta\nu(r)}\phi(\Delta\nu) d(\Delta\nu)\nonumber\;,\\
		&=&\tau_{\nu_0}(r)\int_{\Delta\nu(\text{phot})}^{\Delta\nu(r)}\phi(\Delta\nu) d(\Delta\nu)\nonumber\;.\\
	\end{eqnarray}
		
	Notice that absorption coefficient and density have been extracted from the integral using the Sobolev approximation.
	The integral can be defined as:
	\begin{equation}
		\Phi(\Delta\nu_\mu)=\int_{\Delta\nu(\text{phot})}^{\Delta\nu(r)}\phi(\Delta\nu)d\Delta\nu\;,
	\end{equation}
	with $\Delta\nu_\mu$ being $\nu_p-\nu_0(1+\mu v(r)/c)$, analogous to Eq.~\ref{deltanu}.
	Besides, once again we assume that the intensity emitted by the photosphere is almost constant in the interval covered by $\Delta\nu_\mu$, so we can set the photospheric intensity $I_{\nu_p}^*\simeq I_{\nu_0}^*$ where the last term is the continuum intensity at the rest frequency.
		
	However, we need to consider the radiation coming from all the angles of the stellar photosphere, and for that reason we focus in the mean intensity $J_{\nu_p}$, interpreted as the intensity per steradian of radiation at the frequency $\nu_p$.
	This is equivalent at the half of the integration of the intensity $I_{\nu_p}$ over $\mu=\cos\theta$, from the minimal possible value $\mu_*$ on the photosphere of the star ($\mu_*=\cos\theta_*=\sqrt{1-(R_*/r)^2}$) to the maximal value at $\theta=0$.
	\begin{equation}\label{meanintensitynu}
		J_{\nu_p}=\frac{1}{2}\int_{\mu_*}^1 I_{\nu_p}d\mu=\frac{I_{\nu_0}^*}{2}\int_{\mu_*}^1 e^{-\tau_{\nu_0}\Phi(\Delta\nu_\mu)}d\mu\;.
	\end{equation}
		
	Since $\phi(\Delta\nu)$ is a normalised function, we can later calculate the mean intensity integrated over all frequencies as:
	\begin{eqnarray}\label{meanintensity}
		\bar{J}&=&\frac{I_{\nu_0}^*}{2}\int_{\mu_*}^1 \int_{\Delta\nu_\mu=-\infty}^{\Delta\nu_\mu=+\infty}\phi(\Delta\nu_\mu)e^{-\tau_{\nu_0}\Phi(\Delta\nu_\mu)}d(\Delta\nu)d\mu\;,\nonumber\\
		&=&\frac{I_{\nu_0}^*}{2}\int_{\mu_*}^1 \int_0^1 e^{-\tau_{\nu_0}\Phi(\Delta\nu_\mu)}d\Phi(\Delta\nu)d\mu\;,\nonumber\\
		&=&\frac{I_{\nu_0}^*}{2}\int_{\mu_*}^1 \frac{1-e^{-\tau_{\nu_0}}}{\tau_{\nu_0}}d\mu\;,\nonumber\\
		\bar{J}(r)&=&\frac{I_{\nu_0}^*}{2}\int_{\mu_*}^1 \frac{1-e^{-\tau_{\nu_0}}}{\tau_{\nu_0}}d\mu\;.
	\end{eqnarray}
		
	Eq.~\ref{meanintensity} gives us the amount of energy per second (power) per unit of surface and per steradian gained at radius $r$ by means of the absorption at rest frequency $\nu_0$.
%		Since we are assuming spherical symmetry, we can state that this intensity is doing effect upon a spherical shell of thickness $dr$, infinitesimal mass $dm=4\pi\rho r^2dr$ and velocity between $v(r)$ and $v(r)+dv$.
	Since this energy is absorbed from the radiation coming from the stellar photosphere, the momentum gained by the wind at radius $r$ is equal to absorption coefficient $\kappa_\nu$ times the flux ($=4\pi\bar J$) divided by the speed of light $c$:

	\begin{eqnarray}\label{glineoriginal}
		g_\text{line}&=&\frac{2\pi}{c}\int_{\mu_*}^1\int_{-\infty}^\infty \kappa_{\nu_p}(\Delta\nu) I_{\nu_p}(\mu)d(\Delta\nu)\mu d\mu\;,\nonumber\\
		&=&\frac{2\pi}{c}I_{\nu_0}^*\int_{\mu_*}^1\int_{-\infty}^\infty \kappa_{\nu_p}(\Delta\nu) e^{-\tau_{\nu_0}\Phi(\Delta\nu)}d(\Delta\nu)\mu d\mu\;,\nonumber\\
		&=&\frac{2\pi}{c}\kappa_l I_{\nu_0}^*\int_{\mu_*}^1\frac{1-e^{-\tau_{\nu_0}}}{\tau_{\nu_0}}\mu d\mu\;.
	\end{eqnarray}
		
	The integral can be solved if we consider the limit case that the star is a single point, and therefore the only valid possible value for $\mu_*$ is 1.
	This simplification is called the \textit{point source limit}, and it was introduced by \citet{cak}, and it constitutes a fundamental part of the CAK theory.
	Hence, in the point source limit with $\mu_*=1$ we have:
	\begin{eqnarray}\label{glinemu1}
		g_\text{line}&=&\frac{2\pi}{c}\kappa_l I_{\nu_0}^*\frac{1-e^{-\tau_s}}{\tau_s}\;,\nonumber\\
		&=&\frac{2\pi}{c}\kappa_l I_{\nu}^*\frac{1-e^{-\tau_s}}{\tau_s}\;.
	\end{eqnarray}

	We have eliminated the subindex 0, because hereafter all frequencies are only referred to rest frequency (or rest wavelength).
	For the same reason, we substitute the subindex $\nu_0$ for the optical depth and we replace it for $s$, to emphasise that we are using Sobolev approximation.
				
	If $\mu=1$, Eq.~\ref{sobolevopticaldepth} becomes:
	\begin{eqnarray}\label{sobolevopticaldepthmu1}
		\tau_s(r)&=&\kappa_l\rho\lambda_0\left(\frac{v}{r}\right)^{-1}\left[\frac{r}{v}\frac{dv}{dr}\right]_{r_s}^{-1}\;,\nonumber\\
		&=&\kappa_l(r)\rho(r)\lambda\left(\frac{dv}{dr}\right)^{-1}\;.
	\end{eqnarray}
	Therefore, for line-acceleration:
	\begin{eqnarray}
		 g_\text{line}&=&\frac{2\pi}{c}\kappa_l I_{\nu_0}^*(1-e^{-\tau_s})\frac{dv/dr}{\kappa_l\rho\lambda_0}\;,\nonumber\\
		&=&\frac{2\pi}{c^2}\nu I_\nu^*\left(\frac{dv}{dr}\right)\frac{1-e^{-\tau_s}}{\rho}\;. 
	\end{eqnarray}
	Finally, considering the relation between the intensity and the luminosity:
	$$I_\nu^*=\frac{L_\nu}{4\pi r^2}\;,$$
	we obtain the following analytical expression:
	\begin{equation}\label{lineaccelerationpuls}
		g_\text{line}(r)=\frac{L_\nu \nu}{4\pi r^2c^2}\left(\frac{dv}{dr}\right)\frac{1-e^{-\tau_s}}{\rho(r)}\;.
	\end{equation}
		
	This expression shows that line-acceleration, besides the classical dependence on the frequency $\nu$ and the luminosity $L_\nu$ coming from the photosphere, has a important dependence on the gradient of velocity, the inverse of the density and overall, on the optical depth of the wind at the point $r$.
	This last dependence is perhaps the most important, because it shows that line-acceleration will present a different behaviour depending on the strength of $\tau_s$.
	For example, for the case of lines with small $\tau_s$ (called \textit{optically thin lines}), we can approximate $e^{-\tau_s}\simeq1-\tau_s$ and then:
	\begin{eqnarray}\label{thinacceleration}
		g_\text{thin line}&=&\frac{L_\nu \nu}{4\pi r^2c^2}\left(\frac{dv}{dr}\right)\frac{\tau_s}{\rho(r)}\;,\nonumber\\
		&=&\frac{L_\nu \nu}{4\pi r^2c^2}\left(\frac{dv}{dr}\right)\frac{\kappa_l\lambda}{dv/dr}\;,\nonumber\\
		&=&\frac{L_\nu }{4\pi r^2c}\kappa_l\;,\\
		&=&\frac{\kappa_l}{c}\mathcal F_\nu\;.
	\end{eqnarray}
		
	This result shows that for optically thin lines, acceleration is given mainly by the photospheric flux and is intrinsically dependent on density by means of absorption coefficient, but it is independent on velocity gradient.
	On the other hand, if the line has $\tau_s\gg1$ (called \textit{optically thick lines}):
	\begin{equation}\label{thickacceleration}
		g_\text{thick line}=\frac{L_\nu \nu}{4\pi r^2c^2}\frac{1}{\rho(r)}\left(\frac{dv}{dr}\right)\;,
	\end{equation}
	where line-acceleration is independent on absorption coefficient, but also on photospheric and hydrodynamic conditions \citep{puls00,puls08}.
		
\subsubsection{Line ensemble}
	However, we are interested in evaluate the resulting acceleration produced by \textit{all the lines} involved in the line-driving process, being them optically thick or thin.
	In order to obtain that expression, the work of \citet{cak} consisted in the search of an expression easy to sum and analyse.
	Combining Eq.~\ref{kappal} and Eq.~\ref{sobolevopticaldepthmu1}, the full expression for Sobolev optical depth is:
	\begin{equation}
		\tau_s(r)=\frac{\pi e^2}{m_ec}\lambda f_l N_l\left(1-\frac{N_u}{N_l}\frac{g_l}{g_u}\right)\left(\frac{dv}{dr}\right)^{-1}\;.
	\end{equation}
		
	This value will vary from line to line, because each spectral line carries its own information about atomic populations and statistical weights.
	But the hydrodynamical components (density and velocity gradient) will be the same for all lines because they depend on $r$ only.
	Hence, it is convenient to separate both components, in order to define a new optical depth independent on atomic information.
	For that purpose, \citet{cak} have introduced the new variable $t$, the \textit{independent optical depth} or \textit{optical depth for an expanding atmosphere}\footnote{Hereafter and during all the CAK procedure, we refer $t$ simply as optical depth only, omitting the word \textit{independent}. To avoid confusions, we will explicitly specify when we refer to the classical optical depth $\tau$.}\citep{abbott82}, defined as:
	\begin{equation}\label{tcak}
		t\equiv \sigma_e v_\text{th}\rho(r)\left(\frac{dv}{dr}\right)^{-1}\;,
	\end{equation}
	with $\sigma_e=0.325$ cm$^2$ g$^{-1}$ \citep{cak,abbott82} being the \textit{electron scattering opacity} and $v_\text{th}$ the mean thermal velocity of the protons:
	\begin{equation}\label{thermalvelocity}
		v_\text{th}=\sqrt{\frac{2k_B T_\text{eff}}{m_H}}\;.
	\end{equation}
		
	The inclusion of thermal velocity is important, because random thermal movements plays a role enhancing the range of frequencies to be absorbed by means of Doppler effect.
	We can define the Doppler enhancement due to thermal motions as:
	\begin{equation}\label{nudoppler}
		\Delta\nu_D=\nu\frac{v_\text{th}}{c}=\frac{v_\text{th}}{\lambda}\;.
	\end{equation}
		
	No less important, the component of the full optical depth $\tau$ depending on atomic information only is read as:
	\begin{eqnarray}\label{betainv}
		\eta_\text{line}&=&\frac{\pi e^2}{m_ec}\lambda f_l N_l\left(1-\frac{N_u}{N_l}\frac{g_l}{g_u}\right)\frac{1}{\sigma_e v_\text{th}\rho}\nonumber\;,\\
		&=&\frac{\pi e^2}{m_ec}\frac{1}{\rho\sigma_e \Delta\nu_D} f_l N_l\left(1-\frac{N_u}{N_l}\frac{g_l}{g_u}\right)\;.
	\end{eqnarray}
		
	This term $\eta_\text{line}$ represents the ratio between line to electron scattering opacity, and it is fulfilled that $\tau_\text{line}=t\,\eta_\text{line}$.
	With these new definitions, line-acceleration can be rewritten from Eq.~\ref{lineaccelerationpuls} to:
	\begin{eqnarray}
		g_\text{line}(r)&=&\frac{L_\nu \nu}{4\pi r^2c^2}\left(\frac{dv}{dr}\right)\frac{1-e^{-\eta_l t}}{\rho(r)}\nonumber\;,\\
		&=&\frac{L_\nu \nu}{4\pi r^2c^2}\sigma_e v_\text{th}\frac{1-e^{-\eta t}}{t}\nonumber\;,\\
		&=&\frac{\sigma_e\mathcal F_\nu}{c}\Delta\nu_D\frac{1-e^{-\eta t}}{t}\nonumber\;.
	\end{eqnarray}
	Here, we have used the relation between luminosity and flux from Eq.~\ref{fluxnu} and we have omitted the subindex $l$ for $\eta$.
	This expression for line-acceleration has the advantage of being written in a very similar way to the standard acceleration due to radiation pressure (i.e., that acceleration produced by electron scattering interactions):
	\begin{equation}\label{electronscatteringacceleration}
		g_\text{e}=\frac{\sigma_e \mathcal F}{c}=\frac{\sigma_eL_*}{4\pi r^2c}\;.
	\end{equation}
		
	Therefore, line-acceleration can be expressed as the radiative acceleration due to electron scattering, multiplied by some factor representing the contribution of all the lines involved in the line-driving process.
	This factor was defined by \citet{cak} as the \textit{force multiplier factor} as:
	\begin{equation}\label{forcemultiplier}
		\mathcal M(t)=\frac{g_\text{line}}{g_\text{e}}=\sum_\text{lines}\frac{\Delta\nu_D\mathcal F_\nu}{\mathcal F}\frac{1-e^{-\eta t}}{t}\;.
	\end{equation}
		
	Notice the fact that the force multiplier is now defined not as function of radius, but as optical depth $t$.
	The great advantage of this procedure is, $\mathcal M(t)$ can be easily parametrised by a simple power law.
	\begin{equation}\label{mcak}
		\mathcal M(t)=k\,t^{-\alpha}\;,
	\end{equation}
	being $k$ and $\alpha$ the \textit{line-force [multiplier] parameters}.
		
	This was the most important result from the revolutionary work done by Castor, Abbott and Klein in 1975, and for that reason is called \textit{CAK theory}.
	The authors calculated an acceptable force multiplier following Eq.~\ref{forcemultiplier}, by means of the sum of the spectral lines of the ions of carbon.
	The posteriori challenge was, the inclusion of more precise atomic data (oscillator strengths, statistical weights, excitation energies) for all the individual ions involved in the line-driving process, together with the calculation of an accurate thermodynamical treatment in order to calculate accurate atomic populations $N_l$ and $N_u$.
		
	This pioneer study opened the door to the possibility of obtaining a solution for equation of motion (Eq.~\ref{eqmotion}), and thereafter every study dedicated to stellar wind on massive stars is totally or partially based on CAK theory.
	However, later studies included relaxations to some of the main assumptions of the CAK theory, together with other considerations not taken into account by \citet{cak}.
	These new improvements leaded to the generation of the \textit{modified CAK (m-CAK) theory}.
	Details about these changes are given in the following section.

\subsection{Posteriori improvements}\label{posterioriimprovements}
	Seven years after the introduction of CAK theory, \citet{abbott82} performed a detailed calculation of the line-force multiplier $\mathcal M(t)$ taking into account the contribution of a full set of atomic line transition data for elements from hydrogen to zinc.
	Moreover, the line-force multiplier was calculated over a fixed grid of optical depths $t$ and also for different values of diluted electron densities $N_e/W$, for a wide range of stellar temperatures.
	The most remarkable result from this study, was the inclusion of an extra exponential dependence on the diluted electron density $N_e/W$ for $\mathcal M(t)$:
	\begin{equation}\label{mcakdelta}
		\mathcal M(t)=k\,t^{-\alpha}\left(\frac{N_{e,11}}{W(r)}\right)^\delta\;,
	\end{equation}
	with $N_{e,11}$ being the electron density in units of $10^{11}$ cm$^{-3}$, $\delta$ our third line-force parameter and $W(r)$ the dilution factor, i.e, the function showing how radiation is 'diluted' through the wind:
	\begin{equation}\label{dilutionfactor}
		W(r)=\frac{1}{2}\left(1-\sqrt{1-\frac{R_*^2}{r^2}}\right)\;.
	\end{equation}
	This expression can be easily obtained in the limit $\tau_{\nu_0}\ll1$ for Eq.~\ref{meanintensity}.
	\begin{figure}[t]
		\begin{center}
			\includegraphics[width=0.75\textwidth]{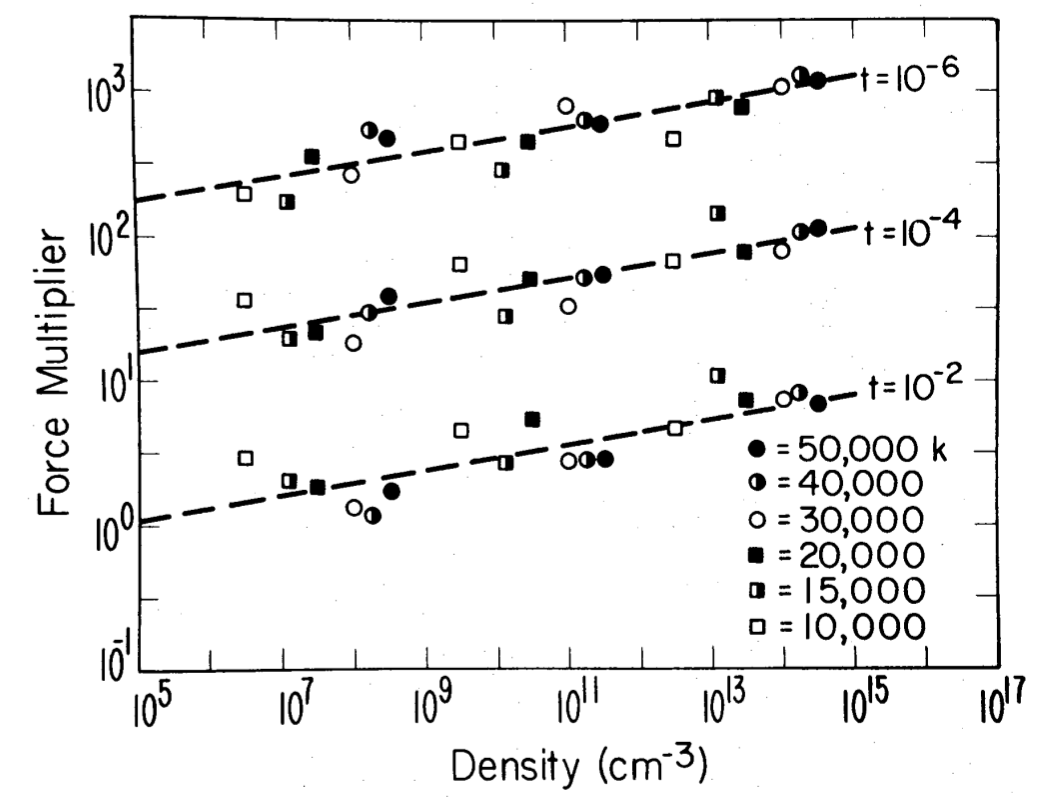}
			\caption[Figure from \citet{abbott82}]{\small{Figures a from \citet{abbott82}, showing the dependence of the force multiplier $\mathcal M(t)$ on electron density.}}\label{abbott82figures}
		\end{center}
	\end{figure}
		
	This dependence $\mathcal M\propto N_{e,11}^\delta$ can be obtained inspecting Fig.~\ref{abbott82figures}, where is clearly seen that force-multiplier increases with electron density in an almost exponential fit, and its explanation lies in the relationship between the number of allowed transitions and the ionisation stage.
	Higher electron densities leads into lower stages of ionisation, as it is shown by Saha equation and its version for ionisation equilibrium in expanded atmospheres given by \citet[][Eq. 5-46]{mihalas78}:
	\begin{equation}\label{mihalasexpanded}
		\frac{N_{i+1}}{N_i}=2\left(\frac{2\pi m_ek_B}{h^2}\right)^{3/2}\frac{T_R\sqrt{T_e}}{N_e/W}\frac{U_{i+1}}{U_i}e^{-\frac{E_i}{k_BT_e}}\;,
	\end{equation}
	being $m_e$ the electron mass, $T_R$ and $T_e$ the radiative and electron temperatures respectively, $U_i$ the partition functions of each ionisation stage and $E_i$ the ionisation energy.
	Lower ionised stages have more lines, which allows the absorption of more radiation by line-driving.
		
	Due to the point-star approximation ($\mu_*=1$ for Eq.~\ref{glineoriginal}) the derived hydrodynamical values for mass-loss rates given by \citet{cak} and \citet{abbott82} were overestimated.
	The explanation for this disagreement lies in the fact that all incoming photons of Fig.~\ref{absandreemit} in reality enter not always with $\theta=0$ (as in the point source approximation) but a range of values for the angle which will reduce the effective value for $\cos\theta$.
	\citet{ppk} and \citet{friend86} relaxed this point source approximation and considered the finite disk shape of the star.
	This modification consisted in the inclusion of a \textit{finite disk correction factor} over the point-source multiplier factor.
	\begin{eqnarray}\label{finitediskcorrection}
		D_f&=&\frac{\mathcal M_\text{disk}(t)}{\mathcal M_\text{point}(t)}\nonumber\;,\\
		&=&\frac{(1+\sigma)^{\alpha+1}-(1-\sigma\mu_*^2)^{\alpha+1}}{(1-\mu_*^2)(\alpha+1)\sigma(1+\sigma)^{\alpha+1}}\;,
	\end{eqnarray}
	with $\alpha$ being the line force parameter.
	With this modified theory (hereafter m-CAK) they solved the equation of momentum and obtained improved theoretical results, in better agreement with the observed mass-loss rates.
	
	Another initial assumption used by \citet{cak} and \citet{abbott82} was the treatment of the atomic populations as being under local thermodynamical equilibrium (LTE).
	Then, ionisation balance were based in Eq.~\ref{mihalasexpanded}, instead taking into account the more complicated statistical relations among all the ions present in the wind.
	Because of the high level of difficulty for a full treatment of atomic populations under non-local thermodynamical equilibrium (hereafter NLTE), there is a shortage of works doing calculations of the line-force parameters, most of them being mostly done in the 90s and thereafter \citep{ppk,puls00,kudritzki02,pauldrach03,noebauer15}.
	As a consequence, it has been difficult to obtain the velocity profiles and mass-loss rates from the m-CAK theory, and thus the massive star researches started to use the $\beta$--law velocity profile.
	Indeed, from the m-CAK simulations performed by \citet{ppk} it was determined a velocity profile following the $\beta-$law \citep{puls08}:
	\begin{equation}
		v(r)=v_\infty\left(1-\frac{R_*}{r}\right)^\beta\;,
		\begin{cases}
			\beta=0.5\text{ for CAK case, point-source limit}\\
			\beta=0.8\text{ for O stars}\\
			\beta=2.0\text{ for BA supergiants}
		\end{cases}
	\end{equation}
	Nevertheless, the value for $\beta$ had to be relaxed in order to fit the spectra, so these theoretical values are not longer valid.
	This simplified description of the velocity field is widely used as input in radiative transfer and NLTE model-atmosphere codes such as FASTWIND \citep{santolaya97,puls05} or CMFGEN \citep{hillier90b,hillier98,hillier01} to calculate NLTE synthetic spectra.
	In this procedure, stellar and wind parameters (terminal velocity and mass-loss rates) are treated as free and are determined by varying them to adjust synthetic profiles to observed ones.
	In the particular case of CMFGEN, the final solution considers a full NLTE treatment and it also provides a radiative acceleration calculated beyond the simplifications undergone from m-CAK (such as Sobolev approximation), but this radiative acceleration is not consistent with the $\beta$--law set as input for hydrodynamics.
	\citet{kudritzki00} argued that the usage of $\beta-$law for the velocity field is only justified \textit{a posteriori} once the fit is achieved.
	However, there are other approaches that coupled the hydrodynamics with comoving frame radiative transfer, see e.g. \citet{sander17} or \citet{kk10,kk17}, that do not use a $\beta-$law velocity profile.
		
	Nevertheless, in spite of the disadvantages associated to m-CAK theory (such as Sobolev approximation and corrections over point source assumptions) and efforts trying to obtain a self-consistent solution beyond it, this remains being a valid reference for the calculation of line-acceleration.
	Moreover, m-CAK theory provides us the great chance to perform a fast self-consistent solution beyond assumptions from $\beta$--law, and for that reason we have chosen this regime to execute our analysis for stellar winds in Chapter~\ref{alfakdelta27}.
	In spite of that, a full parallel analysis of self-consistent solutions in a complete NLTE regime beyond m-CAK prescription will be done in Chapter~\ref{lambert}.

%_____SELF-CONSISTENT SOLUTIONS UNDER M-CAK THEORY__________________________________________
\chapter[Solutions in the frame of m-CAK Theory]{Self-consistent Solutions in the frame of m-CAK Theory}\label{alfakdelta27}
	In this chapter we go in details about the solutions obtained for line-driven winds in the frame of the m-CAK theory described on Section~\ref{mcaktheory}.
	Solution of equation of motion is obtained using the line-acceleration $g_\text{line}$ with the derived values of the line-force multiplier parameters $k$, $\alpha$ and $\delta$.
	This procedure has the enormous advantage of saving a lot of computational effort, and therefore it allows the execution of a large set of models in a short time.
	However, the price to be paid is the adoption of several assumptions whose consequences are discussed.
	Due to scarce works involving NLTE (non-local thermodynamic equilibrium) calculations of the line-force parameters \citep{ppk,puls00,kudritzki02,pauldrach03,noebauer15}, it was difficult to obtain from the m-CAK hydrodynamics the velocity profiles and mass-loss rates, thus, the community of massive star researchers started to use the $\beta$-law velocity profile instead of the proper hydrodynamics.
	This simplified description of the velocity field is widely used as input in radiative transfer and NLTE model-atmosphere codes such as FASTWIND \citep{santolaya97,puls05} or CMFGEN \citep{hillier90a,hillier98,hillier01} in order to calculate synthetic spectra.
	In those procedures, stellar and wind parameters (terminal velocity and mass-loss rates) are treated as free parameters and are determined by adjusting them to fit synthetic line profiles with observed ones.
	However, there are other approaches that coupled the hydrodynamics with comoving frame radiative transfer, see e.g. \cite{sander17} or \cite{kk10,kk17}, that do not use a $\beta$-law velocity profile.

	Calculations of line-force wind parameters coupled with hydrodynamics are necessary to derive self-consistent values of the velocity profiles and the mass-loss rates.
	Moreover, these line-force parameters depend non-linearly on the stellar parameters, chemical abundances, and atomic data via the wind driven mechanism.
	Therefore, to obtain the line-force parameters it is necessary to calculate the \textit{total acceleration} produced by the contribution of hundreds of thousands lines involved in the absorption and re-emission processes (i.e., line-acceleration, $g_\text{line}$) which requires reliable atomic data, as they are essential to perform \textit{line-statistics} calculations.
	
	The number of contributing lines to the line-driven acceleration depends on the wind opacity and it is strongly coupled to the wind density and velocity profiles.
	To solve this highly non-linear problem an iterative procedure is required to satisfy both: line-statistics and m-CAK hydrodynamics.
	
	In this chapter, we calculate self-consistent solutions to obtain accurate m-CAK line-force parameters $(k,\alpha,\delta)$ and wind properties of hot massive stars.
	The hydrodynamics is provided by the code \textsc{HydWind} \citep{michel04}, whereas abundances have been adopted from \citet{asplund09}. 
	Final self-consistent line-force values must correspond to an unique solution obtained when line-force parameters, velocity profile and mass-loss rate simultaneously converge.
	Hence, we present here a new set of m-CAK self-consistent line-force parameters for $T_\text{eff} \ge 32$ kK and $\log g\ge3.4$ (this is, as we will discuss later, the range of validity for our self-consistent solutions), with the corresponding velocity profile  and mass-loss rate.
	These line-force parameters are compared with previous numerical studies.
	Furthermore, with these new results we calculate synthetic spectra with FASTWIND and contrast them with observations.
	We show that applying our procedure we obtain a very good fit to the observed line profiles.
	Finally, we derive:
	
	i) an alternative recipe for the mass-loss rate which only depends on the stellar parameters and the abundance;
	
	ii) a dependency of $v_{\inf}/v_{esc}$ on $\log g$, which was not previously known.
			
	We have to mention that this chapter corresponds to the manuscript \citet{alex19}, also referred as Paper I.
	
%_____Calculation of k, alpha and delta_______________________________________________________________	
\section{Equation of motion with line-force parameters $(k,\alpha,\delta)$}
	Using the expression for line-acceleration from the force-multiplier $\mathcal M(t)$ (Eq.~\ref{mcakdelta}), we re-write (Eq.~\ref{eqmomentum}) as:
	\begin{equation}\label{eqmomentumcak}
		 v\frac{d v}{dr}=-\frac{1}{\rho}\frac{dP}{dr}-\frac{GM_*(1-\Gamma_e)}{r^2}+g_\text{e}\,kt^{-\alpha}\left(\frac{N_{e,11}}{W(r)}\right)^\delta\;,
	\end{equation}
	being $k$, $\alpha$ and $\delta$ the already mentioned line-force parameters (see Eq.~\ref{mcakdelta}).
	Assuming an isothermal ideal gas $P=a^2\rho$, equation of momentum is transformed into equation of motion (Eq.~\ref{eqmotion}):
	\begin{equation}\label{eqmotioncak}
		\left(1-\frac{a^2}{v^2}\right)v(r)\frac{dv}{dr}=\frac{2a^2}{r}-\frac{GM_*(1-\Gamma_e)}{r^2}+g_\text{e}\,k(\sigma_e v_\text{th}\rho)^{-\alpha}\left(\frac{N_{e,11}}{W(r)}\right)^\delta\left(\frac{dv}{dr}\right)^{\alpha}\;.
	\end{equation}
	
	Rewriting $dv/dr$ as $v'$, we can define momentum equation as a function $F(r,v,v')=0$.
	Besides, we use Eq.~\ref{electronscatteringacceleration} to modify $g_\text{e}$ and equation of continuity Eq.~\ref{eqcontinuity} to express density in terms of the mass-loss rate.
	Thus, $F(r,v,v')$ reads:
%	\begin{equation}
%		\;.
%	\end{equation}
%	
%	From \citet{michel04}, we employ the following change of variables:
%	\begin{equation}\label{mcakchangeofvariables}
%		u=-\frac{R_*}{r}\;,\;\; w=\frac{ v}{a}\;\text{ and }\;\;\ w'=\frac{dw}{du}\;,
%	\end{equation}
	\begin{eqnarray}\label{funceqmotioncak}
		F(r,v,v')&=&\left(1-\frac{a^2}{v^2}\right)r^2vv'+GM_*(1-\Gamma_e)-2a^2r-\frac{\sigma_eL_*}{4\pi c}\,k\left(\frac{\sigma_e v_\text{th}\dot M}{4\pi r^2v}\right)^{-\alpha}\left(\frac{N_{e,11}}{W(r)}\right)^\delta\left(\frac{dv}{dr}\right)^{\alpha}\nonumber\;,\\
		&=&\left(1-\frac{a^2}{v^2}\right)r^2vv'+GM_*(1-\Gamma_e)-2a^2r-\frac{\sigma_eL_*}{4\pi c}k\left(\frac{4\pi}{\sigma_e v_\text{th}\dot M}\right)^{\alpha}\left(\frac{N_{e,11}}{W}\right)^\delta\left(r^2vv'\right)^{\alpha}\nonumber\;,\\
		&=&\left(1-\frac{a^2}{v^2}\right)r^2vv'+GM_*(1-\Gamma_e)-2a^2r-h(r)C\left(r^2vv'\right)^{\alpha}=0\nonumber\;,\\
	\end{eqnarray}
	with the constant $C$ being:
	$$C=\frac{\sigma_eL_*}{4\pi c}k\left(\frac{4\pi}{\sigma_e v_\text{th}\dot M}\right)^{\alpha}\;,$$
	and the function\footnote{Here, we use the letter $h$ instead $g$ in order to avoid confusions with acceleration.} $h(r)$ being:
	$$h(r)=\left(\frac{N_{e,11}}{W}\right)^\delta=\left[\frac{N_{e}}{2\times10^{11}}\left(1-\sqrt{1-\frac{R_*^2}{r^2}}\right)\right]^\delta\;.$$
	
	Despite the fact that the ionisation density $h(r)=N_{e,11}/W$ is not strictly constant, $h(r)C$ was assumed as constant under the classical m-CAK formulation because the exponent $\delta$ has values typically in the order of $\lesssim0.15$ \citep{abbott82,ppk}.
	This is the so-called \textit{fast solution}. However, for cases when $\delta$ takes higher values (in the order of $\gtrsim0.3$) $h(r)$ is not longer constant and a new hydrodynamical solutions may arise: the so-called \textit{$\delta$-slow solution} \citep{michel04,michel04b}.
	
	Due to the exponent $\alpha$, Eq.~\ref{funceqmotioncak} is a non-linear differential equation in which we will look for a monotonically increasing function $v(r)$.
	This solution must be unique, and starting from subsonic velocities ($v<a$) close to the stellar photosphere reaching a supersonic asymptotic terminal value $v_\infty\gg a$ at large radius.
	\citet{cak} demonstrated that there are different cases for $v(r)$ satisfying Eq.~\ref{funceqmotioncak}, but no one of them fulfil the conditions previously mentioned.
	However, it is possible to couple one of the solutions starting from subsonic region to one starting from the infinite, which match on a specific point called \textit{critical} or \textit{singular point} \citep[see Section 8.7.2 of][for details]{stellarwind}.
	This critical point $r_c$ is determined once the eigenvalue (mass-loss rate) of the equation of motion is numerically calculated, being then the respective eigenfunction, the formal solution for the hydrodynamics of the wind \citep{friend86,ppk,kppa89}.
	For this calculation, it is necessary to provide a location for this critical point as an standard method to solve the equation.
	On this work however, hydrodynamics is calculated using the code \textsc{HydWind} \citep{michel04}, where the equation of motion is calculated by means of finite-difference method, modified to handle singular points \citep{nobili88}.
	This code has the advantage of obtaining, according to the initial trial solution, different solutions with other critical points in addition to the standard one.
				
	Although in Section~\ref{mcaktheory} we introduced $k$, $\alpha$ and $\delta$ merely as constant parameters to fit the force-multiplier, it is possible to find physical meanings for the resulting line-acceleration and hydrodynamics depending on their final values.
	Then, physical interpretation of the line-force parameters \citep[see, e.g.,][]{puls00} are:
	\begin{itemize}
		\item The $k$ parameter, which takes values between 0 and 1, is directly proportional to the effective number of driving lines, and is related to the fraction of the photospheric flux which would have been blocked by all lines if they were optically thick and overlapping effects were not considered.
		Higher values of $k$ are obtained at higher densities and, therefore, higher mass-loss rates.
		In addition to the dependency on $\rho(r)$, $k$ presents also a strong dependence with metallicity and temperature due to the large number of driving lines: a lower temperature implies lower ionization stages, and thus more lines, therefore a higher $\mathcal M(t)$.
		More lines (above a given threshold line-strength) are also present for higher metallicities.

		The overlapping of two or more spectral lines produces an overestimation in the calculated value of $k$. On the other hand, $k$ is underestimated when multi-scattering effects are not taken into account (i.e., the summation in $\mathcal M(t)$ considers only direct photospheric radiation, and not radiation reprocessed in the wind).
		As was pointed out by \citet{puls87}, the inclusion of both effects might cancel, at least for O stars, and
		the \textit{effective} $k$ becomes moderately reduced. In this work, we have not considered these effects, therefore, our $k$ values should be maximum.
		\item The $\alpha$ parameter, which usually takes values between 0.45 and 0.75, is related to the exponent of the line-strength distribution function, and quantifies also the ratio of the line acceleration from optically thick lines to the total one \citep[for details, see][]{puls08}.
		Higher values of $\alpha$ implies both high mass-loss rates and terminal velocities in the resulting hydrodynamics.
		\item The $\delta$ parameter represents the change in the ionisation throughout the wind.
		According with classical literature \citep{stellarwind}, it takes lower values, rarely higher than 0.1.
		However, it has been found that, high values of $\delta$ ($ \gtrsim 0.25$) makes the wind 'slow', yielding a different wind solution \citep{michel11} because $C$ from Eq.~\ref{funceqmotioncak} cannot be longer treated as constant.
		Besides, according with \citet{puls00} $\delta$ takes an "exact" value of 1/3 for neutral hydrogen as a trace element.
	\end{itemize}
	
	These interpretations are coming from previous studies; hence, our work will consist in the analysis about whether these statements agree with our results or not, and of what new considerations we can include for discussion.
	Besides, some studies have pointed out that the line-force parameters are a function of radius \citep{schaerer94} or can be considered in a piecewise constant structure \citep{kudritzki02}. 
	Nevertheless, in this work we will consider $k$, $\alpha$ and $\delta$ as constants throughout all the wind because their variation is negligible over the final calculated wind parameters compared with uncertainties on stellar parameters (see Section~\ref{constantakd}).
	
%_____METHODOLOGY_______________________________________________________________________________
	\section{Calculation of the $\mathcal M(t)$ factor \label{methods}}
	When we talk about the calculation of the line-force parameters $k$, $\alpha$ and $\delta$, we actually mean the calculation of the line-force multiplier $\mathcal M(t)$ using Eq.~\ref{forcemultiplier}.
	To do that, we perform a script called \textsc{Alfakdelta27}, including the following improvements:
	
	i) a larger atomic line list, 
	
	ii) a quasi-NLTE approach for the ionisation equilibrium, 
	
	iii) a NLTE radiative stellar flux and 
	
	iv) an optical depth range consistent with the wind structure.
	
	Then we test it for one single-step (i.e., without iterations) first and after we execute the whole iteration procedure until the convergence of line-force parameters, velocity profile and mass-loss rate is achieved.
	
%_____Selection of atomic lines database________________________________________________________________	
	\subsection{Selection of atomic database}
		To calculate the line-acceleration and obtain a proper value of $\mathcal M(t)$, \citet{abbott82} established that it is necessary to sum the contribution of hundreds of thousands of spectral lines participating in the line-acceleration processes.
		Indeed, that work was pioneer in the inclusion of a larger atomic database taking into account more ions than those previously considered by \citet{cak}.
		Therefore, aiming to get the most accurate value of $\mathcal M(t)$, we decided to employ around $\sim 900\,000$ line transitions, whose atomic data were obtained (and modified in format) from the atomic database list used by the code CMFGEN\footnote{Atomic data used here are that one which were updated by DJH in 2016 (\url{http://kookaburra.phyast.pitt.edu/hillier/cmfgen_files/atomic_data_15nov16.tar.gz}).} \citep{hillier90a,hillier98}.
		We have chosen this database because it is the most complete one available, specially containing an extensively complete number of atomic transitions for heavy elements like iron and nickel (which contributes with $\sim80\%$ of all the spectral lines).
		Specifically, we have extracted information related to energy levels, degeneracy levels, partition functions and oscillator strengths $f_l$, which are necessary to calculate the absorption coefficient $\eta_\text{line}$ (see Eq.~\ref{betainv}) of each line in terms of lower ($l$) and upper ($u$) level populations $n_l$ and $n_u$, and their statistical weights $g_l$ and $g_u$. 
		
		Before continuing, it is important to remark the differences on notation and definitions compared with other studies.
		The most important atomic information about transitions for the calculation of $\mathcal M(t)$ is the oscillator strength $f_l$, defined as the dimensionless quantity that express the probability of absorption of reemission between the both energy levels corresponding to the line in question.
		However, because a change in the notation some authors express the factor $\eta$ as:
		\begin{eqnarray}\label{betainvgf}
			\eta_\text{line}&=&\frac{\pi e^2}{m_ec}\frac{1}{\rho\sigma_e \Delta\nu_D} f_l g_l\left(\frac{N_l}{g_l}-\frac{N_u}{g_u}\right)\nonumber\;,\\
			&=&\frac{\pi e^2}{m_ec}\frac{1}{\rho\sigma_e \Delta\nu_D}\left(gf\right)\left(\frac{N_l}{g_l}-\frac{N_u}{g_u}\right)\;.
		\end{eqnarray}
		Where the former oscillator strength is now multiplied by the lower statistical weight, creating then the so called $gf$-factor \citep[see, for example, notation used in][]{puls00}.
		Moreover, some other authors use the \textit{Einstein A coefficient} instead of the oscillator strength, which are related by the formula:
		\begin{equation}\label{oscillator}
			A_{ul}=\frac{8\pi^2\nu^2e^2}{\epsilon_0m_ec^3}\frac{g_l}{g_u}f\;[\text{s}^{-1}]\;.
		\end{equation}
		Even when Einstein $A$ coefficient do not appear explicitly on m-CAK notation, some authors such as \citet{noebauer15} have derived their oscillator strength from them (private communication in 2018).
			
		Elements and ionisation stages considered in this work are listed in Table~\ref{atomicdata}.
		Following \citet{abbott82}, we consider ions up to ionisation stage VI only.
		The total number of lines per element is also specified.
		\begin{table}
			\centering
			\scalebox{0.9}{
			\begin{tabular}{rlrrlr}
				\hline\hline
				Elem. & Ions & Nº lines & Elem. & Ions & Nº lines\\
				\hline
				H & I & 599 & He & I$-$II & 1\,342\\
				Li & I$-$III & 273 & Be & I$-$IV & 76\\
				B & I$-$V & 85 & C & I$-$IV & 25\,421\\
				N & I$-$VI & 8\,691 & O & I$-$VI & 6\,851\\
				F & I$-$VI & 187 & Ne & I$-$VI & 30\,880\\
				Na & I$-$VI & 8\,138 & Mg & I$-$VI & 7\,136\\
				Al & I$-$VI & 5\,613 & Si & I$-$VI & 2\,839\\
				P & I$-$VI & 3\,331 & S & I$-$VI & 15\,455\\
				Cl & I$-$VI & 534 & Ar & I$-$VI & 27\,376\\
				K & I$-$VI & 287 & Ca & I$-$VI & 37\,556\\
				Sc & I$-$VI & 322 & Ti & I$-$VI & 791\\
				V & I$-$VI & 920 & Cr & I$-$VI & 779\\
				Mn & I$-$VI & 688 & Fe & I$-$VI & 278\,923\\
				Co & I$-$VI & 489 & Ni & I$-$VI & 492\,341\\
				\hline
			\end{tabular}}
			\caption{\small{Atomic elements and ionisation stages used to calculate $\mathcal M(t)$. Range of frequency of the spectral lines goes from UV to IR range.}}
			\label{atomicdata}
		\end{table}
			
%_____Ionisation equilibrium____________________________________________________________________________
	\subsection{Ionisation equilibrium}
		Line-acceleration is calculated over the contribution of numerous transitions for every element at every ionisation stage present in the wind.
		\citet{abbott82} determined the ionization degrees using the Saha's equation for extended atmospheres \citep{mihalas78}, namely:
		\begin{equation}\label{saha}
			\small{\left(\frac{N_{i+1}}{N_i}\right)_\text{LTE}=2\left(\frac{2\pi m_ek_B}{h^2}\right)^{3/2}\frac{T_R\sqrt{T_e}}{N_e/W}\frac{U_{i+1}}{U_i}e^{-\frac{E_i}{k_BT_e}}\;\;,}
		\end{equation}
		being $T_R$, $T_e$ the radiation and electron temperatures, respectively, and $E_i$ the ionisation energy from stage $i$ to $i+1$.
		More precise treatment called \textit{approximate NLTE} (hereafter quasi-NLTE) has been used by, e.g., \citet{mazzali93} and \citet{noebauer15}.
		Here the ionisation balance is determined by the application of the modified nebular approximation \citep{abbott85}.
		Following this treatment, the ratio of number densities for two consecutive ions can be expressed in term of its LTE value, multiplied by correction effects due to dilution of radiation field and recombinations:
		\begin{equation}\label{recombinated}
			\small{\frac{N_{i+1}}{N_i}\approx\left(\frac{N_e}{W}\right)^{-1}[\zeta_i+W(1-\zeta_i)]\sqrt{\frac{T_e}{T_R}}\left(\frac{N_{i+1}N_e}{N_i}\right)_\text{LTE}\,,}
		\end{equation}
		where $\zeta_i$ represents the fraction of recombination processes that go directly to the ground stage. Eq. (\ref{recombinated}) is an alternative description to the one given by \citet{puls05}, who included a different radiative temperature dependence in the wind, which is specially important in the far UV region of the spectrum that is not optically thick.
			
		Modifications in the treatment of atomic populations $X_i$, being $i$ the excitation level, are also based on the work of \citet{abbott85}.
		It is necessary to make distinction between metastable levels (with no permitted electromagnetic dipole transitions to lower energy levels) and all the other ones:
		\[ \left(\frac{X_i}{X_1}\right) =
		\begin{cases}
		\left(\frac{X_i}{X_1}\right)_\text{LTE}       & \quad \text{metastable levels,}\\
		\\
		W(r)\left(\frac{X_i}{X_1}\right)_\text{LTE}  & \quad \text{others.}
		\end{cases}
		\]
		
		Atomic partition functions, $U_i$ (necessary for Saha's equation and the calculation of atomic populations), are calculated following the formulation of \citet{cardona10}, i.e.:
		\begin{equation}\label{cardonapartition3}
			U_i=U_{i,0}+G_{jk}e^{-\varepsilon_{jk}/T}+\frac{m}{3}(n^3-343)e^{-\hat E_{n*jk}/T}\;\;,
		\end{equation}
		where $U_{i,0}$ are the constant partition functions, $\hat E_{n*jk}$ is the mean excitation energy of the last level of the ion, $n$ is the maximum excitation stage to be considered, while $G_{jk}$, $\varepsilon_{jk}$ and $m$ are parameters tabulated by \citet{cardona10}.
			
		The advantage of this treatment is that it provides values for atomic partition functions explicitly as function of temperature and implicitly of electron density, giving a more accurate ionization balance.
		Following \citet{noebauer15}, the temperature will be treated as a constant ($T_R=T_e=T_\text{eff}$).
		Then, for a specific value of $(T_\text{eff},N_e)$, the ratio between number densities of ionization stage $j$ and  $i$ (for a specific $Z$-element) is calculated by a matrix (hereafter ionization matrix) given by:
		\begin{equation}
			D_{Z,i,j}=\frac{N_j}{N_i}=\prod_{i\leq k< j}\frac{N_{k+1}}{N_k}\;\;.
		\end{equation}
			
		In reference to the abundances of the different chemical elements, these were adopted from the solar abundances given by \citet{asplund09}.
		However, these can be easily modified to evaluate stars with non-solar metallicity (see Section~\ref{mcakresults}).

		At this point, it is necessary to remark that previous authors  \citep{abbott82,noebauer15} have considered the diluted-electron density $N_e/W$ as constant throughout the wind.
		Nevertheless, to calculate $\delta$, $\mathcal M(t)$ must be evaluated considering changes in the ionisation stages, and therefore, $N_e(r)/W(r)$. Since, the calculation of electron density depends on  the ionisation stages of each specie which in turn are functions of $N_e$, we deal with a coupled non-linear problem.
		To obtain a solution, we use the following formula to calculate (as initial value) the electron number density:
		\begin{equation}
		\label{eqNe0}
			N_{e,0}=\frac{\rho(r)}{m_H}\,\frac{X_\text{H}+2X_\text{He}}{X_\text{H}+4X_\text{He}}\, ,
		\end{equation}
		being $m_H$ the hydrogen atom mass, and $X_\text{H}$ and $X_\text{He}$ the abundances of hydrogen and helium, respectively.
			
		We used this initial electron density to start calculating the ionisation matrix and to re-calculate both atomic populations and electron density iteratively:
		\begin{eqnarray}
			N_e(r)&=&\left(X_\text{H}\frac{D_{1,1,2}}{1+D_{1,1,2}}+X_\text{He}\frac{(D_{2,1,2}+2D_{2,1,3})}{1+D_{2,1,2}+D_{2,1,3}}\right)\nonumber\\
			& &\times \, \frac{\rho(r)}{X_\text{H}+4X_\text{He}}\,.
		\end{eqnarray}
			
		\begin{figure}
			\centering
			\includegraphics[width=0.6\textwidth]{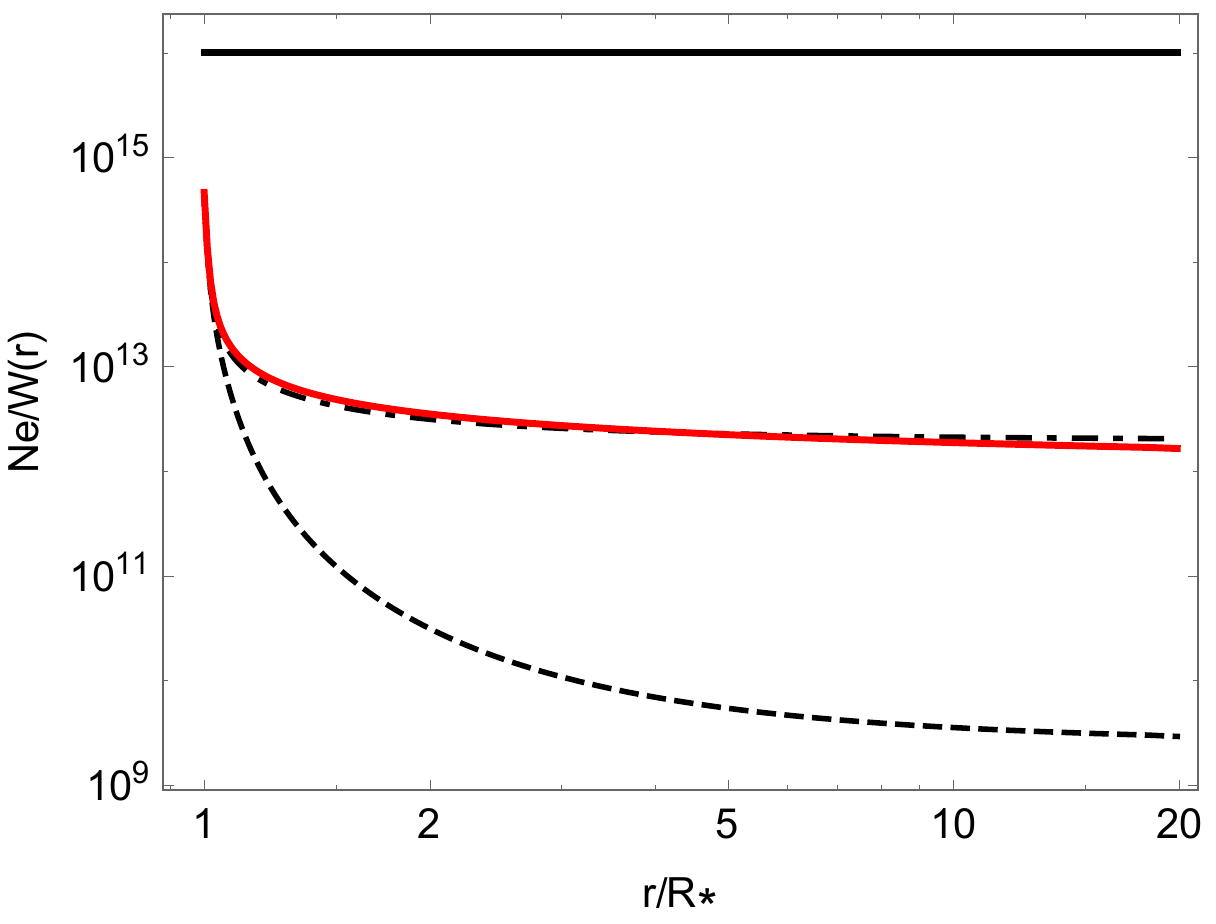}
			\caption{\small{Final value of $N_e/W(r)$ as function of stellar radius even when $N_{e,0}$ is set as constant input (black solid line), after one iteration (single dashed line), after two iterations (dashed-dotted line) and after five iterations (red solid line).}}
			\label{ne-conv}
		\end{figure}
		Convergence of $N_e$ is easily obtained after just a few iterations (see Fig.~\ref{ne-conv}).
		It is important to remark, that alternatively we can start iterations using $N_{e,0}/W$ as a constant value following \citet{abbott82} and \citet{noebauer15} instead of starting using Eq.~\ref{eqNe0}, and anyway the final converged value for $N_e(r)$ is the same.

%_____Radiaton field__________________________________________________________________________________
	\subsection{Radiation field}
		Together with an accurate treatment of atomic populations and electron density, Eq.~\ref{forcemultiplier} requires as an input the radiation field in the terms of $F_\nu/F$.
		
		Simplest expression for the radiation field comes from the black-body Planck's law:
		\begin{equation}\label{plancknu}
			B_\nu(T)=\frac{2h\nu^3}{c^2}\frac{1}{e^{-h\nu/k_\text{B}T}-1}\;.
		\end{equation}
		\begin{figure}[htbp]
			\centering
			\includegraphics[width=0.9\textwidth]{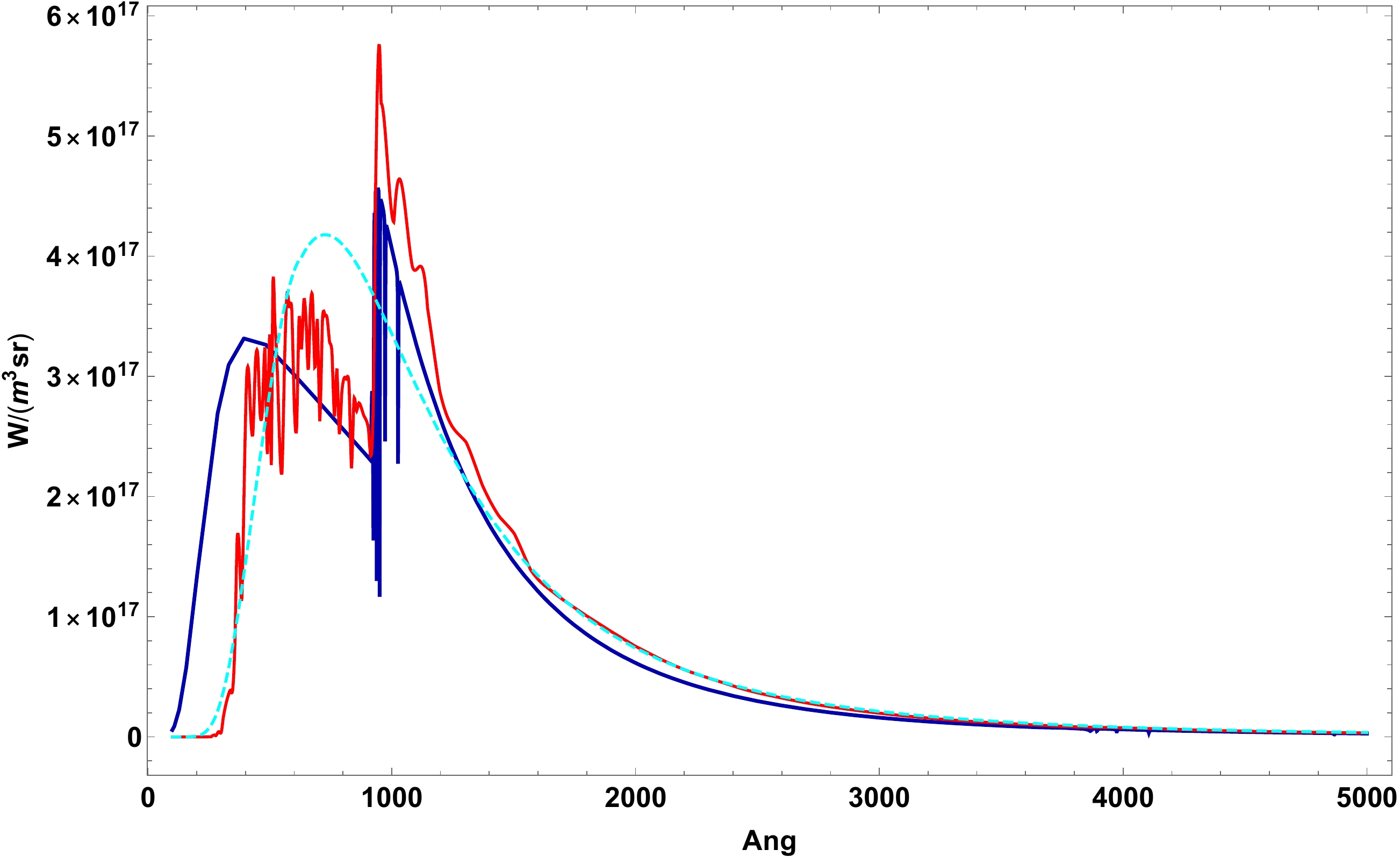}
			\caption{\small{Comparison of radiation fields $F_\nu$ obtained by black-body Planck's radiation law (cyan, dashed) and those ones obtained from the models of Kurucz (red) and \textsc{Tlusty} (dark blue), implemented for a star with $T_\text{eff}=40\,000$ K and $\log g=4.0$.}}
			\label{bb_vs_kurucz_vs_tlusty}
		\end{figure}
		
		But this back-body scenario assumes a full LTE, which is not valid for stellar atmospheres because of the transport of energy and matter; it is required then to solve the equation of radiative transfer in order to incorporate effects due to opacity and hence obtain an accurate radiation field.
		Since this calculation is beyond the scope of the present work, we proceed to simply employ the already calculated flux fields for different stellar models.
		Some of the most common radiation field models used by the stellar wind community are those performed by \citet[][in LTE only]{kurucz79} and the more modern \textsc{Tlusty} models \citep[][in both LTE and NLTE]{hubeny95,lanz03}.
		Differences among these models are presented in Fig.~\ref{bb_vs_kurucz_vs_tlusty}, where black-body radiation Planck's law is shown together with a Kurucz and a \textsc{Tlusty} model.
			
		The usage of already performed stellar models for the flux field was also implemented before.
		For example, \citet{abbott82} used the radiation fields from Kurucz' models, whereas \citet{noebauer15} from a black-body. In this work, we use the radiation field calculated by the NLTE line-blanketed plane-parallel code \textsc{Tlusty}.
		
		The overlap effects among ten of thousands of spectral lines are not considered when we sum the contributions to the force-multiplier $\mathcal M(t)$ across the wind.
		However, line blanketing effects are partially considered as we are using \textsc{Tlusty} radiation field in the calculations of $\mathcal M(t)$.

%_____Determination of the optical depth range____________________________________________________________
	\begin{figure}[h!]
		\centering
		\includegraphics[width=0.48\textwidth]{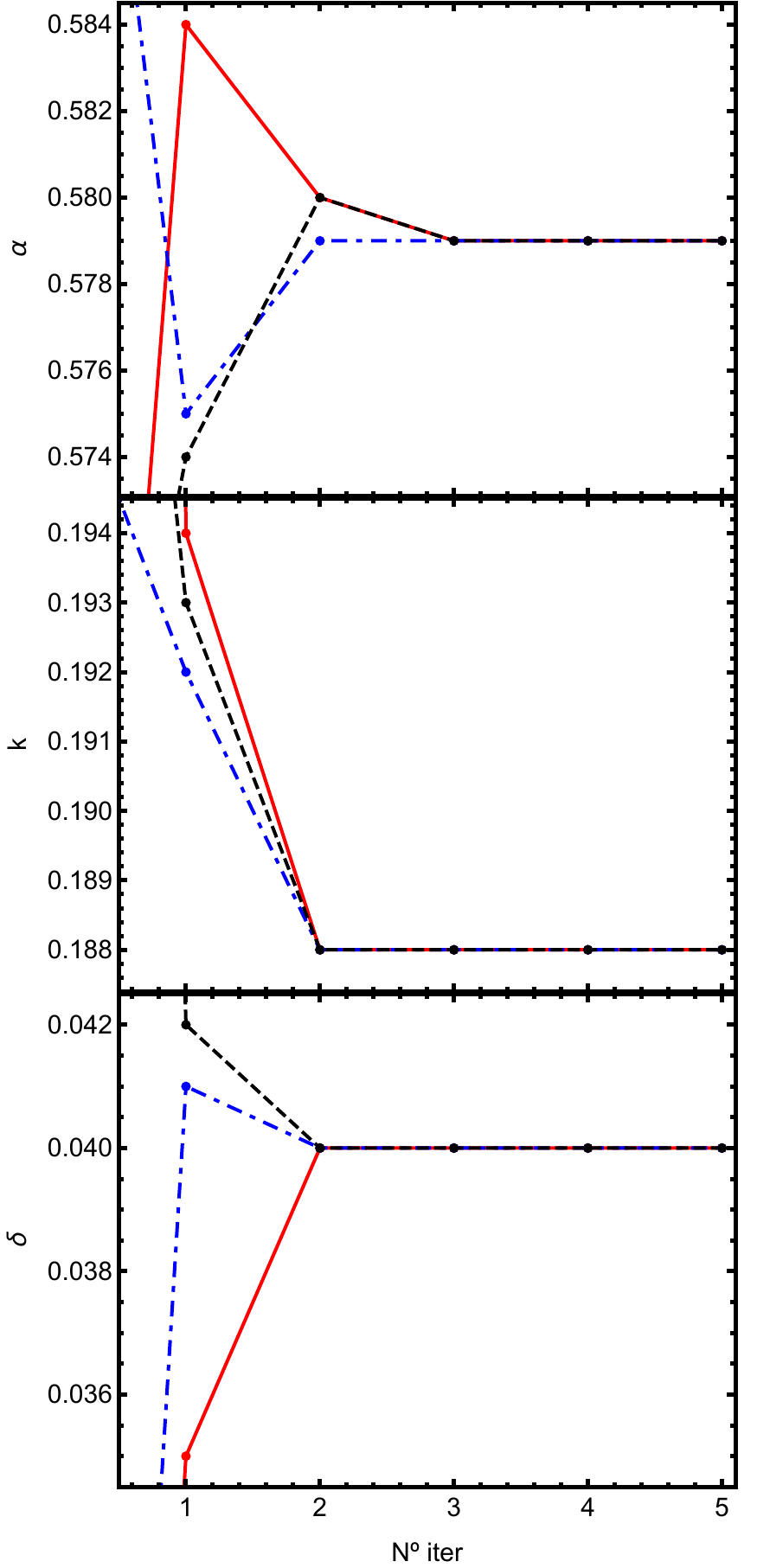}
		\includegraphics[width=0.5\textwidth]{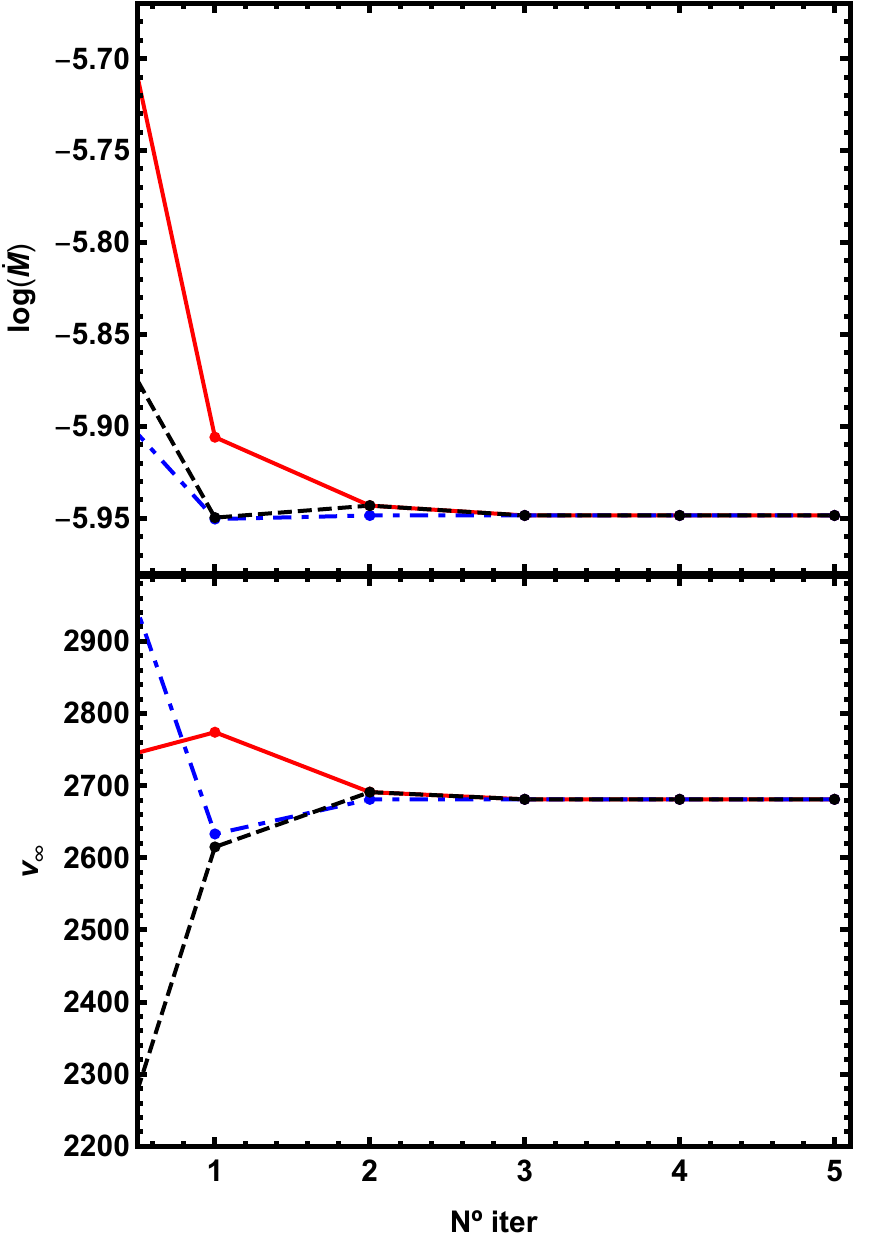}
		\caption{\small{Left: Values of $\alpha$, $k$ and $\delta$ as a function of the iteration number, starting from different initial values. Different initial values (iteration 0, not shown) converge to the same final self-consistent solution. Right: idem as left, but for the mass-loss rate and terminal velocity.}}
		\label{iters}
	\end{figure}

	\subsection{Determination of the optical depth range}
		Previous studies by \citet{abbott82} and \citet{noebauer15} have considered a fixed range for the optical depth $t$ to fit the force multiplier (Eq.~\ref{mcak}).
		However, given the definition of $t$ (Eq.~\ref{tcak}), it is clear that the optical depth range is constrained by the physical properties of the stellar wind (density and velocity profiles).
		For this reason, calculations presented in this work are constrained inside the wind, characterised by this range of $t$.
		
		Because m-CAK theory is based upon Sobolev approximation \citep[][see also Section~\ref{sobolevapprox} of the present Thesis]{sobolev60}\footnote{See also Section 8.4 of \citet[][\textit{Introduction to Stellar Winds}]{stellarwind}} in this work we will use as upper and lower limits for the optical depth $t$, its values at the sonic point and at infinity (usually $r\sim 100\,R_*$), respectively.
		It is important to remark that although $t$ decreases outwards it never reaches zero\footnote{This condition lies in the fact that, at larger distances, both density $\rho$ and velocity gradient $dv/dr$ decrease as $\sim r^{-2}$, cancelling each other.}.
		Therefore, it is always possible to define a proper $t$ range.

%_____Iterative procedure______________________________________________________________________________
	\subsection{Iterative procedure}
		Velocity profile and mass-loss rate from hydrodynamics are required to calculate the line-acceleration $g_\text{line}$.
		At the same time, line-force parameters fitted from $g_\text{line}$, are necessary to solve the m-CAK hydrodynamic equations and obtain the mass-loss rate and velocity profile.
		Therefore a self-consistent iterative procedure must be implemented to solve this coupled non-linear problem.
		
		Our procedure is the following:
		\begin{enumerate}[label=\roman*.]
			\item Using a $\beta-$law profile with a given mass-loss rate, initial values for the line force parameters $(k_0,\alpha_0,\delta_0)$ are calculated. 
		 	\item A numerical solution of the equation of motion (Eq.~\ref{eqmotioncak}) is obtained with \textsc{HydWind \footnote{This code solves the m-CAK equation of motion with an eigenvalue that depends on the mass-loss rate. At the location of the singular point, both solution branches (singular point to stellar surface and singular point to infinity) are smoothly merged to obtain the velocity profile, see \citet{ppk,friend86,michel04} for details.}}, getting an improved hydrodynamics: $v(r)$ and $\dot{M}$.
			\item A new force multiplier is calculated.
			\item New line-force parameters $(k_i,\alpha_i,\delta_i)$ are fitted from  $\mathcal M(t)$
			\item Steps ii - iv are iterated until convergence.
		\end{enumerate}
		
		Convergence is usually obtained after $\sim4-5$ iterations (see left panel of Fig.~\ref{iters}), independently on the initial values.
		Our criterion for convergence is when  two consecutive iterations ($i$, $i+1$) get a value  for $\|\Delta p\|= \|p^{i+1}-p^{i}\| \,\leq 10^{-3}$, where $p$ is a line-force parameter and this condition should be satisfied for each one of these parameters.
		
		Right panel of Fig.~\ref{iters} shows the convergence of the mass-loss rate (top panel) and the terminal velocity (lower panel) as function of the procedure's iterations.
		Both values depend non-linearly on the stellar and line-force parameters.
		
%_____One single iteration test_________________________________________________________________________	
	\begin{table}[htbp]
		\centering
		\scalebox{0.9}{
		\begin{tabular}{ccclcccc}
			\hline
			\hline
			& & & & \multicolumn{2}{c}{previous studies} & \multicolumn{2}{c}{present work}\\
			\cline{5-6} \cline{7-8}
			& $T_\text{eff}$ & $N_e/W$ & $\delta$ & $k$ & $\alpha$ & $k_1$ & $\alpha_1$ \\
			& [kK] & [cm$^{-3}$] &\\
			\hline
			A & 30 & $1.0\times10^8$ & 0.12 & 0.093 & 0.576 & 0.062 & 0.661\\
			A & 30 & $1.0\times10^{11}$ & 0.12 & 0.156 & 0.609 & 0.097 & 0.611\\
			A & 30 & $1.0\times10^{14}$ & 0.12 & 0.571 & 0.545 & 0.487 & 0.450\\
			A & 40 & $1.8\times10^8$ & 0.12 & 0.051 & 0.684 & 0.072 & 0.639\\
			A & 40 & $1.8\times10^{11}$ & 0.12 & 0.174 & 0.606 & 0.120 & 0.609\\
			A & 40 & $1.8\times10^{14}$ & 0.12 & 0.533 & 0.571 & 0.289 & 0.552\\
			N & 42 & $1.0\times10^{15}$ & 0.0 & 0.381 & 0.595 & 0.376 & 0.572\\
			A & 50 & $3.1\times10^8$ & 0.092 & 0.089 & 0.640 & 0.148 & 0.611\\
			A & 50 & $3.1\times10^{11}$ & 0.092 & 0.178 & 0.606 & 0.196 & 0.595\\
			A & 50 & $3.1\times10^{14}$ & 0.092 & 0.472 & 0.582 & 0.289 & 0.566\\
			\hline
		\end{tabular}}
		\caption{\small{Comparison of $k$ and $\alpha$ parameters from Abbott (A) and Noebauer \& Sim (N), with our one single-step results.}}
		\label{initresults}
	\end{table}

	\subsection{A single-step test}
		To compare our line-force parameters with the results obtained by \citet{abbott82} and \citet{noebauer15}, we use just one single-step.
		Following these authors, $\delta$ and $N_e/W$ are set as input and the optical depth range is fixed between $-6<\log t<-1$. The selection of a fixed interval of $\log t$ does not  require any velocity field structure.
		Furthermore we have considered Kurucz' and  black-body fluxes to reproduce  \citet{abbott82} and \citet{noebauer15} calculations, respectively.
		Then, starting from a $\beta$--law and a $\dot{M}$, we calculate $k_1$ and $\alpha_1$ (single-step).
		These results are shown in Table~\ref{initresults}.
		The  coefficients of determination, $R$-Squared, for $\alpha$ and $k$ (respectively) between previous and our single-iteration results are: 
		
		i) $R_{\alpha}^2=0.87$ and $R_k^2=0.93$ for $T_{\rm{eff}} \geq 40\,000$ K; 
		
		ii) $R_{\alpha}^2=0.4$ and $R_k^2=0.81$ for $T_{\rm{eff}} \geq 30\,000$ K.
		
		We conclude that our calculations reproduced previous results, now using a modern atomic database and abundances.
		
%_____LTE RESULTS__________________________________________________________________________________
\section{m-CAK Results}\label{mcakresults}
	This section is focused on the resulting values obtained for both line-force and wind parameters following the methodology previously described.
	\begin{table}[t]
		\centering
		\resizebox{\textwidth}{!}{
		\begin{tabular}{cccc|rc|ccc|cccc}
			\hline
			\hline
			$T_\text{eff}$ & $\log g$ & $R_*/R_\odot$ & $Z_/Z_\odot$ & $\log t_\text{in}$ & $\log t_\text{out}$ & $k$ & $\alpha$ & $\delta$ & $ v^\text{SC}_\infty$ & $\dot M_\text{SC}$\ & $\dot M_\text{SC}/\dot M_\text{Vink}$\\
			$[\text{kK}]$ & & & & & & & & & [km s$^{-1}$] & [$10^{-6}M_\odot\,\text{yr}^{-1}$]\\
			\hline
			45 & 4.0 & 12.0 & 1.0 & $-0.31$ & $-4.53$ & 0.167 & 0.600 & 0.021 & $3\,432\pm240$ & $2.0\pm_{0.5}^{0.65}$ & 1.00\\
			45 & 4.0 & 12.0 & 0.2 & $-0.77$ & $-4.85$ & 0.142 & 0.493 & 0.017 & $2\,329\pm210$ & $0.38\pm_{0.11}^{0.15}$ & 0.74\\
			45 & 3.8 & 16.0 & 1.0 & $0.28$ & $-4.07$ & 0.135 & 0.648 & 0.022 & $3\,250\pm300$ & $6.4\pm_{1.3}^{1.6}$ & 0.84\\
			45 & 3.8 & 16.0 & 0.2 & $-0.06$ & $-4.28$ & 0.114 & 0.545 & 0.014 & $2\,221\pm230$ & $1.7\pm_{0.45}^{0.6}$ & 0.88\\
			42 & 3.8 & 16.0 & 1.0 & $-0.10$ & $-4.36$ & 0.137 & 0.629 & 0.027 & $3\,235\pm300$ & $3.4\pm_{0.7}^{0.9}$ & 0.94\\
			42 & 3.8 & 16.0 & 0.2 & $-0.55$ & $-4.73$ & 0.108 & 0.534 & 0.019 & $2\,313\pm230$ & $0.73\pm_{0.21}^{0.3}$ & 0.79\\
			42 & 3.6 & 20.4 & 1.0 & 0.70 & $-3.80$ & 0.122 & 0.671 & 0.039 & $2\,738\pm230$ & $11\pm_{2.5}^{3.5}$ & 0.74\\
			42 & 3.6 & 20.4 & 0.2 & 0.37 & $-4.09$ & 0.091 & 0.586 & 0.022 & $2\,043\pm200$ & $3.1\pm_{0.75}^{1.2}$ & 0.82\\
			40 & 4.0 & 12.0 & 1.0 & $-0.88$ & $-4.97$ & 0.164 & 0.581 & 0.027 & $3\,300\pm220$ & $0.66\pm_{0.15}^{0.19}$ & 1.17\\
			40 & 4.0 & 12.0 & 0.2 & $-1.43$ & $-5.44$ & 0.133 & 0.492 & 0.038 & $2\,329\pm160$ & $0.11\pm_{0.03}^{0.05}$ & 0.76\\
			40 & 3.6 & 20.4 & 1.0 & $0.42$ & $-3.96$ & 0.118 & 0.659 & 0.044 & $2\,813\pm290$ & $6.6\pm_{1.4}^{1.8}$ & 0.89\\
			40 & 3.6 & 20.4 & 0.2 & $-0.05$ & $-4.40$ & 0.091 & 0.572 & 0.025 & $2\,116\pm230$ & $1.7\pm_{0.4}^{0.5}$ & 0.90\\
			40 & 3.4 & 18.0 & 1.0 & $1.30$ & $-3.14$ & 0.099 & 0.715 & 0.094 & $1\,548\pm240$ & $14.5\pm_{3.5}^{5.0}$ & 0.73\\
			40 & 3.4 & 18.0 & 0.2 & $1.90$ & $-3.50$ & 0.073 & 0.650 & 0.047 & $1\,334\pm230$ & $4.7\pm_{1.3}^{2.4}$ & 0.92\\
			38 & 3.8 & 16.0 & 1.0 & $-0.63$ & $-4.79$ & 0.130 & 0.610 & 0.036 & $3\,153\pm240$ & $1.2\pm_{0.25}^{0.3}$ & 1.10\\
			38 & 3.8 & 16.0 & 0.2 & $-1.18$ & $-5.28$ & 0.091 & 0.542 & 0.033 & $2\,473\pm300$ & $0.25\pm_{0.06}^{0.08}$ & 0.89\\
			36 & 4.0 & 12.0 & 1.0 & $-1.45$ & $-5.50$ & 0.132 & 0.580 & 0.036 & $3\,314\pm200$ & $0.21\pm_{0.05}^{0.065}$ & 1.17\\
			36 & 4.0 & 12.0 & 0.2 & $-1.97$ & $-5.97$ & 0.101 & 0.517 & 0.068 & $2\,402\pm140$ & $0.036\pm_{0.01}^{0.014}$ & 0.78\\
			36 & 3.6 & 20.4 & 1.0 & $-0.29$ & $-4.55$ & 0.104 & 0.644 & 0.062 & $2\,809\pm240$ & $2.2\pm_{0.5}^{0.7}$ & 1.12\\
			36 & 3.6 & 20.4 & 0.2 & $-0.87$ & $-5.09$ & 0.071 & 0.581 & 0.033 & $2\,534\pm220$ & $0.5\pm_{0.13}^{0.17}$ & 1.00\\
			36 & 3.4 & 18.0 & 1.0 & $1.78$ & $-3.77$ & 0.091 & 0.686 & 0.116 & $1\,708\pm170$ & $4.4\pm_{1.0}^{1.6}$ & 1.13\\
			36 & 3.4 & 18.0 & 0.2 & $0.41$ & $-4.21$ & 0.072 & 0.607 & 0.048 & $1\,566\pm160$ & $1.0\pm_{0.25}^{0.4}$ & 1.01\\
			34 & 3.8 & 16.0 & 1.0 & $-1.27$ & $-5.37$ & 0.103 & 0.604 & 0.043 & $3\,093\pm210$ & $0.34\pm_{0.07}^{0.1}$ & 1.12\\
			34 & 3.8 & 16.0 & 0.2 & $-1.93$ & $-5.94$ & 0.069 & 0.555 & 0.028 & $2\,791\pm180$ & $0.074\pm_{0.018}^{0.025}$ & 0.95\\
			34 & 3.6 & 20.4 & 1.0 & $-0.61$ & $-4.82$ & 0.095 & 0.637 & 0.074 & $2\,732\pm180$ & $1.2\pm_{0.3}^{0.4}$ & 1.25\\
			34 & 3.6 & 20.4 & 0.2 & $-1.29$ & $-5.46$ & 0.058 & 0.590 & 0.031 & $2\,642\pm180$ & $0.25\pm_{0.05}^{0.07}$ & 1.03\\
			32 & 3.4 & 18.0 & 1.0 & $0.37$ & $-4.30$ & 0.078 & 0.675 & 0.159 & $1\,653\pm190$ & $1.3\pm_{0.3}^{0.5}$ & 1.67\\
			32 & 3.4 & 18.0 & 0.2 & $-0.70$ & $-4.15$ & 0.053 & 0.610 & 0.052 & $1\,847\pm140$ & $0.23\pm_{0.05}^{0.075}$ & 1.16\\
			\hline
		\end{tabular}}
		\caption{\small{Self-consistent line-force parameters $(k,\alpha,\delta)$ for adopted standard stellar parameters, together with the resulting terminal velocities and mass-loss rates ($\dot M_\text{SC}$). Ratios between self-consistent mass-loss rates and  Vink's recipe values \citep[re-scaled to match metallicity from][]{asplund09} using $ v_\infty/v_{\rm{esc}}=2.6$ are shown in the last column. Error margins for mass-loss rates and terminal velocities are derived over a variation of $\pm500$ for effective temperature, $\pm0.05$ for logarithm of surface gravity and $\pm0.1$ for stellar radius.}}
		\label{standardtable}
	\end{table}
	\begin{figure}[htbp]
		\centering
		\includegraphics[width=0.55\linewidth]{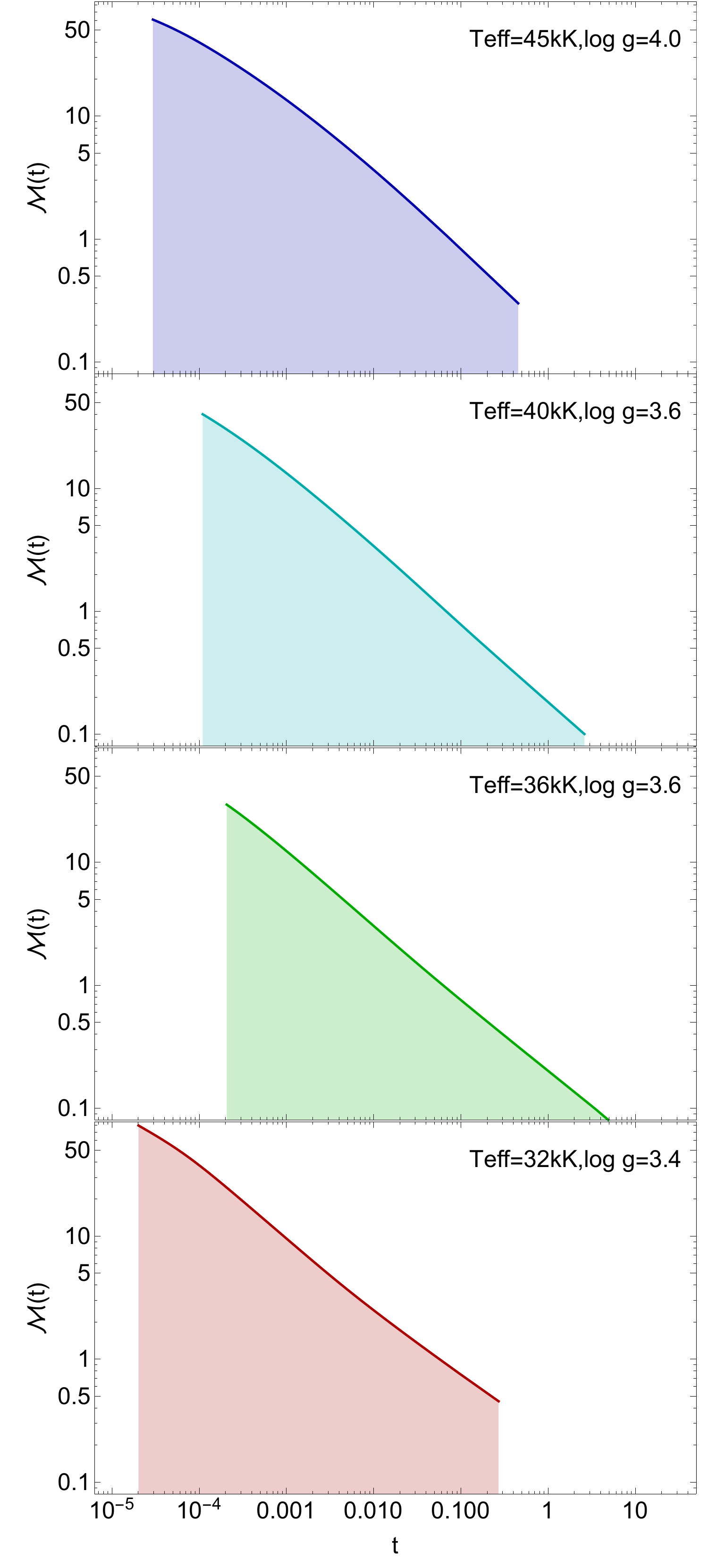}
		\caption{\small{Force-multiplier $\mathcal M(t)$ as function of $t$ for some stellar models presented on Table~\ref{standardtable} with $T_\text{eff}=45\,000$ K and $\log g=4.0$ (blue), $T_\text{eff}=40\,000$ K and $\log g=3.6$ (cyan), $T_\text{eff}=36\,000$ K and $\log g=3.4$ (green) and $T_\text{eff}=32\,000$ K and $\log g=3.4$ (red). Coloured areas below curves indicate the range of $t$ where the fits for $(k,\alpha,\delta)$ have been adjusted.}}
		\label{fastplots}
	\end{figure}
	\begin{figure}[htbp]
		\centering
		\includegraphics[width=0.55\linewidth]{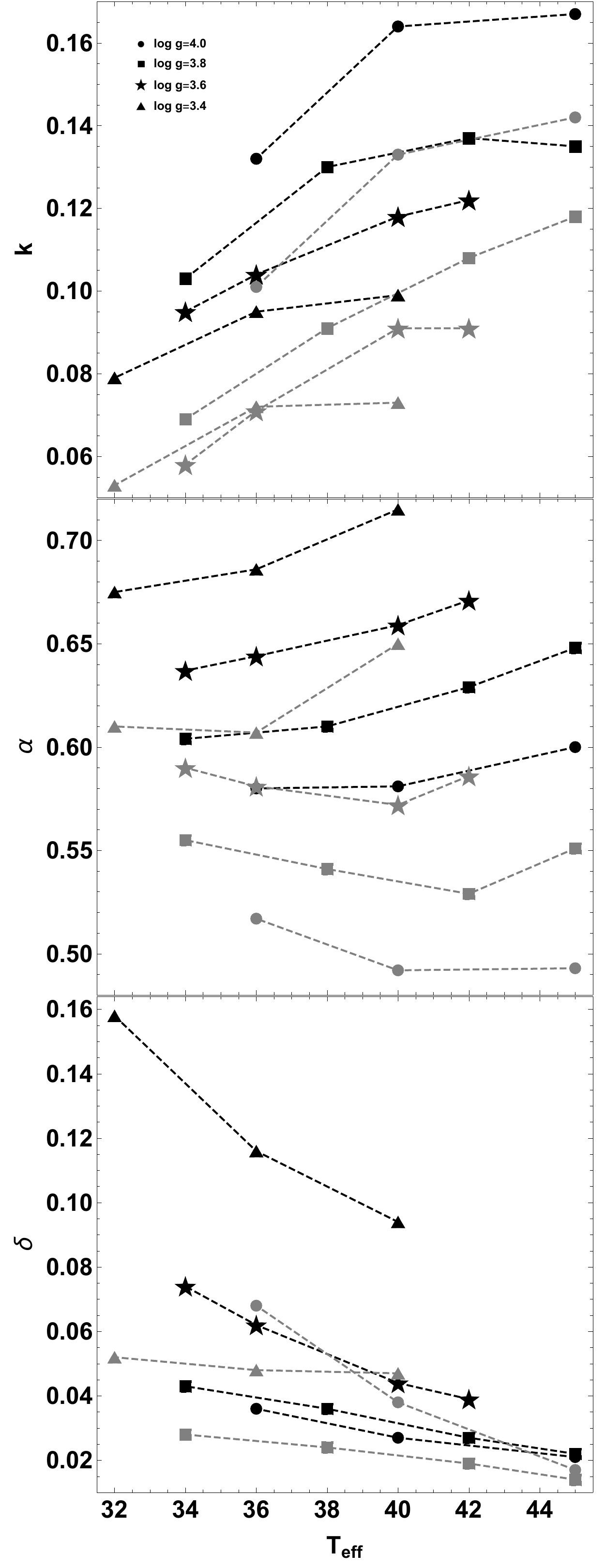}
		\caption{\small{Behaviour of line-force parameters $(k,\alpha,\delta)$ as a function of the effective temperature (in kK), for different surface gravities and metallicities. Circles represent models with $\log\,g=4.0$, squares: $\log\,g=3.8$, stars: $\log\,g=~3.6$, and triangles:  $\log g=3.4$. Black dashed lines are for models with solar metallicity and grey dashed lines for  $Z=Z_{\odot}/5$.}}
		\label{behaviours}
	\end{figure}

%_____Self-consistent calculations_______________________________________________________________________
	\subsection{Self-consistent calculations}
		The following results are computed  self-consistently with the methodology detailed in Section~\ref{methods}.

		Self-consistent solutions for a grid of models are presented in Table~\ref{standardtable}.
		The effective temperature ranges from 32 kK to 45 kK and $\log\, g$ from $3.4$ to $4.0$ dex.
		This grid considers different stellar radii and two abundances: 1 and 1/5 of the solar value.
		This table shows the stellar parameters, the calculated $t$-range, and the fitted m-CAK line-force.
		In addition, we calculated the corresponding wind solution using \textsc{HydWind}, and their error margins were derived considering variations of $\Delta T_{\rm{eff}}=\pm500$, $\Delta\log g=\pm0.05$, and $\Delta R_*=\pm0.1R_\odot$ in the stellar radius, keeping constant the line-force parameters.
		
		Convergence has been checked for each solution.
		Figure~\ref{fastplots} shows the final resulting $\mathcal M(t)$ given by the last iteration for different four models from Table~\ref{standardtable} at their respective ranges of $t$.
		Due to the quasi-linear behaviour of the logarithm of the  force-multiplier, parameters $k$ and $\alpha$ are easily fitted and their values can be considered constant throughout the wind (see Sect.~\ref{constantakd}).
		To fit $\delta$ in the $\mathcal M(t)$--$N_e/W$ plane, it is necessary to perform an extra calculation of $\mathcal M(t)$ using a slightly different value for the diluted-electron density.
		Last column of this table shows the ratio between our mass-loss rate and the one calculated using Vink's recipe \citep{vink01}, with $v_{\infty}/v_{esc}=2.6$ and re-scaled to current abundances \citep{asplund09}.
		The mean value of $\dot M_\text{SC}/\dot M_\text{Vink} = 0.98 \pm 0.2$.
		As we have not included in our procedure multi-line nor line-overlapping processes, we support \citet{puls87} conclusion that these effects are somewhat canceled, because we do not observe relevant discrepancies in the mass-loss rates when a comparison with Vink's recipe is performed.
		
		In Fig.~\ref{behaviours}, we observe clear trends for the behaviour of the $(k,\alpha,\delta)$ parameters with $T_{\rm{eff}}$,  $\log\,g$, and $Z$.
		While $k$ increases, $\delta$ decreases as function of the effective temperature, for both metallicities.
		It is interesting to remark the influence of the surface gravity on the resulting line-force parameters, values for $k$ and $\delta$ decrease as the gravity decreases.
		Notice that globally our line-force parameter results are similar to the values obtained in previous works \citep{puls00,kudritzki02,noebauer15}. However, we found an important dependence on $\log g$ as a result of the hydrodynamic coupling in the self-consistent procedure.
		
		On the other hand, the behaviour of $\alpha$ depends on the metallicity, it increases with effective temperature for solar abundance, but for low abundance and low gravities, it slowly decreases with temperature.
		Moreover, the change in $\alpha$ is more significant for $\log g$ than for $T_{\rm{eff}}$: a difference in $\Delta \log g \pm 0.2$ dex produces a $\Delta \alpha \sim0.04$, whereas variations on $\Delta T_\text{eff} =\pm\, 2\,000\,$K, might produce $\Delta \alpha \sim 0.02$.
				
\begin{figure}[h!]
			\centering
			\includegraphics[width=0.48\linewidth]{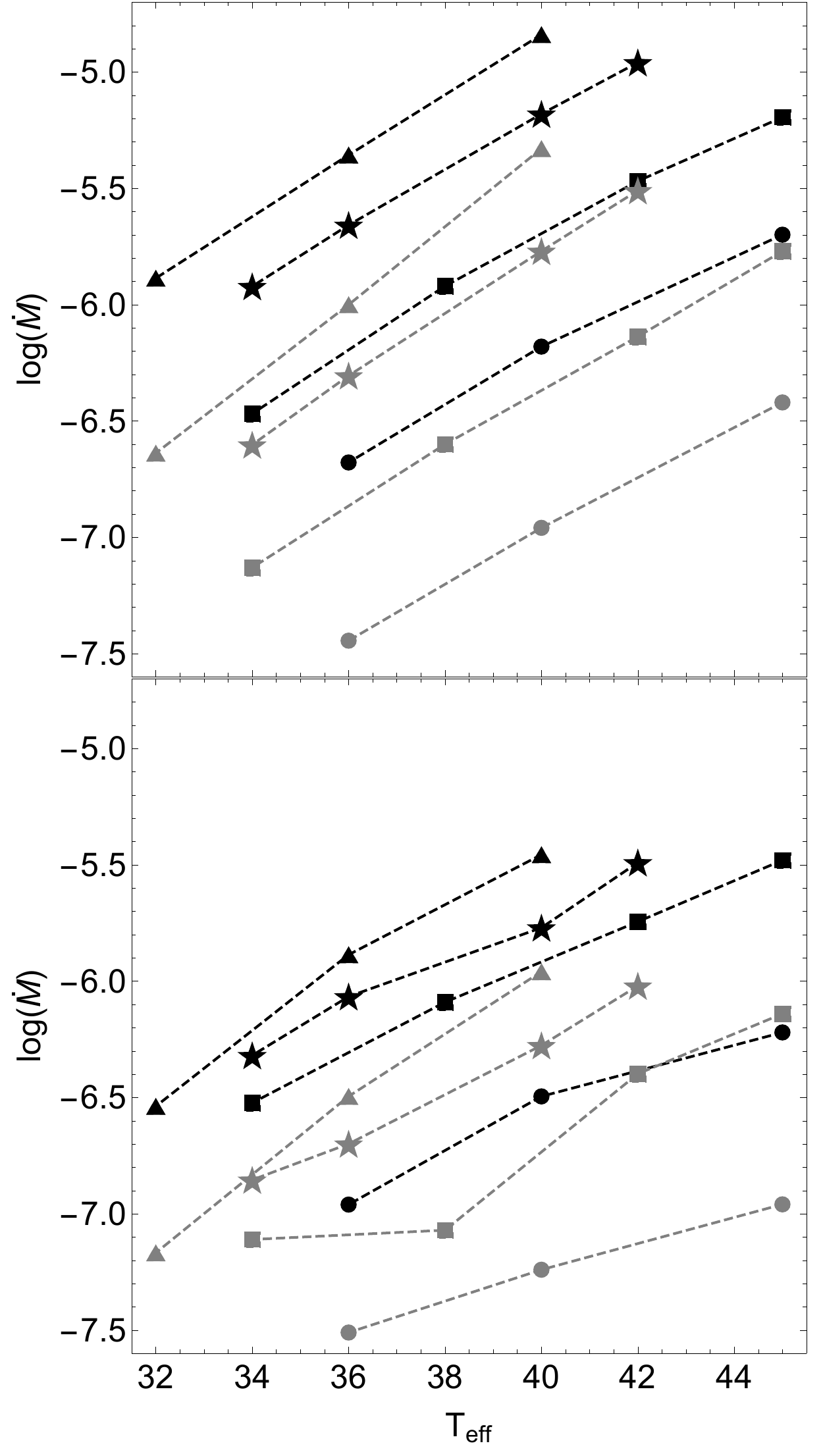}
			\includegraphics[width=0.48\linewidth]{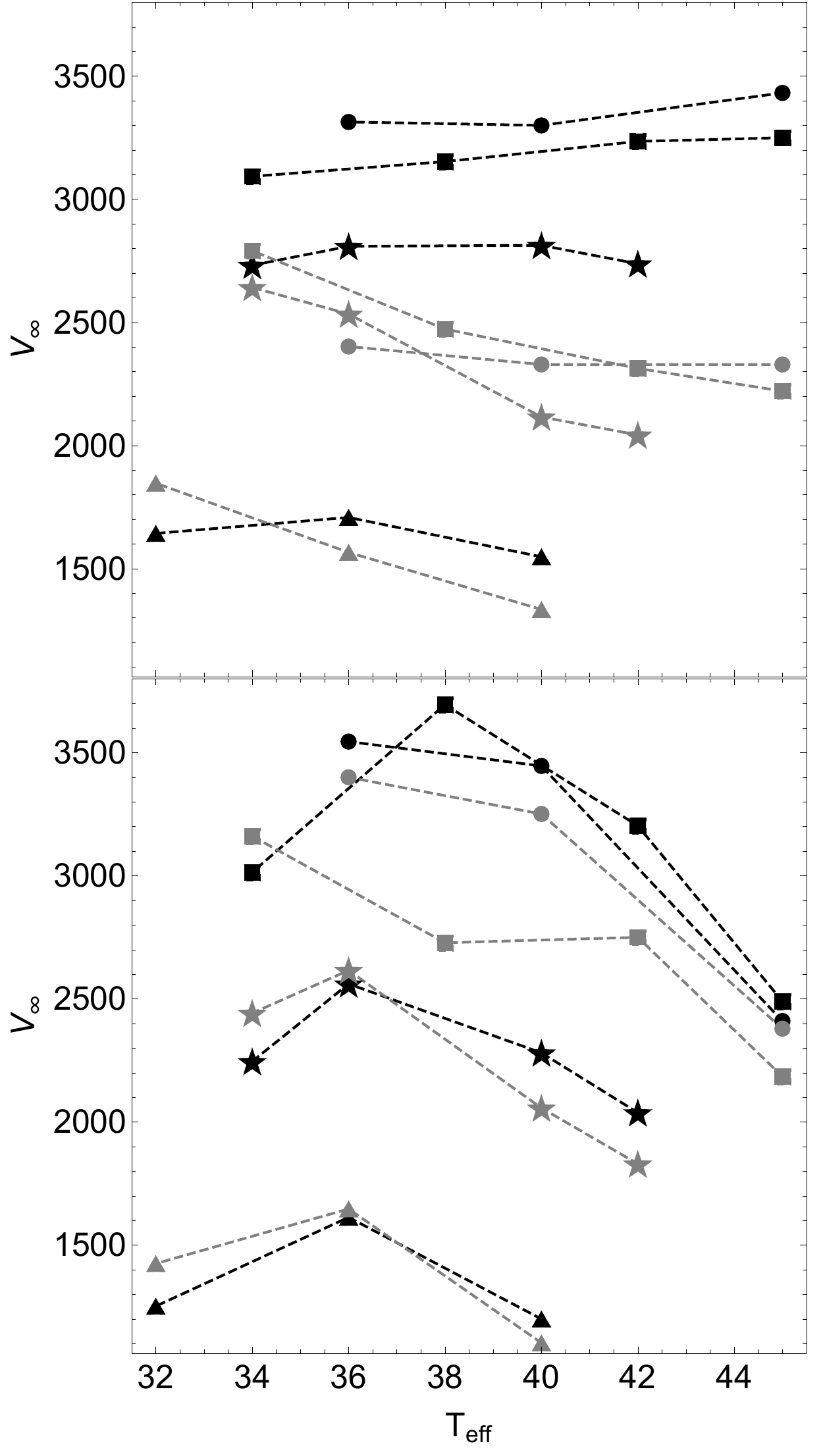}
			\caption{\small{Left: Behaviour of mass-loss rate as a function of effective temperature (in kK) for different abundances and gravities. Top panel is for self-consistent calculations and bottom panel is for Abbott's procedure, now including the finite-disk correction factor. Symbol description is the same as Fig.~\ref{behaviours}. Right: same as left panel, but for terminal velocities.}}
			\label{windbehaviours}
		\end{figure}

		Left panel of Figure~\ref{windbehaviours} shows the results for the mass loss rates as a function of the effective temperature, for different gravities and metallicities.
		Upper panel shows the results from our self-consistent procedure and 
		bottom panel shows the result using Abbott's methodology (a single iteration) to calculate line-force parameters and apply them in our hydrodynamic code \textsc{HydWind} (hereafter Abbott's procedure).
		We found that $\dot{M}$ increases with effective temperature and metallicity and decreases with gravity.
		This behaviour is similar to the one obtained using Abbott's procedure, but the self-consistent calculated mass-loss rates are about $30\%$ larger. 
		
		From the mass-loss results tabulated in Table~\ref{standardtable}, a simple linear relationship for solar-like metallicities (with a  coefficient of determination or $R$--squared, $R^2=0.999$) reads:
		\begin{align}\label{mdotz10eq1}
			\log \dot M_{Z=1.0}=&10.443\times\log\left(\frac{T_\text{eff}}{1000\text{ K}}\right)\nonumber\\
			&-1.96\times\log g\nonumber\\
			&+0.0314\times(R_*/R_\odot)\nonumber\\
			&-15.49\nonumber,\\
		\end{align}
		and for metallicity $Z/Z_\odot=0.2$ the linear relationship reads (also with $R^2=0.999$):
		\begin{align}\label{mdotz02eq1}
			\log \dot M_{Z=0.2}=&11.668\times\log\left(\frac{T_\text{eff}}{1000\text{ K}}\right)\nonumber\\
			&-2.126\times\log g\nonumber\\
			&+0.04\times(R_*/R_\odot)\nonumber\\
			&-17.63\nonumber,\\
		\end{align}
		where $\dot M$ are given in $10^{-6}M_\odot\,\text{yr}^{-1}$.
		
		These relationships could be considered analogous to that given by \citet{vink00} to obtain theoretical mass-loss rates for solar-like metallicities. However, the advantage of our description is that it depends only on \textit{stellar parameters} and we do not need to consider the value of $v_{\infty}/v_{esc}$. 
		It is important to remark however, that this formula has been derived for the following ranges:
		\begin{itemize}
			\item $T_\text{eff}=32-45$ kK
			\item $\log g=3.4-4.25$
			\item $M_*/M_\odot\ge25.0$
		\end{itemize}
		
		Concerning terminal velocities, see right panel of Fig.~\ref{windbehaviours}, self-consistent calculations (top panel) show that $v_\infty$ is almost constant with respect to the effective temperature, but it decreases as a function of $\log g$ and $Z$.
		On the other hand, Abbott's procedure results do not show the same behavior and exhibit a maximum in the $T_{\rm{eff}}$ interval.

%Analysis of errors___________________________________________________________________________________
	\subsection{Range of validity for line-force parameters}\label{constantakd}
		\begin{table}[htbp]
			\centering
			\scalebox{0.9}{
			\begin{tabular}{ccrc|ccc|rr}
				\hline\hline
				$T_\text{eff}$ & $\log g$ & $\log t_\text{in}$ & $\log t_\text{out}$ & $k$ & $\alpha$ & $\delta$ & $|\Delta v_\infty|$ & $|\Delta\dot M|$\\
				& & & & & & & [km s$^{-1}$] & [$10^{-6}M_\odot\,\text{yr}^{-1}$]\\
				\hline
				45\,000 & 4.0 & $-0.31$ & $-2.03$ & 0.099 & 0.686 & 0.037 & $780$ & $0.23$\\
				& & $-0.31$ & $-2.87$ & 0.107 & 0.650 & 0.029 & $600$ & $0.30$\\
				& & $-0.31$ & $-3.71$ & 0.120 & 0.638 & 0.027 & $420$ & $0.21$\\
				& & $-0.31$ & $-4.55$ & 0.167 & 0.600 & 0.021 & $0$ & $0$\\
				\hline
				40\,000 & 4.0 & $-0.87$ & $-2.50$ & 0.099 & 0.633 & 0.040 & $521$ & $0.09$\\
				& & $-0.87$ & $-3.32$ & 0.099 & 0.634 & 0.036 & $610$ & $0.07$\\
				& & $-0.87$ & $-4.14$ & 0.107 & 0.621 & 0.026 & $594$ & $0.07$\\
				& & $-0.87$ & $-4.96$ & 0.164 & 0.581 & 0.027 & $0$ & $0$\\
				\hline
				40\,000 & 3.6 & 0.08 & $-1.44$ & 0.095 & 0.666 & 0.090 & $247$ & $0.58$\\
				& & 0.08 & $-2.28$ & 0.098 & 0.680 & 0.075 & $75$ & $0.13$\\
				& & 0.08 & $-3.12$ & 0.101 & 0.692 & 0.067 & $323$ & $0.92$\\
				& & 0.08 & $-3.96$ & 0.118 & 0.659 & 0.044 & $0$ & $0$\\
				\hline
				36\,000 & 3.6 & $-0.29$ & $-2.00$ & 0.084 & 0.637 & 0.112 & $520$ & $0.58$\\
				& & $-0.29$ & $-2.85$ & 0.092 & 0.648 & 0.078 & $114$ & $0.15$\\
				& & $-0.29$ & $-3.70$ & 0.089 & 0.668 & 0.075 & $267$ & $0.01$\\
				& & $-0.29$ & $-4.55$ & 0.104 & 0.644 & 0.062 & $0$ & $0$\\
				\hline
				32\,000 & 3.4 & $0.37$ & $-1.49$ & 0.066 & 0.630 & 0.251 & $631$ & $0.77$\\
				& & $0.37$ & $-2.43$ & 0.075 & 0.636 & 0.221 & $457$ & $0.57$\\
				& & $0.37$ & $-3.37$ & 0.079 & 0.662 & 0.179 & $168$ & $0.11$\\
				& & $0.37$ & $-4.31$ & 0.078 & 0.675 & 0.159 & $0$ & $0$\\
				\hline
			\end{tabular}}
			\caption{\small{Influence of the optical depth interval on the line-force parameters for some reference models given in Table~\ref{standardtable}. Absolute values of the differences in the resulting wind parameters with respect to the reference solution are presented.}}
			\label{variationofcak}
		\end{table}

		It is important to remember that the range of optical depths used to calculate our self-consistent  line-force parameters is defined along almost all the atmosphere of the star, i.e., downstream from the sonic point.
		This procedure improves  the criterion used by \citet{abbott82}, who arbitrarily defined the parameters at $t=10^{-4}$. 
		This value sometimes lays outside the optical depth range here defined, as it was shown in Fig.~\ref{fastplots}.

		To analyse the change on the line-force parameters due to the selection of the $t$-range, we define four different intervals inside the whole range of $t$, and compute these parameters in each range. Table~\ref{variationofcak} summarises these calculations.
		 Regarding the uncertainties of our procedure in the terminal velocities, these are of the same order as the uncertainties due to the errors in the determination of the stellar parameters in the range $32\,000$ K $ < T_{\rm{eff}} < 40\,000$ K, while, the uncertainties in $\dot{M}$ are much lower than the ones produced by variations of stellar parameters. These small uncertainties indicate that it is a good approximation to consider  line-force parameters as constants throughout the wind.
		Due to the fact that the entire $t$-range represents the physical conditions of almost all the wind, we recommend to use the complete optical depth range to derive the line-force parameters.
		
		For $T_\text{eff} < 30\,000$ K, we found that $\log\,\mathcal M(t)$ is not longer linear with respect to $\log t$ and the corresponding line-force parameters can be approximated to a linear piecewise description.
		Due to this reason, we establish that our set of self-consistent solutions describes stellar winds for effective temperatures and $\log g$ in the range $32\,000-45\,000$ K and $3.4-4.0$ dex, respectively.

% Calculated spectra
	\section{Synthetic spectra}\label{syntheticspectra}	
		In order to know whether our calculations reproduce realistic physical features observed in hot stars, we calculated synthetic spectra for three O-type stars using FASTWIND.
		We selected some stars in the range of the considered $T_{\rm{eff}}$, trying to cover the extreme cases of temperature and $\log g$.
		We chose first the O4 I(n)fp star $\zeta$-Puppis (HD 66811) because it has been extensively studied \citep{puls96,repolust04,puls06,sota11,bouret12,noebauer15}.
		Because \textsc{HydWind} allows the option to include a rotational velocity, self-consistent solutions for $\zeta$-Puppis consider a $v\sin i=210$ km s$^{-1}$ which is a value in agreement with previous authors.
		Mentioned authors have also adopted independently different set of stellar and wind parameters, which are summarised in Table~\ref{zpuppispar}.
		Here, the wind parameters were determined by \citet{repolust04}.
		\citet{puls06} has used Repolust's parameters and derived clumped mass-loss rates from H$\alpha$, IR and radio, using  analytical expressions for the corresponding opacities, whereas \citet{bouret12} used CMFGEN.
		Both calculations include clumping, so these results correspond to a clumped mass-loss rate.\footnote{FASTWIND uses the clumping factor $f_\text{cl}\ge 1$ (with $f_\text{cl} = 1$ representing the smooth limit), where $f_\text{cl} = 1/f$ if the inter-clump medium was void \citep[][]{sundqvist18}. On the other hand, CMFGEN-clumping is represented by the so-called volume filling factor $f$, which scales homogeneous and clumped mass-loss rates under the relationship $\dot M_\text{hom}=\dot M_\text{clump}/\sqrt{f}$ (notice that this $f$ takes values between 0 and 1).}
		On the other hand, the mass-loss rate given by \citet{noebauer15} was obtained using their Monte-Carlo radiation hydrodynamics (MCRH) method assuming a homogeneous media ($f_\text{cl}=1.0$).
			
		Particularly, we compare our results with those given by \citet{puls06}, who did an exhaustive analysis of the clumping throughout the wind.
		Two different values for  mass-loss rate are given by these authors, because they considered different stellar radii depending on the assumed distance for $\zeta$-Puppis: 
		
		i) the "conventional" ($d=460$ pc) and 
		
		ii) the one given by \citet[][$d=730$ pc]{sahu93}.
		
		We examine here the "conventional" case with $R_*/R_\odot=18.6$.
		We can observe from Table~\ref{zpuppispar} (last row), that our new calculated  mass-loss rate agree quite well with the value from \citet{puls06}.
		\begin{figure}[t!]
			\centering
			\includegraphics[width=\linewidth]{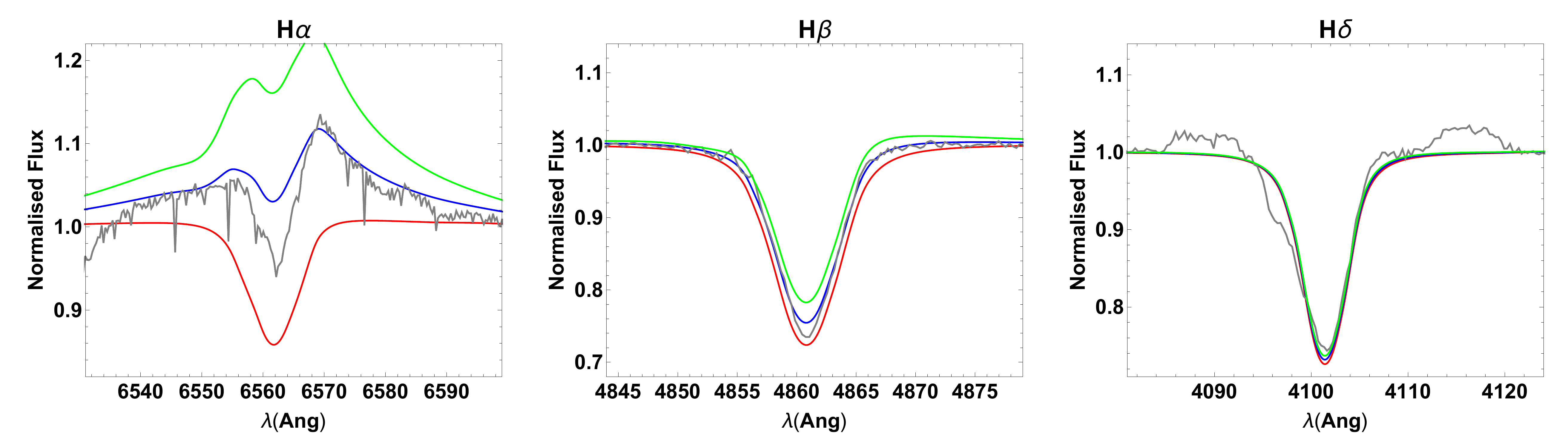}
			\includegraphics[width=\linewidth]{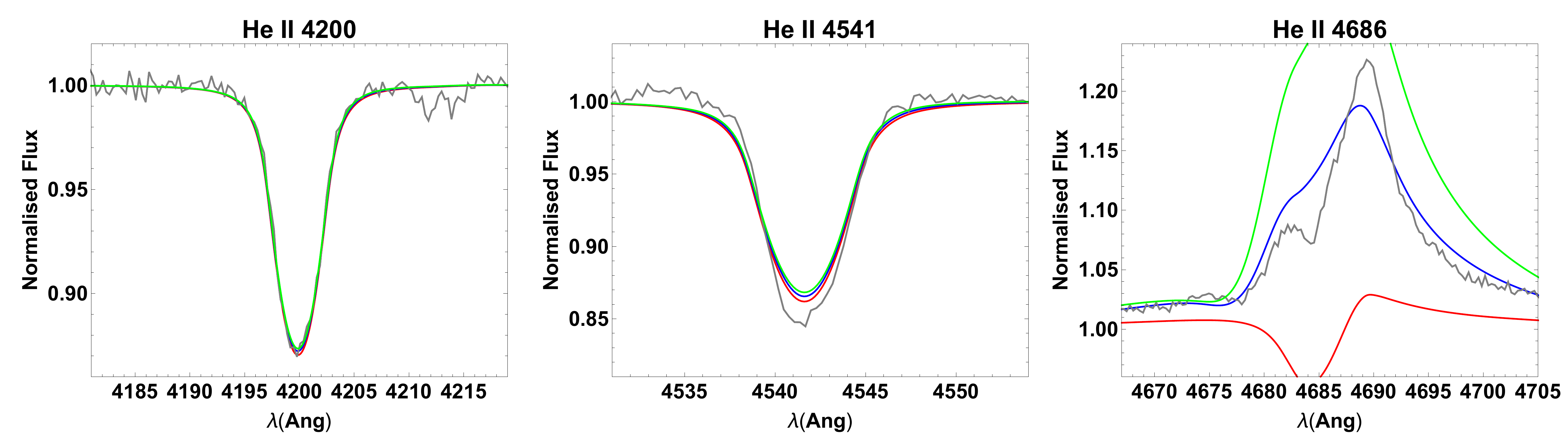}
			\includegraphics[width=\linewidth]{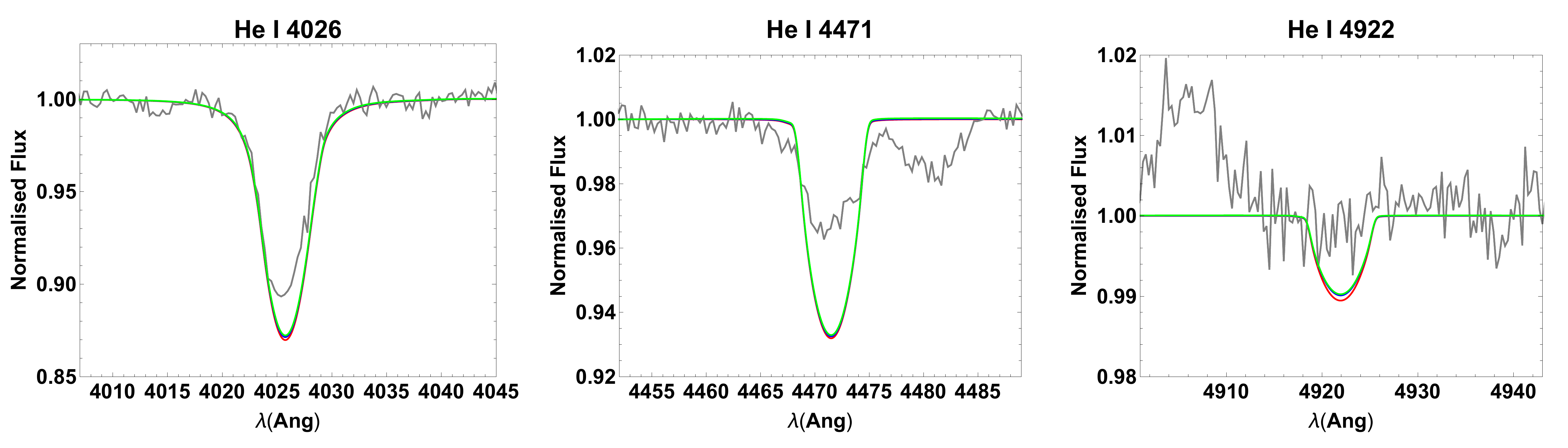}
			\caption{\small{Resulting FASTWIND spectra for $\zeta$-Puppis with $T_\text{eff}=39$ kK, $\log g=3.6$, $R_*/R_\odot=18.6$ and $\dot M=4.6\times10^{-6}$ $M_\odot$~yr$^{-1}$. Clumping factors are $f_\text{cl}=1.0$ (red, homogeneous), $f_\text{cl}=5.0$ (blue) and $f_\text{cl}=9.0$ (green).}}
			\label{comp413540}
		\end{figure}
		\begin{figure}[t!]
			\centering
			\includegraphics[width=\linewidth]{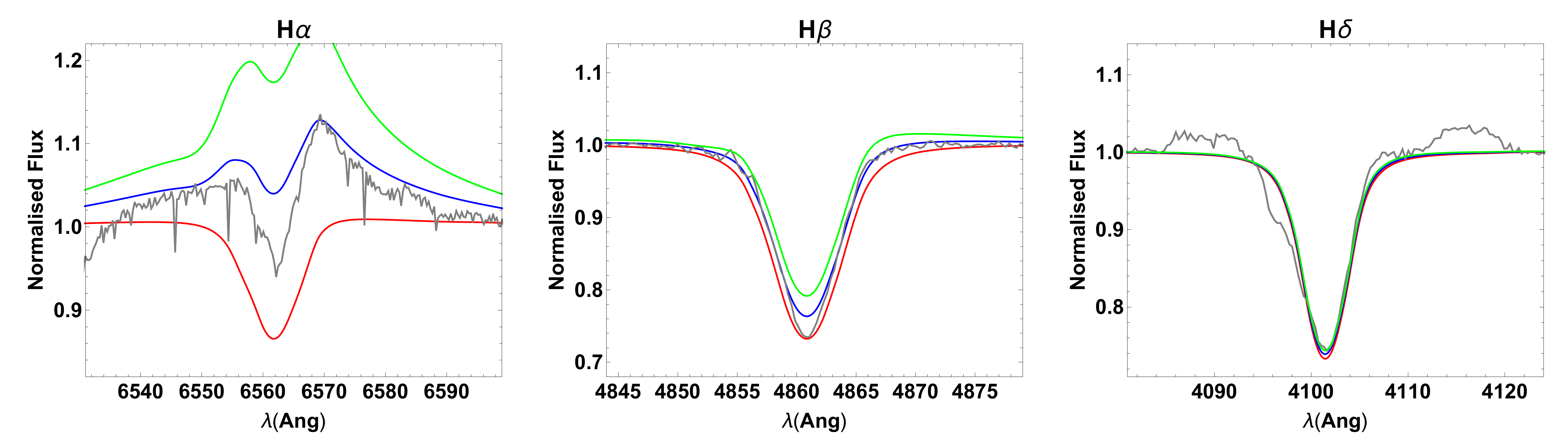}
			\includegraphics[width=\linewidth]{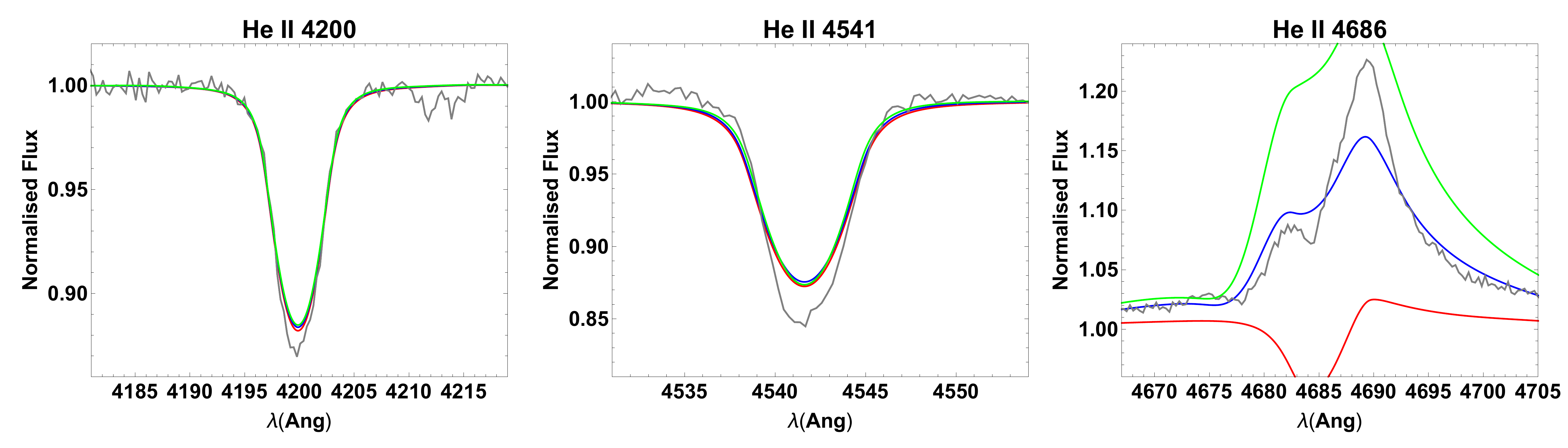}
			\includegraphics[width=\linewidth]{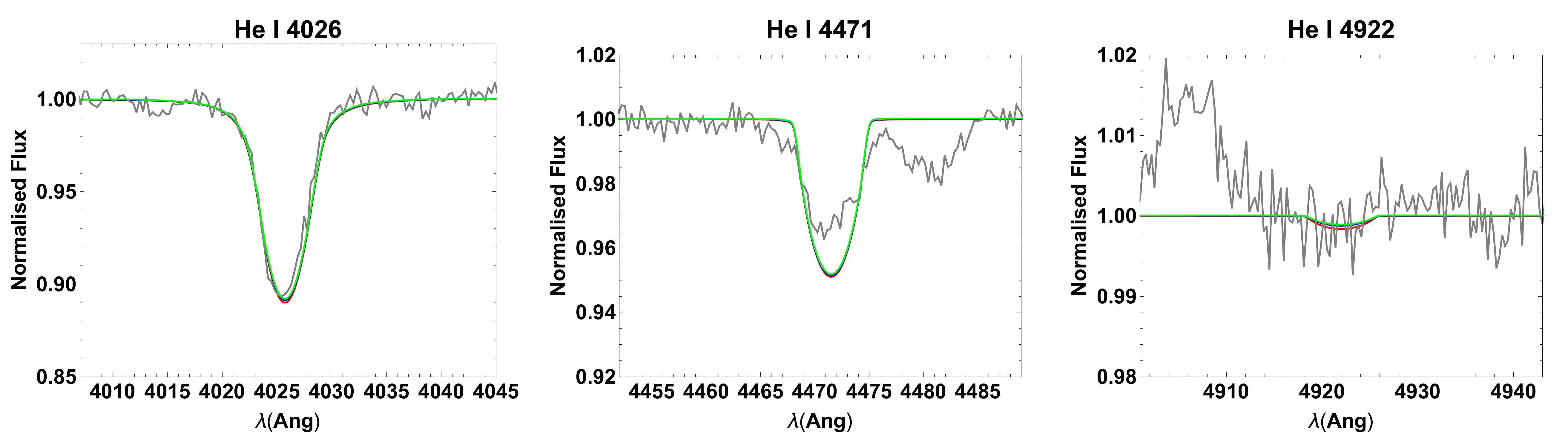}
			\caption{\small{Resulting FASTWIND spectra for $\zeta$-Puppis with $T_\text{eff}=40$ kK, $\log g=3.64$, $R_*/R_\odot=18.6$ and $\dot M=5.2\times10^{-6}$ $M_\odot$~yr$^{-1}$. Clumping factors are $f_\text{cl}=1.0$ (red, homogeneous), $f_\text{cl}=5.0$ (blue) and $f_\text{cl}=9.0$ (green).}}
			\label{comp000809}
		\end{figure}
		\begin{figure}[t!]
			\centering
			\includegraphics[width=\linewidth]{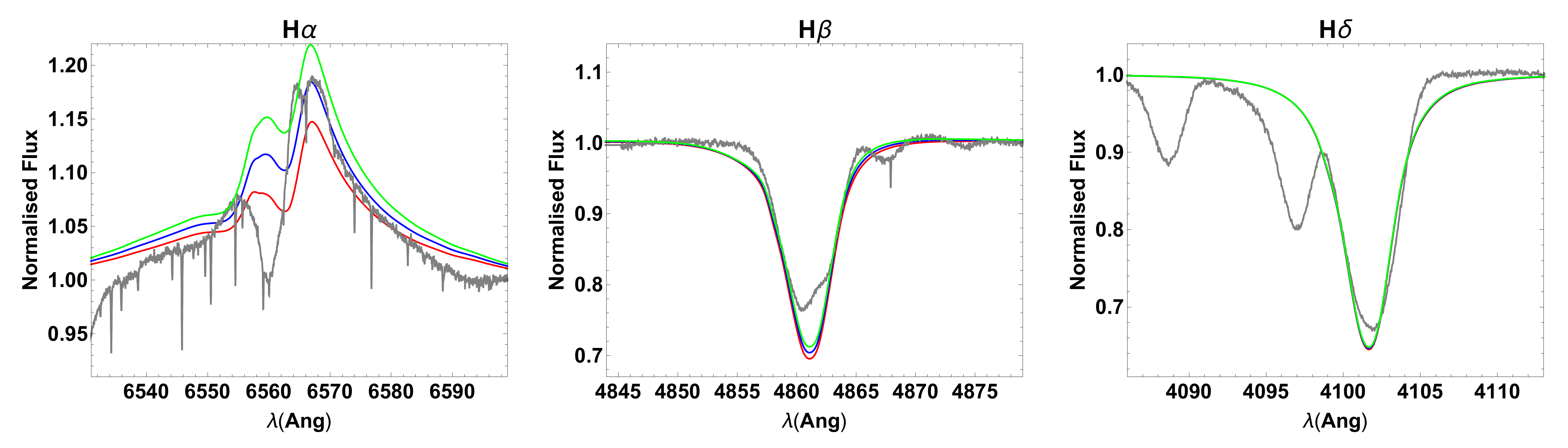}
			\includegraphics[width=\linewidth]{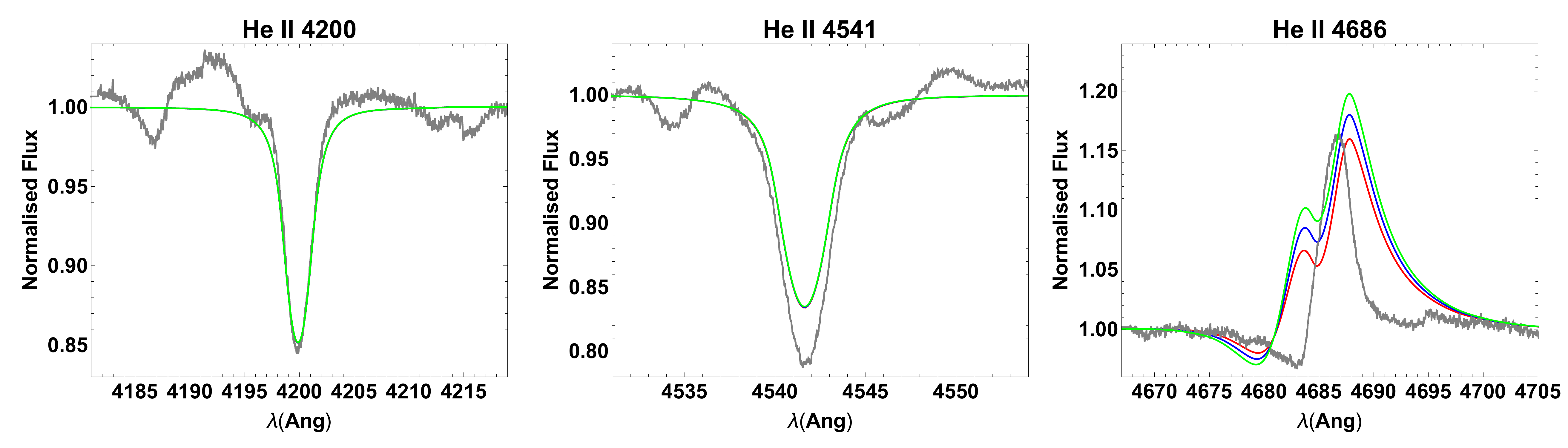}
			\includegraphics[width=\linewidth]{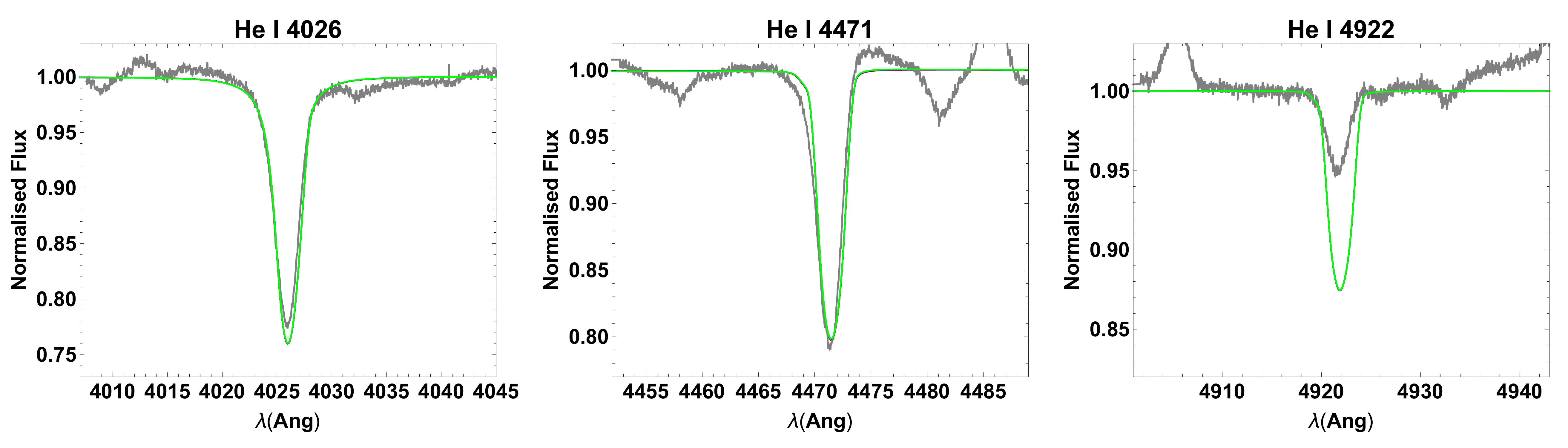}
			\caption{\small{Resulting FASTWIND spectra for HD 163758 with $T_\text{eff}=34.5$ kK, $\log g=3.41$, $R_*/R_\odot=21.0$ \citep[see][]{bouret12} and $\dot M=3.3\times10^{-6}$ $M_\odot$~yr$^{-1}$. Clumping factors are $f_\text{cl}=5.0$ (red), $f_\text{cl}=6.0$ (blue) and $f_\text{cl}=7.0$ (green).}}
			\label{comp537366}
		\end{figure}
		\begin{figure}[t!]
			\centering
			\includegraphics[width=\linewidth]{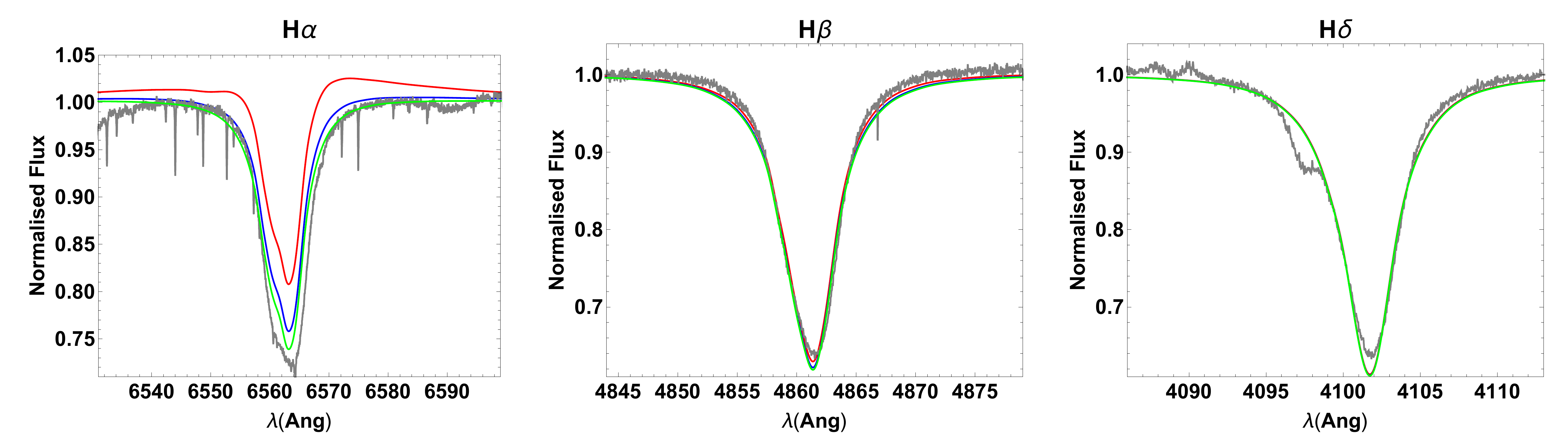}
			\includegraphics[width=\linewidth]{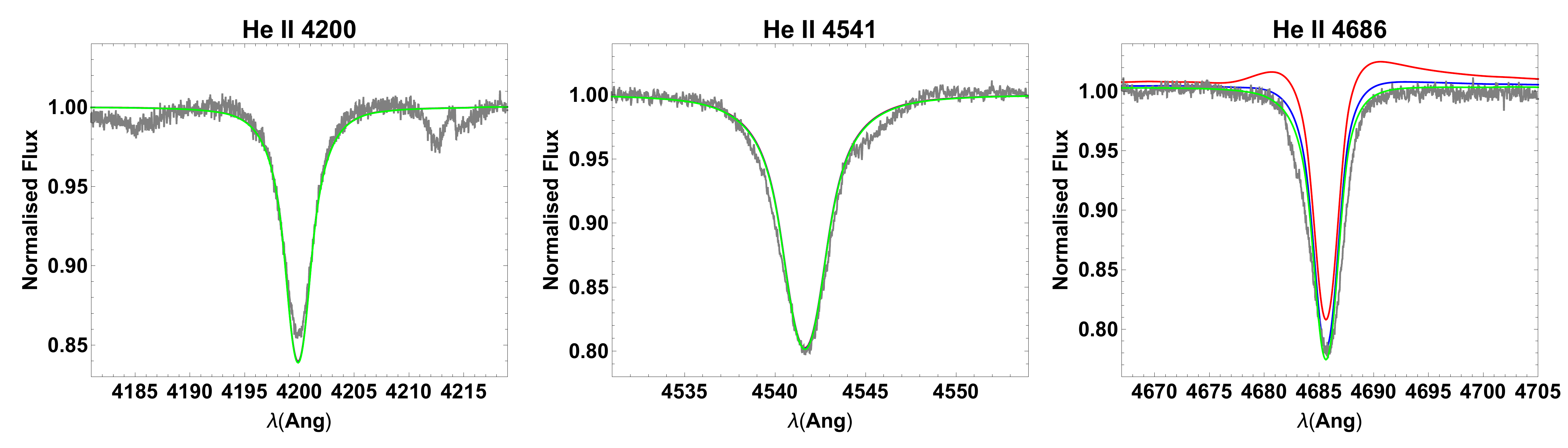}
			\includegraphics[width=\linewidth]{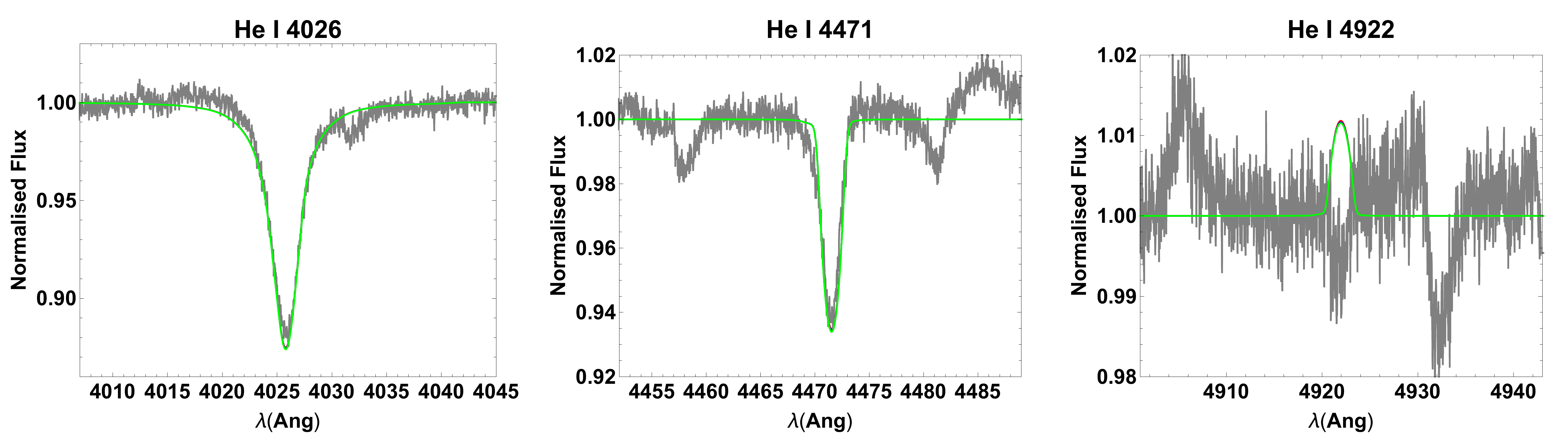}
			\caption{\small{Resulting FASTWIND spectra for HD 164794 with $T_\text{eff}=43.8$ kK, $\log g=3.92$, $R_*/R_\odot=13.1$ \citep[stellar parameters taken from][]{krticka15} $\dot M=2.3\times10^{-6}$ $M_\odot$~yr$^{-1}$. Clumping factors are $f_\text{cl}=5.0$ (red), $f_\text{cl}=2.0$ (blue) and $f_\text{cl}=1.0$ (homogeneous, green).}}
			\label{comp858687}
		\end{figure}

		\begin{table}[htpb]
			\resizebox{\textwidth}{!}{
			\begin{tabular}{lccccc|ccccc}
				\hline
				\hline
				\multicolumn{6}{c}{previous studies} & \multicolumn{5}{c}{present work}\\
				Reference & $T_\text{eff}$ & $\log g$ & $R_*/R_\odot$ & $\dot M$ & $v_\infty$ & $k$ & $\alpha$ & $\delta$ & $\dot M_\text{SC}$ & $v^\text{SC}_{\infty}$\\
				& $[\text{kK}]$ & & & [$10^{-6}M_\odot\,\text{yr}^{-1}$] & [km\,s$^{-1}$] & & & & [$10^{-6}M_\odot\,\text{yr}^{-1}$] & [km\,s$^{-1}$]\\
				\hline
				\small{\citet{noebauer15}} & 42 & 3.6 & 19.0 & $45.0$ & 881 & 0.120 & 0.678 & 0.041 & $11.0\pm_{3.0}^{3.5}$ & $2\,500\pm280$\\
				\small{\citet{bouret12}} & 40 & 3.64 & 18.7 & $2.0$ & $2\,300$ & 0.120 & 0.655 & 0.039 & $5.2\pm_{1.2}^{1.6}$ & $2\,700\pm300$\\
				\multirow{2}{*}{\small{\citet{puls06}}} & 39 & 3.6 & 29.7 & $8.5$ & $2\,250$ & 0.115 & 0.654 & 0.044 & $9.3\pm_{2.2}^{2.9}$ & $3\,200\pm350$\\
				& 39 & 3.6 & 18.6 & $4.2$ & $2\,250$ & 0.114 & 0.658 & 0.049 & $4.6\pm_{1.1}^{1.3}$ & $2\,570\pm300$\\
				\hline
			\end{tabular}}
			\caption{\small{Stellar and wind parameters for $\zeta$-Puppis from previous studies compared with our  self-consistent results. Line-force parameters are also listed.}}
			\label{zpuppispar}
		\end{table}
			
		Figure~\ref{comp413540} shows the observed spectra (kindly provided by D. J. Hillier) and the resulting synthetic spectra for $\zeta$-Puppis.
		Stellar parameters are taken from \citet[][see Table~\ref{zpuppispar}]{puls06} and wind parameters from our self-consistent procedure ($\dot M_\text{SC}=4.6\times10^{-6}$ $M_\odot\,\text{yr}^{-1}$).
		We calculated three synthetic spectra with different clumping factors: $f_\text{cl}=1.0$ (homogeneous), $f_\text{cl}=5.0$ and $f_\text{cl}=9.0$.
		The best fit is for $f_\text{cl}=5.0$, which is the same clumping factor found by \citet{puls06} with their $\dot M=4.2\times10^{-6}$ $M_\odot$~yr$^{-1}$.
		Moreover, we also include the synthetic spectra obtained with the self-consistent solution (see Fig.~\ref{comp000809}), calculated using the stellar parameters given by \citet[][see Table~\ref{zpuppispar}]{bouret12} and \citet{najarro11}.
		The best fit is achieved when we use a clumping factor of $f_\text{cl}=5.0$.
		These results suggest that the real stellar parameters lies in the neighbourhood given by \citet{puls06} and \citet{najarro11}.
		
		The observed spectrum for  HD 163758 (O9 I) has been obtained from UVES-POP database\footnote{\url{http://www.eso.org/sci/observing/tools/uvespop/field_stars_uptonow.html}}.
		We calculated the synthetic spectra for this star (see Fig.~\ref{comp537366}) using stellar parameters from \citet{bouret12} and wind self-consistent parameters (see Table~\ref{hd163758par}) with different clumping factors, the best fit is for $f_\text{cl}=6.0$. 
		\begin{table}[htpb]
			\centering
			\resizebox{\textwidth}{!}{
			\begin{tabular}{lccccc|ccccc}
				\hline
				\hline
				\multicolumn{6}{c}{previous studies} & \multicolumn{5}{c}{present work}\\
				Name & $T_\text{eff}$ & $\log g$ & $R_*/R_\odot$ & $\dot M$ & $v_\infty$ & $k$ & $\alpha$ & $\delta$ & $\dot M_\text{SC}$ & $v^\text{SC}_{\infty}$\\
				& $[\text{kK}]$ & & & [$10^{-6}M_\odot\,\text{yr}^{-1}$] & [km\,s$^{-1}$] & & & & [$10^{-6}M_\odot\,\text{yr}^{-1}$] & [km\,s$^{-1}$]\\
				\hline
				HD 163758 & 34.5 & 3.41 & 21.0 & 1.6 & 2\,100 & 0.087 & 0.679 & 0.112 & $3.3\pm_{0.8}^{1.1}$ & $2\,040\pm280$\\
				HD 164794 & 43.8 & 3.92 & 13.1 & 2.9 & 3\,090 & 0.141 & 0.614 & 0.020 & $2.3\pm_{0.5}^{0.6}$ & $3\,304\pm400$\\
				\hline
			\end{tabular}}
			\caption{\small{Idem Table~\ref{zpuppispar}, but for HD 163758 and HD 164794. Stellar and wind parameters are  from \citet{bouret12} and \citet{krticka15} respectively.}}
			\label{hd163758par}
		\end{table}
		
		Last spectrum corresponds to the O3.5 V star HD 164794, also obtained from UVES-POP database.
		Stellar parameters were extracted from \citet{krticka15}, as shown in Table~\ref{hd163758par}.
		Contrary to previous cases, the best fit is obtained for the homogeneous model ($f_\text{cl}=1.0$, see Fig.~\ref{comp858687}).

		In view of these first results, our self-consistent iterative procedure takes us quickly into the neighborhood of the solution that reproduces the observed wind spectra for O-type stars.

%_____DISCUSSION__________________________________________________________________________________
\section{Discussion}\label{discussion}
	We have developed a self-consistent methodology to calculate the line-force parameters and derived consequently mass-loss rates and velocity profiles.
	We found that mass-loss rate is about $30\%$ larger than the one obtained using Abbott's procedure (non self-consistent calculation).

%Terminal velocity________________________________________________________________________________________
	\subsection{Terminal velocity}
		It is well known the scaling relation for the terminal velocity in the frame of the line-driven wind theory. This  relation \citep{puls08} reads:
		\begin{equation}
			\label{cinfinit}
			v_{\infty} \approx 2.25 \, \sqrt{\frac{\alpha}{1-\alpha}} \, v_{\rm{esc}}\,.
		\end{equation}
		This is an approximation of the formula found by \citet[][their Eqs. 62 to 65]{kppa89}.

\begin{figure}[htbp]
			\centering
			\includegraphics[width=0.6\linewidth]{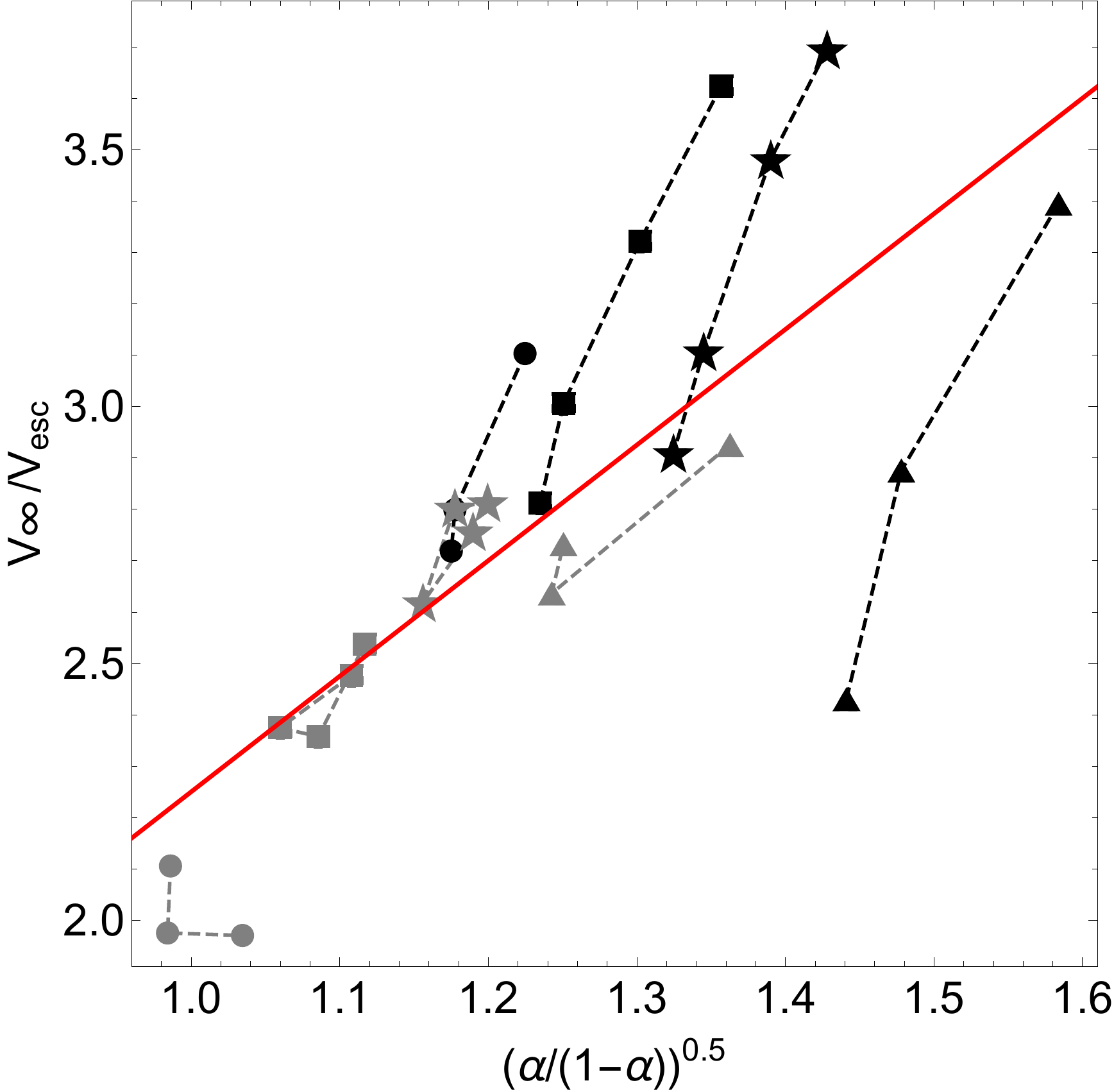}
			\caption{\small{$v_{\infty}/v_{\rm{esc}}$ versus $\sqrt{\alpha/(1-\alpha)}$. For each set of $\log\,g$ values there is a linear dependence for $Z_\odot$. Slope 2.25 of Eq.~\ref{cinfinit} is also displayed. For sub-solar abundance there is a unique linear relationship (see text for details). Symbol description is the same as in Fig.~\ref{behaviours}}}
			\label{vinfvesc}
		\end{figure}

		In Fig.~\ref{vinfvesc} we have plotted $v_{\infty}/v_{\rm{esc}}$ versus $\sqrt{\alpha/(1-\alpha)}$ using the results from Table~\ref{standardtable}.
		Contrary to the expected result (Eq.~\ref{cinfinit}) for solar abundances, we find a different linear behaviour, which strongly depends on the value of $\log\,g$. This is a new result that comes from applying our self-consistent procedure.
		The m-CAK equation of momentum shows an interplay between the gravity ($\log\,g$) and the line-force term.
		This balance of forces defines the location of the singular point and therefore fixes the value of $\dot{M}$.
		As a consequence, the velocity profile depends also on the value of $\log\,g$.
		This result cannot be obtained from Eq.~\ref{cinfinit} which is an oversimplification of this non-linear coupling.
		However, Eq.~\ref{cinfinit} presents a fair fit when $Z$=$Z_{\odot}/5$, where the dependence of the slope on $\log g$ is weak since the radiation force is driven by few ions.

		The dependence of $v_{\infty}/v_{\rm{esc}}$ on $\log\,g$ yield that stars with  solar abundances present an intrinsic variations of $v_{\infty}/v_{\rm{esc}}$  in the range $2.4 - 3.7$, as shown in Fig.~\ref{vinfvesc}.
		This range might explain the scatter observed in the hot side of the bi-stability jump shown by \citet[][in their Fig. 12]{markova08}.

%Mass-loss rate______________________________________________________________________________________
	\subsection{Mass-loss rate}
		In this section we want to compare our theoretical results with the ones obtained from line-profile fittings for homogeneous (unclumped) winds with a $\beta$--law, and the mass-loss (recipe) from \citet{vink00}.
		\begin{table}[htpb]
			\centering
			\resizebox{\textwidth}{!}{
			\begin{tabular}{lccc|ccc|cc}
				\hline
				\hline
				Model&$T_\text{eff}$ & $\log g$ & $R_*/R_\odot$ & $k$ & $\alpha$ & $\delta$ & $v_\infty$ & $\dot M$\\
				& [kK] & & & & & & [km s$^{-1}$] & [$10^{-6}M_\odot\,\text{yr}^{-1}$]\\
				\hline
				Self-Consistent& 43.5 & 4.0 & 11.9 & 0.159 & 0.603  & 0.032 & $3\,342\pm240$ & $1.55\pm_{0.3}^{0.45}$ \\
				$\beta=1.0$ & 43.5 & 4.0 & 11.9  & 0.118 & 0.647 & 0.021 & $4\,187\pm290$ & $1.45\pm_{0.25}^{0.35}$ \\
				\hline
				Self-Consistent & 39 & 3.6 & 19.45 & 0.116 & 0.657 & 0.079 & $2\,412\pm210$ & $5.8\pm_{1.3}^{2.0}$ \\
				$\beta=0.8$ & 39 & 3.6 & 19.45  & 0.039 & 0.815 & 0.062 & $6\,789\pm570$ & $4.2\pm_{0.7}^{0.9}$ \\
				\hline
			\end{tabular}}
			\caption{\small{Comparison of self-consistent with $\beta$--law (single-step) models for the two stars analyzed by \citet{bouret05}. Self-consistent models reproduce better the line-fitted wind parameters obtained by these authors ($\beta$=$1$: $v_{\infty}=3000\,$ km s$^{-1}$, $\dot{M}=1.8\times10^{-6}\,$ $M_\odot\,\text{yr}^{-1}$, and $\beta$=$0.8$: $v_{\infty}=2300\,$ km s$^{-1}$, $\dot{M}=6\times10^{-6}\,$ $M_\odot\,\text{yr}^{-1}$).}}
			\label{tablebouret05atlas}
		\end{table}

		Table~\ref{tablebouret05atlas} shows our results for two O-type star reported by \citet{bouret05}: HD 96715, $T_\text{eff}=43.5$ kK, $\log g=4.0$, and  HD 1904290A, $T_\text{eff}=39$ kK, $\log g=3.6$.
		These results were obtained for the self-consistent solution together with the ones after just one iteration starting from a $\beta$--law.
		It is observed that models starting from a $\beta$--law largely overestimate the terminal velocity and slightly underestimate the mass-loss rate.
		Self-consistent calculations find a fairly good agreement to both: the observed mass-loss rate and terminal velocity.
		For the mass-loss rate in this Figure, we have included the result calculated using \citet{vink00} recipe.
		It is clear that our self-consistent method gives values of $\dot{M}$ much closer to the observed ones.
		\begin{table}[htpb]
			\centering
			\newcolumntype{d}{D{,}{,}{-1} }
			\resizebox{\textwidth}{!}{
			\begin{tabular}{lllc|ccc|cc|cc}
				\hline
				\hline
				Field star & $T_\text{eff}$ & $\log g$ & $R_*/R_\odot$ & $k$ & $\alpha$ & $\delta$ & $v^{SC}_\infty$ & \multicolumn{1}{c}{$\dot M$} & $\frac{\dot M_\text{SC}}{\dot M_\text{obs}}$ & $\frac{\dot M_\text{SC}}{\dot M_\text{Vink}}$\\
				& [kK] & & & & & & [km s$^{-1}$] & \multicolumn{1}{c}{$[10^{-6}M_\odot\,\text{yr}^{-1}]$} & &\\
				\hline
				HD 169582 & 37 & 3.5 & 27.2 & 0.102 & 0.668 & 0.063 & $3\,017\pm700$ & $7.1\pm_{2.4}^{3.6}$ & 1.10 & 1.26\\
				CD-43 4690 & 37 & 3.61 & 14.1 & 0.105 & 0.653 & 0.058 & $2\,310\pm540$ & $1.5\pm_{0.55}^{0.9}$ & 1.22 & 1.16\\
				HD 97848 & 36.5 & 3.9 & 8.2 & 0.123 & 0.601 & 0.034 & $2\,532\pm470$ & $0.17\pm_{0.06}^{0.09}$ & 0.89 & 0.95\\
				HD 69464 & 36 & 3.51 & 20.0 & 0.099 & 0.664 & 0.076 & $2\,412\pm580$ & $3.2\pm_{1.2}^{1.9}$ & 1.14 & 1.30\\
				HD 302505 & 34 & 3.6 & 14.1 & 0.092 & 0.643 & 0.077 & $2\,331\pm460$ & $0.68\pm_{0.26}^{0.42}$ & 1.24 & 0.98\\
				\hline
				HD 148546 & 31 & 3.22 & 24.4 & 0.073 & 0.718 & 0.243 & $1\,300\pm350$ & $5.3\pm_{2.5}^{4.7}$ & 0.94 & 2.24\\
				HD 76968a & 31 & 3.25 & 21.3 & 0.071 & 0.711 & 0.248 & $1\,212\pm300$ & $3.5\pm_{1.7}^{3.3}$ & 1.43 & 2.11\\
				HD 69106 & 30 & 3.55 & 14.2 & 0.068 & 0.644 & 0.149 & $1\,455\pm300$ & $0.21\pm_{0.09}^{0.16}$ & 1.48 & 1.78\\
				\hline
			\end{tabular}}
			\caption{\small{Resulting self-consistent wind parameters ($v^\text{SC}_\infty$ and $\dot M_\text{SC}$) calculated for stars analyzed by \citet{markova18}. Error margins presented here for wind parameters are undergone from uncertainties of $\pm1\,000$ for $T_\text{eff}$ and $\pm0.1$ for $\log g$. Last two columns show the ratio between self-consistent and observed mass-loss rates and the ratio between self-consistent and  Vink's  mass-loss rates.}}
			\label{tablemarkova18}
		\end{table}
		\begin{figure}[htbp]
			\centering
			\includegraphics[width=0.6\linewidth]{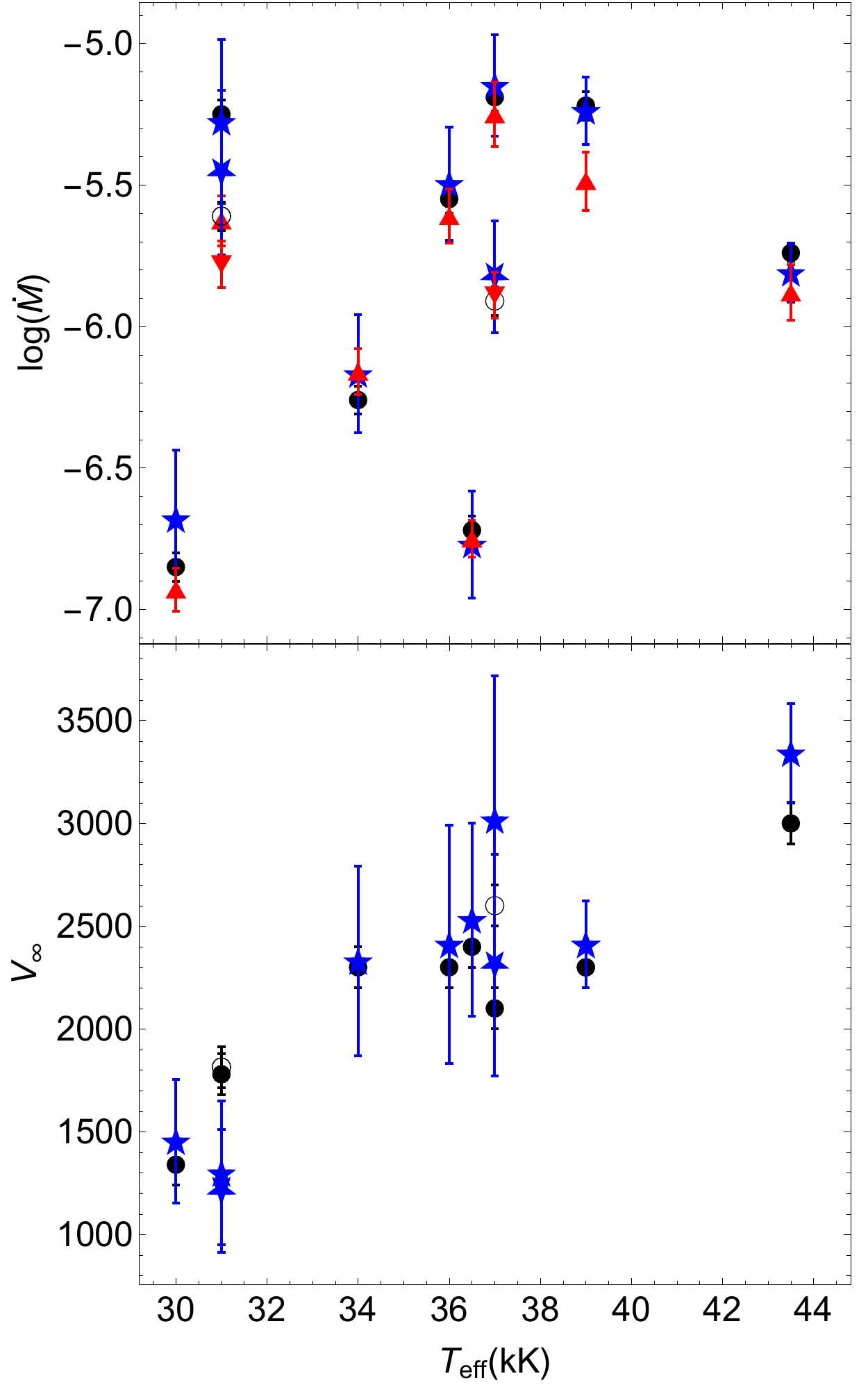}
			\caption{\small{Comparison of mass-loss rates (upper panel) and terminal velocities (lower panel) as a function of the effective temperature. Blue stars correspond to results from this work, black disks to \citet{bouret05} and \citet{markova18} results, and red triangles to theoretical values from \citet{vink00}. The same colour code but with modified symbols (inverted blue stars, unfilled black circles and inverted red triangles) are used to represent Markova's stars with the same effective temperature but higher surface gravity.}}
			\label{masslossbouret05}
		\end{figure}
			
		We also apply our self-consistent procedure to objects analysed by means of FASTWIND adopting unclumped winds.
		For that purpose, we also examine some field Galactic O-type stars from \citet{markova18}.
		Table~\ref{tablemarkova18} summarises our results, where we found a fair agreement between observed and calculated mass-loss rates (see Fig.~\ref{masslossbouret05}).
		These results confirm that our methodology delivers the proper mass-loss rate for the ranges in  $T_\text{eff}$ and $\log g$ given above.
		Below these thresholds, mass-loss rates present larger values compared with both: observational and  Vink's theoretical values.
		This is probably due to the fact that the line-force multiplier is not longer a linear function of $t$ (in the $\log$-$\log$ plane, see Fig.~\ref{fastplots}), and the line-force parameters are not constant throughout the wind.
						
		However, it is important to remark that uncertainties of $\Delta T_{\rm{eff}} \sim \pm1\,000$ K and $\Delta \log g \sim \pm0.1$ dex, produce uncertainties in the  mass-loss rates up to a factor of 2 (see blue error bar in top panel of Fig.~\ref{masslossbouret05}), which can be considered as the upper threshold for the mass-loss rate.
		Hence, even though our self-consistent hydrodynamics gives confident values for $\dot M$, these good results are strongly dependent on the assumed stellar parameters.

%_____CONCLUSIONS________________________________________________________________________________
\section{Conclusions for self-consistent m-CAK solutions}\label{mcakconclusions}
	In the present Chapter we have presented a method to calculate a self-consistent line-force parameters coupled with the hydrodynamics in the frame of the radiation driven wind theory.
	Thanks to this procedure, we achieve a unique well-converged solution that does not depend on the chosen initial values.
	This is important because it reduces the number of free parameters (now $\beta$, $v_\infty$ and $\dot{M}$ are no more input parameters) to be determined by fitting synthetic spectra against observed ones.

	Our calculations contemplate the contribution to the line-force multiplier from more than $\sim 900\,000$ atomic transitions, a NLTE radiation flux from the photosphere and a quasi-LTE approximation for the occupational numbers.
	We have to notice that for $T_{\rm{eff}} > 30\,000$ K the line force parameters can be confidently used as constants throughout the wind. 

	The set of solutions given in Table~\ref{standardtable} differs from previous line-force parameter calculations performed by \citet{abbott82} and \citet{noebauer15}.
	With these new values, we found a different scale relation for the terminal velocity that is steeper than the usually accepted one.
	This new relation might explain the observed scatter found in the terminal velocity from massive stars located at the hot side of the bi-stability jump \citep{markova08}.

	Concerning the wind parameters derived from modelling O-type stars with homogeneous winds, our mass-loss rates are in better agreement with the predicted ones given by \citet{vink00} formula. 
	
	For the calculation of synthetic spectra for O-type stars ($\zeta$-Puppis, HD 163758 and HD 164794), we conclude that our procedure's values for mass-loss rate and hydrodynamics reproduce the observed line-profiles when an adequate value for the clumping factor is chosen.
	
	Even knowing the limitations of the m-CAK theory, this remains an extremely useful framework to get an approach about the real parameters of stellar winds on massive stars.
	In spite of the approximations assumed under this theory, we obtain  reliable values for mass-loss rates and self-consistent hydrodynamics in a short period of time with a great CPU time saving (compare with big efforts made by, e.g., \citealt{mokiem05} or \citealt{fierro18}).

	Our new self-consistent procedure can be used to derive accurate mass-loss rates and: 
	\begin{enumerate}[label=\roman*.]
		\item Build evolutionary tracks, where a high precision on terminal velocities is not required.
		\item Derive truly clumping factors via line-profile fittings.
	\end{enumerate}

%_____SELF-CONSISTENT SOLUTIONS UNDER W-LAMBERT PROCEDURE_________________________________
\chapter[Solutions Under Lambert Procedure]{Self-consistent Solutions Under $W$-Lambert Procedure}\label{lambert}
	During the previous two Chapters, we have focused our analysis on self-consistent solutions for line-driving winds based on m-CAK theory, which has demonstrated to be a fast and confident enough prescription to give theoretical values for wind parameters, specially for mass-loss rates.
	However, self-consistent solutions calculated by \textsc{Alfakdelta27} are based on two main assumptions: 
	
	i) Sobolev approximation which is a fundamental part of m-CAK theory and
	
	ii) more important, the quasi-NLTE treatment given by atomic populations.
	
	Despite the fact that quasi-NLTE scenario provided us reliable results, a more complete analysis is required for a complete NLTE treatment in stellar winds to make a parallel between both approaches.
	
	The goal of this chapter will be finding a self-consistent solution for the stellar winds using the radiative transfer CMFGEN \citep{hillier90a,hillier90b,hillier98} as a tool.
	This code, together with providing us with a synthetic spectrum for a specific set of stellar and wind parameters taking into account all the statistical equilibrium relationships between all the atomic populations existing in the atmosphere, it also provides us with an output for radiative acceleration $g_\text{rad}$ calculated from the solution of the radiative transfer equation.
	Since this output can be expressed as a function of radius only $g_\text{rad}(r)$, it can be adopted later to solve Eq.~\ref{eqmomentum} (equation of momentum) analytically using the so-called $W$-Lambert equation.
	By means of introducing this new velocity field calculated from equation of momentum to re-execute CMFGEN, it is possible to perform again an iterative procedure capable of reaching a new self-consistent solution, this time with the full NLTE line-acceleration given by CMFGEN and the analytical solution for the equation of momentum.
	The price to pay in this case is a bigger computational effort and hence more time, so the analysis made during this Chapter is limited only to the well-known star $\zeta$-Puppis.
	Another price to pay is the fact that, as we mentioned in Section~\ref{generalhydrodynamics}, since Eq.~\ref{eqmotion} does not have an implicit dependence on density (and then mass-loss rate) and line-acceleration is obtained directly from the output of CMFGEN, mass-loss rate is an extra free parameter to be set, together with effective temperature, surface gravity and abundances.
	Nevertheless, results presented here could be the basis for extended studies on other massive stars.
	
	The discussion of the results and analysis done in this Chapter will be compared with the study implemented by \citet{sander17}, who performed a similar procedure for a full NLTE self-consistent solution using the radiative transfer code \textsc{PoWR} \citep{grafener02,hamann03}.
	And later, the comparison will focus between this full NLTE treatment with the quasi-NLTE performed for \citet{alex19}.
	
%_____Lambert Equation_______________________________________________________________________________
\section{$W$-Lambert equation}
	As we previously mentioned, equation of momentum (Eq.~\ref{eqmomentum}) can be solved analytically with the help of a mathematical tool called the \textit{Lambert $W$-function} (also named \textit{product logarithm}), which is defined by the inverse of the function:
	\begin{equation}\label{invertlambert}
		z(w)=w e^w\;\;,
	\end{equation}
	with $z$ being any complex number.
	That means:
	\begin{equation}
		W(z)e^{W(z)}=z\;\;,
	\end{equation}
	\citep{corless93,corless96}.
	
\begin{figure}[htbp]
		\centering
		\includegraphics[width=0.65\linewidth]{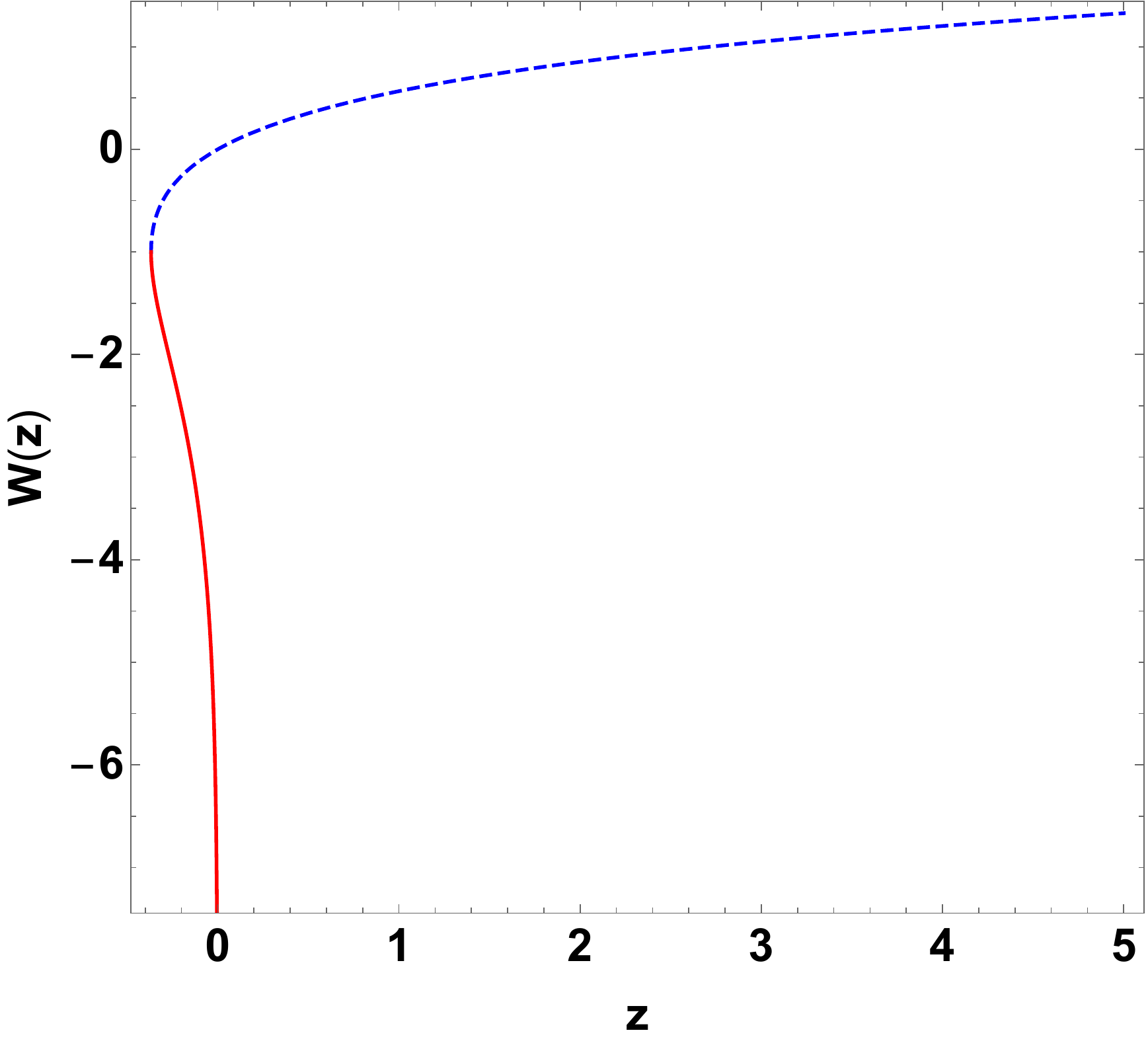}\\
		\caption{\small{Real branches of $W_k(z)$-Lambert function: $k=0$ (dashed blue) and $k=-1$ (solid red).}}
		\label{lambertfunc}
	\end{figure}

	Lambert $W$-function presents infinite solutions depending of all the possible non-zero values that $z$ may take.
	If we limit our search only for real values of $z$, we find that $W(z)$ function is multivalued because $f(w)=we^w$ is not injective.
	Because of this reason, we split Lambert function into two sections, corresponding them to the branches $W_0$ for $W(z)\ge-1$ and $W_{-1}$ for $W(z)\le-1$ (see Figure~\ref{lambertfunc}), which are coupled in the lowest value for $z$:
	\begin{equation}
		W_0\left(-\frac{1}{e}\right)=W_{-1}\left(-\frac{1}{e}\right)=-1\;.
	\end{equation}
	
	From Eq.~\ref{invertlambert} and Figure~\ref{lambertfunc}, it is clearly seen that the domains for $z$ and the codomains for the two branches are\footnote{Notation for open and closed intervals following definition given by the Encyclopaedia of Mathematics:\\
\url{https://www.encyclopediaofmath.org/index.php/Interval and segment}}:
	\begin{equation}
		z\in\left[-\frac{1}{e},0\right[\;\text{ for }W_{-1}\in[-1,-\infty[\;,
	\end{equation}
	\begin{equation}
		z\in\left[-\frac{1}{e},+\infty\right[\;\text{ for }W_0\in[-1,+\infty[\;.
	\end{equation}
	
	This means, branch $W_{-1}$ diverges when $z$ approaches zero or, in other words, $z$ is an asymptote when $W_{-1}$ tends to $-\infty$.
	Because of this condition, $W_{-1}$ function can be used to solve any function with the same asymptotic behaviour as $z$, such as the velocity field of a stellar wind.
	Then, the analytical solution for the equation of momentum with $g_\text{line}(r)$ given is obtained once we reformulate Eq.~\ref{eqmomentum} in terms of the Lambert function, as it is demonstrated in Section~\ref{solutionmomentum}.
	
%_____METHODOLOGY_______________________________________________________________________________
\section{Lambert-procedure}\label{lambertprocedure}
	To evaluate whether Lambert-procedure can be successfully applied into hydrodynamic models, we take as starting point a CMFGEN model for $\zeta$-Puppis with the parameters shown in Table~\ref{initialmodel}.
	These both (stellar and wind) parameters are chosen because they were used to fit $\zeta$-Puppis \citep[][Section 6.5]{bouret12} together with being also determined by \citet{marcolino17} on their spectral analysis in infrared.
	\begin{table}[htbp]
		\centering
		\begin{tabular}{cc}
			\hline\hline
			$T_\text{eff}$ & 41\,000 K\\
			$\log g$ & 3.6\\
			$R_*$ & 17.9 $R_\odot$\\
			$\dot M$ & $2.7\times10^{-6}\,M_\odot$ yr$^{-1}$\\
			$v_\infty$ & 2\,300 km s$^{-1}$\\
			$\beta$ & 0.9\\
			$f_\infty$ & 0.1\\
			\hline
		\end{tabular}
		\caption{\small{Stellar and wind parameters of our initial CMFGEN model. $R_*$ is considered for optical depth $\tau=2/3$, whereas parameter $f_\infty$ corresponds to the infinite filling factor (see Section~\ref{masslossandclumping} for details).}}
		\label{initialmodel} % is used to refer this table in the text
	\end{table}
	\begin{figure}[t!]
		\centering
		\includegraphics[width=0.62\linewidth]{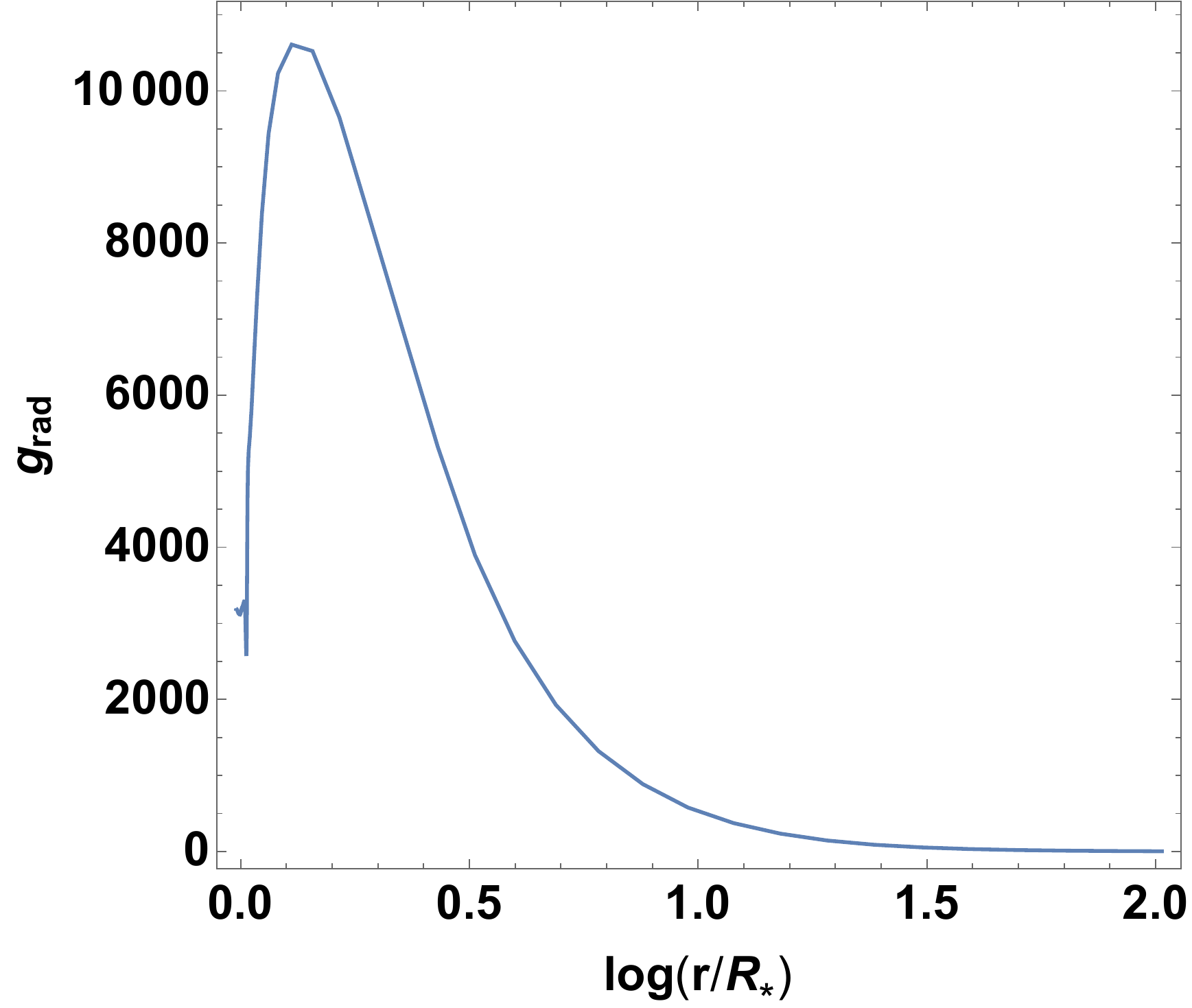}\\
		\caption{\small{Radiative acceleration (arbitrary units) as a function on radius for the initial CMFGEN model with $\beta$--law.}}
		\label{initialgline}
	\end{figure}

	In the previously mentioned Section 6.5 of \citet{bouret12}, they also check the consistency of the line-force.
	Radiative acceleration for this initial model is plotted in Fig.~\ref{initialgline}.
	This $g_\text{rad}(r)$ was internally calculated by CMFGEN starting from the initial parameters shown in Table~\ref{initialmodel}.
	However, the values tabulated there do not lead to recover the same wind parameters; hence, it is not a self-consistent result.
	A truly self-consistent solution must satisfy the equation of momentum: 
	\begin{equation}\label{findgline}
		g_\text{rad}=v\frac{dv}{dr}+\frac{1}{\rho}\frac{dp}{dr}+\frac{GM_*}{r^2}\;.
	\end{equation}
	
	Mass-loss rate $M_\infty$ is linked with density and velocity profiles by means of the equation of continuity (Eq.~\ref{eqcontinuity})
	
\begin{figure}[t!]
		\centering
		\includegraphics[width=0.62\linewidth]{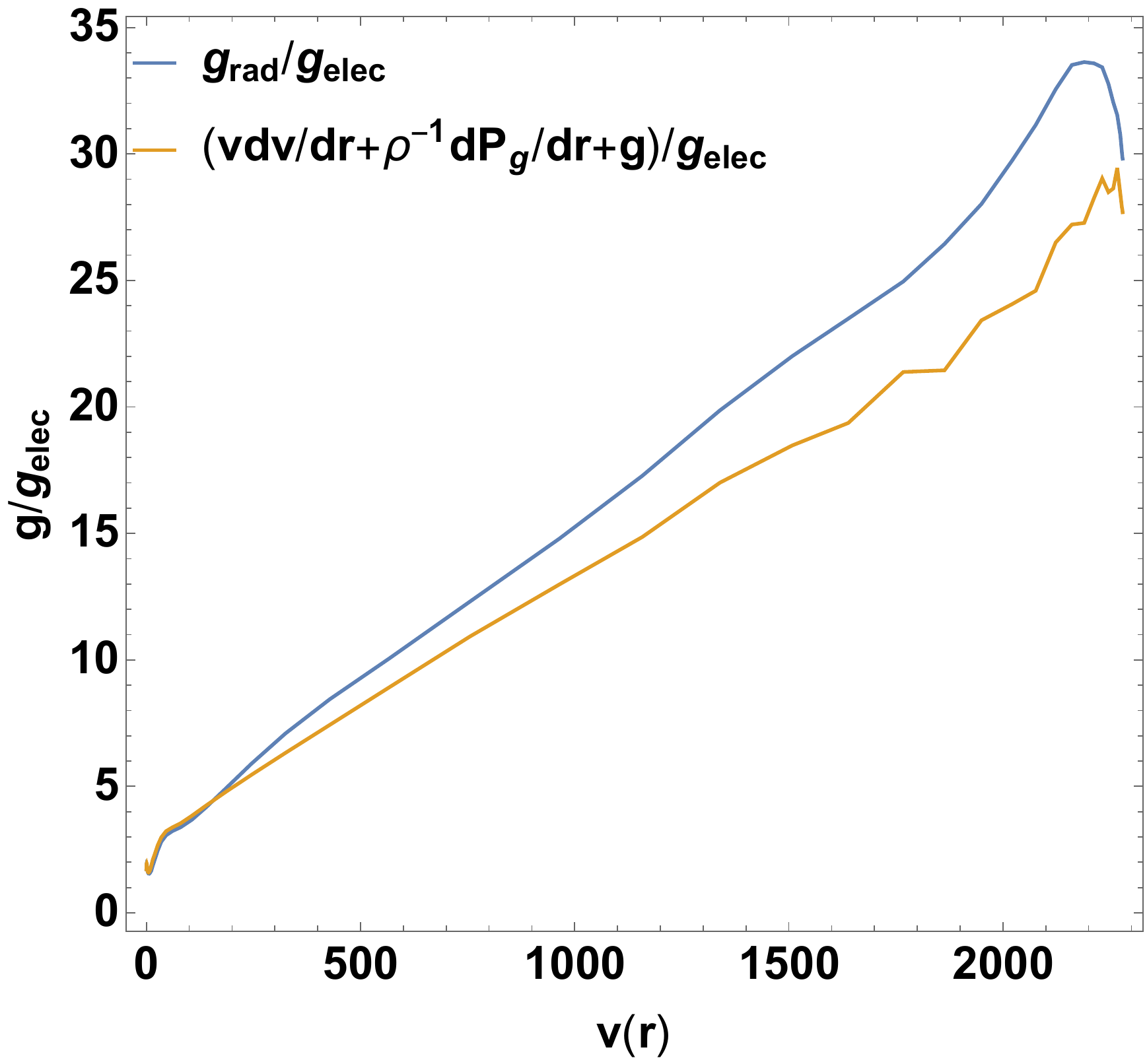}\\
		\caption{\small{Comparison of left-hand side of Eq.~\ref{findgline} (blue) and right-hand side (orange) for the hydrodynamics obtained from the initial CMFGEN model (Table~\ref{initialmodel}). Accelerations were rescaled by $g_\text{elec}$ (Eq.~\ref{gelec}) for illustrative reasons.}}
		\label{initialglinev}
	\end{figure}

	Figure~\ref{initialglinev} shows both left-hand and right-hand sides of Eq.~\ref{findgline} for the initial CMFGEN model tabulated in Table~\ref{initialmodel}, where it is clearly seen that they do not match.
	This discrepancy yields in the fact that velocity field is not calculated from the line-acceleration itself, but it is consequence of using a $\beta$--law.
	It would be possible to argue that this discrepancy might be a specific situation only, but it was previously noticed also by \citet{bouret12}; hence, the lacking of consistency between hydrodynamics and radiative acceleration for CMFGEN seems to be a general rule.
	
	As a consequence of this, we must perform a prescription capable to equalise both sides of Eq.~\ref{initialgline}.
	Following Lambert-procedure described below in this section is done aiming to evaluate:
	
	i) whether there is an hydrodynamic solution capable to properly couple line-acceleration and velocity/density fields,
	
	ii) whether this solution is stable or not and
	
	iii) whether this solution does reproduce a spectrum in agreement with the observations.

%_____Calculation of g_line_____________________________________________________________________________
\subsection{Calculation of line-acceleration $g_\text{line}(r)$}
	Despite the fact that line-acceleration used to solve equation of momentum is an output of a converged CMFGEN model, it is important to describe in general terms how is this calculated inside the code.
	For that purpose, it is necessary to do a brief summary about radiative transfer, and how this is related with the resulting $g_\text{line}(r)$.
	To avoid confusions, we remark the fact that what CMFGEN gives us exactly, corresponds to the \textit{total radiative acceleration} $g_\text{rad}$, i.e., the sum of both the acceleration produced by the line-driving process and that one produced by the continuum by means of Thomson scattering.
	However, these two terms are easily separated, as it is shown in Eq.~\ref{grad}.
	
	Radiative transfer equation, which describes the gaining or losing of radiative energy trough a path due to emission and absorption, is given by the formula:
	\begin{equation}\label{radiativetransferdif}
		\frac{dI_\nu}{ds}=-\chi_\nu I_\nu+\eta_\nu\;,
	\end{equation}
	with $\chi_\nu=\kappa_\nu\rho$ being the opacity, given $\kappa_\nu$ as defined in Eq.~\ref{kappanu}, and $\eta_\nu$ the emissivity and $ds$ is an infinitesimal element of path\footnote{We have presented here the simplest expression for the differential radiative transfer equation (i.e., along a simple segment $ds$) because of didactic reasons. The full expression introduced by \citet{mihalas75} for a spherical coordinate system is:
	\begin{equation}
		\mu\frac{\partial}{\partial r}I_\nu+\frac{1-\mu^2}{r}\frac{\partial}{\partial\mu}I_\nu-\frac{\nu_0 v(r)}{cr}\left[1-\mu^2+\mu^2\left(\frac{d\ln v}{d\ln r}\right)\right]\frac{\partial}{\partial\nu}I_\nu=-\chi_\nu I_\nu+\eta_\nu\;.
	\end{equation}
	
	In this case, the variables $\chi_\nu$, $\eta_\nu$ and $I_\nu$ are in function of the radius $r$ and the angle $\mu=\cos\theta$.} \citep{hillier90a}.
	Using the differential form of optical depth (Eq.~\ref{opticaldepth1}), $d\tau_\nu=\chi_\nu ds=\kappa_\nu\rho ds$, we rewrite radiative transfer equation as:
	\begin{eqnarray}\label{radiativetransfertau}
		\frac{dI_\nu}{\chi_\nu ds}&=&-I_\nu+\frac{\eta_\nu}{\chi_\nu}\nonumber\;,\\
		\frac{dI_\nu}{d\tau_\nu}&=&-I_\nu+S_\nu\;,
	\end{eqnarray}
	with $S_\nu$ being the \textit{source function}, the ratio between the emission and absorption coefficients.
	Notice that the formal solution of this equation is:
	\begin{equation}
		I_\nu(\tau_\nu)=I_\nu(0)e^{-\tau_\nu}+\int_0^{\tau_\nu}e^{\tau'_\nu-\tau_\nu}S_\nu(\tau'_\nu)d\tau'_\nu\;,
	\end{equation}
	which is the general expression for Eq.~\ref{radiationsobolev} where, thanks to the Sobolev approximation, we were focused in the absorption only and then we neglected possible any emission source.
	
	Full solution for the intensity of radiation then, depends on the value of the source function $S_\nu$ which in turn depends on the case if we are assuming LTE conditions or not.
	For local thermodynamic equilibrium, it is fulfilled that the source function is equal to the Planck function $B_\nu$ \citep[][Section 9.4]{carroll96}.
	However, for expanding atmospheres this is not the case, and thus $S_\nu$ must be calculated taking into account all the atomic processes among the different ions (collisions, recombinations, bound-free and bound-bound transitions, etc.).
	Since many of these interactions depend on the intensity of radiation, this becomes a coupled problem and then codes such as CMFGEN, \textsc{FastWind} or \textsc{PoWR} are necessary to solve the statistical equilibrium equations.
	Henceforth, solving this coupled problem for the radiative transfer it is possible later to calculate the flux mean opacity by means of the integration over all the frequencies:
	\begin{equation}\label{meanopacity}
		\frac{1}{\bar\chi}=\frac{\int_0^\infty\chi_\nu\frac{\partial S_\nu}{\partial T} d\nu}{\int_0^\infty\frac{\partial S_\nu}{\partial T} d\nu}\;.
	\end{equation}
	
	Radiative acceleration is then evaluated using the formula:
	\begin{equation}\label{glinecalc}
		g_\text{rad}(r)=\frac{\bar\chi_\text{f}L_*}{4\pi c\rho r^2}\;\;,
	\end{equation}
	where $\chi_\text{f}$ is the flux mean opacity.
	
	This radiative acceleration given by CMFGEN corresponds to the \textit{total} acceleration due to radiative processes, i.e., it considers not only the effects of absorption and reemission of photons by line transitions, but also electron scattering.
	However, acceleration by electron scattering is implicitly included in the momentum equation by means of the Eddington factor $\Gamma_e$.
	\begin{equation}\label{gelec}
		g_\text{elec}=\frac{GM\Gamma_e}{r^2}\;\;.
	\end{equation}
	
	Then, acceleration due to line-effects only (i.e., line-acceleration) corresponds to:
	\begin{equation}
		g_\text{line}=g_\text{rad}-g_\text{elec}\;\;.
	\end{equation}

%_____Solution of equation of momentum______________________________________________________________________
\subsection{Solution of equation of momentum}\label{solutionmomentum}
	The equation of momentum for a stationary, one-dimensional, non-rotating, isothermal, outflowing wind in spherical coordinates is given by:
	\begin{equation}\label{motion1}
		 v\frac{d v}{dr}=-\frac{1}{\rho}\frac{dp}{dr}-\frac{GM_*(1-\Gamma_e)}{r^2}+g_\text{line}(r)\;\;,
	\end{equation}
	where $p$ is the gas pressure, $ v$ is the wind velocity and $GM_*(1-\Gamma_e)/r^2$ is the gravitational effective acceleration.
		
	According with \citet{muller08} and \citet{araya14}, Eq.~\ref{eqmotion} can be expressed in a dimensionless way by making the following change of variables:
	\begin{equation}\label{changeofvariables}
		\hat r=\frac{r}{R_*}\;,\;\;\hat v=\frac{ v}{a}\;\text{ and }\;\;\hat v_\text{crit}=\frac{ v_\text{esc}}{a\sqrt{2}}=\frac{1}{a}\sqrt{\frac{GM_*(1-\Gamma_e)}{R_*}}\;,
	\end{equation}
	being $a$ the isothermal speed of sound given by:
	\begin{equation}\label{sound_plus_vturb}
		a^2=\frac{k_BT_\text{eff}}{\mu m_H}+\frac{1}{2} v_\text{mic}^2
	\end{equation}
	
	This formula differs from those used by \citet{sander17}, because we are considering temperature field $T(r)$ and mean particle $\mu(r)$ as constants.
	Micro-turbulence velocity $ v_\text{mic}$ is included, which is using to be in the order of $\sim10$ km~s$^{-1}$.
	After that, defining the dimensionless line acceleration:
	\begin{equation}
		\hat g_\text{line}(r)=\frac{R_*}{a^2}g_\text{line}(r)\;\;,
	\end{equation}
	the equation of momentum reads:
	\begin{equation}\label{motion2}
		\hat v\frac{d\hat v}{d\hat r}=-\frac{1}{\rho}\frac{dp}{d\hat r}-\frac{\hat v_\text{crit}^2}{\hat r^2}+\hat g_\text{line}(\hat r)\;.
	\end{equation}
	
	With the use of equation of state for an ideal gas ($p=a^2\rho$), Eq.~\ref{motion2} becomes independent of density (and therefore independent of mass-loss rate) equation of motion:
	\begin{equation}\label{momentum3}
		\left(\hat v-\frac{1}{\hat v}\right)\frac{d\hat v}{d\hat r}=-\frac{\hat v_\text{crit}^2}{\hat r^2}+\frac{2}{\hat r}+\hat g_\text{line}(\hat r)\;\;,
	\end{equation}
	By the integration of both sides along the atmosphere, Eq.~\ref{momentum3} can be solved analytically in order to obtain $\hat v(\hat r)$.
	
\subsubsection{Subsonic v/s supersonic region}
	Assuming a monotonic behaviour of the velocity field throughout the atmosphere ($d\hat v/d\hat r>0$), it is clearly seen that Eq.~\ref{momentum3} becomes zero when $v=a$.
	This condition is fulfilled at the \textit{sonic (or critical) point} $\hat r_c$:
	\begin{equation}\label{sonicpoint}
		-\frac{\hat v_\text{crit}^2}{\hat r_c^2}+\frac{2}{\hat r_c}+\hat g_\text{line}(\hat r_c)=0\;\;,
	\end{equation}
	
	Because of the monotonic behaviour of $ v(r)$ (and therefore $\hat v$), the sonic point becomes a boundary between two regions.
	The first of them is where $\hat v<1$ (velocity field below sound speed), which makes both sides of Eq.~\ref{momentum3} be less than zero, it is called the \textit{subsonic region}.
	The second one, where $\hat v>1$, is then called the \textit{supersonic region}.
	Differentiation between both is not only made due to mathematical analysis but also physical reasons.
	In a one dimensional fluid moving at velocities below sound speed perturbations are propagated both inwards and outwards, whereas perturbations occurred on a supersonic fluid only are propagated in the direction of the flux.
	
	This means that we actually are searching \textit{two solution branches} that merged at the critical point.
	If we focus at first in the supersonic region, going from the sonic point towards infinite and where velocity field scales from sound speed to the asymptotic terminal velocity, the equation to be solved by Lambert $W$-function (branch $W_{-1}$) is obtained by the integration of Eq.~\ref{momentum3}.
	%For that purpose we proceed to solve Equation~\ref{momentum3} by the integration along all the stellar atmosphere, making the division around $\hat r_c$ and using each branch of Lambert $W$-function per each regime.
	\begin{eqnarray}\label{lambertintegrated}
		\hat v^2e^{-\hat v^2}&=&-\left(\frac{\hat r_c}{\hat r}\right)^4\exp\left[-1-2\,\hat v_\text{crit}^2\left(\frac{1}{\hat r}-\frac{1}{\hat r_c}\right)\right]\nonumber\\
		 &\times&\exp\left[-2\int_{\hat r_c}^{\hat r}\hat g_\text{line}\,d\hat r\right]\;\text{, with }\hat r>\hat r_c\;\;.
	\end{eqnarray}
	
	Therefore, if we define the right-hand side of Eq.~\ref{lambertintegrated} as $x(\hat r)$, expression for velocity field is directly calculated from the Lambert $W$-function:
	\begin{equation}\label{vsqrtw}
		\hat v(\hat r)=\sqrt{-W_j(x(\hat r))}\;,
	\end{equation}
	with $j$ being the branch of the Lambert $W$-function (0 or $-1$).
	\begin{figure}[t!]
		\centering
		\includegraphics[width=0.6\linewidth]{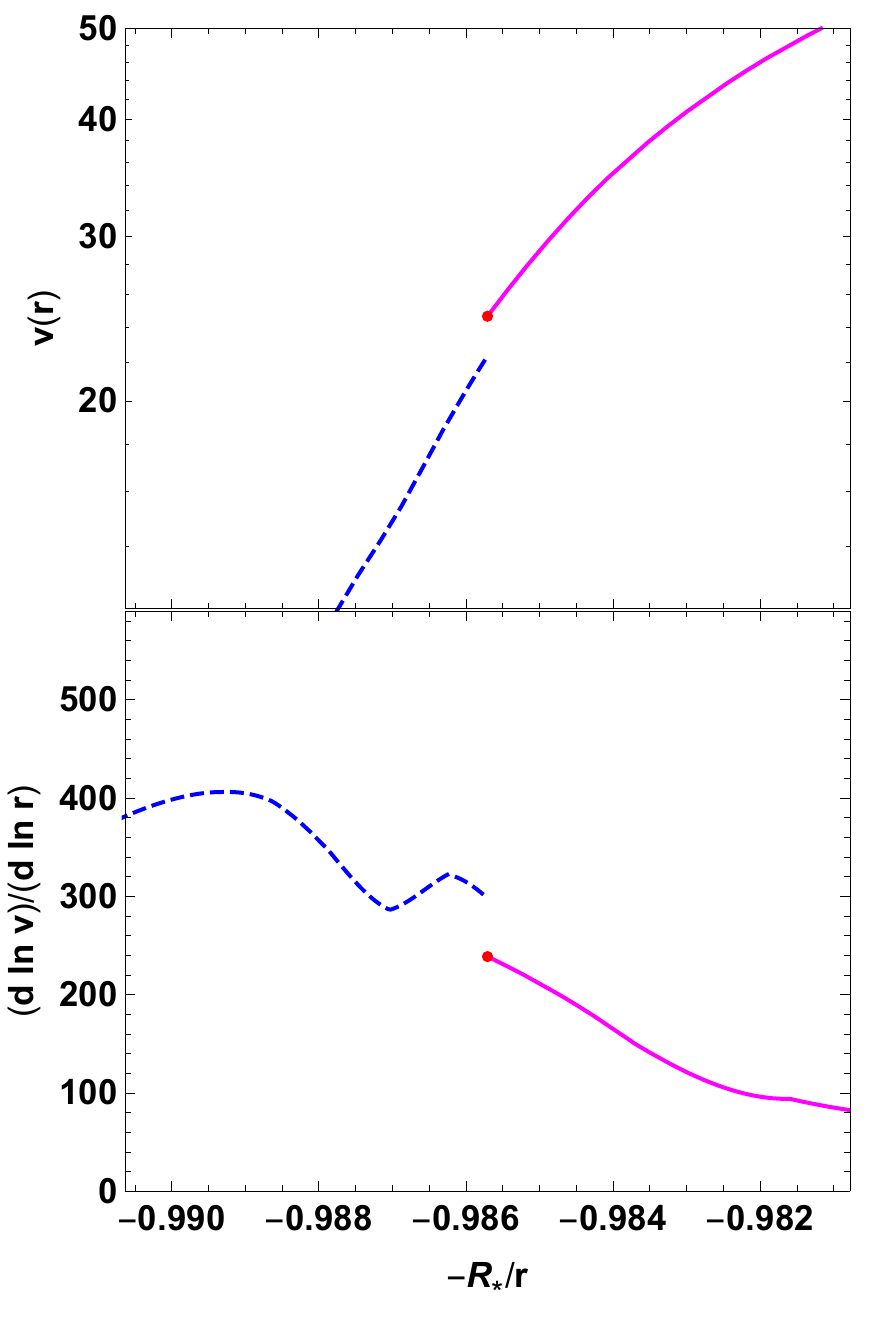}
		\caption{\small{Zoom over the sonic point (red point), where subsonic and supersonic regions are coupled during the Lambert-procedure. Magenta solid line represents the new velocity profile for the supersonic region obtained with Lambert $W$-function, whereas blue dashed line represents the old velocity profile for the subsonic region.}}
		\label{coupling}
	\end{figure}
	\begin{figure*}[t!]
		\centering
		\includegraphics[width=\linewidth]{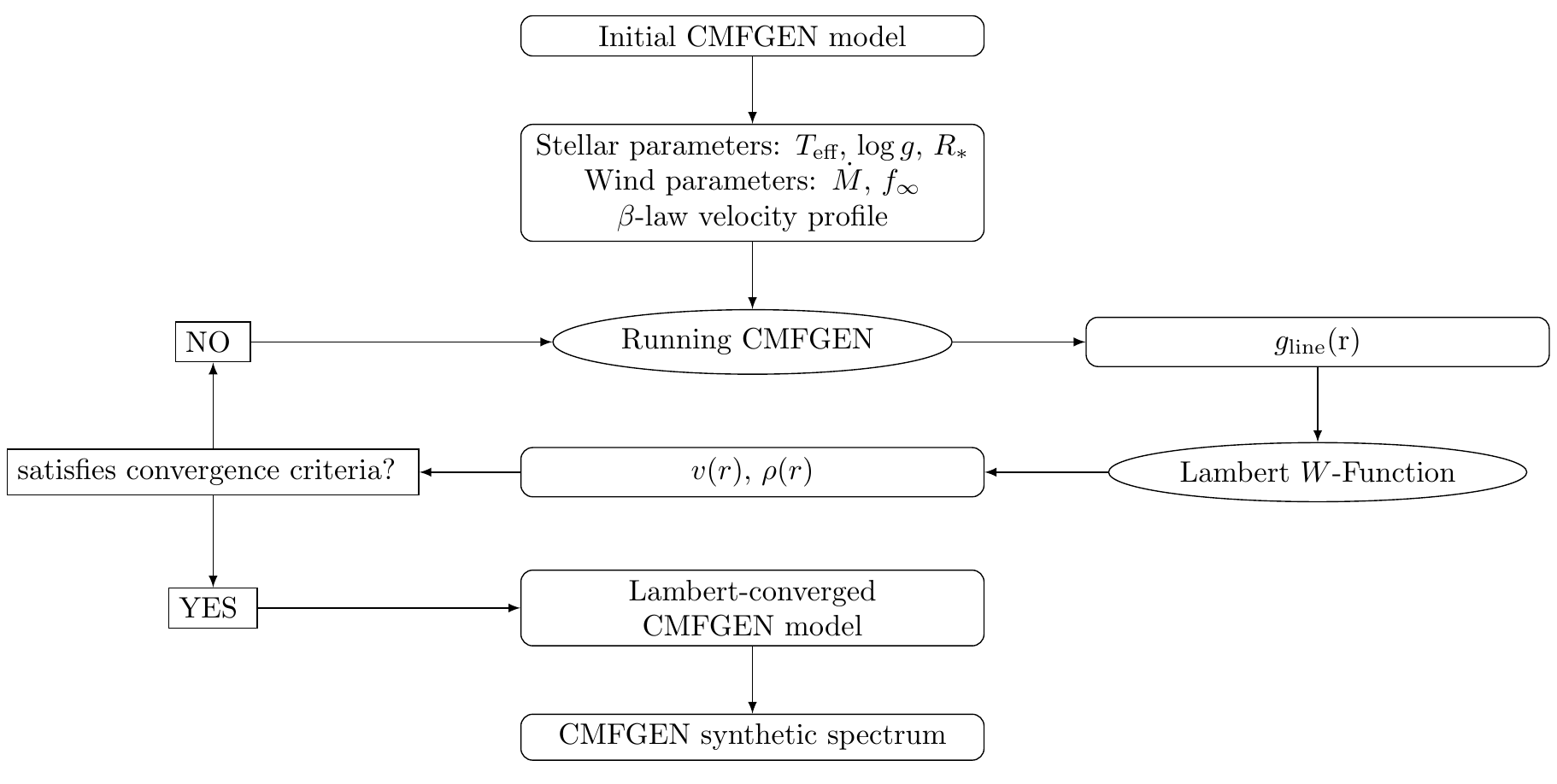}
		\caption{\small{Flowchart of the Lambert-procedure. Notice that stellar parameters are the same during the entire procedure, as well as mass-loss rate and clumping factor.}}
		\label{flowchart}
	\end{figure*}
	
	Because of the behaviour of the branches of the Lambert $W$-function, which are also merged in a specific point (see Fig.~\ref{lambertfunc}), it would be expectable to obtain an analytical expression for the subsonic region using the branch $W_0$.
	However, when this idea was executed on CMFGEN big errors were produced for the acceleration and velocity profiles on the lowest part of the wind (hydrostatic region, below the photosphere).
	Therefore, we decide to use Lambert-procedure to calculate hydrodynamics above the sonic point only, making later a good coupling between both supersonic and subsonic regions without introduce big errors in the hydrostatic part.
	This coupling must ensure a smooth transition between subsonic and supersonic zones, reason why those conditions must be fulfilled:
	\begin{equation}\label{fitv}
		\lim_{r\rightarrow a-} v(r)=\lim_{r\rightarrow a+} v(r)\;\;,
	\end{equation}
	and
	\begin{equation}\label{fitdv}
		\lim_{r\rightarrow a}\frac{d v(r)}{dr}\Biggr\rvert_{r<a}=\lim_{r\rightarrow a}\frac{d v(r)}{dr}\Biggr\rvert_{r>a}\;\;.
	\end{equation}
	
	The second limit is easier to evaluate if we use the logarithm derivatives by means of the equivalence:
	\begin{equation}
		\frac{d\ln v}{d\ln r}=\frac{r}{v(r)}\frac{dv}{dr}\;.
	\end{equation}
	
	Then:
	
	\begin{equation}\label{fitdlnv}
		\lim_{r\rightarrow a}\frac{d \ln v(r)}{d\ln r}\Biggr\rvert_{r<a}=\lim_{r\rightarrow a}\frac{d\ln v(r)}{d\ln r}\Biggr\rvert_{r>a}\;\;.
	\end{equation}
	
	As consequence of these conditions, subsonic region must be readapted too, in function of saving a smooth transition and avoid instabilities.
	Fig.~\ref{coupling} shows the coupling of $v(r)$ and $d\ln v/d\ln r$ around the sonic point for a velocity profile after solving hydrodynamics with Lambert $W$-function.
	The gaps displayed are produced because the sonic point given by finding the root of Eq.~\ref{sonicpoint} may not be equal to the old $v(r_c)=a$ given by the input velocity profile \citep[differences on these two sonic points were also previously referred by][see their Fig. 1]{sander17}, and because Lambert $W$-function may produce a more or less steep wind close to the sonic point.
	In any case, our self-consistent solution must be rescaled in the subsonic region in order to cancel these gaps.
	
	Aiming this, it is important an accurate determination of features on the inner part of the wind, such as photospheric radius and the sonic point location itself.
	We do not use Lambert $W$-function to obtain a new velocity field for the subsonic region because CMFGEN is too sensitive to modifications close to the photosphere (the hydrostatic part of the wind).
	Steep modifications affect largely features such as opacity and atomic populations, which leads to large errors in the re-calculation of line-acceleration.
	That is the main reason why we apply Lambert $W$-function only to supersonic region, whereas wind below sonic point is slightly modified by rescaling the velocity profile in order to satisfy accurately Eq.~\ref{fitv} and Eq.~\ref{fitdlnv}.

%_____Convergence of models__________________________________________________________________________	
\subsection{Convergence of models}
	The convergence of the Lambert-iterations can be checked by evaluating both either velocity field $ v(r)$ or line-acceleration $g_\text{line}(r)$.
	In order to accept a Lambert-model as well converged, we impose as condition to satisfy the following relationships:
	\begin{equation}\label{thresholdv}
		\Biggr\lvert\log\frac{ v(r)_n}{ v(r)_{n-1}}\Biggr\rvert\le0.01\;\;,
	\end{equation}
	and:
	\begin{equation}\label{thresholdg}
		\Biggr\lvert\log\frac{ g_\text{line}(r)_n}{g_\text{line}(r)_{n-1}}\Biggr\rvert\le0.01\;\;,
	\end{equation}
	where $n$ is the $n$-th Lambert-iteration executed\footnote{Notice that, if $f(r)_n=f(r)_{n-1}$ the value of the absolute expression would be zero, because the ratio between both functions would be 1.}.
	This condition is applied for line-acceleration too.
	Reason to choice this value as threshold yields in the fact that no significative differences on resulting spectra are observed between models with hydrodynamic differences smaller that this value.
	
	The threshold previously presented is applied not only for external part of the wind (i.e., terminal velocity $v_\infty$) but for the entire range on radius $r$ from the photosphere outwards.
	This, because the part which is most sensitive to changes is where the coupling is done (around the sonic point).
	A good convergence around this zone allows us the obtention of a stable Lambert-model with a smooth transition between subsonic and supersonic regions.

	Later, existence of a unique solution must be confirmed.
	Wind hydrodynamics expressed under $\beta$--law depends on two parameters: terminal velocity $v_\infty$ and the $\beta$ exponent.
	If the Lambert-procedure is able to generate a well converged hydrodynamic solution, it should be the same for all the initial velocity profiles (i.e., initial $v_\infty$ and $\beta$ parameters) chosen.
	\begin{table}
		\centering
		\begin{tabular}{c c c}
			\hline\hline
			Model & $v_\infty$ (km s$^{-1}$) & $\beta$\\
			\hline
			\texttt{T41blaw01} & $2\,300$ & $0.9$\\
			\texttt{T41blaw02} & $2\,700$ & $0.9$\\
			\texttt{T41blaw03} & $2\,300$ & $1.1$\\
			\texttt{T41blaw04} & $2\,000$ & $0.9$\\
			\texttt{T41blaw05} & $2\,300$ & $0.7$\\
			\hline
		\end{tabular}
		\caption{\small{Hydrodynamic parameters of initial models.}}
		\label{tableini} % is used to refer this table in the text
	\end{table}
	
	Table~\ref{tableini}\footnote{Nomenclature for the name CMFGEN models executed in this procedure is as follows: \texttt{T41} means a model with $T_\text{eff}=41$ kK, \texttt{blaw} means a model with $\beta$--law whereas \texttt{lamb} means a converged Lambert-model.} shows hydrodynamic parameters for each initial model.
	All these models have the same value for mass-loss rate ($M=2.7\times10^{-6}$ $M_\odot$~yr$^{-1}$) and clumping filling factor ($f_\infty=0.1$), i.e., only initial conditions for velocity field are different.
	As consequence, the final Lambert-converged hydrodynamics (if there is one) should be the same.
	Once convergence of all these initial models be demonstrated, we can start to evaluate what is the result when different values for $\dot M$ and $f_\infty$ are implemented.
	
	A scheme of Lambert-procedure is presented in Fig.~\ref{flowchart}.
	Typically, Lambert-procedure converges after 5 or 6 Lambert-iterations, each one of them corresponding to a CMFGEN model executed with 30 inner iterations.

%_____RESULTS_____________________________________________________________________________________
\section{Results}\label{lambertresults}
	Through this section we will analyse the convergence of the Lambert procedure.
	First part is focused on the results starting from the initial model with the parameters given in Table~\ref{initialmodel}, together with the check of the convergence of the alternative models tabulated in Table~\ref{tableini} over the same final solution.
	Second part is a more extensive analysis evaluating the consequences of using different initial values for mass-loss rate and clumping.
	\begin{figure}[t!]
		\centering
		\includegraphics[width=0.49\linewidth]{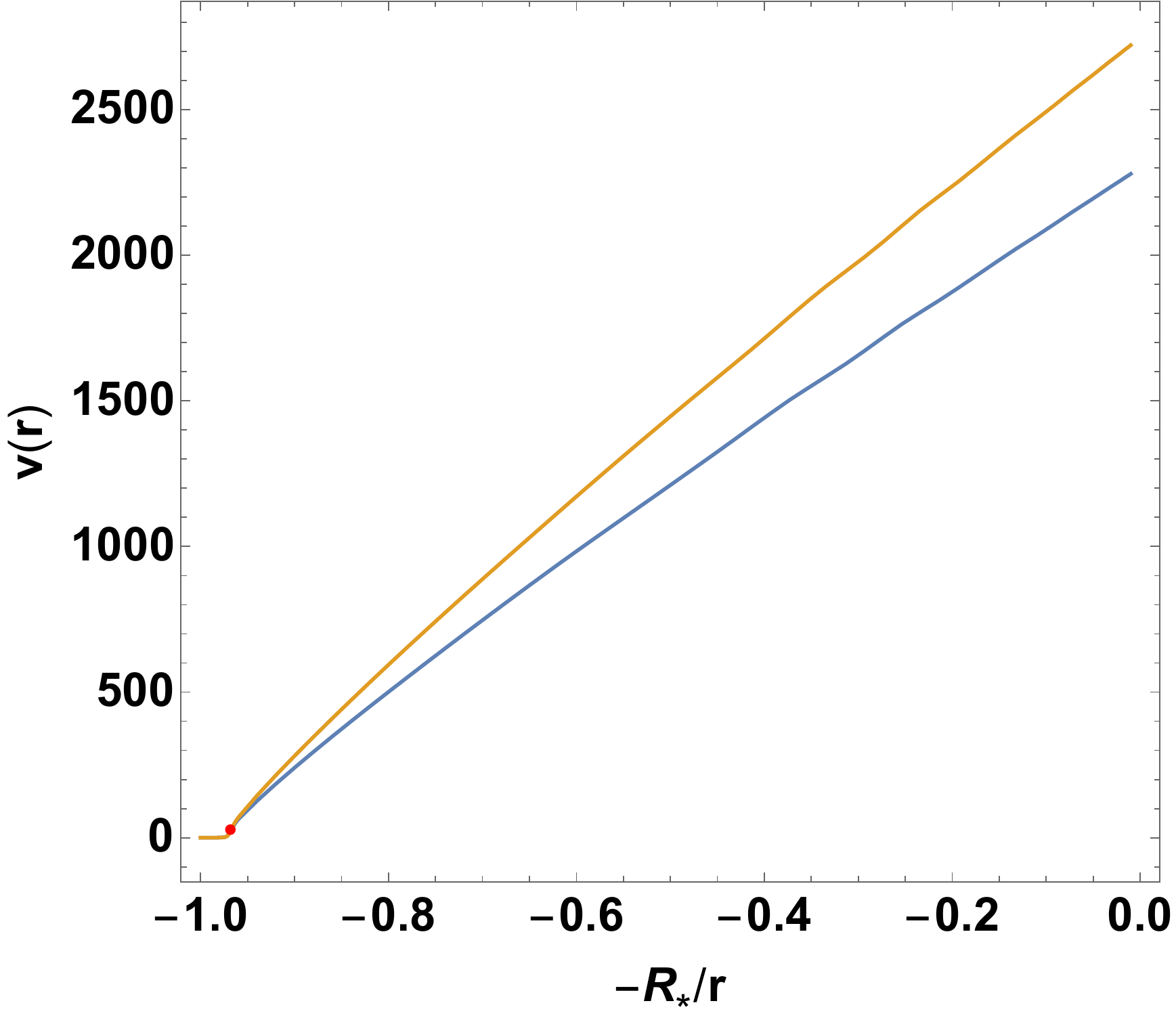}
		\includegraphics[width=0.49\linewidth]{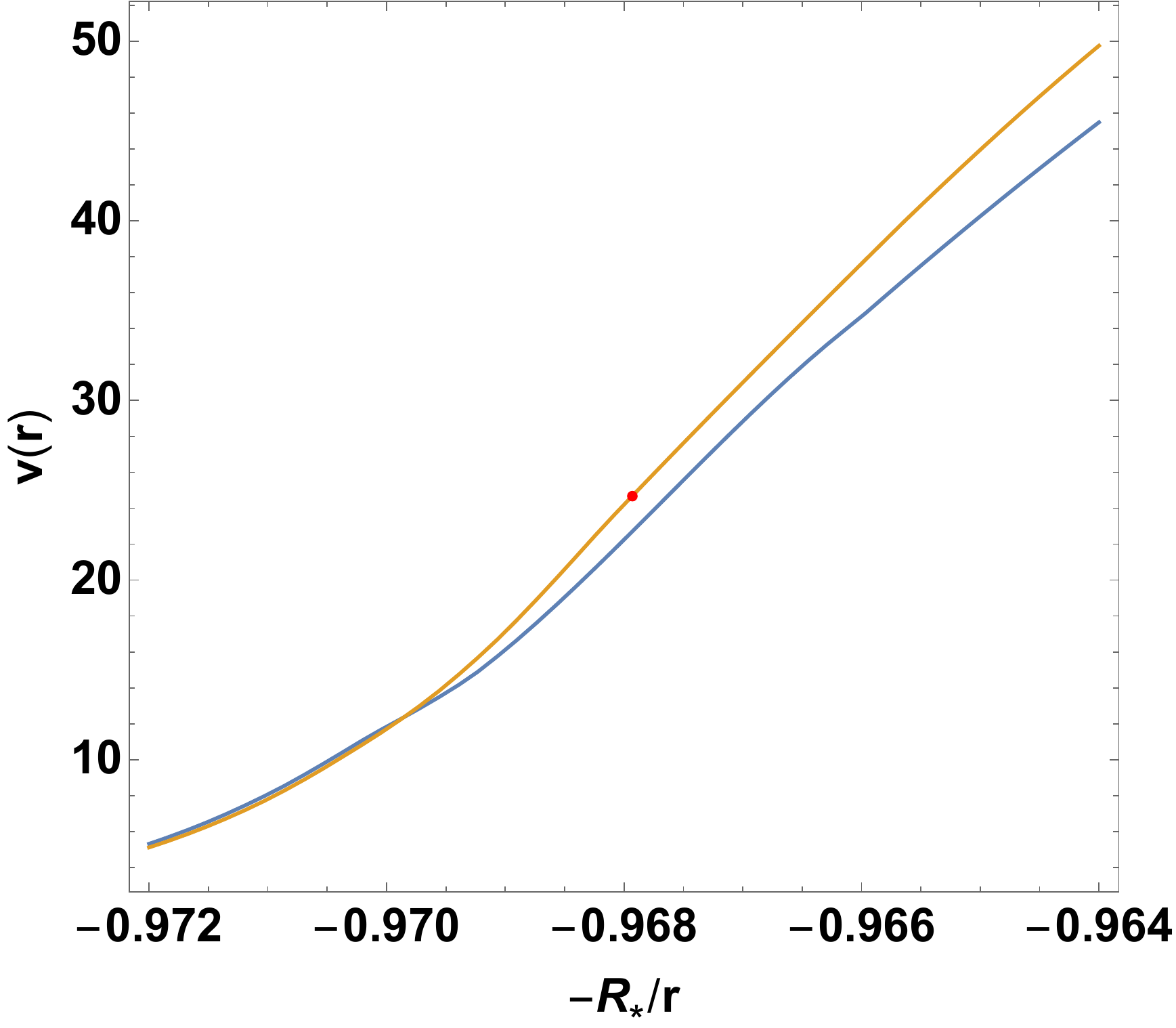}
		\caption{\small{Differences between velocity field with $\beta$--law (blue) and the resulting after Lambert-iterations (orange) for a CMFGEN model with $\dot M=2.7\times10^{-6}$ and $f_\infty=0.1$. A zoom over the transition region is included, displaying the sonic point with a red dot.}}
		\label{oldvsnewvel}
	\end{figure}
	\begin{figure}[t!]
		\centering
		\includegraphics[width=0.6\linewidth]{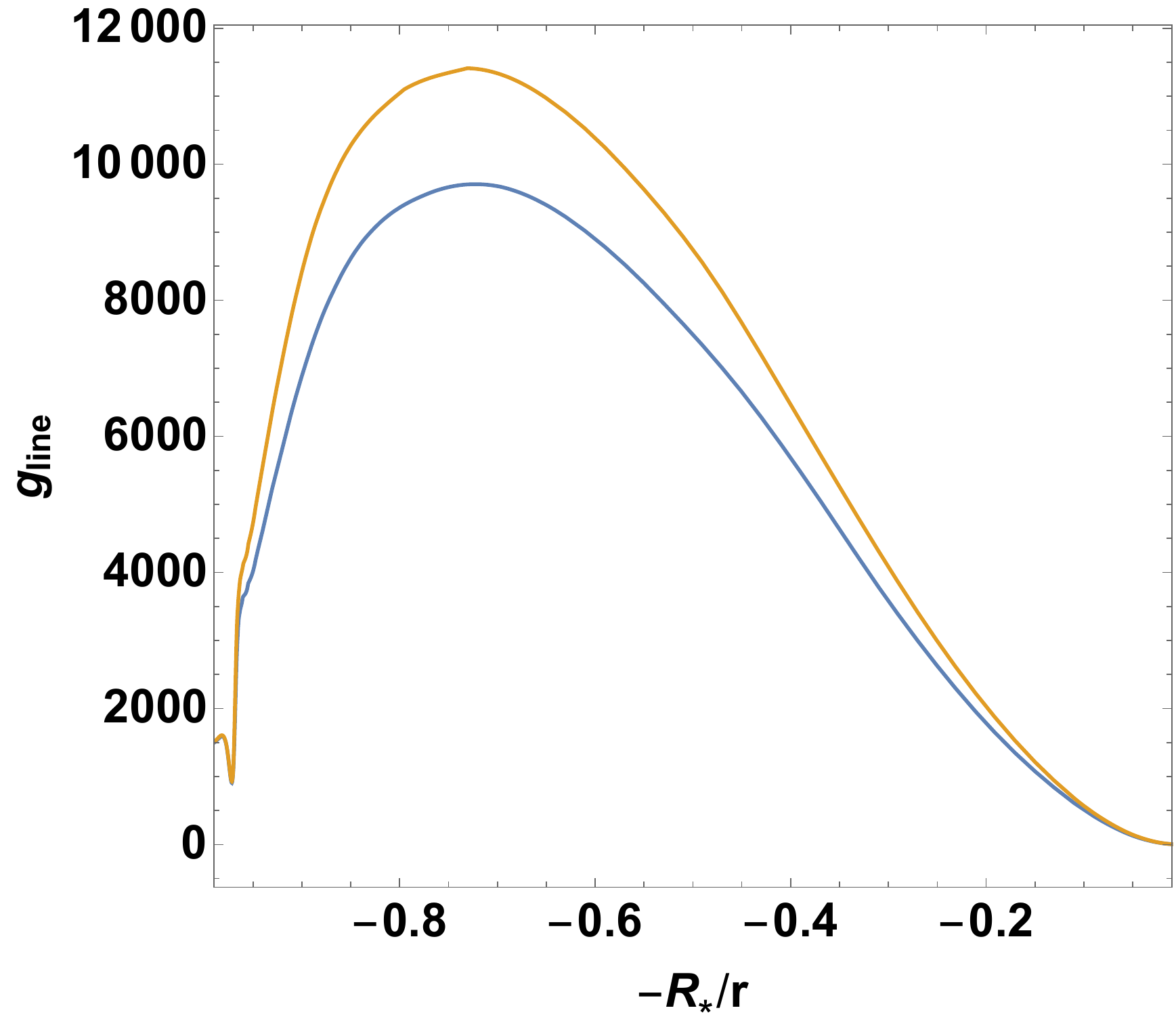}
		\caption{\small{Same as Fig.~\ref{oldvsnewvel}, but comparing line-accelerations instead velocity fields.}}
		\label{oldvsnewgline}
	\end{figure}

%_____Initial converged Lambert-hydrodynamics_____________________________________________________________
\subsection{Initial converged Lambert-hydrodynamics}
\begin{figure}[t!]
		\centering
		\includegraphics[width=0.6\linewidth]{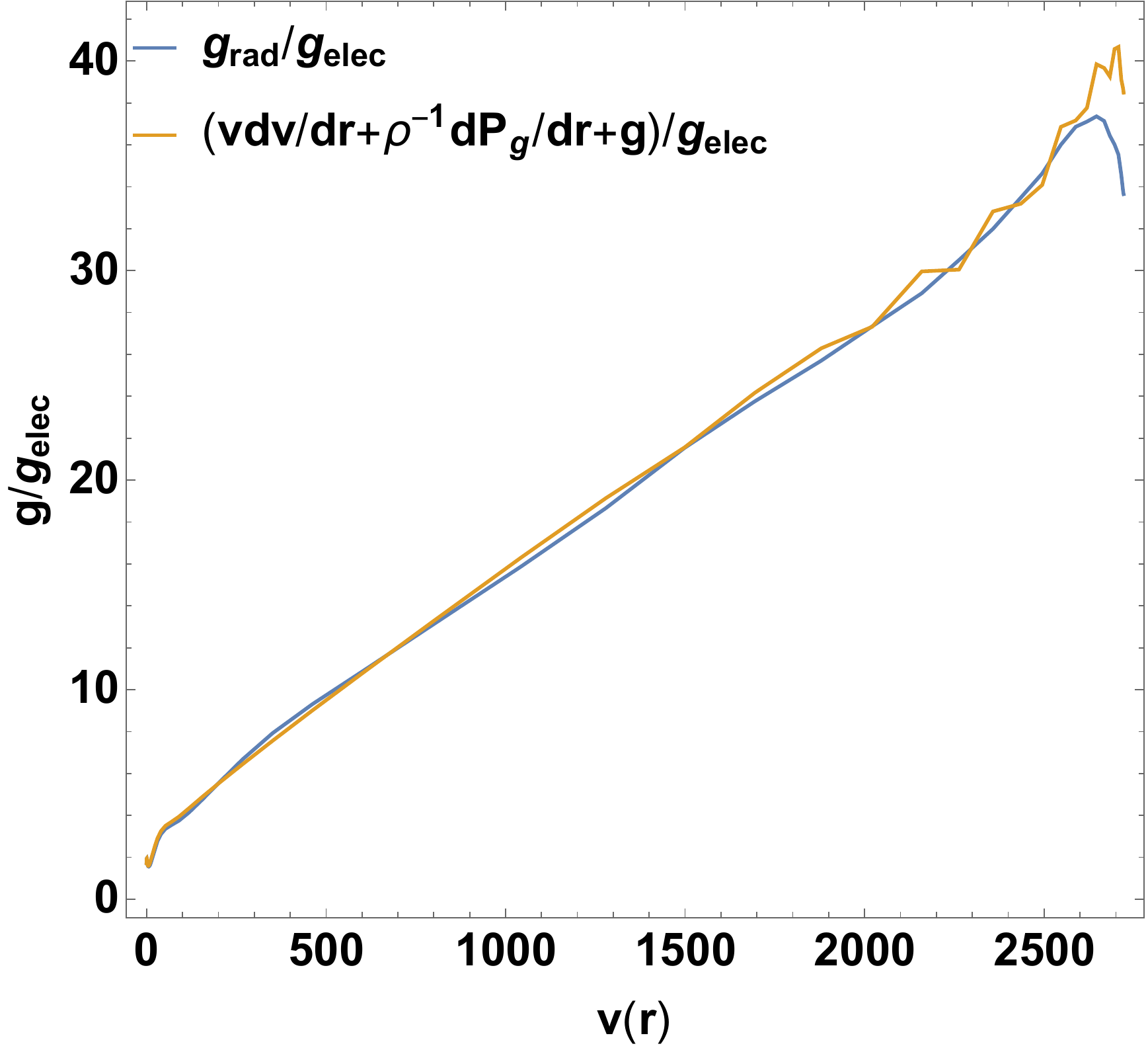}\\
		\caption{\small{Radiative acceleration (divided by acceleration due to electron scattering) as a function on wind velocity for the final CMFGEN Lambert-model (blue), compared with the expected value when equation of momentum is solved (orange).}}
		\label{finalglinev}
	\end{figure}

	Applying Lambert-procedure over the starting CMFGEN model with the parameters shown in Table~\ref{initialmodel}, a new CMFGEN model has been created with a new hydrodynamics and a new line-acceleration.
	These both new features are related each other by means of the equations previously discussed in Section~\ref{lambertprocedure}, i.e., wind hydrodynamics and line-acceleration are self-consistent between them.
	Comparisons between initial and converged self-consistent velocity fields are shown in Fig.~\ref{oldvsnewvel}, whereas comparisons between line-accelerations are shown in Fig.~\ref{oldvsnewgline}.
	Both features were evaluated to satisfy the threshold condition from Eq.~\ref{thresholdv}.
	Because this time the model is self-consistent, both left-hand and right-hand sides of Eq.~\ref{findgline} are in agreement as it is observed in Fig.~\ref{finalglinev} (compare with previous Fig.~\ref{initialglinev}).

	Concerning to the converged self-consistent new hydrodynamic, as first comment we observe that resulting terminal velocity has increased in a factor of $\sim1.2$ from 2\,300 to 2\,740 km~s$^{-1}$.
	Resulting line-acceleration is also higher for the Lambert-model, but this result may be consequence on the chosen initial parameters ($T_\text{eff}$, $\log g$, $\dot M$) only.
	Besides, due to the rescaling in the subsonic region, there are not significant differences in the resulting velocity profile below the sonic point, just above $\sim12-13$ km s$^{-1}$.
	As we previously stated in Section~\ref{lambertprocedure}, this is the most effective method to ensure a good coupling between two zones, and also it estabilised the subsonic regions (which is very sensitive to sharp changes in CMFGEN).
%	As consequence, in spite it is observed differences on the velocity field below the sonic point, these ones are significant until half sound speed only.
	\begin{figure}[htbp]
		\centering
		\includegraphics[width=0.6\linewidth]{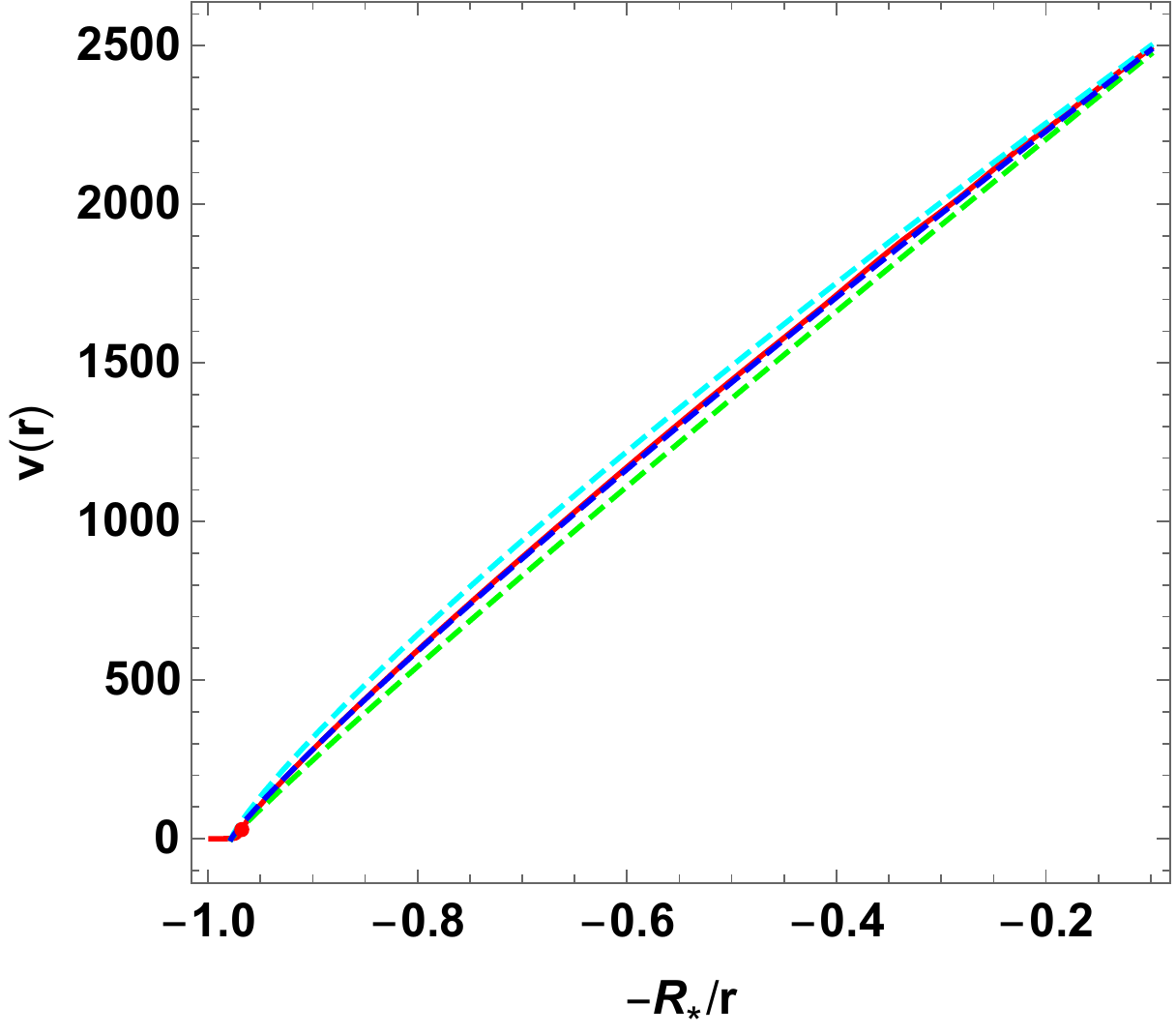}\\
		\includegraphics[width=0.6\linewidth]{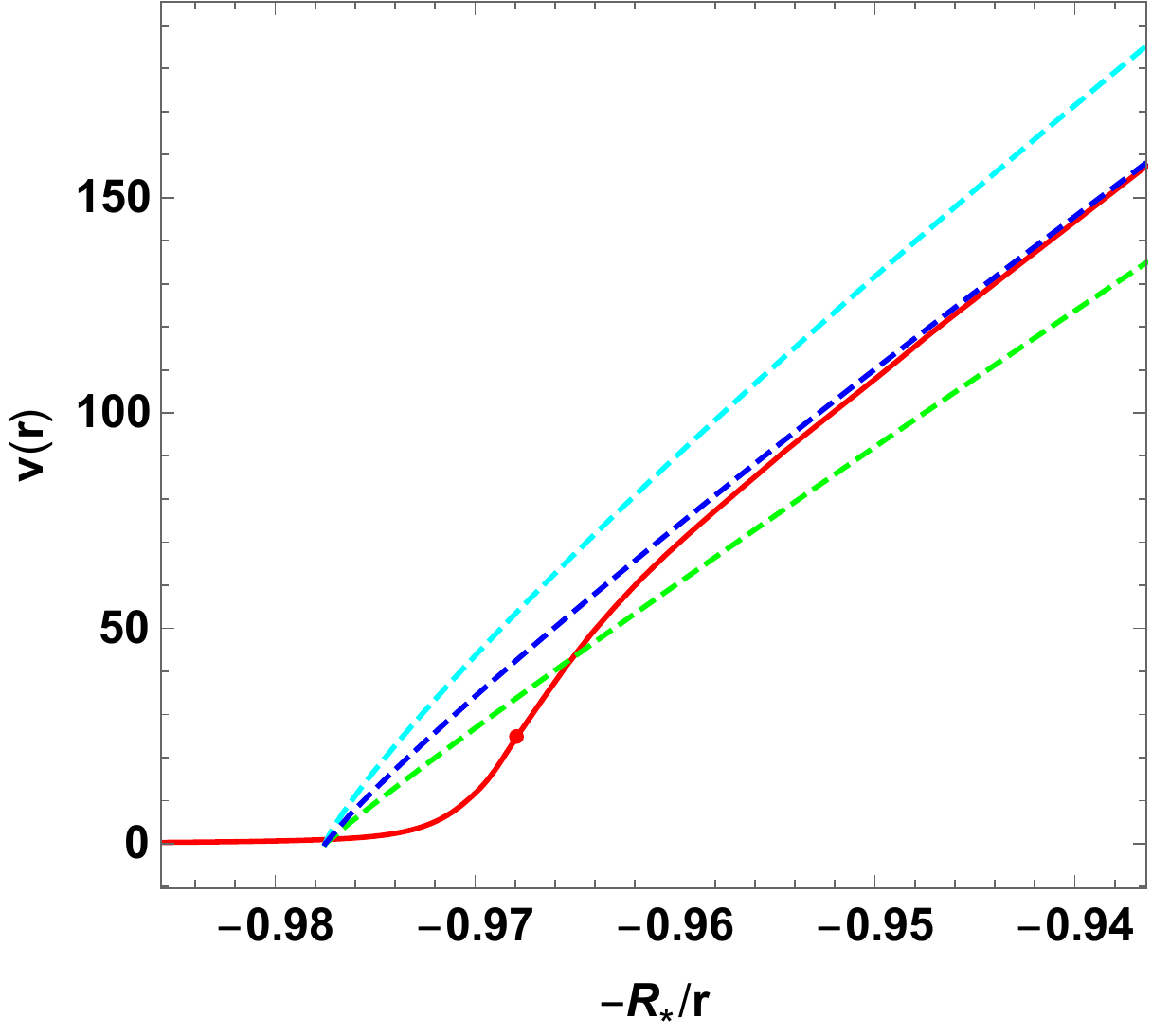}
		\caption{\small{Velocity profile of the converged Lambert-model (red), fitted with a $\beta$-law using $\beta=0.85$ (cyan), $\beta=0.95$ (green) and $\beta=0.9$ (blue). Lower panel shows a zoom around the sonic point.}}
		\label{oldvsnewvelbeta}
	\end{figure}
	\begin{figure}[htbp]
		\centering
		\includegraphics[width=0.48\linewidth]{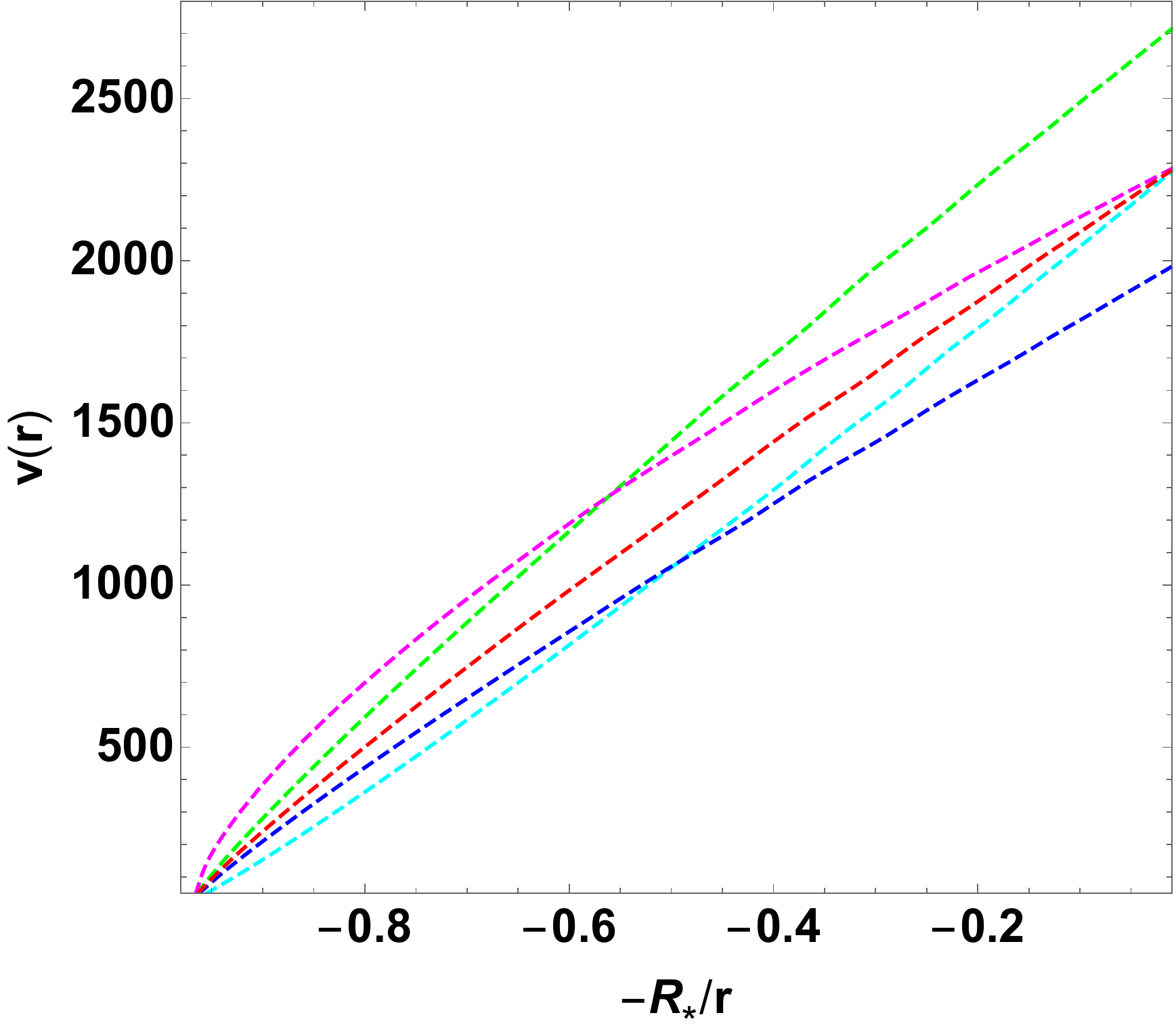}
		\includegraphics[width=0.48\linewidth]{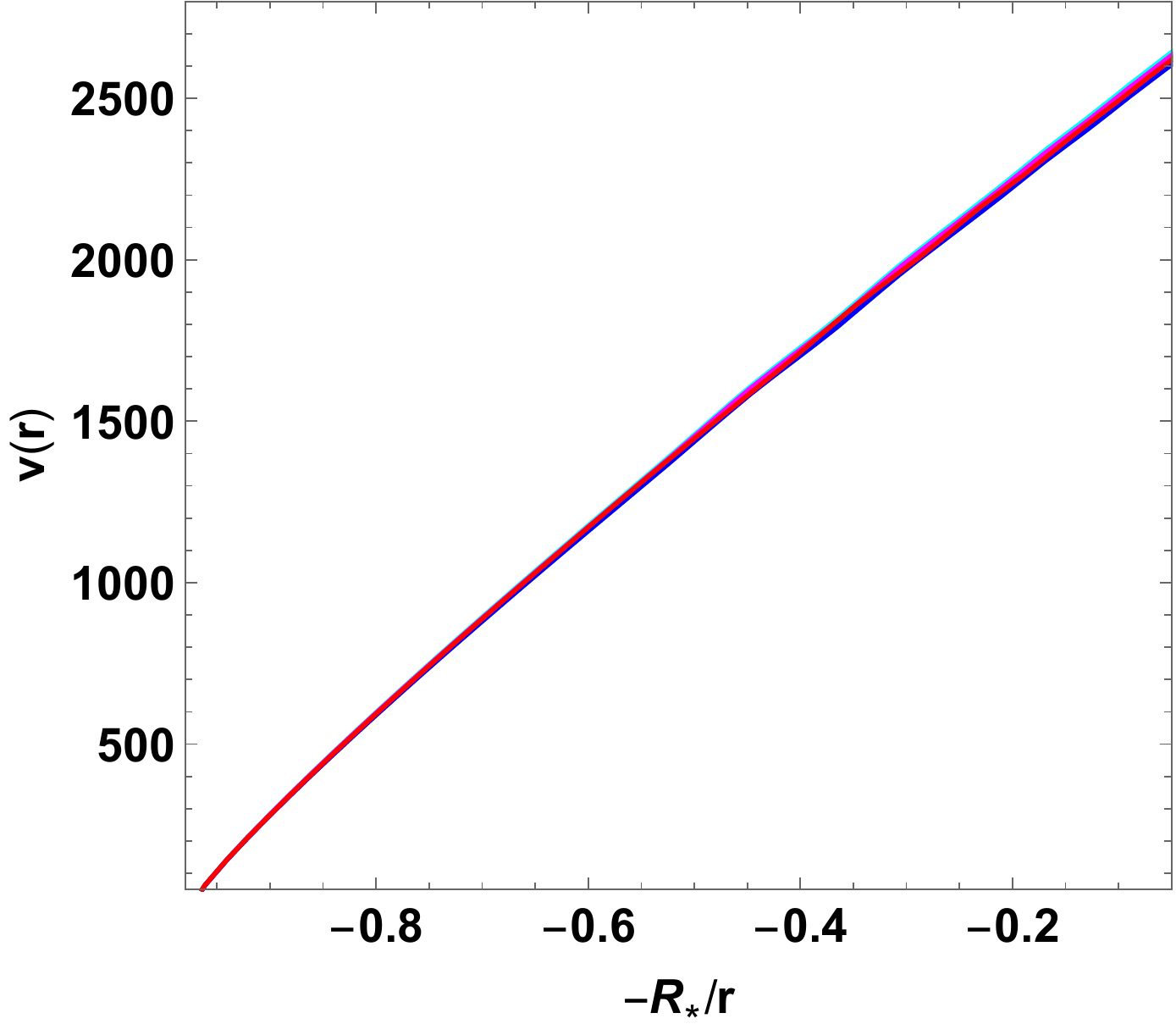}
		\caption{\small{Left panel: velocity profiles (in units of km~s$^{-1}$) corresponding to the models shown in Table \ref{tableini}: \texttt{T41blaw01} (red), \texttt{T41blaw02} (green), \texttt{T41blaw03} (cyan), \texttt{T41blaw04} (blue), \texttt{T41blaw05} (magenta). Right panel: converged Lambert-profiles for each initial model.}}
		\label{hydros_alt}
	\end{figure}
	
	Moreover, it is interesting the fact that hydrodynamics can be characterised not only with a new terminal velocity, but also they can be fairly approximated with a new $\beta$ factor (which may differ from the initially set as input).
	This effect is clearly seen in Fig.~\ref{oldvsnewvelbeta}, where velocity profile with $\beta=0.9$ overlaps the self-consistent $v(r)$, although that does not mean that Lambert procedure gives a new value for $\beta$, because resulting hydrodynamics are not fitted but calculated from respective line-acceleration.
	However, it is possible to find a $\beta$ value (hereafter quasi-$\beta$) capable to closely fit the new velocity field.
	This approximate fit with a $\beta$-law is expectable because the wind of $\zeta$-Puppis lies in the range of the so-called \textit{fast solutions} \citep{michel12,alex19}, but we cannot assure that we could find the same quasi-$\beta$ for stars at lower temperatures with $\delta$-slow solutions \citep{michel11}.

	To be sure that this well converged Lambert-solution is independent on initial values for $\beta$ and $ v_\infty$, we proceed to repeat the process for every model signalled in Table~\ref{tableini}, and whose graphs are shown in Fig.~\ref{hydros_alt}: left panel represents the initial velocity profiles with $\beta$-law, whereas right panel shows the final converged Lambert velocity profiles (where the overlap demonstrates all the initial models converging into the same solution).

%_____Mass-loss rate as free parameter_________________________________________________________________
\subsection{Mass-loss rate as free parameter}\label{masslossandclumping}
	Previously, when we transform initial momentum equation, Eq.~\ref{eqmotion}, into Eq.~\ref{momentum3}, the explicit term for $\rho(r)$ was replaced by $v(r)$ by means of the equation of continuity, Eq.~\ref{eqcontinuity}, and thus we obtained a formula independent on density.
	As consequence, Lambert-procedure cannot calculate a new mass-loss rate because $\dot M$ must be set as an input.
	For this reason, this wind parameter is considered as free when we are calculating our Lambert-procedure.

\begin{figure}[t]
		\centering
		\includegraphics[width=0.6\linewidth]{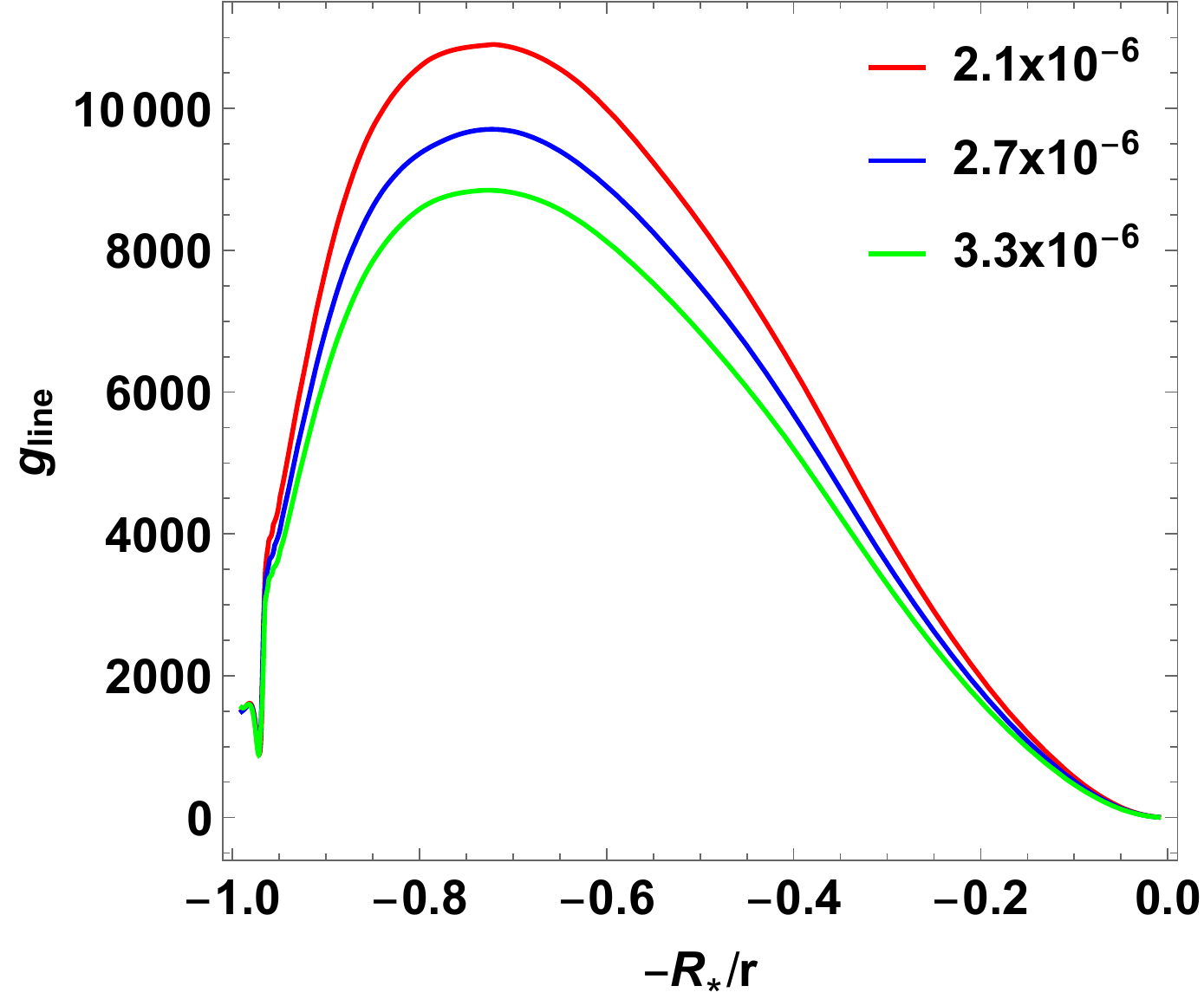}
		\caption{\small{Line-accelerations from initial CMFGEN models with different values for $\dot M$. Every other stellar and wind parameter is the same as in Table~\ref{initialmodel}.}}
		\label{mdotsinitial}
	\end{figure}

	However, the dependence on density, and therefore the mass-loss rate, is implicitly included in Eq.~\ref{momentum3} inside the $g_\text{line}$ term.
	From Eq.~\ref{glinecalc} remains clear that line-acceleration is inversely proportional to mass-loss rate $\dot M$, and hence also inversely proportional to the density $\rho(r)$.
	Because this density (implicitly included inside line-acceleration) is directly proportional to $\dot M$ which keeps constant during Lambert-procedure, these differences in $g_\text{line}(r)$ leads to different final Lambert-hydrodynamics.
	Higher mass-loss rates produces slower line-accelerations (Fig.~\ref{mdotsinitial}), which undergoes into velocity profiles with lower terminal velocities.
	
	Thus, the mass-loss rate is not the only wind parameter to be considered as free.
	Line-acceleration is affected not only by the general matter density but also for the small scale inhomogeneities (i.e., clumping) present through the wind.
	Clumping is implemented by CMFGEN in terms of the \textit{volume filling factor} $f$, which assumes a void interclump medium and the clumps to be small compared to the photons mean free path.
	The filling factor is such that $\rho=\rho_0/f$, where $\rho_0$ is the homogeneous (unclumped) wind density \citep{bouret05}.
	Volume filling factor is defined in terms on the velocity field:
	\begin{equation}
		f( v(r))=f_\infty+(1-f_\infty)e^{- v(r)/ v_\text{cl}}\;\;.
	\end{equation}
	
\begin{figure}[t]
		\centering
		\includegraphics[width=0.6\linewidth]{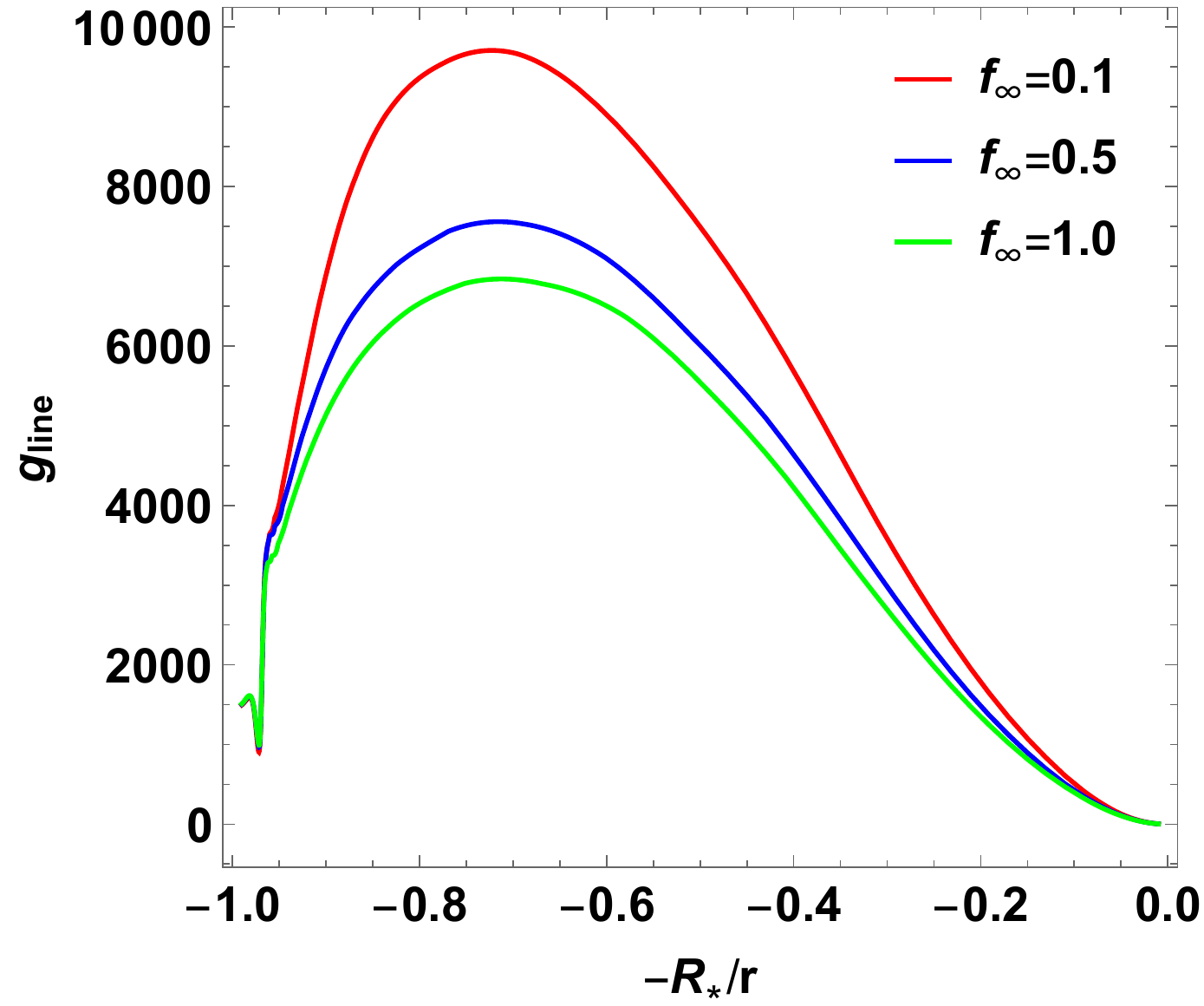}
		\caption{\small{Line-accelerations from initial CMFGEN models with different values for $f_\infty$. Every other stellar and wind parameter is the same as in Table~\ref{initialmodel}.}}
		\label{clumpinitial}
	\end{figure}

	Usually, clumping can be expressed by the \textit{[infinite] filling factor} $f_\infty$ only.
	%, because this is the parameter that determines the intensity of the clumping throughout the wind whereas the effect of $ v_\text{cl}$ (the point in velocity field where clumping starts from) is less relevant.
	Any stronger clumping factor, expressed with a smaller filling factor $f_\infty$, gives a stronger line-acceleration (Fig.~\ref{clumpinitial}), which will also lead into faster hydrodynamics with higher terminal velocities.
	This effect is not a consequence derived from Eq.~\ref{glinecalc} because density profile keeps unchanged in all these cases.
	We could argue that the presence of the overdensities (which produce the overestimation for mass-loss rates) is the responsible of the increasing on $g_\text{line}$ but as we pointed out previously, line-acceleration is inversely proportional to $\rho(r)$ and hence with higher clumping factor $g_\text{line}$ should be smaller.
	Therefore, the reason why line-acceleration becomes higher when clump inhomogeneities intensifies remains not completely clear.
	
	Since neither mass-loss rate nor filling factor are altered when Lambert-iterations are executed, different combinations of $(\dot M,f_\infty)$ will lead into different converged new Lambert-hydrodynamics, with their corresponding new terminal velocities.
	Some of these new self-consistent hydrodynamics generated by Lambert-iterations are shown in Table~\ref{lambertsolutions}.
	The homogeneous equivalence for mass-loss rate ($\dot M/\sqrt{f_\infty}$) is shown in order to make clearer the influence of chosen clumping into the resulting Lambert-hydrodynamics.
	The most important result derived from Table~\ref{lambertsolutions} (even when it could be considered an obvious consequence), is the direct relationship between line-acceleration and the resulting terminal velocity for each self-consistent solution.
	The higher $g_\text{line}$, the larger final $v_\infty$, and therefore a self-consistent solution with an specific terminal velocity can be obtained using the adequate set of $\dot M$ and $f_\infty$.
%	As a secondary result, we could add that higher line-accelerations produce steeper velocity profiles (i.e., lower values for the quasi-$\beta$), but differences seems to be negligible compared with effects on terminal velocity.
	\begin{table*}[htbp]
		\centering
		\begin{tabular}{cccccc}
			\hline\hline
			Name & $\dot M_\text{clump}$ & $f_\infty$ & $\dot M_\text{hom}$ & $ v_{\infty,\text{Lamb}}$\\
			& ($M_\odot$ yr$^{-1}$) & & ($M_\odot$ yr$^{-1}$) & (km s$^{-1}$)\\
			\hline
			\texttt{T41lamb01} & $2.7\times10^{-6}$ & 0.1 & $8.5\times10^{-6}$ & 2\,740\\%lamb022
			\texttt{T41lamb02} & $2.7\times10^{-6}$ & 1.0 & $2.7\times10^{-6}$ & 2\,120\\%lamb040
			\texttt{T41lamb03} & $2.1\times10^{-6}$ & 0.1 & $6.6\times10^{-6}$ & 3\,010\\%lamb046
			\texttt{T41lamb04} & $3.5\times10^{-6}$ & 0.1 & $1.1\times10^{-5}$ & 2\,490\\%lamb066
			\texttt{T41lamb05} & $2.85\times10^{-6}$ & 0.5 & $4.0\times10^{-6}$ & 2\,280\\%lamb070
			\texttt{T41lamb06} & $2.7\times10^{-6}$ & 0.5 & $3.8\times10^{-6}$ & 2\,290\\%lamb077
			\texttt{T41lamb07} & $2.85\times10^{-6}$ & 0.1 & $9.0\times10^{-6}$ & 2\,670\\%lamb078
			\texttt{T41lamb08} & $2.85\times10^{-6}$ & 0.3 & $5.2\times10^{-6}$ & 2\,380\\%lamb089
			\texttt{T41lamb09} & $3.5\times10^{-6}$ & 0.2 & $7.8\times10^{-6}$ & 2\,340\\%lamb093
			\texttt{T41lamb10} & $2.85\times10^{-6}$ & 0.2 & $6.37\times10^{-6}$ & 2\,470\\%lamb094
			\texttt{T41lamb11} & $3.5\times10^{-6}$ & 0.3 & $6.4\times10^{-6}$ & 2\,220\\%lamb095
			\hline
		\end{tabular}
		\caption{\small{Converged models, given a set of different values for mass-loss rate and clumping filling factor for stellar parameters $T_\text{eff}=41$ kK, $\log g=3.6$ and $R_*=17.9R_\odot$.}}
		\label{lambertsolutions}
	\end{table*}
	
	However, it is important to emphasise that this family of numerically well converged Lambert-hydrodynamics given different combinations for $\dot M$ and $f_\infty$ are not valid for \textit{any} possible mass-loss rate, but it is constrained within a range of plausible values.
	This comes from the fact that the error associated in the equation of momentum for a CMFGEN model, defined as:
	\begin{equation}\label{errorcmfgen}
		\text{Err}(r)=\frac{200\left(v\frac{dv}{dr}+\frac{1}{\rho}\frac{dP}{dr}+\frac{P}{\rho}\frac{dT}{dr}-g_\text{tot}\right)}{\biggr|v\frac{dv}{dr}\biggr|+\biggr|\frac{1}{\rho}\frac{dP}{dr}\biggr|+\biggr|\frac{P}{\rho}\frac{dT}{dr}\biggr|+|g_\text{tot}|}\;,
	\end{equation}
	grows considerably when the initial hydrodynamics introduced departs too far away from the previous model, specially in the so-called \textit{acceleration zone} (where velocity field takes values around $\sim25-100$ km s$^{-1}$).
	For example, when we consider mass-loss rates below $1.0\times10^{-6}$ $M_\odot$ yr$^{-1}$, resulting hydrodynamic is so fast in the supersonic region that does not allow a smooth coupling; hence, we can consider this value as the lower plausible limit for Lambert-solutions.
	A similar scenario is seen when the introduced mass-loss rate is too large as more than $6.0\times10^{-6}$ $M_\odot$ yr$^{-1}$: resulting Lambert-hydrodynamics is so slow that is "compressed" in the supersonic region.
	The existence of a minimum error should be used as a tracer to find the most accurate value for the mass-loss rate; nevertheless this is not possible at all, because the erratic behaviour of $\text{Err}(r)$ does not allow the performance of a good constraint.
	This implies, the best free mass-loss rate (and clumping filling factor) must be determined by the spectral fit.

%_____Discussion_____________________________________________________________________________
\section{Discussion}
	Discussion about Lambert-solutions will be centred on their capacity to be combined with spectral fitting analysis in order to derive the best self-consistent wind parameters, and the comparison of these new Lambert-solutions with previous self-consistent studies \citep[see, e.g.,][and Chapter~\ref{alfakdelta27}]{sander17}.
	
%_____Fitting observational spectra______________________________________________________________________
\subsection{Fitting observational spectra}
	Despite the find of several plausible self-consistent hydrodynamics for (this case) $\zeta$-Puppis, the \textit{real physical} solution must be unique, and just one of these solutions can be the accurate one.
	For that reason, the following step is to evaluate the resulting spectra from all these Lambert-hydrodynamic solutions, checking which combination of $(\dot M,f_\infty)$ provides an accurate fit for the observed spectrum.
	Observed spectra for $\zeta$-Puppis (HD 66811) were taken with FEROS\footnote{Fiber-fed Extended Range Optical Spectrograph: \url{https://www.eso.org/public/chile/teles-instr/lasilla/mpg22/feros/?lang}} for the visible range and with IUE\footnote{International Ultraviolet Explorer: \url{https://archive.stsci.edu/iue/}} for the ultraviolet, and corresponds to those ones used previously by \citet{bouret12}.
	
\begin{figure*}[t!]
		\centering
		\includegraphics[width=\linewidth]{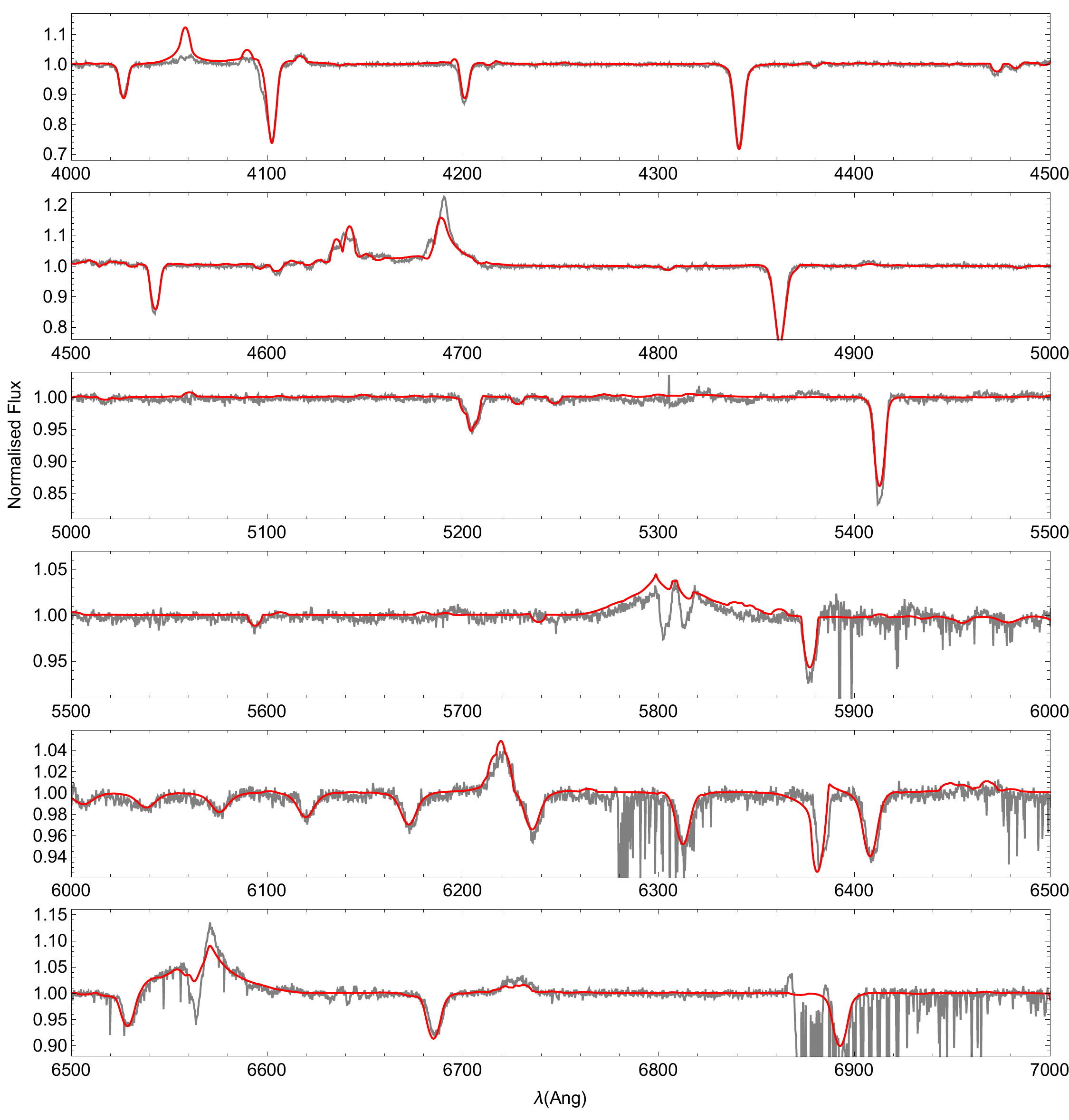}
		\caption{\small{Fit of \texttt{T41lamb09} (see Table~\ref{lambertsolutions}) over the observed FEROS spectrum for the visible range from 4\,000 to 7\,000 $\AA$.}}
		\label{visiblefit}
	\end{figure*}

\begin{figure*}[t!]
		\centering
		\includegraphics[width=\linewidth]{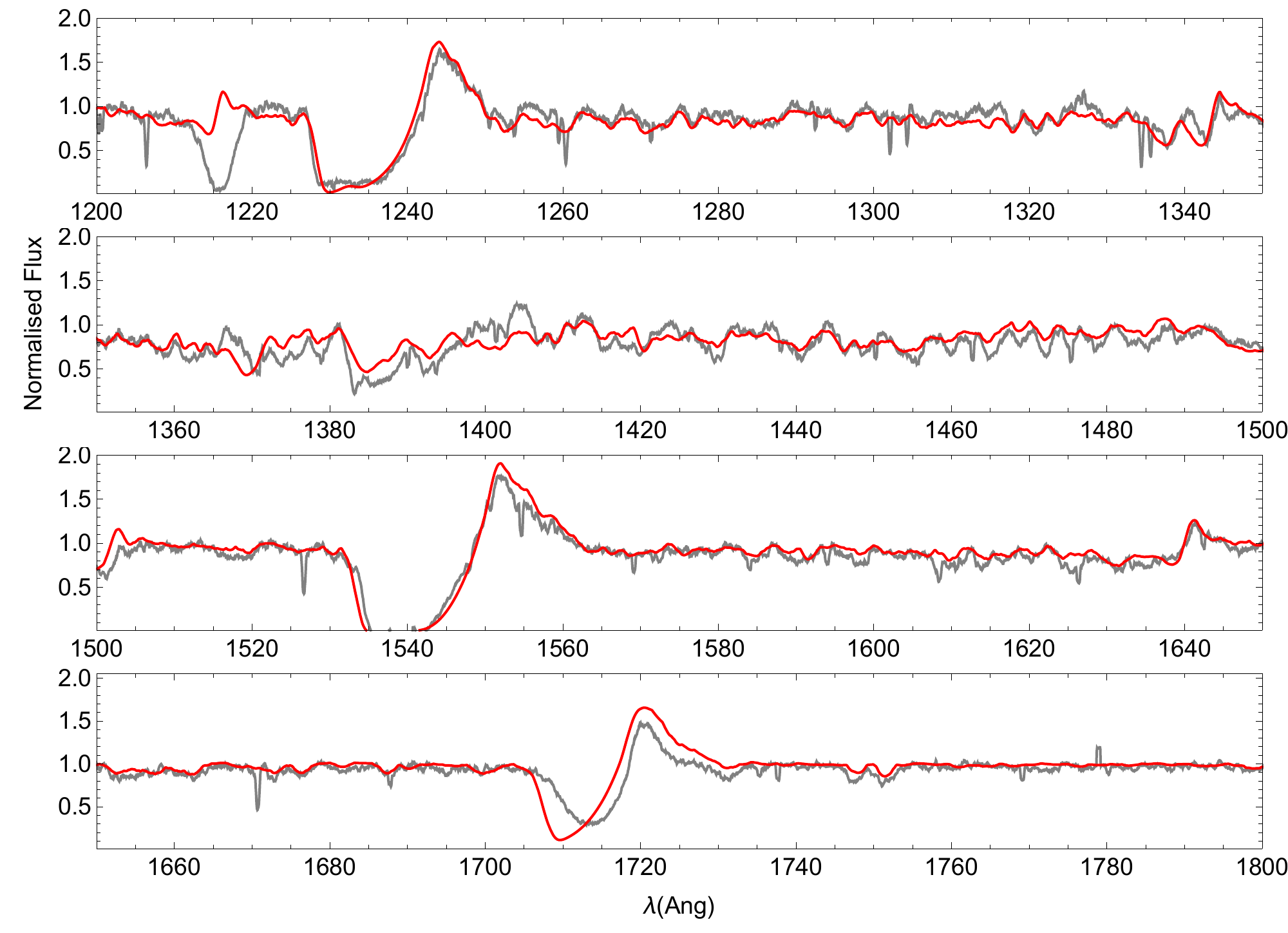}
		\caption{\small{Fit of \texttt{T41lamb09} (see Table~\ref{lambertsolutions}) over the observed IUE spectrum for the ultraviolet range from 1\,200 to 1\,800 $\AA$.}}
		\label{uvfit}
	\end{figure*}

	The two main criteria to be satisfied on the spectral fitting are: the blue side of the P-Cygni profile C IV 1548 for the ultraviolet range, and H$\alpha$ for the optical.
	C IV 1548 is a good indicator of $v_\infty$ in O, B and WR stars \citep{prinja90}, and the blue side of this P-Cygni profile can be reproduced using CMFGEN \citep{alex14}.
	Because of the direct proportionality between line-acceleration and self-consistent terminal velocity, it is possible to find a set of $(\dot M,f_\infty)$ capable to fit $v_\infty$, which has a value around $\sim2\,300$ km s$^{-1}$.
	This, together with the constraint coming from the fitting of the H$\alpha$ profile (which is proportional to the homogeneous mass-loss rate, tabulated in Table~\ref{lambertsolutions}) lead us to an unique combination of $\dot M$ and $f_\infty$ capable to satisfy both criteria.
	Following this, we proceed then to analyse the spectra of our Lambert-solutions, and it is found that the best fit (done by-eye) corresponds to the model \texttt{T41lamb09}, with $\dot M=3.5\times10^{-6}$ $M_\odot$~yr$^{-1}$ and $f_\infty=0.2$ (see Table~\ref{lambertsolutions}).
	The synthetic and observed spectra are shown in Fig.~\ref{visiblefit} and Fig.~\ref{uvfit}.

	First comment to be done is, given the initial set of stellar parameters shown in Table~\ref{initialmodel}, the self-consistent solution provided us a mass-loss rate enhanced by a factor of $\sim1.3$ but using a value for clumping half than before.
	This implies an almost similar value for the "homogeneous" mass-loss rate ($\dot M/\sqrt{f_\infty}$): $7.8\times10^{-6}$ $M_\odot$ yr$^{-1}$ for us, $8.5\times10^{-6}$ $M_\odot$ yr$^{-1}$ for \citet{bouret12}.
	In spite of this result, since this initial Lambert-converged solution was obtained starting from a set of stellar parameters different to those finally determined for the star, a more deep analysis must be done starting from the stellar parameters used by \citet{bouret12} to fit the spectra of $\zeta$-Puppis, in order to include a comparison of spectral fits, i.e., new initial stellar parameters are $T_\text{eff}=40$ kK, $\log g=3.64$ and $R_*=18.7\,R_\odot$.
	\begin{table*}[t!]
		\centering
		\begin{tabular}{cccccc}
			\hline\hline
			Name & $\dot M_\text{clump}$ & $f_\infty$ & $\dot M_\text{hom}$ & $ v_{\infty,\text{Lamb}}$\\
			& ($M_\odot$ yr$^{-1}$) & & ($M_\odot$ yr$^{-1}$) & (km s$^{-1}$)\\
			\hline
%			\texttt{lamb164} & $3.0\times10^{-6}$ & 0.2 & $6.7\times10^{-6}$ & 2\,300 & 0.95\\
			\texttt{T40lamb01} & $3.0\times10^{-6}$ & 0.2 & $6.7\times10^{-6}$ & 2\,260\\%lamb175
			\texttt{T40lamb02} & $3.0\times10^{-6}$ & 0.1 & $9.5\times10^{-6}$ & 2\,400\\%lamb184
			\texttt{T40lamb03} & $3.3\times10^{-6}$ & 0.1 & $6.7\times10^{-6}$ & 2\,320\\%lamb196
			\texttt{T40lamb04} & $3.0\times10^{-6}$ & 0.15 & $7.7\times10^{-6}$ & 2\,310\\%lamb197
			\texttt{T40lamb05} & $2.7\times10^{-6}$ & 0.1 & $8.5\times10^{-6}$ & 2\,620\\%lamb200
			\hline
		\end{tabular}
		\caption{\small{Converged models, given a set of different values for mass-loss rate and clumping filling factor for stellar parameters $T_\text{eff}=40$ kK, $\log g=3.64$ and $R_*=18.7R_\odot$.}}
		\label{lambertsolutionst40}
	\end{table*}
	
\begin{figure*}[t!]
		\centering
		\includegraphics[width=\linewidth]{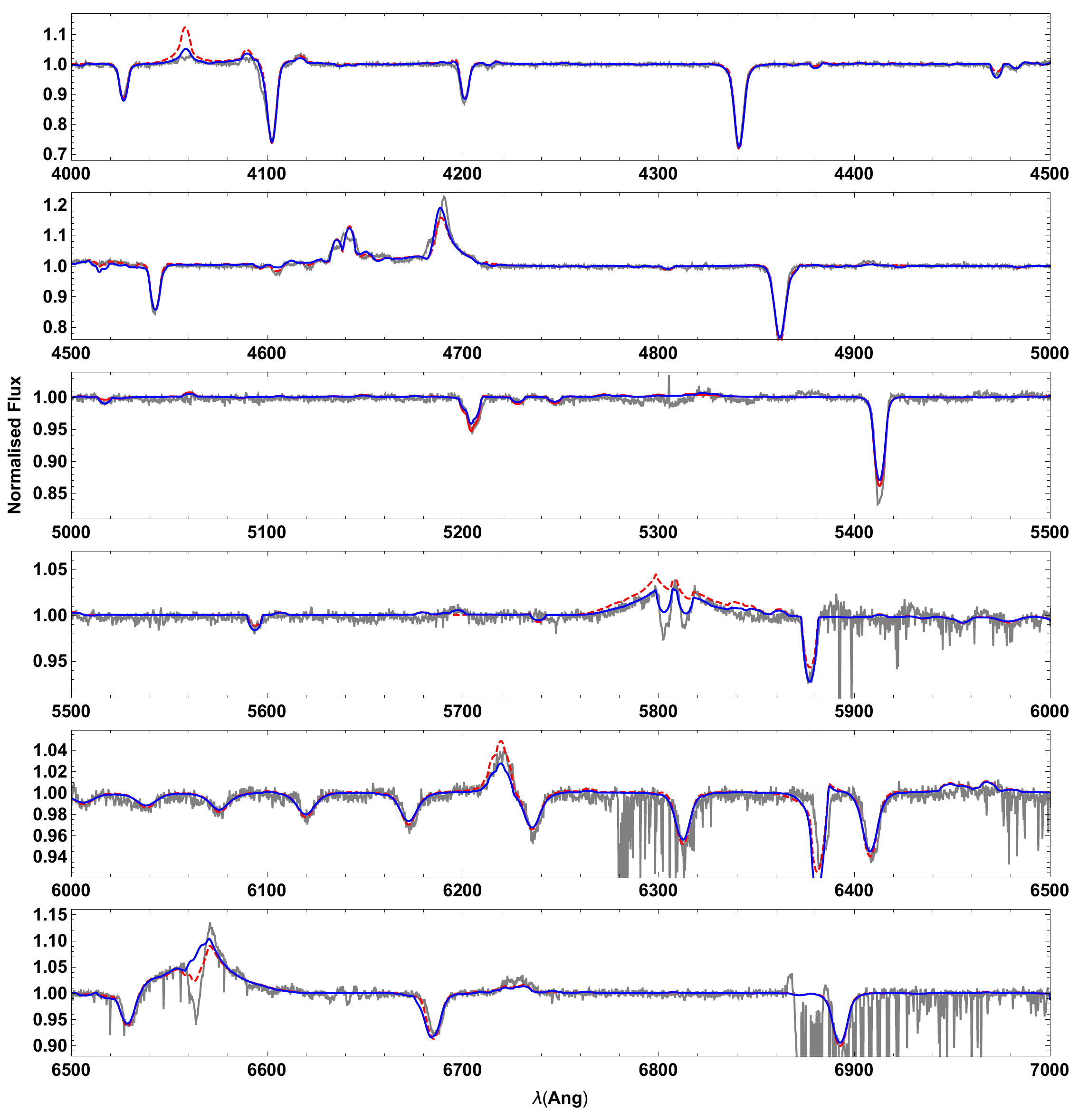}
		\caption{\small{Fit of \texttt{T40lamb04} from Table~\ref{lambertsolutionst40} (blue solid line) over the observed FEROS spectrum for the visible range from 4\,000 to 7\,000 $\AA$. Former spectrum for \texttt{T41lamb09} (red dashed line) is included behind, for comparison purposes.}}
		\label{visiblefit093197}
	\end{figure*}

\begin{figure*}[t!]
		\centering
		\includegraphics[width=\linewidth]{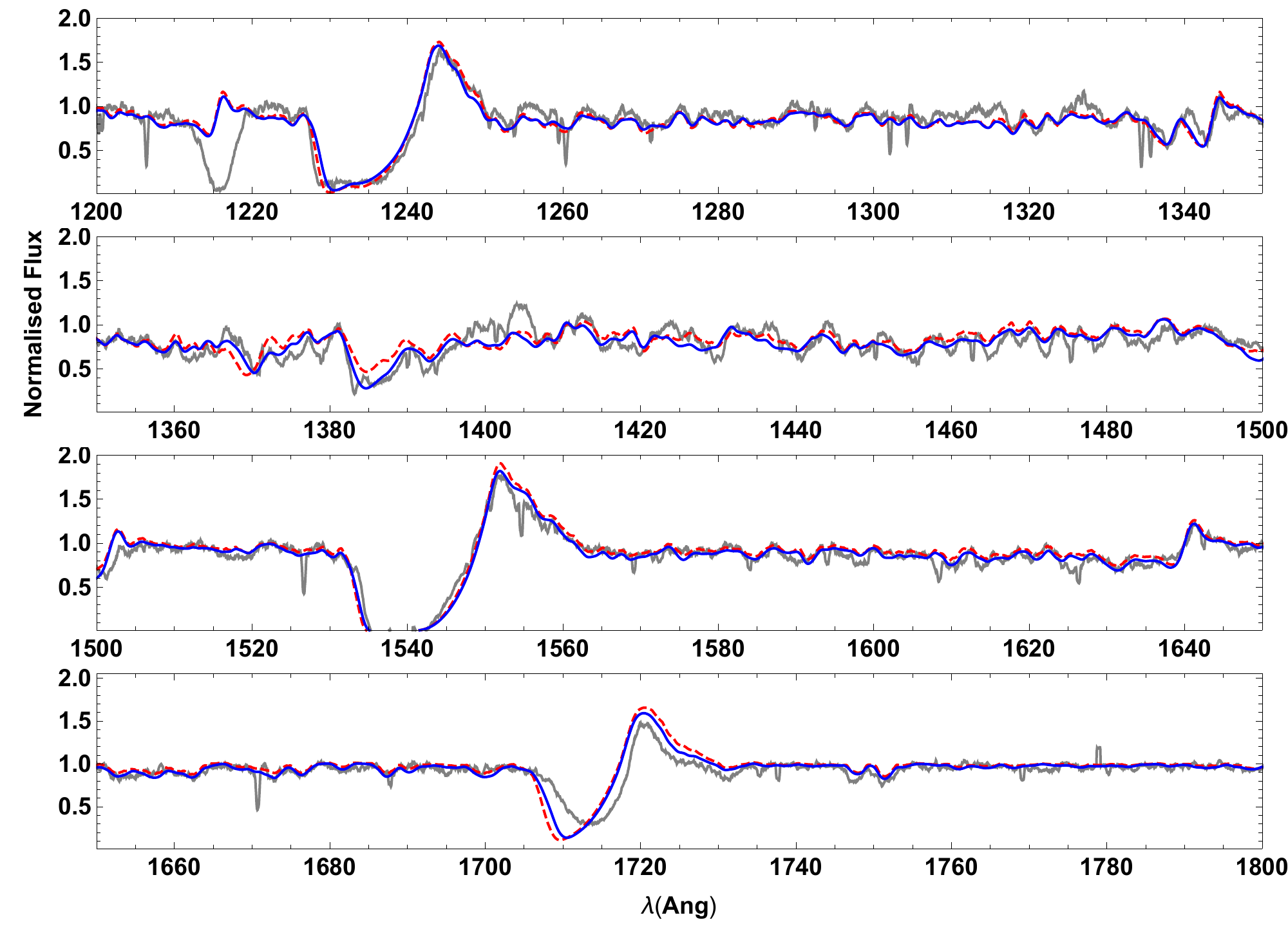}
		\caption{\small{Fit of \texttt{T40lamb04} from Table~\ref{lambertsolutionst40} (blue solid line) over the observed IUE spectrum for the ultraviolet range from 1\,200 to 1\,800 $\AA$. Former spectrum for \texttt{T41lamb09} (red dashed line) is included behind, for comparison purposes.}}
		\label{uvfit093197}
	\end{figure*}

	Summary of Lambert-converged CMFGEN models using this new set of stellar parameters are presented on Table~\ref{lambertsolutionst40}.
	In this case, we found that the best model corresponds to \texttt{T40lamb04}, whose spectra for optical and ultraviolet are presented in Fig.~\ref{visiblefit093197} and Fig.~\ref{uvfit093197} respectively.
	Because this new model presents a slightly better fit for some lines such as H$\alpha$ and He II 4684, we will consider this new self-consistent solution given by  $T_\text{eff}=40$ kK, $\log g=3.64$ and $R_*=18.7\,R_\odot$ as stellar parameters to be our best fit for $\zeta$-Puppis.

%_____Comparison with Sander_________________________________________________________________________
%\vspace{5mm}
\subsection{Comparison with previous self-consistent studies}
	The search for a full self-consistent solution (coupling line-acceleration, hydrodynamics and radiative transfer) for the wind of massive stars has been approached previously by other studies \citep{puls00,kudritzki02}.
	In particular, we emphasise the work done by \citet{sander17}, where another iterative procedure was implemented in order to obtain a self-consistent solution for the wind hydrodynamics of $\zeta$-Puppis under the non-LTE regime, using the radiative transfer code PoWR \citep{grafener02,hamann03}.
	In that study, radiative acceleration was also obtained from the output of the radiative transfer solution (PoWR for them, CMFGEN for us), and that acceleration was re-used to obtain a new hydrodynamic profile.
	
	Although their study shares a similar philosophy with ours, there are important differences.
	First, our new velocity profile is calculated by means of the Lambert $W$-function, whereas their $v(r)$ is recalculated by updating the stratification of the wind.
	The advantage of $W$-Lambert procedure yields in the fact that this is a mathematical tool which ensures the existence of a unique solution at the end, i.e., final converged Lambert-model does not depend on the initial velocity profile used ($ v_\infty$ and $\beta$).
	Besides, our procedure considers iterative changes only in the velocity profile, letting the stellar parameters, the mass-loss rate and the clumping factors as free.
	This could be considered as a disadvantage taking into account that \citet{sander17} did their procedure in order to obtain a final result for both stellar and wind parameters; however, the relaxation imposed by us allows the search by eye inspection of the best spectral fit, reducing the number of free parameters (hydrodynamic will depend now from the initial stellar parameters and the free $\dot M$), whereas Sander's prescription does not allow a relaxation because every parameter is being recalculated.
	Finally, the methodology introduced by us is focused on the line-acceleration $g_\text{line}$ instead radiative acceleration $g_\text{rad}$, which gives the chance to compare these new results with previous formulations based on m-CAK framework (Chapter~\ref{alfakdelta27}).
	\begin{table*}[htbp]
		\centering
		%\resizebox{\textwidth}{!}{
		\begin{tabular}{cccc}
			\hline\hline
			& \citet{sander17} & Chapter~\ref{alfakdelta27} & This work\\
			\hline
			RT code & \textsc{PoWR} & FASTWIND & CMFGEN\\
			Hydro method & – & \textsc{HydWind} & Lambert $W$-function\\
			$T_\text{eff}$ (kK) & 40.7 & 40 & 40\\
			$\log g$ & 3.63 & 3.64 & 3.64\\
			$R_*/R_\odot$ & 15.9 & 18.7 & 18.7\\
			$v_\infty$ (km s$^{-1})$ & 2\,046 & 2\,700 & 2\,310\\
			$\dot M$ ($M_\odot$ yr$^{-1}$) & $1.6\times10^{-6}$ & $5.2\times10^{-6}$ & $3.0\times10^{-6}$\\
			$f_\infty$ ($=D_\infty^{-1}=f_\text{cl}^{-1}$) & 0.1 & 0.2 & 0.15\\
			$\dot M_\text{hom}$ ($M_\odot$ yr$^{-1}$) & $5.1\times10^{-6}$ & $1.1\times10^{-5}$ & $7.7\times10^{-6}$\\
			\hline
		\end{tabular}%}
		\caption{\small{Summary of self-consistent solutions performed by \citet{sander17}, \citet{alex19} and this present study. These three models assume the same abundances as \citet{bouret12}. Radiative transfer code used to perform the respective synthetic spectra is also indicated, together with the code/methodology used to calculate the hydrodynamics. Equivalencies between filling factor and clumping factor $f_\text{cl}$ (used by FASTWIND) and $D_\infty$ (used by \textsc{PoWR}) were made assuming the interclump medium is void, following \citet{sundqvist18}.}}
		\label{zpuppissolutions}
	\end{table*}
	
	For those reasons, we select these two previous studies to be compared with the results given by the Lambert-procedure.
	Comparison of these three self-consistent solutions for $\zeta$-Puppis is summarised in Table~\ref{zpuppissolutions}.
	As an initial comment, we remark the discrepancy between the wind parameters derived from \citet{sander17} and those derived by us, even when differences on stellar parameters are not so significative.
	First, for the case of the terminal velocity, their value lies below typical values obtained by spectral fitting: 2\,250 km s$^{-1}$ by \citet{puls06} and 2\,300 km s$^{-1}$ by \citet{bouret12}.
	This is an important detail, because our self-consistent wind parameters under Lambert-procedure were calculated satisfying the criterium to find an accurate terminal velocity by fitting C IV 1548 line (i.e., self-consistent mass-loss rate would have been different if we were looking for solutions with $v_\infty=2\,046$ km s$^{-1}$).
	However, since it is derived from Tables~\ref{lambertsolutions} and~\ref{lambertsolutionst40} that mass-loss rate is inversely proportional to the resulting terminal velocity for a self-consistent hydrodynamics, it is clear that a solution with lower $v_\infty$ it would have generated an even higher $\dot M$ than the one tabulated on Table~\ref{zpuppissolutions}.
	Second, for mass-loss rates we found our self-consistent value doubles the Sander's one. 
	Although we employ a less deep clumping factor, our "homogeneous" $\dot M$ is a $\sim50\%$ higher.
	Nevertheless, because the fit on H$\alpha$ (Fig.~\ref{visiblefit093197}) is more accurate using our wind parameters than those given by \citet[][see their Fig. 9]{sander17}, we consider our self-consistent solution as more reliable.
	
	The comparison with the self-consistent solutions under m-CAK theory performed by \citet{alex19} is more extended, because both studies share the same stellar parameters.
	Therefore, following analysis is focused in the differences between this current methodology using Lambert-function with former m-CAK methodology, which had three main aspects.
	In Chapter~\ref{alfakdelta27} we:
	\begin{itemize}
		\item used a simplified "quasi-NLTE" scenario for the treatment of atomic populations, following formulations performed by \citet{mazzali93} and \citet{puls00}; whereas in this work we solve the proper statistical equilibrium equations when CMFGEN is executed.
		\item used a flux field calculated by \textsc{Tlusty} \citep{lanz03}, which uses the plane-parallel approximation; whereas our flux field is calculated within CMFGEN run.
		\item did not consider effects from clumping upon the resulting $g_\text{line}$; whereas, from Fig.~\ref{clumpinitial}, it is clear that line-acceleration changes when mass-loss rate keeps constant but clumping factor is modified.
		\item includes the effects of rotation inside the execution of \textsc{HydWind}, whereas Lambert-procedure have not included any influence of rotation.
	\end{itemize}
	
	Some of these factors might explain the differences between wind parameters for $\zeta$-Puppis given by Paper I and current study.
	However, it is important to remark again that current results are semi-theoretical: hydrodynamic is self-consistent with line-acceleration and velocity profile is dependent on mass-loss rate, but this latter is set as a free parameter to be constrained by the spectral fit.
	\begin{figure}[t!]
		\centering
		\includegraphics[width=0.65\linewidth]{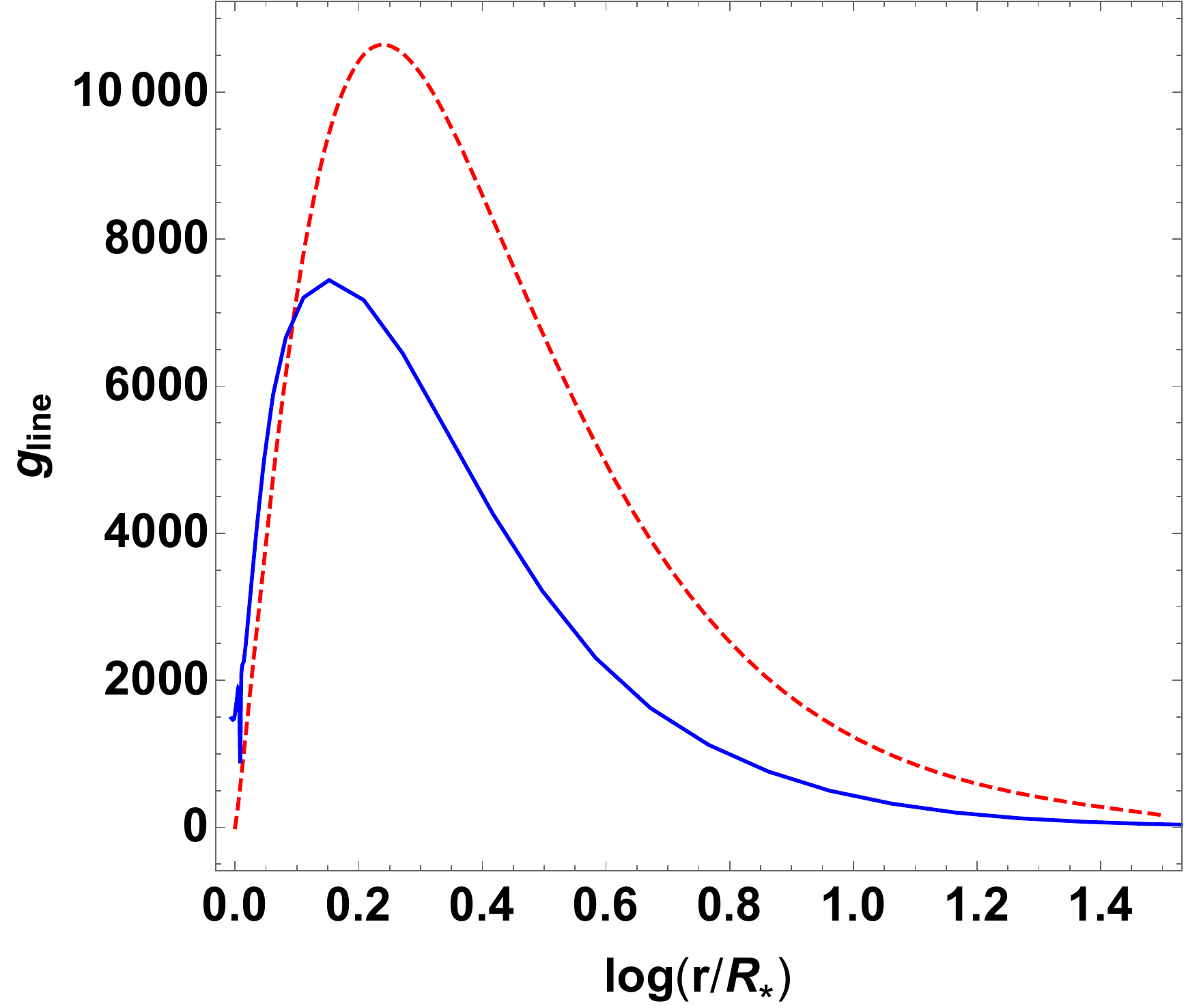}
		\caption{\small{Line-accelerations given by the self-consistent Lambert-procedure (continue blue), compared with the one given by the self-consistent .}}
		\label{gline_cmfcak}
	\end{figure}

	Differences on the self-consistent wind parameters between Chapter~\ref{alfakdelta27} and this present chapter are related with the differences on the resulting self-consistent line-acceleration.
	Fig.~\ref{gline_cmfcak} shows the comparison of both $g_\text{line}$, the obtained for the self-consistent Lambert-solution \texttt{T40lamb04} and the one obtained by the m-CAK self-consistent solution for $\zeta$-Puppis presented on the Chapter~\ref{alfakdelta27} by means of:
	\begin{equation}
		g_\text{line}=\frac{\sigma_e\mathcal F}{c}k\,t^{-\alpha}\left(\frac{N_{e,11}}{W(r)}\right)^\delta\;,
	\end{equation}
	with $k=0.120$, $\alpha=0.655$ and $\delta=0.039$ \citep[see Table 5 of][]{alex19}.
	Line-acceleration obtained with m-CAK prescription is higher than $g_\text{line}$ obtained with the Lambert-procedure, which could be explain then the resulting mass-loss rate and terminal velocity presented on the Paper I.
	
	Therefore, it is possible to claim that m-CAK theory predicts a higher line-acceleration than a fully calculated hydrodynamics.
	The explanation may lie in the three points previously stated: m-CAK prescription for the calculation of the line-force parameters (and hence $g_\text{line}$) uses different atomic populations and different flux field, together with a different treatment for the clumping.
	
	This could lead us to conclude that differences between these two prescriptions (Chapter~\ref{alfakdelta27} and this one) lies on the line-acceleration only, but the trend observed on Tables~\ref{lambertsolutions} and~\ref{lambertsolutionst40} shows that this is not true.
	Even if the resulting $g_\text{line}$ calculated by CMFGEN with parameters $\dot M=5.2\times10^{-6}$ $M_\odot$ yr$^{-1}$ and $f_\infty=0.2$, the resulting self-consistent hydrodynamics would have a terminal velocity far below $\sim2\,200$ km s$^{-1}$.
	Hence, the wind parameters predicted by the m-CAK prescription are not recoverable by the Lambert $W$-function.
	
	The influence of rotational effects is interesting, because it is well known that rotation enhances the values for mass-loss rate on the equator of the star.
	Hydrodynamics calculated with \textsc{HydWind} in \citet{alex19} used a value for the normalised stellar angular velocity\footnote{Normalised stellar angular velocity is defined as:
	$$\Omega=\frac{v_\text{rot}}{v_\text{crit}}\;,$$
	with $v_\text{crit}$ as defined in Eq.~\ref{changeofvariables}.} of $\Omega=0.39$ in order to reproduce the known value of $v\sin i=210$ km s$^{-1}$ for $\zeta$-Puppis \citep{bouret12}, whereas Lambert-procedure do not consider rotational effects because CMFGEN presents a 1D geometry.
	According to \citet{venero16}, mass-loss rates increase their values in a factor of $\sim1.2-1.3$ for $\Omega=0.4$, which leads us to think that self-consistent value for $\dot M$ obtained on Chapter~\ref{alfakdelta27} would decrease to $\sim4\times10^{-6}$ $M_\odot$ yr$^{-1}$ if $\Omega=0$, closer to the $\dot M$ determined by the Lambert-procedure.
	
	However, it is important to remind that both m-CAK and Lambert prescriptions, despite both obtain self-consistent solutions, work under different philosophies.
	On Chapter~\ref{alfakdelta27}, m-CAK prescription calculates its own self-consistent mass-loss rate, which is later tested by spectral analysis in order to \textit{check how near or far} falls from the real solution.
	On the other hand, Lambert-procedure presented on this work calculates a self-consistent solution for the wind hydrodynamics letting the mass-loss rate as a free parameter, which is later \textit{constrained by the spectral fitting}.
	Besides, let us consider the fact that we are working with two different radiative transfer codes (FASTWIND and CMFGEN) which have discrepancies between them \citep{massey13}.
	Therefore, it is expectable that both methodologies do not find exactly the same values, but plausible ones deserving to discuss them.

%_____Conclusions for Lambert-procedure______________________________________________________________________
\section{Conclusions for Lambert-procedure}
	In the present Chapter we have presented a methodology to calculate self-consistent hydrodynamics beyond the m-CAK prescription presented on \citet[][Chapter~\ref{alfakdelta27}]{alex19}.
	For that purpose, we have used the CMFGEN radiative transfer code and the mathematical Lambert $W$-function.
	This function allows us to analytically solve the equation of motion (Eq.~\ref{eqmomentum}), using the line-acceleration $g_\text{line}$ given by the solution of the radiative transfer equation on CMFGEN, to provide us a new velocity profile.
	This procedure is iterated (Lambert-procedure) until the convergence.
%	This methodology presents an alternative to Paper I, where some assumptions such as quasi-NLTE treatment for atomic populations were considered.
%	On the other side, 
	Hydrodynamic solution given by Lambert-procedure is valid only in the supersonic region of the wind, whereas subsonic region needs to be rescaled in order to obtain a continuous solution.
	
	Lambert-procedure has proved to converge into the same hydrodynamic solution, independent of the initial velocity profile (i.e., terminal velocity and $\beta$-value) chosen (Fig.~\ref{hydros_alt}).
	On the other hand, because Eq.~\ref{eqmotion} is explicitly independent of density, mass-loss rate is not recovered by the self-consistent iterations and then it needs to be set as a free parameter.
	However, dependence on density (and hence dependence on mass-loss rate too because of equation of continuity Eq.~\ref{eqcontinuity}) is \textit{implicitly included} in the modified equation of momentum by means of the line-acceleration term (Fig.~\ref{mdotsinitial}).
	Line-acceleration also depends on the clumping factor (Fig.~\ref{clumpinitial}), thus the final self-consistent hydrodynamic obtained by the Lambert-procedure depends on the initial values for $\dot M$ and $f_\infty$ introduced to the CMFGEN models.
	Therefore, given a specific set of stellar parameters (effective temperature, surface gravity, stellar radius and abundances), a range of self-consistent hydrodynamics is found for different mass-loss rates and different clumping effects\footnote{This does not mean that there are different possible hydrodynamics for a specific set of stellar parameters. We reproduce a range of solutions for different combinations of $(\dot M,f_\infty)$ because Lambert-procedure is not capable to calculate a proper density for the wind, but it is clear that only one mass-loss rate and only one clumping factor are the correct ones for the initial set of stellar parameters.}.
	
	The correct mass-loss rate and clumping factor for the self-consistent hydrodynamics is determined by spectral fitting.
	Particularly, we look for a solution capable to accurately fit the blue wing of C IV $\lambda$1548 in the ultraviolet range (indicator for terminal velocity) and the emission line of H$\alpha$ (indicator for mass-loss rate).
	Following this recipe, we have found a self-consistent hydrodynamic for the set of stellar parameters presented on Table~\ref{initialmodel}, together with another self-consistent hydrodynamic for the set of stellar parameters determined by \citet{bouret12}.
	Best-fit spectrum found for $\zeta$-Puppis using a self-consistent solution, is obtained for this stellar parameters.
%	For the initial set of stellar parameters tabulated on Table~\ref{tableini}, we have found a solution with $\dot M=3.5\times10^{-6}$ $M_\odot$ yr$^{-1}$ and $f_\infty=0.2$, and the respective spectral fits are shown in Fig.~\ref{visiblefit} and Fig.~\ref{uvfit}.
	
%	In order to properly compare the results with the last two studies concerning self-consistent hydrodynamics involving $\zeta$-Puppis \citep{sander17,alex19}, we change the stellar parameters to $T_\text{eff}=40$ kK, $\log g=3.64$ and $R_*/R_\odot=18.6$, the same determined by \citet{bouret12} and used on the Paper I.
%	This time, wind parameters found are $\dot M=3.0\times10^{-6}$ $M_\odot$ yr$^{-1}$ and $f_\infty=0.15$, with the respective spectral fit in Fig.~\ref{visiblefit093197} and Fig.~\ref{uvfit093197}.
%	These wind parameters are far different for those fitted by \citet[][$\dot M=2.0\times10^{-6}$ $M_\odot$ yr$^{-1}$ and $f_\infty=0.05$]{bouret12}, but they produce a close similar spectral fit (compare with their Figures A.4 and A.5).
%	
	Compared with \citet{sander17}, where all stellar and wind parameters are recalculated inside the iterative process letting then no free parameters, Lambert-procedure presented in this work has the advantage of letting some parameters such as the stellar ones and mass-loss rate as free.
	Hence, these parameters can be tuned independently in order to look for the self-consistent hydrodynamics fitting better the spectral observations.
%	We observe our mass-loss rate is higher by a factor of 2, whereas their terminal velocity is far below the classical value $\sim2\,300$ km s$^{-1}$ found for $\zeta$-Puppis \citep{puls06,bouret12}.
	Therefore, following the studied correlations found for self-consistent solutions, such as the inverse proportionality between initial mass-loss rate and final terminal velocity (see Table~\ref{lambertsolutions} and Table~\ref{lambertsolutionst40}), it is possible to find the best combination of $(\dot M,f_\infty)$.
	
	As a result of the comparison with Chapter~\ref{alfakdelta27}, we find that both our mass-loss rate and our terminal velocity are lower, despite stellar parameters and the atomic information are the same for both cases.
	Even though these differences could be partially explained by the differences on the self-consistent line-accelerations found (indicating then that differences on final wind parameters would lie on the difference of the methodologies to calculate $g_\text{line}$), a more detailed analysis is needed in order to support this hypothesis.
	
	Finally, despite the big computational effort required in an iterative loop involving CMFGEN, Lambert-procedure has demonstrated to be a confident methodology to find hydrodynamically self-consistent solutions for a stellar wind.
	Follow-up research would be focused on the expansion of this prescription for more stars beyond $\zeta$-Puppis.

%_____FITTING OBSERVABLE SPECTRA WITH SELF-CONSISTENT SOLUTIONS_______________________________________
\chapter[Spectra with Self-consistent Solutions]{Fitting Spectra of Massive Stars with First Self-consistent Solutions}\label{rob}
	In this chapter, we will apply the self-consistent methodology, developed, calculated and analysed using the m-CAK prescription in Chapter~\ref{alfakdelta27}.
	To test the accuracy of these spectral fits is a necessary step, because one of the goals of using a new prescription capable of self-consistently describing the hydrodynamics of hot massive stars is to determine both stellar and wind parameters with the help of spectral fitting.
	
	For that purpose, we use the observed spectral data from a set of hot massive stars obtained with the \textsc{Hermes} spectrograph\footnote{High-Efficiency and high-Resolution Mercator Echelle Spectrograph: \url{https://fys.kuleuven.be/ster/instruments/the-hermes-spectrograph}}.
	Particularly, we analyse the O type stars HD 57682, HD 195592, 9 Sge and HD 192639.
	This set of stars have been previously used in previous studies \citep{grunhut17,debecker10,martins15,bouret12}, who had already constrained some of the stellar parameters such as effective temperature and stellar mass; therefore, they represent a starting point in our search of stellar and wind parameters.
	However, different from Figures~\ref{comp413540} to~\ref{comp858687}, where the spectral fit was focused on the search for the best clumping filling factor keeping the stellar parameters fixed, here we will proceed modifying \textit{all the stellar parameters} in order to find the best fit to each spectrum.
	In some cases, the new found stellar parameters may lie close to those found in previous studies, whereas in others they are remarkably different.
	The consequences and conclusions are presented at the end of this chapter.
	
	Most of this work was done during a two months internship at the Royal Observatory of Belgium in Brussels, under the supervision of Dr. Alex Lobel, as part of the \textbf{Physics Of Extremely Massive Stars} (POEMS) project\footnote{\url{http://stelweb.asu.cas.cz/~kraus/POEMS/Secondments.html}}.
	It is important to remark that, this research work is still in progress.
	For future work, many other stellar spectra should be included in our study, for strengthening confidence in our self-consistent m-CAK procedure.

\section{Methodology}\label{hermesmethodology}
	Firstly, it is important to emphasise the differences among the parameters that will be constrained by the spectral fit.
	Because of the self-consistent procedure, the wind parameters (mass-loss rate and terminal velocity) are depending on the set of initial stellar parameters and cannot be individually modified.
	\begin{itemize}
		\item \textit{Stellar parameters} ($T_\text{eff}$, $\log g$, $R_*$ and abundances) are set at the beginning of the m-CAK prescription, and they directly determine the final self-consistent values for mass-loss rate and terminal velocity.
		Therefore, their modifications are the most complicated point because it requires the execution of a new iterative procedure from the beginning.
		For this reason, we will call them \textit{first-order modifications} (FOM) and they are the most important parameters to be fine-tuned during the spectral fitting.
		\item Parameters such as the clumping factor $f_\text{cl}$ and the turbulence velocity $v_\text{turb}$ are set at the beginning of the execution of FASTWIND but are not included in the self-consistent procedure.
		We will call them \textit{second-order modifications} (SOM), and they are modified after FOMs once the self-consistent solution has been achieved.
		\item Final parameters such as macro-turbulence $v_\text{macro}$ and the rotational velocity $v_\text{rot}$ are set at the end once the output of FASTWIND is obtained.
		We will call them \textit{third-order modifications} (TOM).
		Inside this group we should include the radial velocity $v_\text{rad}$, which fits the core of individual spectral lines.
	\end{itemize}

	\begin{figure}[t!]
		\centering
		\includegraphics[width=0.95\linewidth]{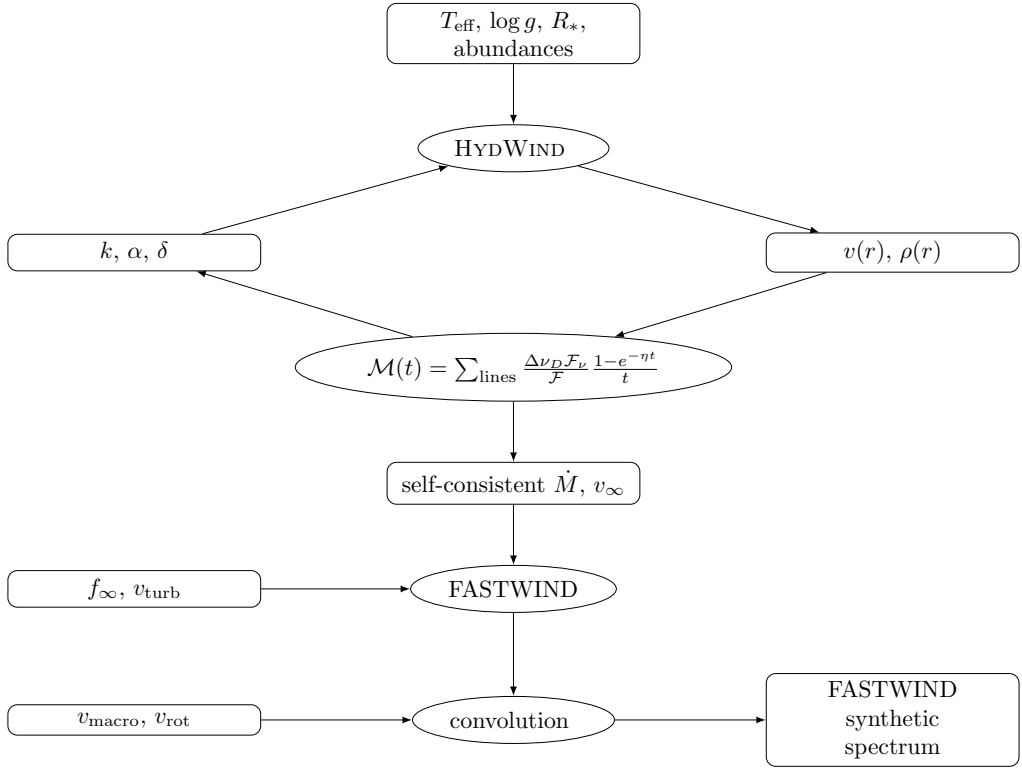}
		\caption{\small{Scheme of the self-consistent m-CAK procedure combined with FASTWIND, showing the stage where the different initial parameters are set.}}
		\label{mcakfastwind_flowchart}
	\end{figure}
	
	The full picture of the self-consistent procedure plus the execution of the FASTWIND code is shown in Fig.~\ref{mcakfastwind_flowchart}.
	Notice that the mass-loss rate and terminal wind velocity are not longer independent, because their values are determined by the FOMs at the beginning.
	A similar situation is valid for the line-force parameters represented in Fig.~\ref{mcakfastwind_flowchart}: final values for $k$, $\alpha$ and $\delta$ are a consequence of the modification of FOMs.
	Hence, tuning the line-force and wind parameters first requires a fine-tuning of the stellar parameters.
	
	It is important to remark, however, the order we presented is related to the nature of the parameters which does not represent an \textit{order step} for doing the fit.
	The order step results from the empirical effects on the lines; for example, the macro-turbulence velocity is only modified at the end (TOM), because its value is directly proportional to the width of He I lines, and it is therefore one of the first parameters to be constrained.
	
	For each star, we use parameters derived from previous studies as starting values.
	In case of lacking reliable information about a particular star, we use the calibration of stellar parameters by \citet{martins05}.
	The most important stellar parameters modified in the self-consistent procedure (effective temperature and stellar radius) are tuned simultaneously in order to keep the stellar luminosity constant, according with Stefan-Boltzmann definition of luminosity:
	\begin{equation}\label{stefan-boltzmann}
		\left(\frac{L_\text{ini}}{L_\text{fin}}\right)=\left(\frac{R_\text{ini}}{R_\text{fin}}\right)^2\left(\frac{T_\text{ini}}{T_\text{fin}}\right)^4=1\;.
	\end{equation}
	
	Moreover, we mention that inside the item \textit{abundances} for the FOM, we consider metallicity, He/H ratio and the individual element abundances.
	With regards to the last item, we consider as default the solar abundances of \citet{asplund09}.
	Due to the fact that spectra of O type stars are dominated by lines of hydrogen, helium, carbon and nitrogen mainly, it is currently not possible to individually fine-tune elements such as the Iron group (iron, cobalt and nickel).
	However, heavier elements.
	Elements contributing more to the line-acceleration $g_\text{line}$ are those having large number of permitted atomic transition lines.
	Because the iron group (hereafter FeG) represents $\sim80\%$ of the total line-force (see Table~\ref{atomicdata}) it is important to investigate if slight changes in the abundances of these elements can affect our final self-consistent solution.
	After running many self-consistent models using different abundance values, it is possible to state that the self-consistent mass-loss rate is directly proportional to abundance of iron, cobalt and nickel:
	\begin{equation}\label{masslossirongroup}
		\dot M\propto\frac{\text{Iron-group}_*}{\text{Iron-group}_\odot}\;\;.
	\end{equation}
	
	Hereafter, we will consider the simultaneous fine-tuning of the three elements Fe, Co and Ni together, represented as the modification of FeG$_*$/FeG$_\odot$.
	
\subsection{FASTWIND spectral lines}
	The group of lines calculated by FASTWIND is presented in Table~\ref{fastwindlines}.
	\begin{table}[h!]
		\centering
		\begin{tabular}{lllll}
			\hline\hline
			H$\alpha$ 6563 & H$\beta$ 4861 & \textcolor{blue}{H$\gamma$ 4340} & H$\delta$ 4101 & \textcolor{blue}{H$\epsilon$ 3970}\\
			He II 4200 & He II 4541 & He II 4686 & \textcolor{blue}{He II 6527} & \textcolor{blue}{He II 6683}\\
			\textcolor{blue}{He I 4387} & He I 4471 & He I 4713 & He I 4922 & \textcolor{blue}{He I 6678}\\
			\hline
		\end{tabular}
		\caption{\small{Lines included in FASTWIND. Blue lines are available for HD 192639 only.}}
		\label{fastwindlines}
	\end{table}
	
	As a first stage, we decide to use H and He lines only\footnote{It would be possible to include lines of silicon and nitrogen, but that requires rebooting the compilation of FASTWIND, together with a big computational effort to run these models. For practical purposes, we will limit this section to the usage of hydrogen and helium only.} for the spectral fitting in the optical and infrared wavelength range, whereas our observational data consist of optical spectra only.
	Thus, the lines to be fitted correspond to those ones available both in FASTWIND and in the observational data.
%	Since this work is still in progress, the parameters found by best fit may change in future as more lines are included in the analysis.
	In the specific case of our first three stars, we had available only nine spectral lines for the analysis, whereas HD 192639 was studied using fifteen spectral lines in FASTWIND (see Table~\ref{fastwindlines}).
	
	Given the large number of free parameters to fine-tune the spectral fitting (even when due to the self-consistent procedure we eliminated three of them, see Fig.~\ref{mcakfastwind_flowchart}), a deeper analysis is necessary to determine how different elements and different lines are affected by the modification.
	After many empirical tests and comparisons letting only one parameter vary while the other are held fixed, we have observed some general trends that help us to determine the best model fit: 
	\begin{itemize}
		\item Macro-turbulent velocity $v_\text{macro}$ is fitted to the width of He I lines, whereas $v_\text{rot}$ must be fitted to the shape of the wings of He II $\lambda$4200 and He II $\lambda$4541.
		\item $T_\text{eff}$ is fitted to He II $\lambda$4200 and He II $\lambda$4541, because these lines present only a significant variation when the effective temperature is modified.
		However, because of the change in the ionisation when the temperature is modified He I will also be affected.
		\item $\log g$ is fitted to the He I lines, particularly He I $\lambda$4471.
		At the same time, the modification of FeG$_*$/FeG$_\odot$ also affects the intensity of helium lines, specially the He I lines (changes of He II are of minor relevance compared to those produced by $T_\text{eff}$), hence both parameters are fitted simultaneously.
		\item $v_\text{turb}$ is fitted to He I $\lambda$4922.
		\item $f_\text{cl}$ is fitted to the Balmer lines, specially fitting the peak of H$\alpha$.
	\end{itemize}
	
	The radial velocity, could be obtained from the SIMBAD database\footnote{\url{http://simbad.u-strasbg.fr/simbad/}}, but because we ignore if the spectrographs were calibrated for $v_\text{rad}$ we fit it manually to make the core of the lines match in wavelength.

\section{Results}
	We present here the spectra of the best fit spectrum for each analysed star.
	Normalised spectra were provided by Dr. Alex Lobel.
	Comments about individual lines are also included.
	
\subsection{HD 57682}
	As a first step, we model the wind of the O 9.5 V star HD 57682.
	\citet{grunhut12} have studied this star, and found line profile variability (LPV) probably due to magnetic fields.
	We aim to fit spectra self-consistently, in order to check whether or not can reproduce part of the line profiles.
		
\begin{figure}[t!]
		\centering
		\includegraphics[width=0.9\textwidth]{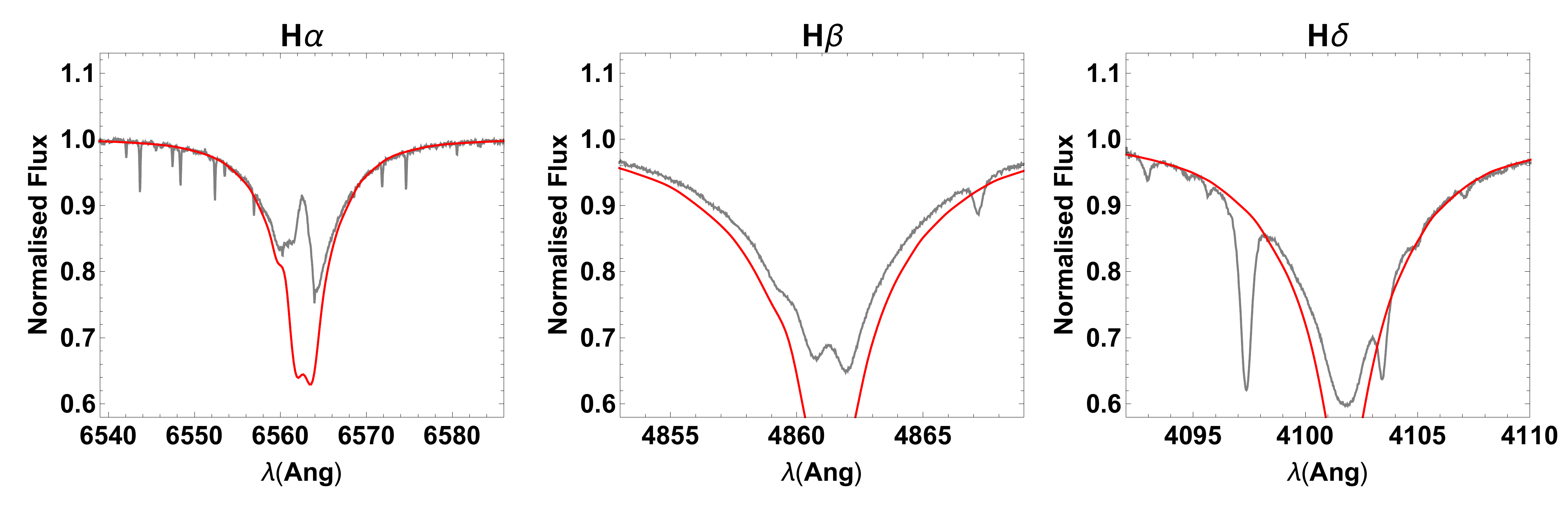}\\
		\includegraphics[width=0.9\textwidth]{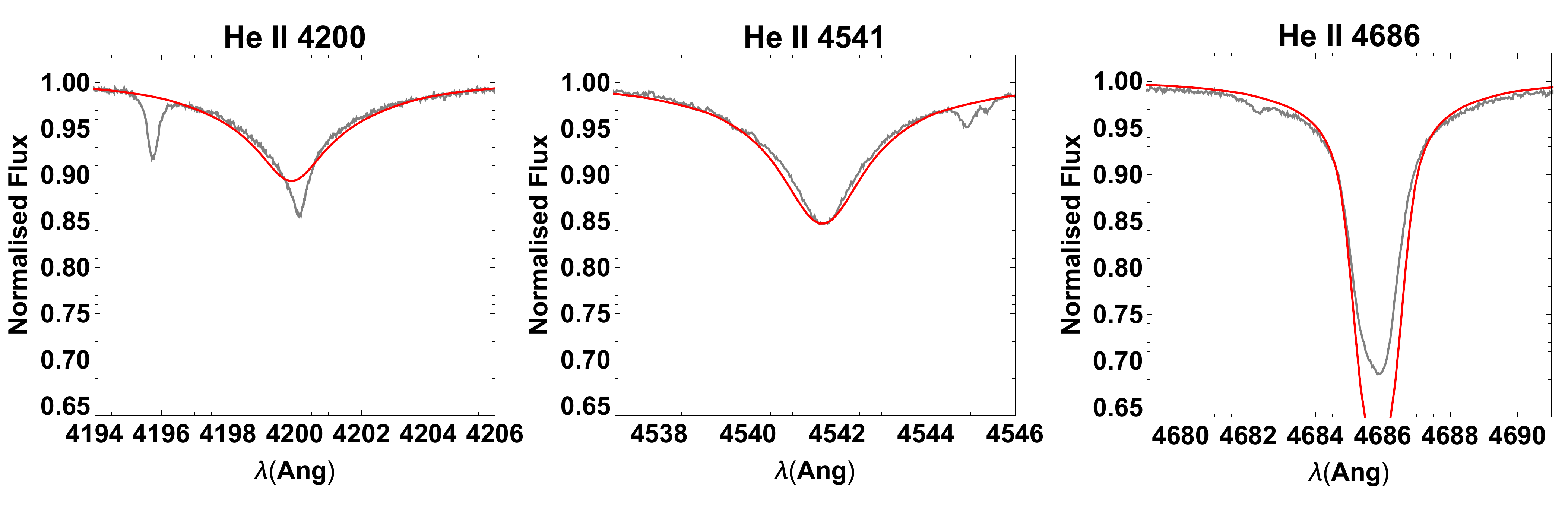}\\
		\includegraphics[width=0.9\textwidth]{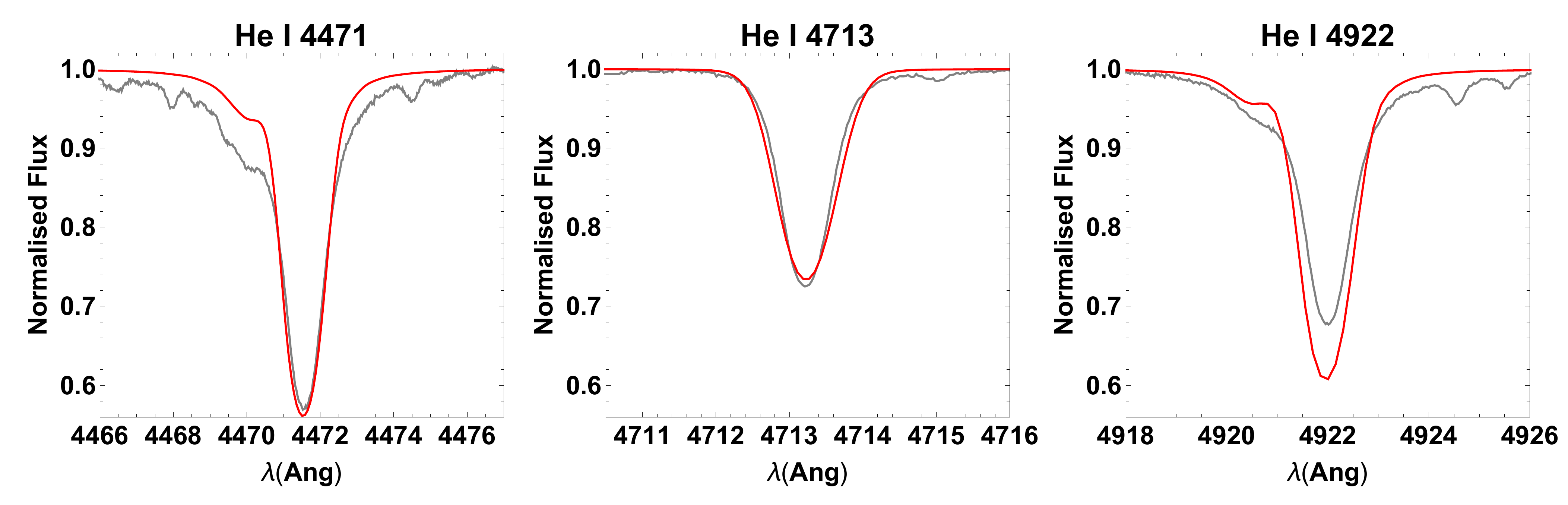}
		\caption{\small{Best fit for the star HD 57682, using $v_\text{rad}=10$ km s$^{-1}$.}}
		\label{fitfast240}
	\end{figure}

	Based on \citet{grunhut12}, we start with the following stellar parameters: $T_\text{eff}=35$ kK, $\log g=4.0$ and $R_*=7\,R_\odot$.
	In this particular case, the self-consistent solution obtained from this set of stellar parameters quickly leads to a good fit.
	The corresponding spectrum is presented in Fig.~\ref{fitfast240}, whereas the obtained parameters are shown in Table~\ref{tablefast240}.
	\begin{table}[h!]
		\centering
		\begin{tabular}{c|cc}
			\hline\hline
			& \multicolumn{2}{c}{Parameters HD 57682}\\%fastmod240
			& This work & \citet{grunhut12}\\
			\hline
			$T_\text{eff}$ (kK) & 35.0 & 35.0\\
			$\log g$ & 3.9 & 4.0\\
			$R_*/R_\odot$ & 9.0 & 7.0\\
			$M_*/M_\odot$ & $23.5$ & 17.0\\%\pm2.6
			$[\text{He/H}]$ & 0.085 & –\\
			$[\text{FeG}_*/\text{FeG}_\odot]$ & 1.0 & –\\
			$(k,\alpha,\delta)$ & $(0.097,0.613,0.022)$ & –\\
			$\dot M$ ($M_\odot$ yr$^{-1}$) & $1.15\times10^{-7}$ & $1.4\times10^{-9}$\\
			$v_\infty$ (km s$^{-1}$) & 2\,970 & 1\,200\\
			$f_\text{cl}$ & 25.0 & –\\
			$v_\text{rot}$ (km s$^{-1}$) & 10 & 10\\
			$v_\text{turb}$ (km s$^{-1}$) & 25 & –\\
			$v_\text{macro}$ (km s$^{-1}$) & 20 & 62\\
			\hline
		\end{tabular}
		\caption{\small{Summary of stellar and wind parameters used to fit HD 57682 (Fig.~\ref{fitfast240}). Values for $v_\text{rot}$ and $v_\text{macro}$ taken from \citet{grunhut17}.}}
		\label{tablefast240}
	\end{table}
	
	Methodology described on Section~\ref{hermesmethodology} helped us to find good fits to the helium lines, especifically to He II $\lambda$4541 (which has shown to be a very good constrainer the rotational velocity), He I $\lambda$4471 and He I $\lambda$4713, but not for He II $\lambda$4200 and He I $\lambda$4922.
	The case of He II $\lambda$4686 is special, because it presents a general (not only for HD 57682) anomalous behaviour making it impossible to fit, either in this work as in previous works.
	For the hydrogen lines, the line-wings are fairly well reproduced with the exception of H$\beta$, but even very high value for the clumping factor does not help to reproduce the abnormal shape of the line cores.
	This is because it is thought that the line cores show the presence of magnetic fields in HD 57682 \citep{grunhut17}, which are beyond the current capacity of the self-consistent fit procedure.
	
	Compared to \citet{grunhut17}, we find that our $v_\text{rot}$ is close to their $v\sin i$ (8 km s$^{-1}$) but their macro-turbulence is too large (65 km s$^{-1}$).

\subsection{HD 195592}
	We proceed obtaining self-consistent parameters for O9.7Ia \citep{sota11} star HD 195592.
	The initial mass $M_*=30\,M_\odot$, radius $R_*=30\,M_\odot$ and luminosity $L_*=3.1\times10^5$ $L_\odot$ were taken from \citet{debecker10}, who derived them from \citet{martins05}.
	This, and the following stars, corresponds to late O supergiants, which are near the lower threshold of validity of our self-consistent fit procedure ($T_\text{eff}\ge30$ kK and $\log g\ge3.4$, see Chapter~\ref{alfakdelta27}).
	\begin{figure}[t!]
		\centering
		\includegraphics[width=0.9\textwidth]{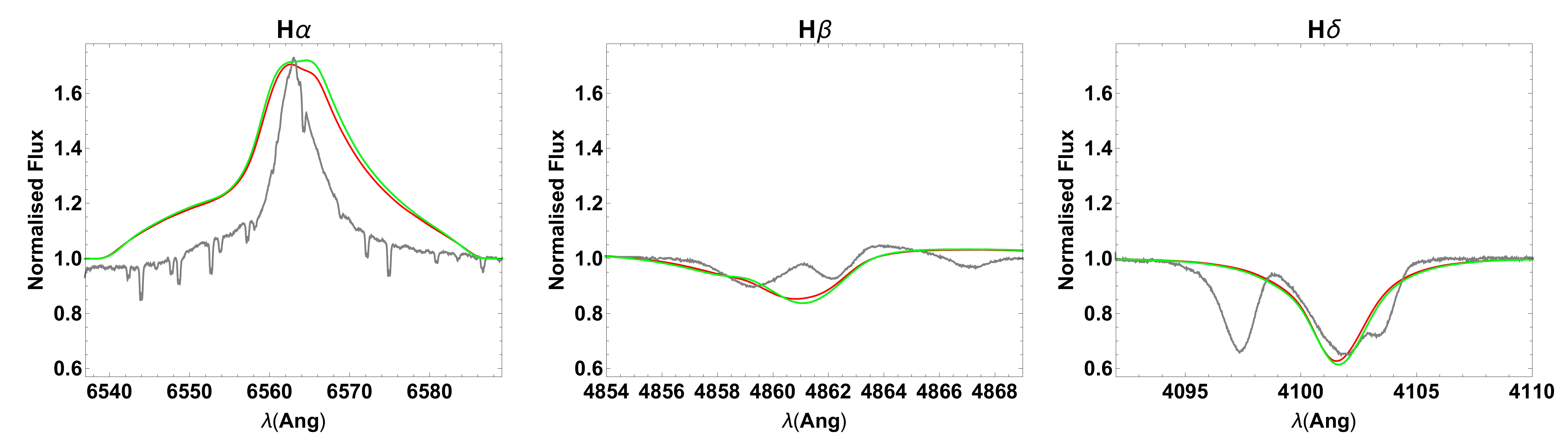}\\
		\includegraphics[width=0.9\textwidth]{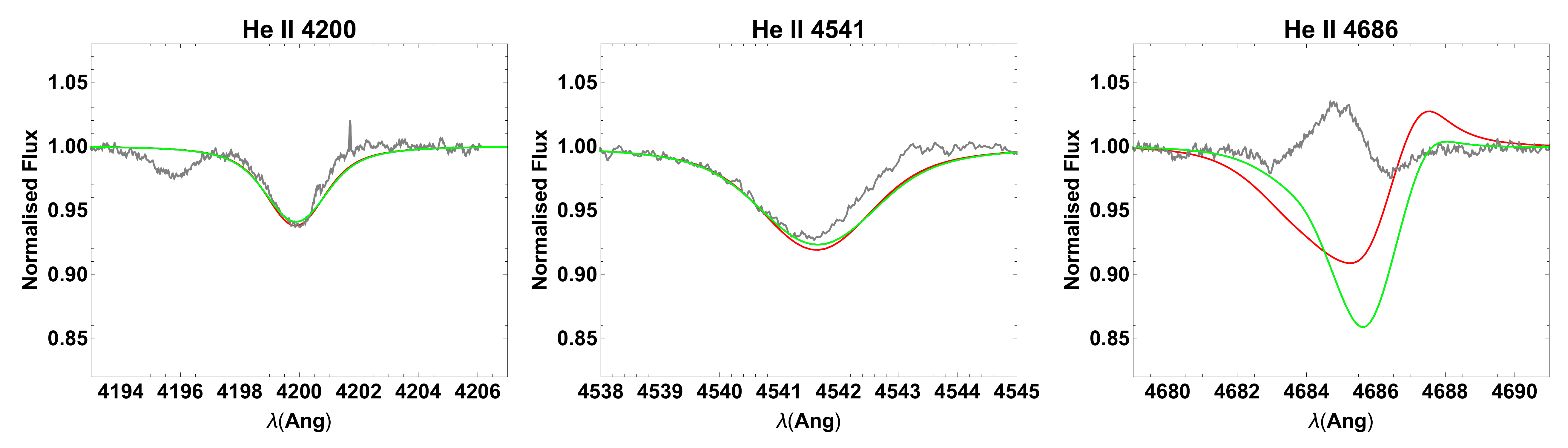}\\
		\includegraphics[width=0.9\textwidth]{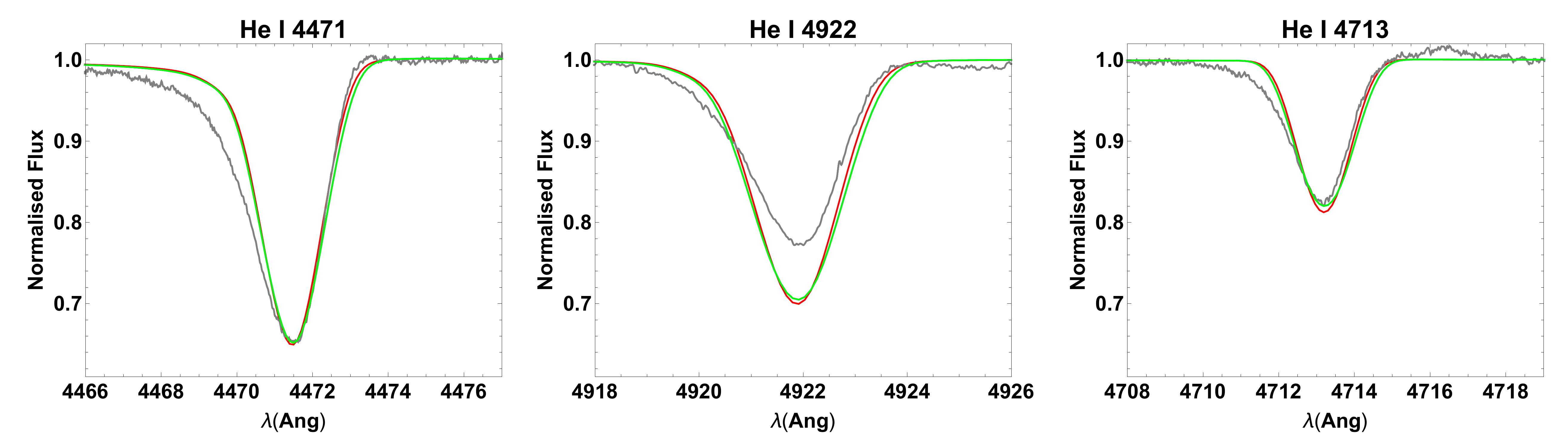}
		\caption{\small{Plot of the models with FeG/FeG$_\odot=1.0$ (red lines) and with FeG/FeG$_\odot=0.7$ (green lines), compared to the observed spectrum of HD 195592.}}
		\label{fitfast123150}
	\end{figure}
	\begin{table}[h!]
		\centering
		\begin{tabular}{c|ccc}
			\hline\hline
			& \multicolumn{3}{c}{Parameters HD 195592}\\
			& \textcolor{red}{\texttt{hd195592fe10}} & \textcolor{green}{\texttt{hd195592fe07}} & \citet{debecker10}\\
			\hline
			$T_\text{eff}$ (kK) & 29.5 & 29.5 & 28.4\\
			$\log g$ & 3.2 & 3.2 & 3.19\\
			$R_*/R_\odot$ & 21.5 & 21.5 & 23\\
			$[\text{He/H}]$ & 0.085 & 0.085 & –\\
			$[\text{FeG}_*/\text{FeG}_\odot]$ & 1.0 & 0.7& –\\
			$(k,\alpha,\delta)$ & $(0.086,0.696,0.260)$ & $(0.074,0.691,0.252)$ & –\\
			$\dot M$ ($M_\odot$ yr$^{-1}$) & $3.6\times10^{-6}$ & $2.4\times10^{-6}$ & –\\
			$v_\infty$ (km s$^{-1}$) & 1\,060 & 1\,050 & –\\
			$f_\text{cl}$ & 2.0 & 4.5 & –\\
			$v_\text{rot}$ (km s$^{-1}$) & 60 & 70 & 60\\
			\hline
		\end{tabular}
		\caption{\small{Summary of stellar and wind parameters used to fit HD 195592.}}
		\label{tablefithd195592}
	\end{table}
		
	Only for this particular case, we present two fits with different sets of parameters (Fig.~\ref{fitfast123150}), in order to show the relevance of the modification of the abundances for the iron group already mentioned.
	Parameters used are tabulated in Table~\ref{tablefithd195592}.
	Fit to helium lines is made by fine-tuning FOM parameters in the same way as the previous case, whereas the hydrogen lines also require to fit clumping factor.
	The model with solar iron abundance has a higher mass-loss rate, so the clumping factor needed to reproduce the H$\alpha$ core is only $f_\text{cl}=2.0$, whereas the model having 70$\%$ of solar abundance fits the observed spectra using a larger clumping factor $f_\text{cl}=4.5$.
	We find that the solar-like model is the best of the two, because it reproduced the emission in He II $\lambda$4686, although the second model could better reproduce its wings.
	The main conclusion we draw is the fact that we can use models with lower/higher iron group abundance values for constraining the other parameters, especially the mass-loss rate.
%	How to perform fits for these constraints is part of our ongoing work.
		
	A remarkable difference compared with our previous fit for HD 57682 is that in this case we are capable to reproduce the core of H$\alpha$ but not its wings because they are too wide.
	This problem is also present for the next stars.
	We therefore concentrate on fitting only the H$\alpha$ core same as for the other lines.
	The consequences of this disagreement are a matter of discussion.

\subsection{9 Sge}
\begin{figure}[t!]
		\centering
		\includegraphics[width=0.9\textwidth]{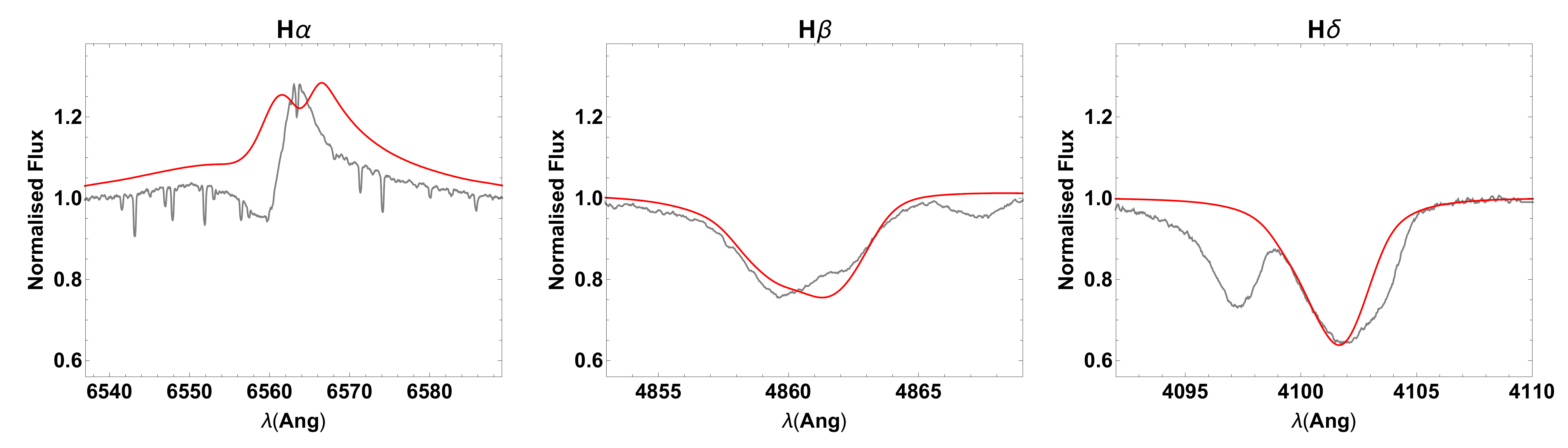}\\
		\includegraphics[width=0.9\textwidth]{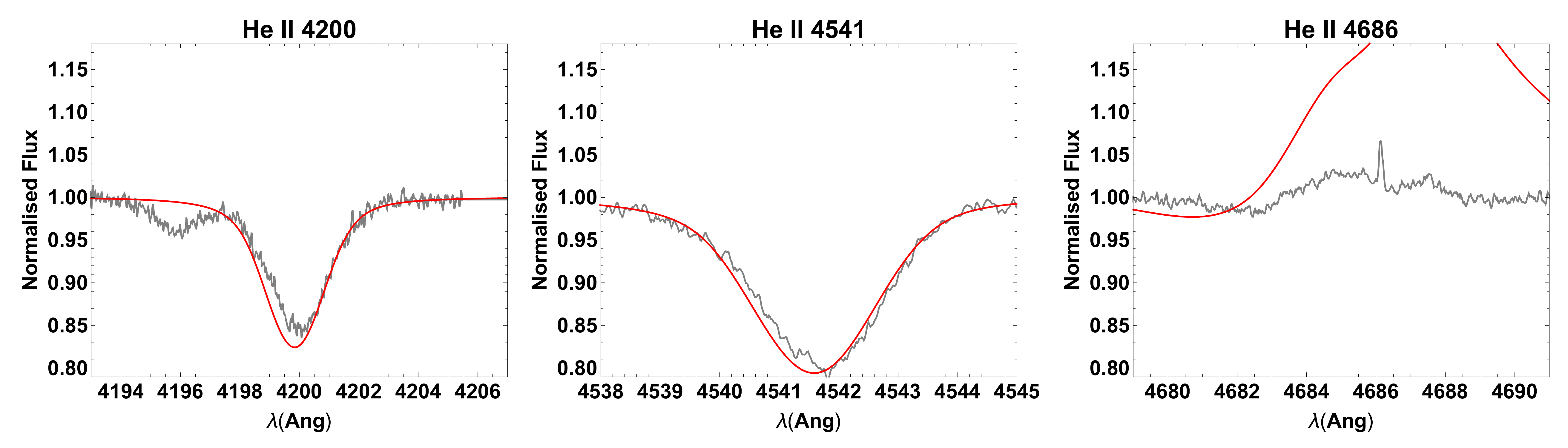}\\
		\includegraphics[width=0.9\textwidth]{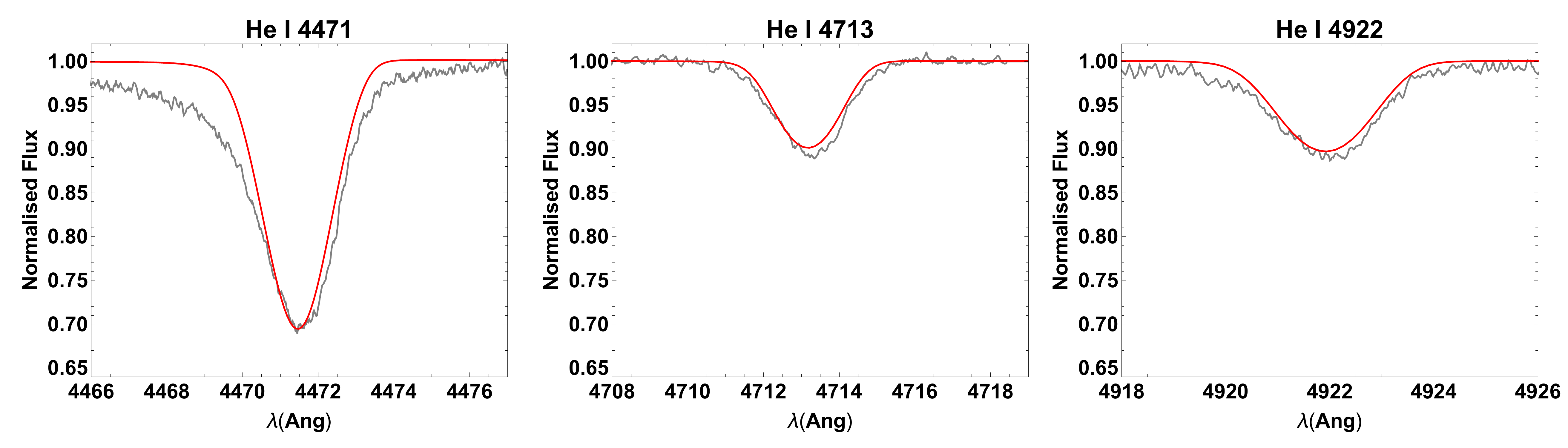}
		\caption{\small{Best fit for the star 9 Sge, using $v_\text{rad}=30$ km s$^{-1}$.}}
		\label{fitfast202}
	\end{figure}

	9 Sge %\footnote{Do not confuse 9 Sge (from Sagitta constellation) with 9 Sgr (from Sagittarius constellation). This latter one is also an O star but belongs to the Main Sequence, spectral type O 3.5 V \citep{sota14}.} 
	(HD 188001) has a spectral type of O 7.5 Iab, according to \citet{sota11}.
	Following this classification, the initial stellar parameters are taken from the catalogue of \citet{martins05}: $T_\text{eff}=34$, $\log g=3.36$ and $R_*=20.8\,R_\odot$, which are also the parameters given in the \textit{VizieR Online Data Catalog}\footnote{\url{http://vizier.u-strasbg.fr/viz-bin/VizieR?-source=J/A+A/620/A89}} \citep{nebot18}.
	Alternatively, \citet{martins15} found $T_\text{eff}=33$, $\log g=3.35$.
	The fitted spectrum and parameters are presented on Fig.~\ref{fitfast202} and Table~\ref{tablefast202} respectively.
	\begin{table}[h!]
		\centering
		\begin{tabular}{c|cc}
			\hline\hline
			& \multicolumn{2}{c}{Parameters 9 Sge}\\%fastmod202
			& This work & \citet{nebot18}\\
			\hline
			$T_\text{eff}$ (kK) & 34.5 & 34.0\\
			$\log g$ & 3.32 & 3.36\\
			$R_*/R_\odot$ & 20.2 & –\\
			$M_*/M_\odot$ & $31.1$ & –\\
			$[\text{He/H}]$ & 0.14 & –\\
			$[\text{FeG}_*/\text{FeG}_\odot]$ & 1.0 & –\\
			$(k,\alpha,\delta)$ & $(0.081,0.698,0.121)$ & –\\
			$\dot M$ ($M_\odot$ yr$^{-1}$) & $4.4\times10^{-6}$ & $8.5\times10^{-7}$\\
			$v_\infty$ (km s$^{-1}$) & 1\,700 & –\\
			$f_\text{cl}$ & 3.0 & –\\
			$v_\text{rot}$ (km s$^{-1}$) & 80 & 60\\
			$v_\text{turb}$ (km s$^{-1}$) & 20 & –\\
			$v_\text{macro}$ (km s$^{-1}$) & 30 & 46\\
			\hline
		\end{tabular}
		\caption{\small{Summary of stellar and wind parameters used to fit 9 Sge (Fig.~\ref{fitfast202}).}}
		\label{tablefast202}
	\end{table}
	
	Similar to HD 195592, the wings of H$\alpha$ are too wide even when we fit the emission intensity, although the model with the same $\dot M$ but without clumping fits the wings better but not the core.
	Besides, He II $\lambda$4684 is the only line that cannot be properly fitted.
	Nevertheless, the fits for the other helium lines are very precise, which is partially because the fine-tuning of $\log g$ has improved with up to decimals.

\subsection{HD 192639}
	Here we present our analysis using self-consistent solution for the supergiant O 7.5 I, HD 192639, observed with \textsc{Hermes} in 2013.
	The initial stellar parameters ($T_\text{eff}=32$ kK and $\log g=3.36$) were taken from standard calibrations done by \citet{martins05} using their theoretical values for effective temperature.
	The initial rotational velocity is set to $v_\text{rot}=80$ km s$^{-1}$.

\begin{figure}[t!]
		\centering
		\includegraphics[width=\textwidth]{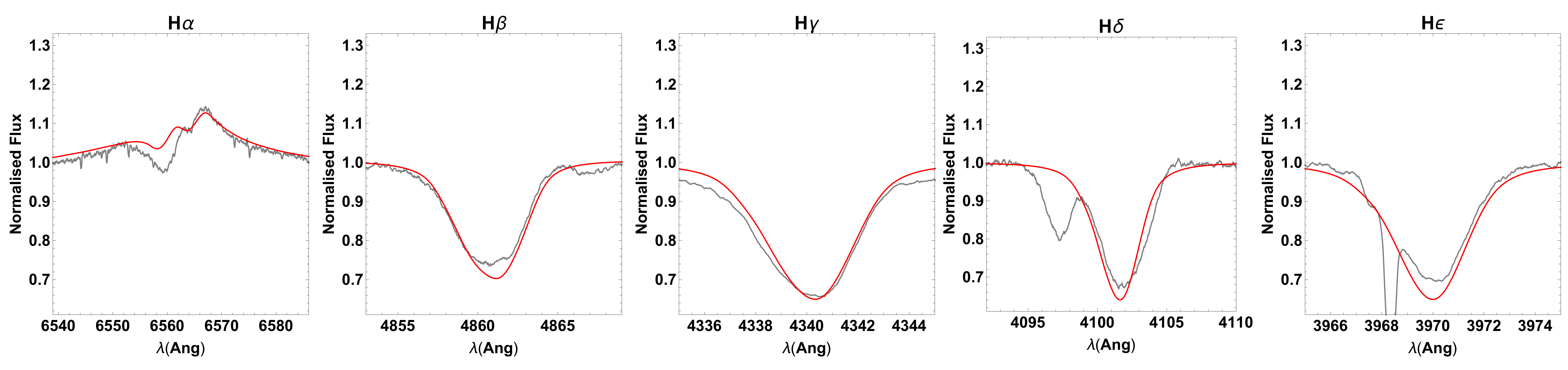}\\
		\includegraphics[width=\textwidth]{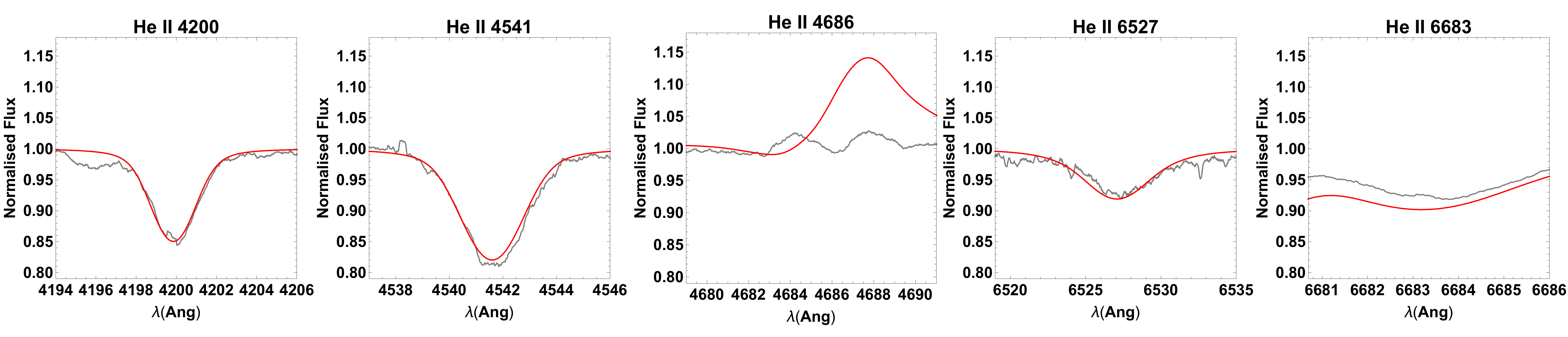}\\
		\includegraphics[width=\textwidth]{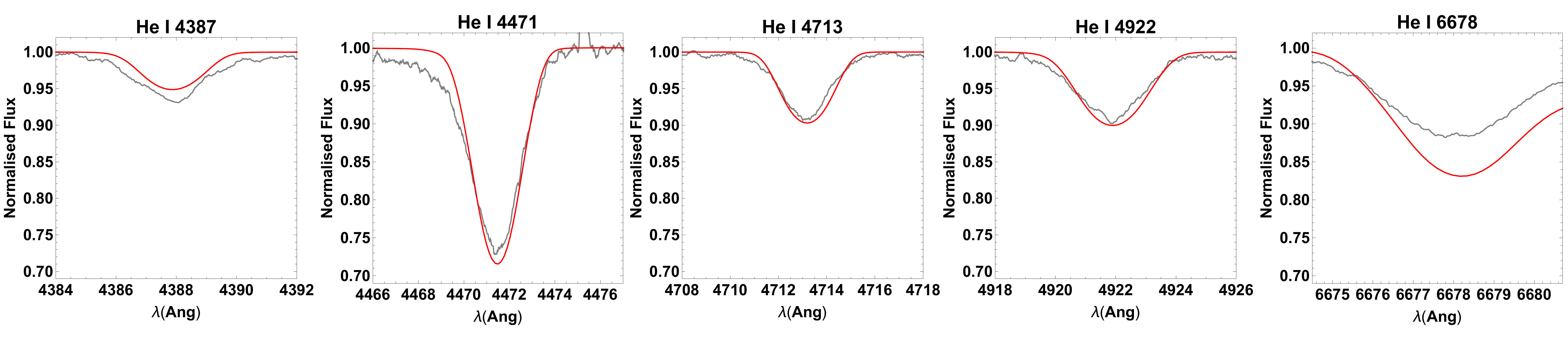}
		\caption{\small{Best fit for the star HD 192639, using $v_\text{rad}=-10$ km s$^{-1}$.}}
		\label{fitfast216}
	\end{figure}

	The stellar and wind parameters obtained for the fit to HD 192639 are given in Table~\ref{tablefast216}, whereas the fit is shown in Fig.~\ref{fitfast216}.
	This time, we have included all hydrogen and helium lines available on FASTWIND.
	The effective temperature is higher than the initial value, whereas the surface gravity is $\sim10\%$ smaller.
	Concerning the spectral fitting, only for this star it was possible to get a good fit to both the wings and the $H\alpha$ emission peak.
	We suppose it is because HD 192639 is a standard star \citep{bouret12,martins15}, therefore its spectrum does not show signs of any unusual behaviour such as magnetic fields or fast rotational velocity.
	\begin{table}[h!]
		\centering
		\begin{tabular}{c|cc}
			\hline\hline
			& \multicolumn{2}{c}{Parameters HD 192639}\\%fastmod216
			& This work & \citet{bouret12}\\
			\hline
			$T_\text{eff}$ (kK) & 33.5 & 33.5\\
			$\log g$ & 3.28 & 3.42\\
			$R_*/R_\odot$ & 19.8 & –\\
			$M_*/M_\odot$ & $27.3$ & –\\
			$[\text{He/H}]$ & 0.14 & 0.15\\
			$[\text{FeG}_*/\text{FeG}_\odot]$ & 1.0 & –\\
			$(k,\alpha,\delta)$ & $(0.073,0.703,0.214)$ & –\\
			$\dot M$ ($M_\odot$ yr$^{-1}$) & $4.05\times10^{-6}$ & $1.2\times10^{-6}$\\
			$v_\infty$ (km s$^{-1}$) & 1\,380 & 1\,900\\
			$f_\text{cl}$ & 1.0 & 20.0\\
			$v_\text{rot}$ (km s$^{-1}$) & 100 & 90\\
			$v_\text{micro}$ (km s$^{-1}$) & 25 & –\\
			$v_\text{macro}$ (km s$^{-1}$) & 30 & 43\\
			\hline
		\end{tabular}
		\caption{\small{Summary of stellar and wind parameters used to fit HD 192639 (Fig.~\ref{fitfast216}).}}
		\label{tablefast216}
	\end{table}
	
\section{Summary and future work}
	We have presented spectral fits combined with our self-consistent m-CAK procedure \citep{alex19} calculating synthetic spectra with the NLTE radiative transfer code FASTWIND.
	Different from previous studies, the parameters we obtain for the wind (mass-loss rate and terminal velocity) are self-consistent with the line-acceleration and the hydrodynamics of the wind.
	For that reason, in almost all the cases they differ for the stellar parameters determined in previous studies.
	Our work presents new best parameters for characterising massive stars.
	
	The methodology presented here has demonstrated be able to accurately fit the helium lines, with the exception of He II $\lambda$4686 (which was not well fitted by previous authors).
%	Thanks to that, we can obtain a value for $\log g$ with up to two relevant decimals
	For the hydrogen lines, the combination of a self-consistent mass-loss rate with an adequate clumping factor allows us to acceptably fit the Balmer lines (with the exception of H$\alpha$):	H$\alpha$ yields a good fit on the line cores but not the wings.
	It is well known that H$\alpha$ is very important for the determination of mass-loss rates but also its shape can reveal other phenomena such as accretion disks or magnetic fields (which are beyond the scope of the FASTWIND and self-consistent procedure presented here).
%	Hence, we rely on the values given by our methodology.
	
	Another important new result, is that we have been able to fit the spectra with stellar parameters below the threshold stated by \citet{alex19}.
	This is an indicator the the lower limits mentioned there should be revisited.
	
	As for future work, we mention the necessity of including all the hydrogen and helium lines available on FASTWIND for the first fitted stars.
	Besides, we have not included yet modifications of CNO abundances for 9 Sge and HD 192639, which are detailed in \citet{martins15}.
	Finally, an appropriate prescription for constraining stellar parameters with reliable error bars is necessary in order to use the new parameters in subsequent studies such as the determination of stellar masses of massive binary systems. 

%_____EVOLUTIONARY PATHS FOR SELF-CONSISTENT SOLUTIONS_______________________________________
\chapter[Evolution of self-consistent solutions]{Evolution of Massive Stars with self-consistent Hydrodynamic Models}\label{tcd}
	The main goal of the present thesis has is development of a methodology capable to couple both, the calculation of line-acceleration and the hydrodynamics of the wind (i.e., a self-consistent solution).
	We have performed a self-consistent prescription based on the m-CAK theory, which has demonstrated to give confident values for wind parameters (mass-loss rate and terminal velocity, see Chapter~\ref{alfakdelta27}).
	The contrast between these m-CAK solutions, under a quasi-NLTE scenario for their atomic populations, and a self-consistent solution under a full NLTE scenario by CMFGEN was presented and discussed on Chapter~\ref{lambert}.
	Even though some differences, self-consistent m-CAK prescription has been capable to provide us accurate spectral fits for massive stars, as it is shown in Chapter~\ref{rob}.
	Thus, self-consistent mass-loss rates can be implemented for future studies, such as the evolution of massive stars.
	
	Mass-loss rate, as we already briefly mentioned on Chapter~\ref{generalintro}, plays a key role in the evolution of massive stars.
	The pioneer work from \citet{meynet94} summarised us that even changes on $\dot M$ by a factor of two may dramatically affect the fate of a star.
	More recent studies \citep{ekstrom12,groh19} have confirmed the same trend.
	Particularly, recent studies as the mentioned above calculated their evolutionary tracks for massive stars using mass-loss values set by Vink's formula \citep{vink01} which, as we pointed out on Chapter~\ref{alfakdelta27}, is not self-consistent.
	The study of the evolution of massive stars that include self-consistent solutions for their stellar winds is still not fully performed.
	
	For that reason, on this chapter we present new evolutionary tracks for a set of massive stars, assuming mass-loss rate given by the self-consistent procedure instead of Vink's formula.
	We use the \textit{Geneva evolutive code} \textsc{Genec} \citep{maeder83,maeder87a,maeder87b} to perform the evolutionary tracks, editing the code to include our prescription.
	The foundations of the work presented in this Chapter was performed during a four months internship at Trinity College Dublin in Ireland, under the supervision of Professor Dr. Jose Groh.
	Following the outline presented on Section~\ref{mcakresults}, this analysis will be done for a case with solar metallicity ($Z_\odot=0.014$) and a 0.2 solar metallicity ($Z=0.003$), both without rotation.
	The resulting evolutive tracks will be compared with those initially given using Vink's recipe, with the respective discussion.

%_____Evolutive Methodology__________________________________________________________________________________________
\section{Methodology}\label{evolutivemethodology}
	Previously, when we presented the results for wind parameters under self-consistent m-CAK prescription, we introduced Eq.~\ref{mdotz10eq1} and Eq.~\ref{mdotz02eq1} as linear relationships to calculate self-consistent mass-loss rate from stellar parameters, in an analogous way to the formulae presented by \citet{vink01}.
	But, even when the coefficient of determination was extremely confident, these expressions were limited to the range of validity given in \citet{alex19}, namely $T_\text{eff}\ge32$ kK and $\log g\ge3.4$.
	Spectral fits shown in Chapter~\ref{rob} have suggested that these lower thresholds could be decreased in order to be valid for more stars, specially O supergiants.
	For that reason, we proceed to calculate more self-consistent solutions for a new set of standard stellar parameters in order to improve these relationships, complementing then the results presented on Table~\ref{standardtable}.
		
\begin{figure}[t!]
		\centering
		\includegraphics[width=0.65\textwidth]{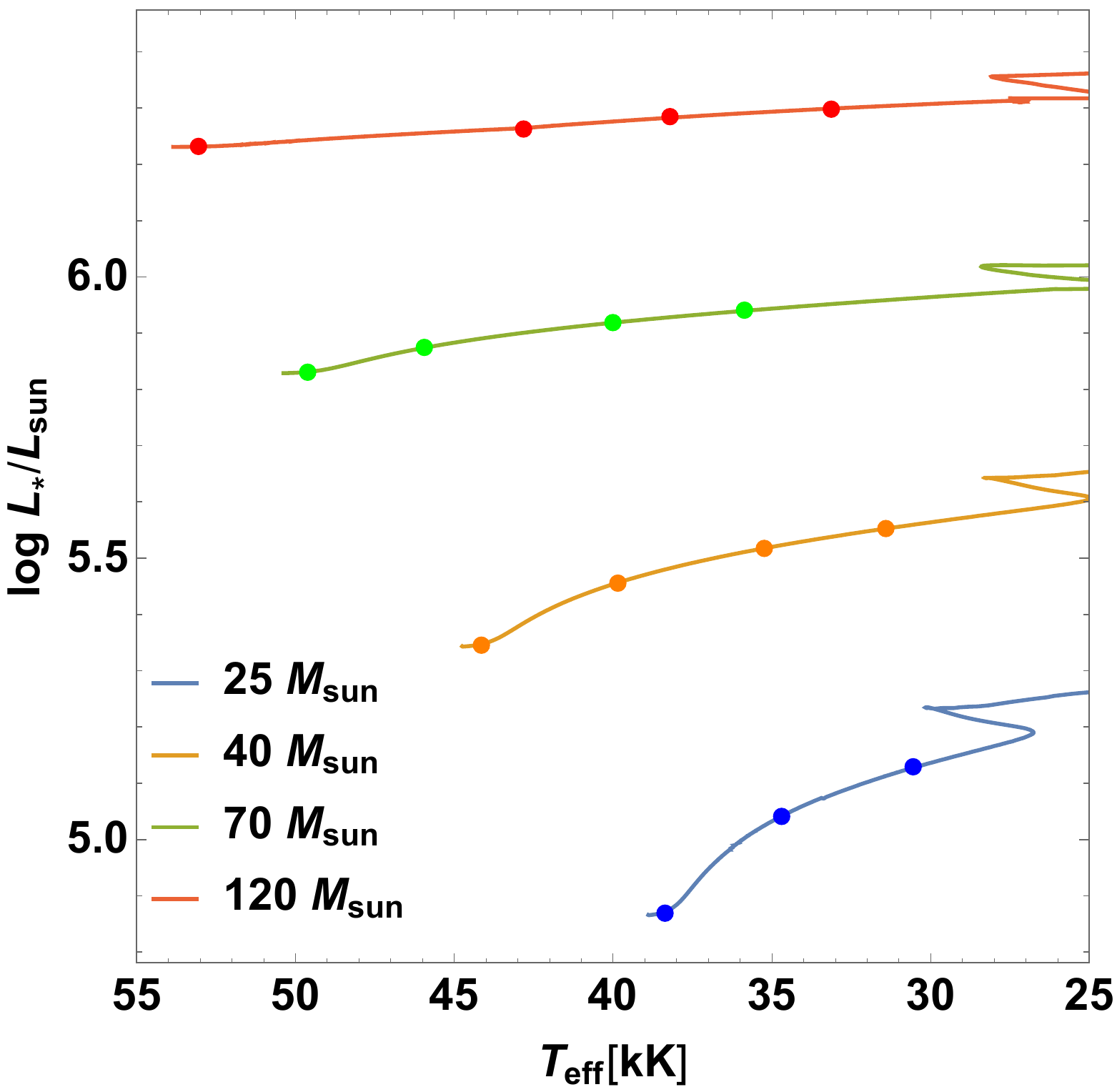}
		\caption{\small{Evolutionary tracks for stars with 120, 70, 40 and 25 $M_\odot$ at solar metallicity ($Z=0.014$).}}
		\label{hrd_z14_ini}
	\end{figure}

	For this time however, standard parameters will be taken for those obtained after running \textsc{Genec}.
	We select the evolutive tracks for non-rotating stars with 25, 40, 70 and 120 solar masses, picking then four representative points for each track (only three for the case of 25 $M_\odot$).
	The points were selected in order to be the most representative on temperature and almost equitable, as it is shown in Fig.~\ref{hrd_z14_ini}.
	The search of new line-force parameters is not limited only for the variations on effective temperature, surface gravity and stellar radius, but also for the abundances.
	Even when metallicity can be considered as constant through the entire evolutionary track, changes in the He to H ratio or in the individual abundance of metal elements affects the resulting line-acceleration\footnote{\textsc{Genec} provides us the change of abundances both on the core of the star and its surface. However, for the calculation of line-acceleration we are interested only on the modification of abundances in the surface of the star.} and therefore affecting the final mass-loss rate.
	These effects produced by abundances are also studied, in order to incorporate them at the final expressions for $\dot M$.
	Given \textsc{Genec} does not provide us surface abundances for important metals such as the iron-group (in order to include a mathematical expression analogous to the relationship shown in Eq.~\ref{masslossirongroup}), our analysis of individual metal elements is constrained to CNO elements only.
	
	\begin{table}[t!]
		\centering
		\resizebox{\textwidth}{!}{
		\begin{tabular}{ccccc|ccc|cccc}
			\hline
			\hline
			$T_\text{eff}$ & $\log g$ & $R_*$ & $M_*$ & $\log L_*$ & $k$ & $\alpha$ & $\delta$ & $ v_\infty$ & $\dot M_\text{SC}$ & $\dot M_\text{SC}/\dot M_\text{Vink}$\\
			$[\text{K}]$ & & $[R_\odot]$ & $[M_\odot]$ & $[L_\odot]$ & & & & [km s$^{-1}$] & [$M_\odot\,\text{yr}^{-1}$]\\
			\hline
%			48\,500 & 4.0 & 18.4 & 123.6 & 6.23 & 0.122 & 0.656 & 0.014 & $5\,148$ & $8.0\times10^{-6}$ & 0.95\\
			54\,000 & 4.15 & 15.0 & 116.0 & 6.24 & 0.126 & 0.660 & 0.014 & $5\,148$ & $8.0\times10^{-6}$ & 0.95\\
			43\,500 & 3.7 & 24.0 & 105.4 & 6.27 & 0.110 & 0.680 & 0.013 & $4\,192$ & $1.2\times10^{-5}$ & 0.80\\
			\hdashline
			39\,000 & 3.45 & 30.7 & 97.4 & 6.28 & 0.102 & 0.694 & 0.031 & $3\,205$ & $1.7\times10^{-5}$ & 1.05\\
			\multicolumn{4}{c}{\small{$[\text{He/H}]=0.11$}} & & 0.102 & 0.694 & 0.029 & $3\,306$ & $1.6\times10^{-5}$ & 1.02\\
			\multicolumn{4}{c}{\small{$[\text{C/C}_\odot]=0.04$}} & & 0.102 & 0.693 & 0.030 & $3\,202$ & $1.6\times10^{-5}$ & 1.02\\
			\multicolumn{4}{c}{\small{$[\text{O/O}_\odot]=0.17$}} & & 0.104 & 0.688 & 0.028 & $3\,148$ & $1.65\times10^{-5}$ & 1.04\\
			\multicolumn{4}{c}{\small{$[\text{He/H}]=0.11$, mod. in CNO abun.}} & & 0.104 & 0.686 & 0.025 & $3\,227$ & $1.6\times10^{-5}$ & 1.02\\
			\hdashline
			34\,000 & 3.2 & 40.8 & 96.3 & 6.3 & 0.071 & 0.729 & 0.187 & $1\,732$ & $2.8\times10^{-5}$ & 1.92\\
			\multicolumn{4}{c}{\small{$[\text{He/H}]=0.14$}} & & 0.074 & 0.722 & 0.169 & $1\,871$ & $2.5\times10^{-5}$ & 1.71\\
			\multicolumn{4}{c}{\small{$[\text{C/C}_\odot]=0.025$}} & & 0.071 & 0.726 & 0.184 & $1\,728$ & $2.7\times10^{-5}$ & 1.85\\
			\multicolumn{4}{c}{\small{$[\text{N/N}_\odot]=6.5$}} & & 0.072 & 0.729 & 0.186 & $1\,738$ & $2.85\times10^{-5}$ & 1.95\\
			\multicolumn{4}{c}{\small{$[\text{O/O}_\odot]=0.06$}} & & 0.070 & 0.724 & 0.190 & $1\,680$ & $2.7\times10^{-5}$ & 1.85\\
			\multicolumn{4}{c}{\small{$[\text{He/H}]=0.14$, mod. in CNO abun.}} & & 0.074 & 0.715 & 0.169 & $1\,818$ & $2.3\times10^{-5}$ & 1.58\\
			\hline
%			45\,500 & 4.05 & 13.1 & 70.0 & 5.82 & 0.132 & 0.629 & 0.014 & $4\,414$ & $2.1\times10^{-6}$ & 0.95\\
			50\,500 & 4.2 & 10.8 & 70.0 & 5.83 & 0.136 & 0.631 & 0.017 & $4\,693$ & $2.3\times10^{-6}$ & 0.90\\
			46\,000 & 4.0 & 13.5 & 66.4 & 5.87 & 0.126 & 0.643 & 0.013 & $4\,360$ & $2.8\times10^{-6}$ & 0.78\\
			40\,500 & 3.7 & 18.4 & 62.4 & 5.92 & 0.107 & 0.662 & 0.015 & $3\,670$ & $3.6\times10^{-6}$ & 0.90\\
			36\,500 & 3.5 & 23.2 & 60.7 & 5.94 & 0.094 & 0.672 & 0.034 & $3\,101$ & $4.0\times10^{-6}$ & 0.82\\
%			\multicolumn{4}{c}{\small{$[N/N_\odot]=11.0$}} & & 0.093 & 0.675 & 0.035 & $3\,133$ & $4.1\times10^{-6}$ & 0.84\\
%			32\,000 & 3.25 & 30.8 & 59.5 & 5.96 & 0.073 & 0.707 & 0.203 & $1\,638$ & $7.5\times10^{-6}$ & 1.85\\
			\hline
%			40\,300 & 4.08 & 9.5 & 40.0 & 5.33 & 0.138 & 0.598 & 0.020 & $3\,457$ & $3.45\times10^{-7}$ & 0.79\\
			45\,000 & 4.25 & 7.8 & 40.0 & 5.34 & 0.158 & 0.593 & 0.017 & $3\,899$ & $4.2\times10^{-7}$ & 0.84\\
			40\,000 & 3.95 & 11.0 & 38.4 & 5.45 & 0.120 & 0.623 & 0.019 & $3\,554$ & $6.1\times10^{-7}$ & 0.70\\
			36\,000 & 3.65 & 15.0 & 37.5 & 5.51 & 0.094 & 0.649 & 0.025 & $3\,010$ & $9.0\times10^{-7}$ & 0.89\\
			32\,000 & 3.45 & 19.3 & 37.0 & 5.55 & 0.081 & 0.662 & 0.089 & $2\,270$ & $1.0\times10^{-6}$ & 1.04\\
%			30\,000 & 3.3 & 22.1 & 36.8 & 5.56 & 0.077 & 0.675 & 0.222 & $1\,350$ & $1.6\times10^{-6}$ & 1.89\\
			\hline
%			35\,000 & 4.1 & 7.4 & 25.0 & 4.87 & 0.102 & 0.593 & 0.032 & $3\,145$ & $4.3\times10^{-8}$ & 0.65\\
			39\,000 & 4.3 & 6.0 & 25.0 & 4.87 & 0.166 & 0.558 & 0.029 & $3\,185$ & $5.4\times10^{-8}$ & 0.69\\
			35\,000 & 3.9 & 9.0 & 24.5 & 5.04  & 0.097 & 0.613 & 0.022 & $2\,972$ & $1.15\times10^{-7}$ & 0.74\\
			31\,000 & 3.6 & 12.7 & 24.3 & 5.12 & 0.068 & 0.639 & 0.054 & $2\,400$ & $1.4\times10^{-7}$ & 0.81\\
%			27\,000 & 3.3 & 17.7 & 24.2 & 5.18 & 0.018 & 0.671 & 0.057 & $2\,201$ & $3.2\times10^{-8}$ & 0.23\\
			\hline
		\end{tabular}}
		\caption{\small{Analogous to Table~\ref{standardtable}, self-consistent line-force parameters $(k,\alpha,\delta)$ for adopted standard stellar parameters, together with the resulting terminal velocities and mass-loss rates ($\dot M_\text{SC}$). Ratios between self-consistent mass-loss rates and  Vink's recipe values \citep[re-scaled to match metallicity from][]{asplund09} using $ v_\infty/v_{\rm{esc}}=2.6$ are shown in the last column. Solid lines separate the groups of stellar parameters for the different evolutionary tracks on Fig.~\ref{hrd_z14_ini}, whereas dashed lines gather models with identical temperature, gravity and radius but different abundances. Error margins for mass-loss rates and terminal velocities have been neglected for practical purposes.}}
		\label{evolstandardtable}
	\end{table}
	
	The new line-force parameters, with their respective new self-consistent mass-loss rate, are presented in Table~\ref{evolstandardtable}.
	We have also included columns to display the stellar mass and the luminosity at each stage.
	It is important to notice that changes on surface abundances for CNO elements and the He to H ratio is just presented for the case of 120 $M_\odot$, because is the only one evolutive track that exhibits this behaviour\footnote{This is a direct consequence of being working with non-rotating models: reactions in the core abruptly modify the abundance structure at some point for only extreme high mass cases such as 120 $M_\odot$. Evolutive tracks that consider rotation exhibit a gradual modification of surface abundances even for stars with masses below 25 $M_\odot$ \citep{ekstrom12}.}.
	In these cases, a self-consistent solution with the abundances unchanged is presented first, followed by self-consistent solutions after the single modification of He to H ratio and the single CNO elements, to finish showing the last self-consistent solution with all the abundance modifications included.
	
\begin{figure}[t!]
		\centering
		\includegraphics[width=0.65\textwidth]{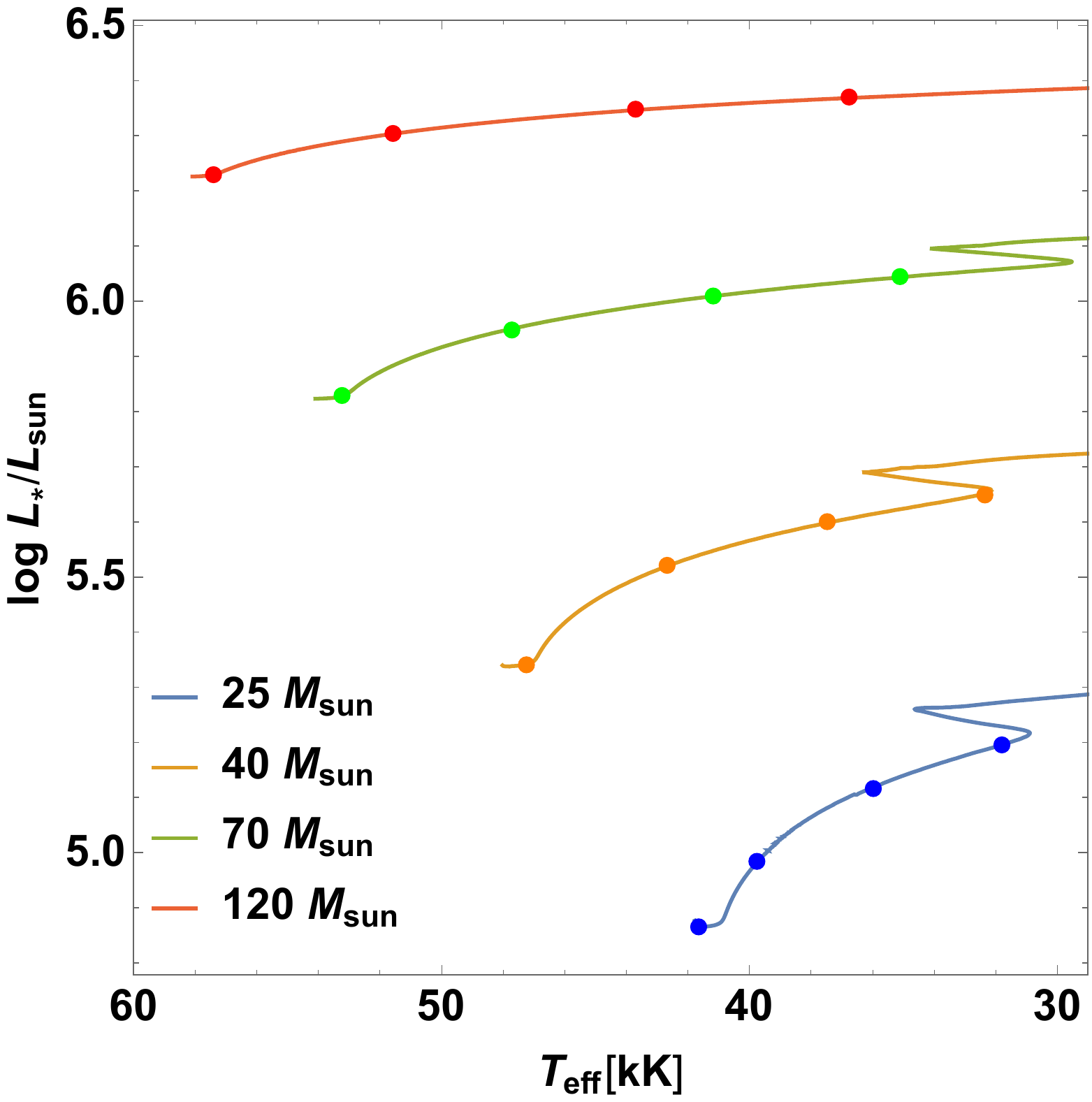}
		\caption{\small{Evolutionary tracks for stars with 120, 70, 40 and 25 $M_\odot$ at low metallicity ($Z=0.003$).}}
		\label{hrd_z03_ini}
	\end{figure}

	Same procedure must be done for the case of metallicity $Z=0.003$.
	The Hertzsprung-Russell diagram with the evolutive tracks and the selected points are shown in Fig.~\ref{hrd_z03_ini}, whereas their respective line-force parameters are presented on Table~\ref{evolstandardtablez03}.
	As a first comment about abundance modifications under low metallicity, we found on the tracks that only significative changes have been observed for nitrogen, which is in agreement with the stated by \citet{groh19}.
	\begin{table}[t!]
		\centering
		\resizebox{\textwidth}{!}{
		\begin{tabular}{ccccc|ccc|cccc}
			\hline
			\hline
			$T_\text{eff}$ & $\log g$ & $R_*$ & $M_*$ & $\log L_*$ & $k$ & $\alpha$ & $\delta$ & $ v_\infty$ & $\dot M_\text{SC}$ & $\dot M_\text{SC}/\dot M_\text{Vink}$\\
			$[\text{K}]$ & & $[R_\odot]$ & $[M_\odot]$ & $[L_\odot]$ & & & & [km s$^{-1}$] & [$M_\odot\,\text{yr}^{-1}$]\\
			\hline
			58\,000 & 4.3 & 12.8 & 120.0 & 6.23 & 0.114 & 0.537 & 0.010 & $3\,790$ & $1.9\times10^{-6}$ & 0.79\\
			\hdashline
			52\,000 & 4.0 & 17.4 & 116.5 & 6.30 & 0.085 & 0.589 & 0.019 & $3\,279$ & $3.8\times10^{-6}$ & 1.25\\
			\multicolumn{4}{c}{\small{$[\text{N/N}_\odot]=0.3$}} & & 0.085 & 0.591 & 0.019 & $3\,299$ & $3.9\times10^{-6}$ & 1.26\\
			\hdashline
			44\,500 & 3.7 & 25.0 & 114.3 & 6.34 & 0.075 & 0.616 & 0.015 & $2\,993$ & $5.2\times10^{-6}$ & 1.29\\
			\multicolumn{4}{c}{\small{$[\text{N/N}_\odot]=2.4$}} & & 0.073 & 0.724 & 0.019 & $3\,048$ & $5.45\times10^{-6}$ & 1.35\\
			\hdashline
			38\,000 & 3.4 & 35.2 & 112.9 & 6.37 & 0.061 & 0.644 & 0.020 & $2\,650$ & $6.0\times10^{-6}$ & 0.95\\
			\multicolumn{4}{c}{\small{$[\text{N/N}_\odot]=2.5$}} & & 0.060 & 0.647 & 0.023 & $2\,582$ & $6.0\times10^{-6}$ & 0.95\\
			\hline
			54\,000 & 4.35 & 9.3 & 70.0 & 5.82 & 0.145 & 0.489 & 0.012 & $3\,162$ & $3.75\times10^{-7}$ & 0.61\\
			48\,000 & 4.0 & 13.6 & 68.6 & 5.95 & 0.094 & 0.559 & 0.025 & $2\,841$ & $1.0\times10^{-6}$ & 1.16\\
			42\,000 & 3.7 & 19.3 & 67.9 & 6.01 & 0.076 & 0.596 & 0.018 & $2\,642$ & $1.6\times10^{-6}$ & 1.54\\
			36\,000 & 3.4 & 27.0 & 67.5 & 6.04 & 0.058 & 0.633 & 0.021 & $2\,446$ & $1.9\times10^{-6}$ & 1.45\\
			\hline
			48\,000 & 4.4 & 6.8 & 40.0 & 5.34 & 0.193 & 0.441 & 0.017 & $2\,490$ & $5.0\times10^{-8}$ & 0.5\\
			42\,000 & 4.0 & 10.4 & 39.6 & 5.51 & 0.107 & 0.526 & 0.026 & $2\,405$ & $1.8\times10^{-7}$ & 1.09\\
			38\,000 & 3.7 & 14.6 & 39.4 & 5.59 & 0.069 & 0.586 & 0.034 & $2\,328$ & $3.1\times10^{-7}$ & 1.55\\
			33\,000 & 3.4 & 20.7 & 39.3 & 5.65 & 0.049 & 0.625 & 0.022 & $2\,292$ & $4.0\times10^{-7}$ & 2.0\\
			\hline
			42\,000 & 4.4 & 5.2 & 25.0 & 4.88 & 0.254 & 0.408 & 0.024 & $2\,050$ & $5.7\times10^{-9}$ & 0.26\\
			40\,000 & 4.2 & 6.5 & 24.9 & 4.98 & 0.141 & 0.475 & 0.046 & $2\,067$ & $1.4\times10^{-8}$ & 0.39\\
			36\,000 & 3.9 & 9.3 & 24.8 & 5.12 & 0.067 & 0.559 & 0.046 & $2\,246$ & $3.2\times10^{-8}$ & 0.51\\
			32\,000 & 3.6 & 13.0 & 24.7 & 5.19 & 0.053 & 0.607 & 0.013 & $2\,431$ & $7.6\times10^{-8}$ & 1.12\\
			\hline
		\end{tabular}}
		\caption{\small{Summary of standard stars extracted from the initial evolutionary paths for low metallicity ($Z=0.003$), analogous to Table~\ref{evolstandardtable}. Notice that in this case the initial ratio was N/N$_\odot=0.2$.}}
		\label{evolstandardtablez03}
	\end{table}
	
%_____Evolutionary tracks at solar metallicity____________________________________________________________________________
\section{Evolutionary tracks at solar metallicity}
	Combining solutions presented on both Table~\ref{standardtable} and Table~\ref{evolstandardtable}, we perform a new fit to obtain the self-consistent formula for mass-loss rate as function of the stellar parameters.
	\begin{align}\label{mdotz10eq2}
		\log \dot M_{Z=Z_\odot}=&\;42.79\times\left[\log\left(\frac{T_\text{eff}}{1000\text{ K}}\right)\right]^{-1}-54.5\times\left[\log\left(\frac{T_\text{eff}}{1000\text{ K}}\right)\right]^{-2}\nonumber\\
		&-73\times(\log g)^{-1}+178.97\times(\log g)^{-2}\nonumber\\
		&-10.29\times(R_*/R_\odot)^{-1}-3.72\nonumber\\
		&-0.227\times[\log(\text{He/H})-\log(\text{He/H})_i]\nonumber\\
		&-0.0114\times\log(\text{C/C}_i)\nonumber\\
		&-0.227\times\log(\text{N/N}_i)\;\;.
	\end{align}
	
	Unlike Eq.~\ref{mdotz10eq1}, the dependence of Eq.~\ref{mdotz10eq2} on the stellar parameters is not linear.
	Moreover, we included the already mentioned dependence on abundance modifications.
	This time, we consider this formula (Eq.~\ref{mdotz10eq2}) valid while $\log g\ge3.2$, whereas for surface gravities below that value, the mass-loss rate will be determined by Vink's formula \citep{vink01}\footnote{It is necessary to remind that \textsc{Genec} uses more than one recipe to calculate the mass-loss rate, depending on the evolutive stage of the star \citep{ekstrom12}. For the particular case of stars with $M_*\ge25$ $M_\odot$ (which is the focus of this chapter), Vink's formula is used from ZAMS to the end of the Main Sequence, whereas for the Wolf-Rayet stage it is used the recipe given by \citet{grafener08}.}.
	
\begin{figure}[t!]
		\centering
		\includegraphics[width=0.65\textwidth]{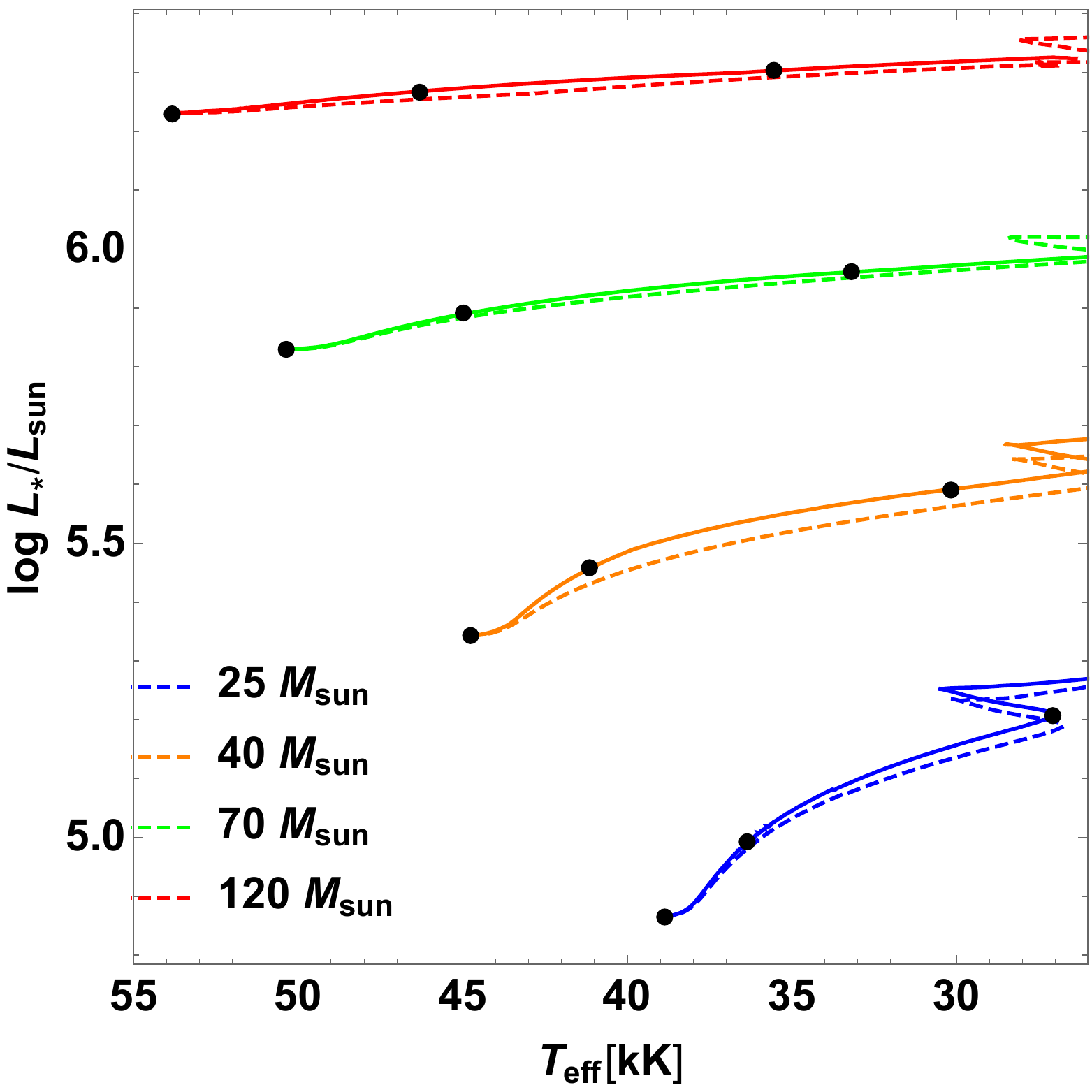}
		\caption{\small{Comparison of evolutionary tracks for stars with 120, 70, 40 and 25 $M_\odot$ at solar metallicity ($Z=0.014$), calculated using our self-consistent $\dot M_\text{SC}$ (solid lines) and using Vink's formula $\dot M_\text{Vink}$ (dashed lines).}}
		\label{hrd_z14_fin}
	\end{figure}

\begin{figure}[t!]
		\centering
		\includegraphics[width=0.51\textwidth]{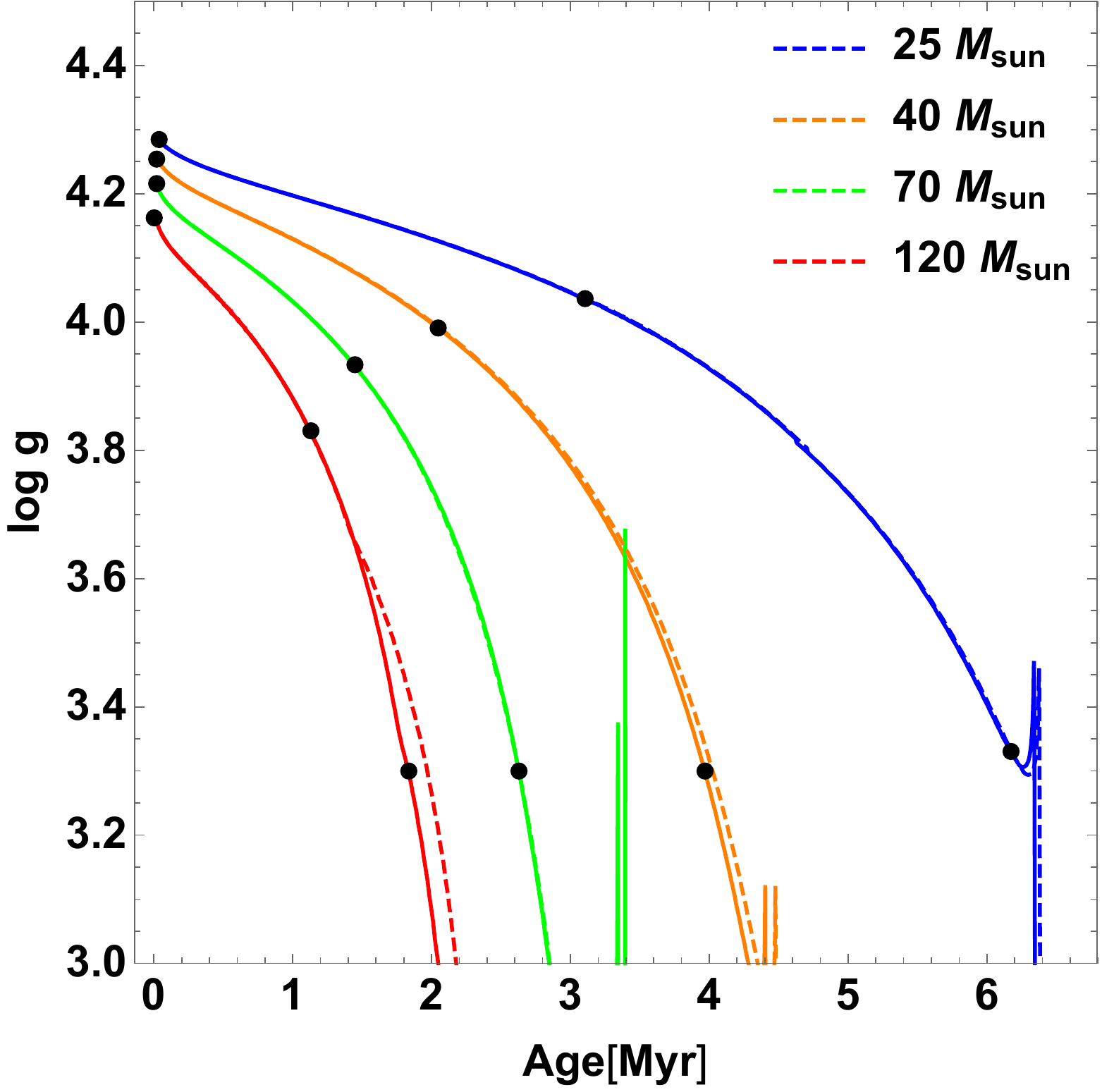}
		\caption{\small{Evolution of $\log g$ of evolutionary tracks for stars with 120, 70, 40 and 25 $M_\odot$ at solar metallicity ($Z=0.014$), calculated using our self-consistent $\dot M_\text{SC}$ (solid lines) and using Vink's formula $\dot M_\text{Vink}$ (dashed lines).}}
		\label{logg_z14_fin}
	\end{figure}

	Resulting new evolutionary tracks are presented on Fig.~\ref{hrd_z14_fin}.
	Our first comment, it is clearly seen that the cases using self-consistent mass-loss rates exhibit slightly more luminous paths.
	Additionally, we show the temporal evolution of the surface gravity on Fig.~\ref{logg_z14_fin}.
	For each track, we have highlighted three points, each one representing a specific stage:
	\begin{itemize}
		\item Zero Age Main Sequence: the beginning of our evolutive tracks.
		\item Abundances $X_\text{He}=X_\text{H}=0.5$ in the core: as a form to represent a middle point on the Main Sequence stage.
		\item Surface gravity $\log g=3.3$: the ending point of our prescription, entering then a linear transition until return to Vink's formula for $\log g=3.2$.
	\end{itemize}
	
%	The synthetic spectra on these three stages for every track by the usage of FASTWIND, in order to elucidate how spectral lines change over time for massive stars.
	
\begin{figure}[t!]
		\centering
		\includegraphics[width=0.49\textwidth]{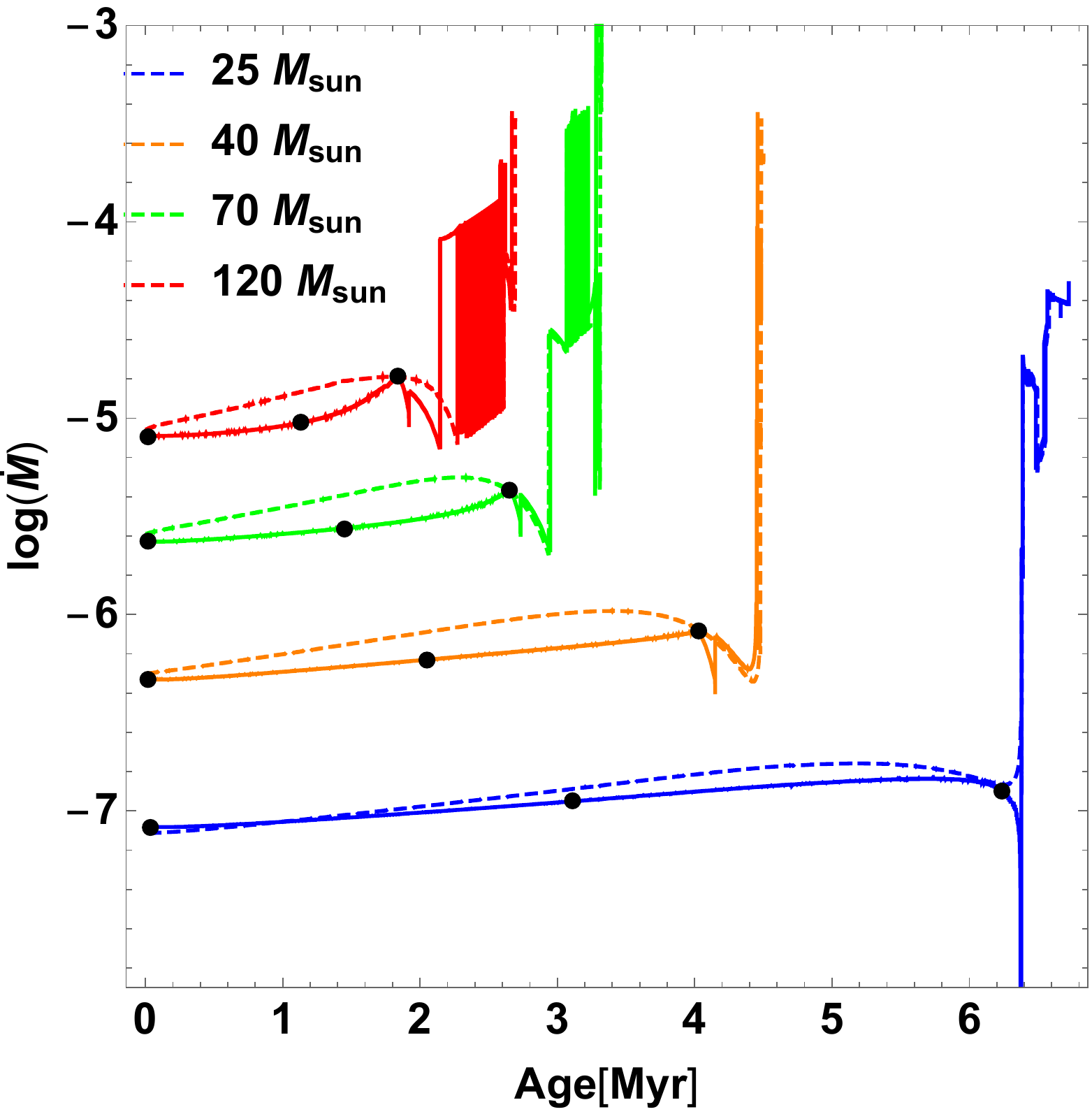}
		\includegraphics[width=0.5\textwidth]{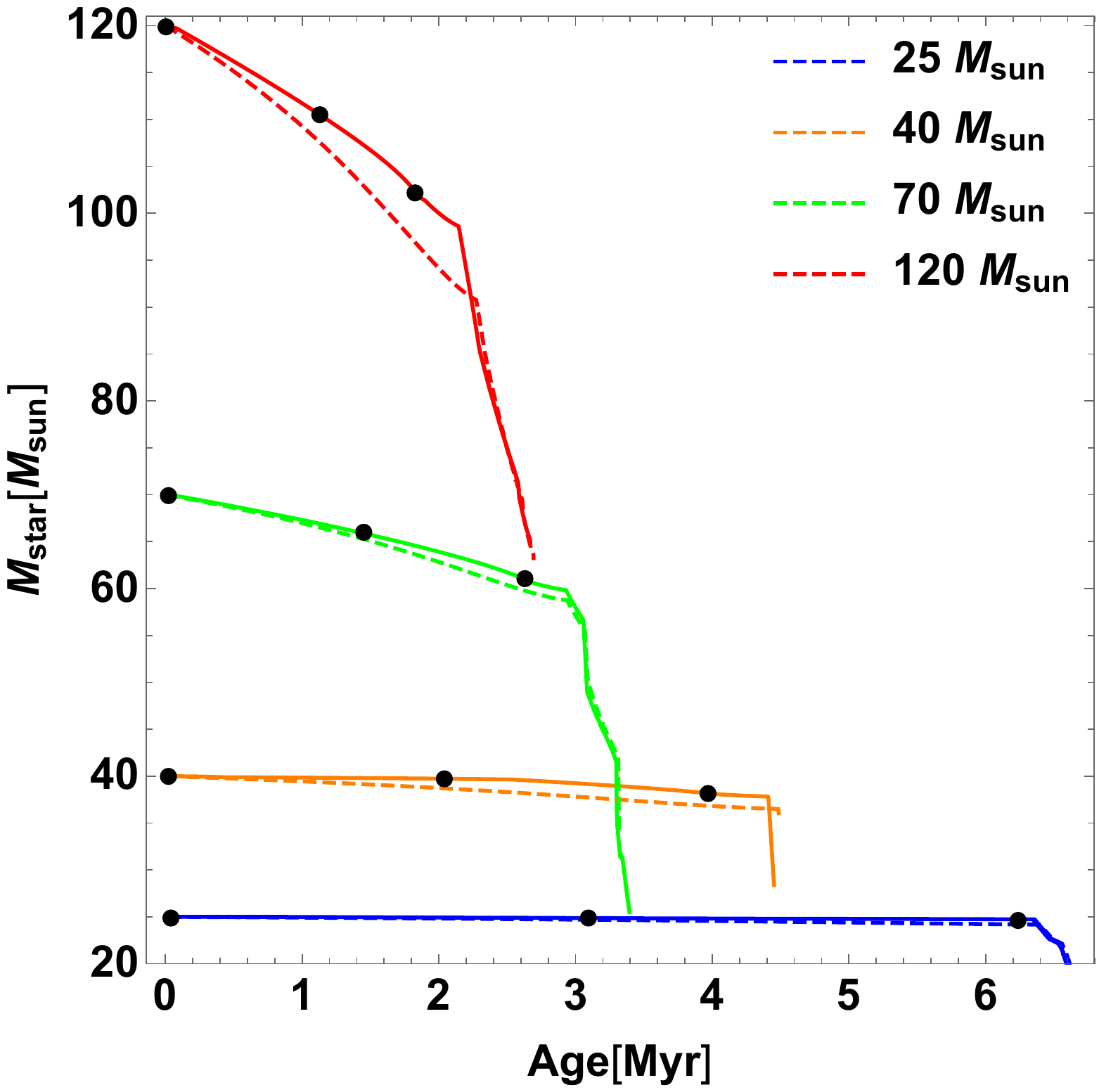}
		\caption{\small{Evolution of $\dot M$ (left panel) and $M_*$ (right panel) of evolutionary tracks for stars with 120, 70, 40 and 25 $M_\odot$ at solar metallicity ($Z=0.014$), calculated using our self-consistent $\dot M_\text{SC}$ (solid lines) and using Vink's formula $\dot M_\text{Vink}$ (dashed lines).}}
		\label{mass_z14_fin}
	\end{figure}

	We include also the study of the behaviour of the stellar masses.
	Figure~\ref{mass_z14_fin} shows the evolution of mass-loss rates and stellar masses.
	Given that the initial self-consistent $\dot M$ is smaller that the $\dot M$ given by Vink's recipe, the resulting evolutionary paths remain with a lower mass-loss rate until the end of the self-consistent prescription and thus, the stellar mass function for stars decreases slower than before\footnote{Result presented on left-panel of Fig.~\ref{mass_z14_fin} seems to be in contradiction with the last column of Table~\ref{evolstandardtable}, where it is shown that $\dot M_\text{SC}>\dot M_\text{Vink}$. However, ratios on the table represent a comparison between self-consistent and Vink's mass-loss rates \textit{for the old evolutionary track using Vink's recipe}, whereas Fig.~\ref{mass_z14_fin} represents the evolution of the mass-loss rate for each different evolutionary tracks, so the $\dot M_\text{SC}$ plotted are valid for the new path and then they do not belong to the parameters tabulated on Table~\ref{evolstandardtable}.}.
	
	The most noticeable new result, is that the usage of self-consistent mass-loss rates produce more luminous paths, increased approximately a $\sim7\%$, and this effect seems to be almost similar for all the stellar masses plotted on Fig.~\ref{hrd_z14_fin}.
	Meanwhile, from Fig.~\ref{logg_z14_fin} it is seen that evolution becomes "faster", i.e., evolutionary tracks reach the third point with $\log g=3.3$ in a slightly shorter time using $M_\text{SC}$.
	Even when these differences are on the edge of the significance, their consequences still need to be understood.

%_____Evolution of spectra at solar metallicity___________________________________________________________________________	
\subsection{Evolution of spectra at solar metallicity}
	\begin{table}[t!]
		\centering
%		\resizebox{\textwidth}{!}{
		\begin{tabular}{ccccc|ccc|cccc}
			\hline
			\hline
			$T_\text{eff}$ & $\log g$ & $R_*$ & $M_*$ & $\log L_*$ & $k$ & $\alpha$ & $\delta$ & $ v_\infty$ & $\dot M_\text{SC}$\\
			$[\text{K}]$ & & $[R_\odot]$ & $[M_\odot]$ & $[L_\odot]$ & & & & [km s$^{-1}$] & [$M_\odot\,\text{yr}^{-1}$]\\
			\hline
			45\,000 & 4.25 & 7.8 & 39.8 & 5.35 & 0.191 & 0.574 & 0.023 & $3\,540$ & $4.7\times10^{-7}$\\
			41\,000 & 3.99 & 10.5 & 39.6 & 5.46 & 0.148 & 0.605 & 0.024& $3\,300$ & $7.3\times10^{-7}$\\
			30\,000 & 3.30 & 22.9 & 40.0 & 5.35 & 0.061 & 0.671 & 0.142 & $1\,750$ & $8.6\times10^{-7}$\\
			\hline
		\end{tabular}%}
		\caption{\small{Stellar and line-force parameters for the set of black points highlighted on Fig.~\ref{logg_z14_fin} for the case with 40 $M_\odot$.}}
		\label{selfconsistentblackpoints}
	\end{table}
	\begin{figure}[t!]
		\centering
		\includegraphics[width=0.9\textwidth]{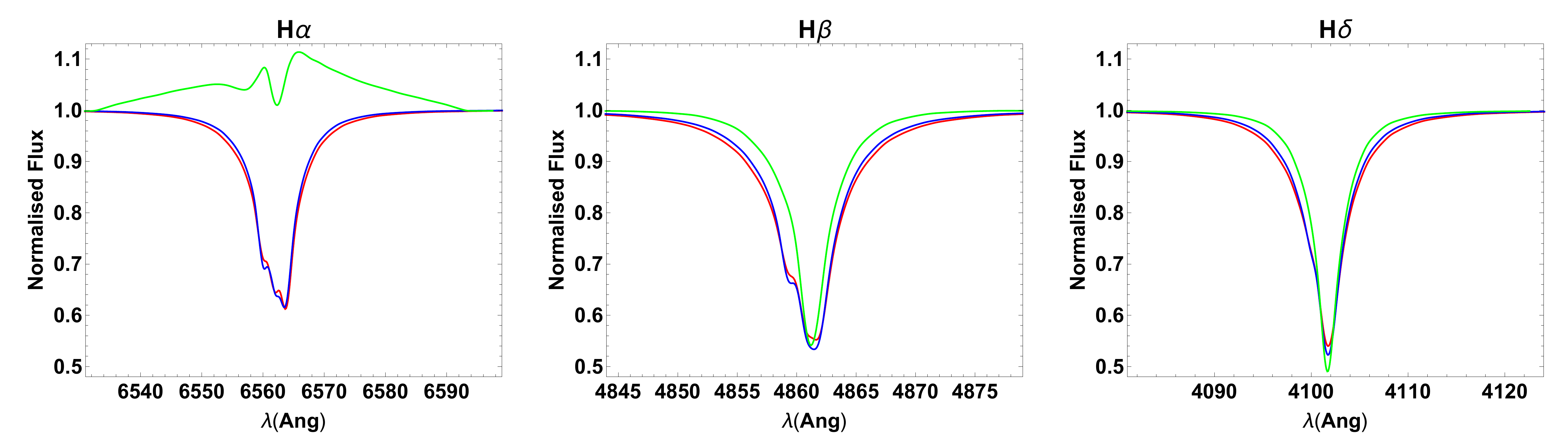}\\
		\includegraphics[width=0.9\textwidth]{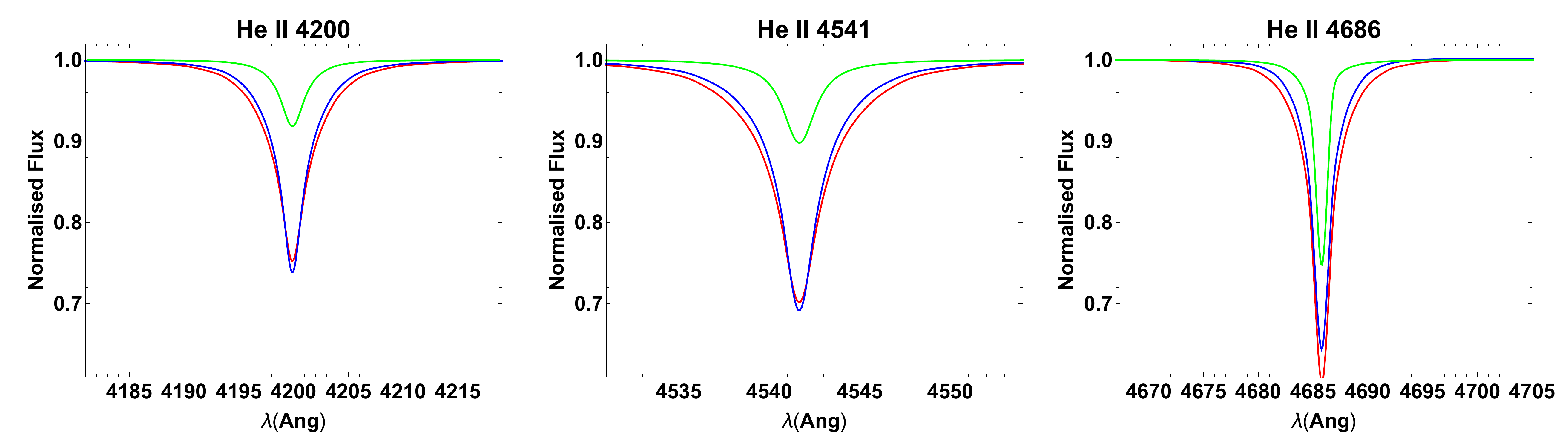}\\
		\includegraphics[width=0.9\textwidth]{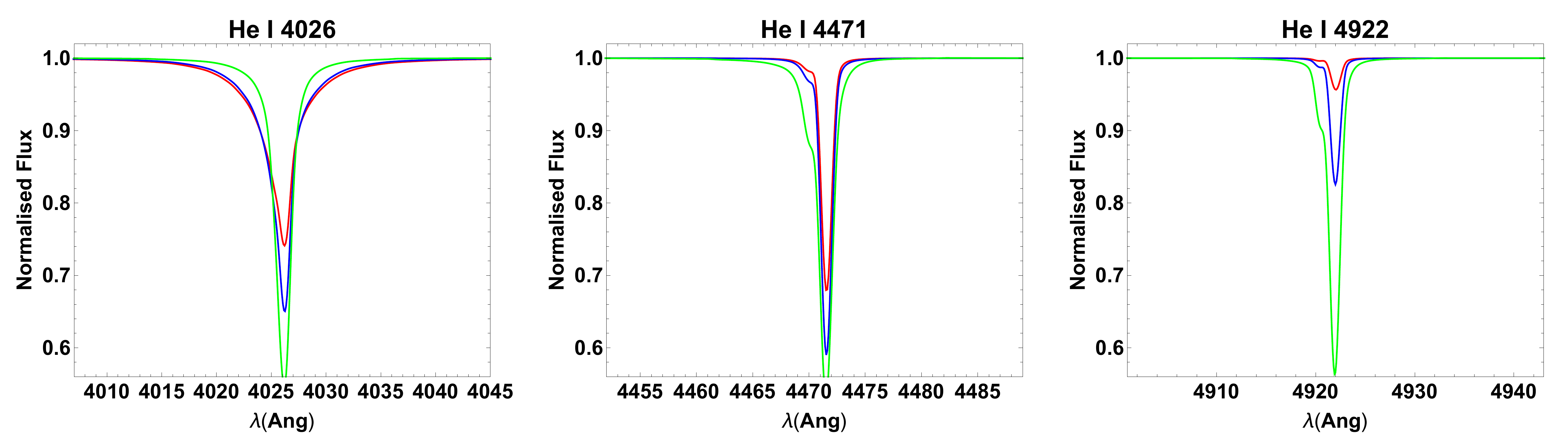}
		\caption{\small{Synthetic spectra created by FASTWIND for the three highlighted stars tabulated on Table~\ref{selfconsistentblackpoints}: $T_\text{eff}=45$ kK (red), $T_\text{eff}=41$ kK (blue) and $T_\text{eff}=30$ kK (green). All models were executed without convolution by rotation and with a clumping factor $f_\text{cl}=5.0$.}}
		\label{spectraP040z14}
	\end{figure}
	
	In order to have an idea about how stellar spectra are changing through the evolutionary tracks, we decide to run self-consistent solutions for each one of the three points previously mentioned.
	Motivation is, to perform an analysis on the evolution of spectra similar to the works previously developed \citep{groh14,liermann15}, but using the self-consistent hydrodynamics and FASTWIND.
	We will present here the evolution of the track for 40 $M_\odot$.
	%Spectra for the other masses is work still in progress.
	
	Self-consistent solutions for the three stages highlighted on the 40 $M_\odot$ evolutionary track are shown in Table~\ref{selfconsistentblackpoints}, whereas Fig.~\ref{spectraP040z14} presents the synthetic spectra.
	It is interesting to observe that, for the first two stages there is no significant differences on the profiles of the hydrogen lines; the most remarkable difference is the absorption intensity of the He I lines, presumably because the lower ionisation at lower temperature (He II/He I ratio becomes smaller as effective temperature decreases).
	This change on the ionisation is visible for the third stage too, specially with the weakness of the He II lines, but besides we observe emission components on H$\alpha$.
	This emission is an effect produced by the reduction of the surface gravity more than an effect produced by the clumping: from the fits previously performed for stars HD 164794 (Fig.~\ref{comp858687}, $\log g=3.92$) and HD 163758 (Fig.~\ref{comp537366}, $\log g=3.41$), it is clear than low surface gravities may produce emission components on H$\alpha$ even for homogeneous models, whereas increase on clumping does not produce emission for large values for $\log g$.
	At the same time, this emission component for H$\alpha$ may be produced by the change on the line-force parameters, specially the increase on the values for $\alpha$ and $\delta$ \citep[see Fig. 2 from][]{araya17}.
	However, a more complete analysis considering the evolutionary tracks for all the studied stellar masses and with a more extended analysis at different clumping values is required in order to confirm or discard this preliminary trends.
	
%_____Evolutionary tracks at low metallicity____________________________________________________________________________
\section{Evolutionary tracks at low metallicity}
	Repeating the procedure shown in previous section, we combine the self-consistent solutions from Table~\ref{standardtable} with those presented on Table~\ref{evolstandardtablez03} in order to create a specific formula for the case $Z/Z_\odot=0.2$:
	\begin{align}\label{mdotz02eq2}
		\log \dot M_{Z=0.2Z_\odot}=&-8.071\times\left[\log\left(\frac{T_\text{eff}}{1000\text{ K}}\right)\right]^{-1}-17.176\times\left[\log\left(\frac{T_\text{eff}}{1000\text{ K}}\right)\right]^{-2}\nonumber\\
		&+33.5\times(\log g)^{-1}-7.444\times(\log g)^{-2}\nonumber\\
		&-8.94\times(R_*/R_\odot)^{-1}-2.336\nonumber\\
		&-0.227\times\log(\text{N/N}_i)\;\;.
	\end{align}
	
\begin{figure}[t!]
		\centering
		\includegraphics[width=0.65\textwidth]{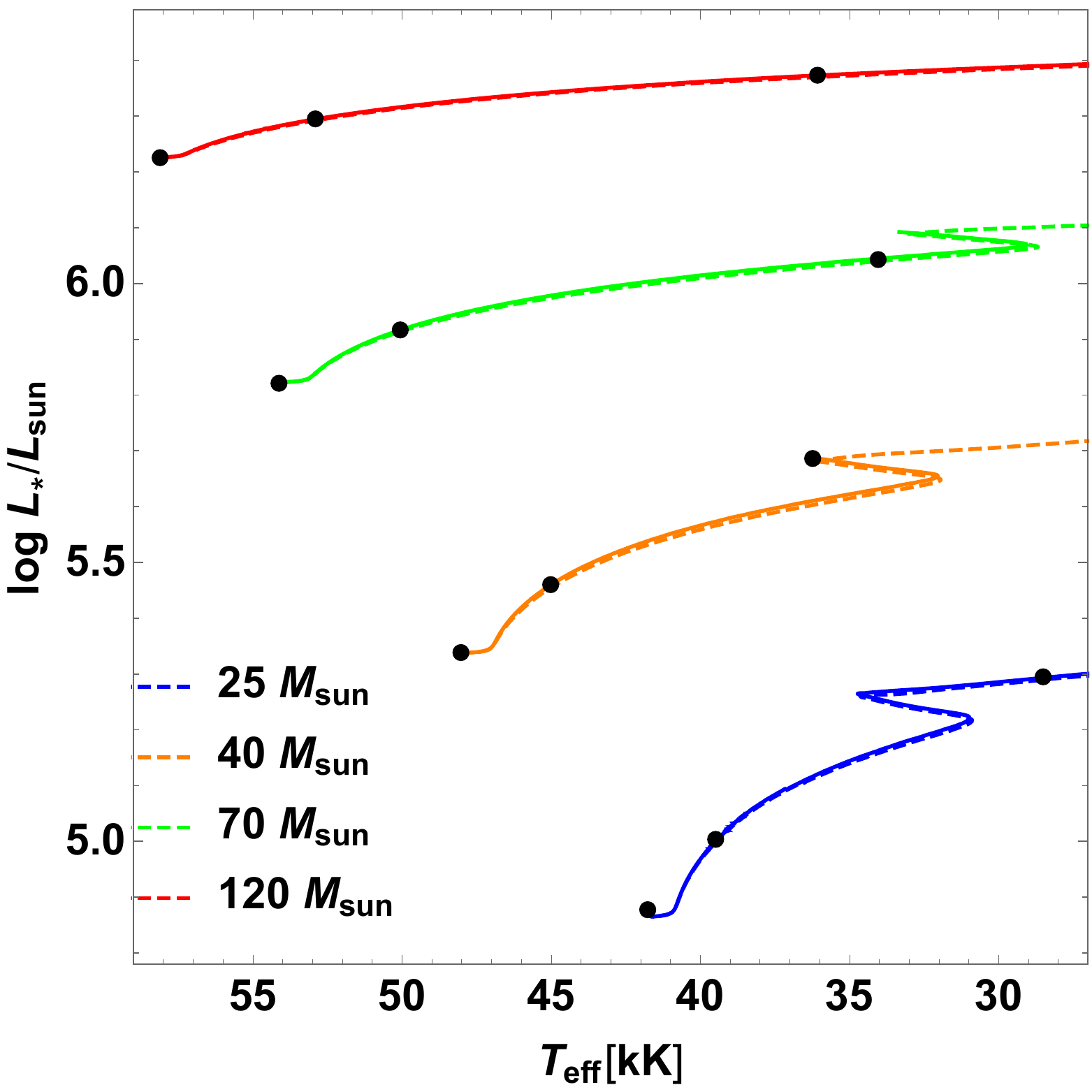}
		\caption{\small{Comparison of evolutionary tracks for stars with 120, 70, 40 and 25 $M_\odot$ at solar metallicity ($Z=0.003$), calculated using our self-consistent $\dot M_\text{SC}$ (solid lines) and using Vink's formula $\dot M_\text{Vink}$ (dashed lines).}}
		\label{hrd_z03_fin}
	\end{figure}

\begin{figure}[t!]
		\centering
		\includegraphics[width=0.51\textwidth]{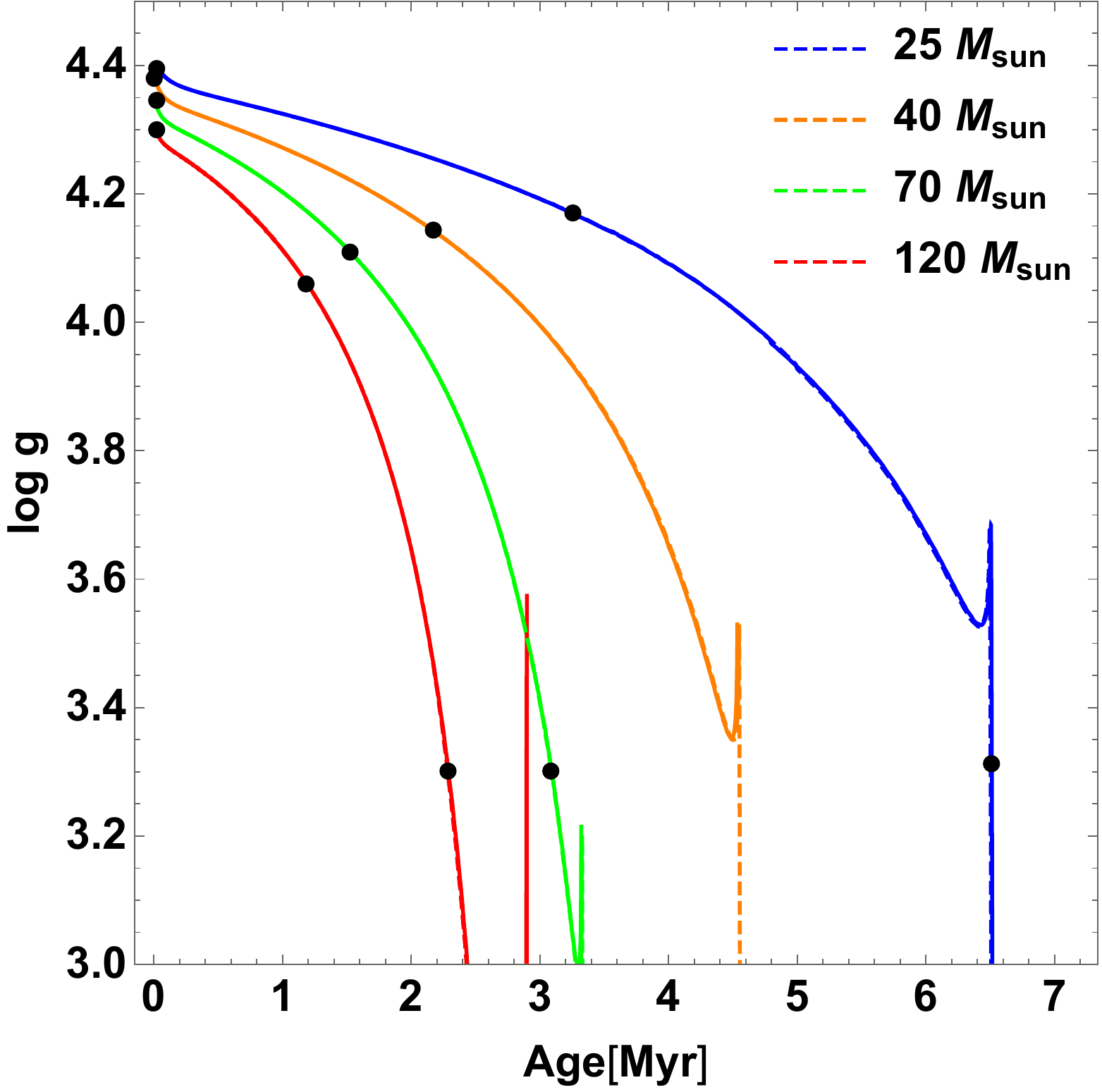}
		\caption{\small{Evolution of $\log g$ of evolutionary tracks for stars with 120, 70, 40 and 25 $M_\odot$ at solar metallicity ($Z=0.003$), calculated using our self-consistent $\dot M_\text{SC}$ (solid lines) and using Vink's formula $\dot M_\text{Vink}$ (dashed lines).}}
		\label{logg_z03_fin}
	\end{figure}

	The new evolutionary tracks are presented on Fig.~\ref{hrd_z03_fin}, whereas the evolution of surface gravities is shown in Fig.~\ref{logg_z03_fin}.
	In this low-metallicity case, even though new paths are slightly more luminous, differences with Vink's tracks is negligible.

	The behaviour of the masses however, shows that mass-loss rates are comparatively far below previous values, in contrast with the case at solar metallicity.
	In spite of this, the effects on the reduction of stellar mass is lower because the value of the mass-loss rate at low metallicity is smaller than the solar metallicity mass-loss rate.
	\begin{figure}[t!]
		\centering
		\includegraphics[width=0.49\textwidth]{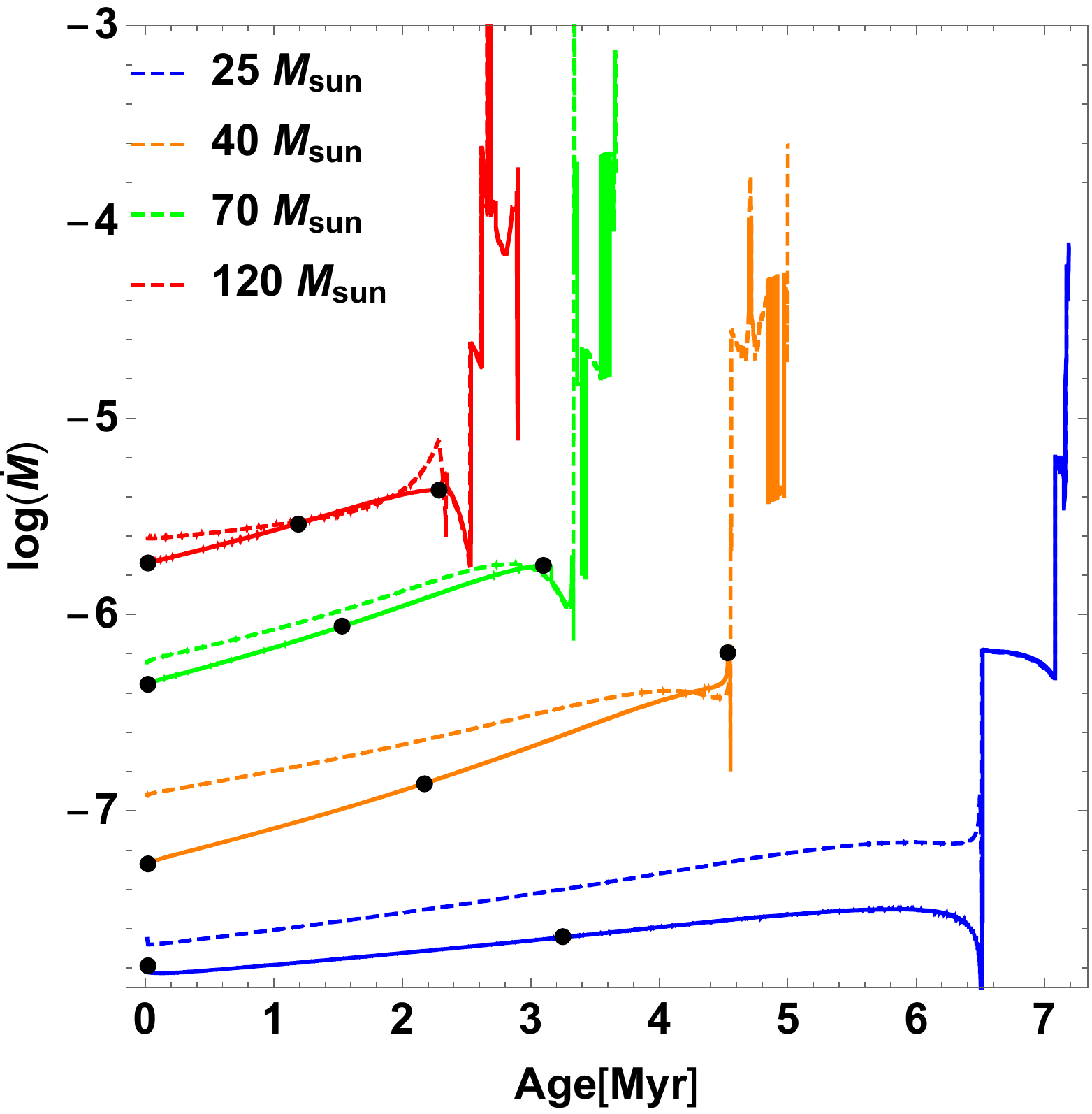}
		\includegraphics[width=0.5\textwidth]{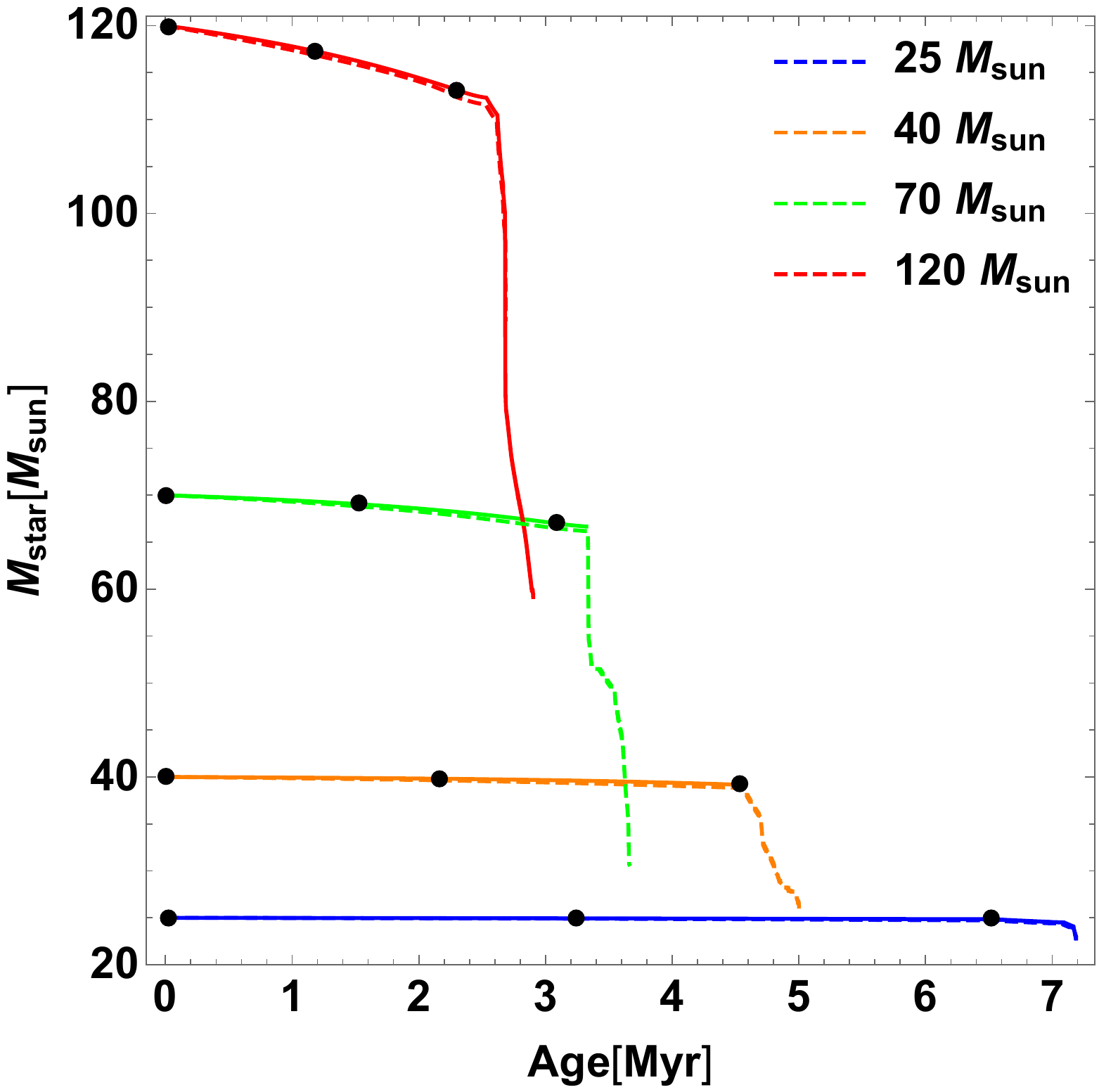}
		\caption{\small{Evolution of $\dot M$ (left panel) and $M_*$ (right panel) of evolutionary tracks for stars with 120, 70, 40 and 25 $M_\odot$ at low metallicity ($Z=0.003$), calculated using our self-consistent $\dot M_\text{SC}$ (solid lines) and using Vink's formula $\dot M_\text{Vink}$ (dashed lines).}}
		\label{mass_z03_fin}
	\end{figure}

%_____Future work_______________________________________________________________________________________________
\section{Summary and future work}
	Given the new self-consistent solutions for stellar winds on massive stars performed on this Thesis \citep[Chapter~\ref{alfakdelta27},][]{alex19}, we have started to use the resulting mass-loss rate to calculate new evolutionary tracks as an alternative to those performed under Vink's recipe.
	Implementing this in \textsc{Genec} code, these new paths over the Hertzsprung-Russell diagram for a set of massive stars at solar and low metallicities are shown in Fig.~\ref{hrd_z14_fin} and Fig.~\ref{hrd_z03_fin}.
	The analysis of these results and their consequences are still a work in progress.
	
	By the obtention of new self-consistent solutions for a new set of stellar parameters, former quick formulae to calculate $\dot M_\text{SC}$ \citep[][Eq.~\ref{mdotz10eq1} and Eq.~\ref{mdotz02eq1}]{alex19} have been improved into Eq.~\ref{mdotz10eq2} and Eq.~\ref{mdotz02eq2}, where it has been included the dependence on the modification of CNO abundances and the He to H ratio.
	In order to ensure our confident on these new formulae, it would be necessary to run more self-consistent solutions to improve the statistics.
%	The best way to do this would include a multi-parameter machine learning procedure, but its implementation and plausability is a matter of discussion yet.
	
	As we mention at the beginning of this chapter, evolutionary tracks for massive stars using self-consistent hydrodynamics is still a not fully performed topic, and therefore it has a lot of potential to research.
	However, as a first stage we will concentrate on the non-rotating case only.
	Self-consistent hydrodynamics under m-CAK prescription are implemented to include the effects of rotation \citep{michel04,araya18}, but a good connection with the correction factor over mass-loss rates described by \citet{maeder00} is needed in order to perform a coherent methodology, a work that is currently beyond the results presented here.

%_____SUMMARY AND CONCLUSIONS_______________________________________________________________________________
\chapter[Summary \& Conclusions]{Summary and Conclusions}\label{conclusions}
	The important role that the massive stars play on the field of Stellar Astrophysics, makes them interesting objects to study.
	In particular, the main link that massive stars have with other topics of Astronomy such as Galactic Astrophysics or Cosmology, is the chemical enrichment and energy output produced by their powerful stellar winds.
%	Therefore, studies on massive stars are necessary to understand the 

	The main feature of the massive stars is their strong mass-loss rates produced by line-driven stellar winds, which largely affects their evolution and therefore their contribution to the galactic chemical enrichment.
	For that reason it is necessary to properly understand the physics behind the stellar wind in order to better constrain values for mass-loss rates.
	For massive stars, line driven winds theory has provided a quite complete theoretical framework capable to predict values for mass-loss rate in the order of the observed values, with small uncertainties \citep{puls08}.
	In the frame of line-driven theory, m-CAK prescription \citep{cak,ppk} have also demonstrated to be versatile and accurate enough to describe the hydrodynamics of the wind and also predict mass-loss rates.
	
	Under the framework of this line-driven wind theory, through this thesis we have performed prescriptions to calculate the line-acceleration in an iterative way combined with the hydrodynamics of the wind, in other words, a \textit{self-consistent solution} for the stellar wind on massive stars.
	Solutions presented here not only satisfy the concordance between the line-acceleration and the hydrodynamics, but also have been calculated using new numerical procedures, new mathematical tools and new atomic information not considered by previous studies, having been self-consistent or not.
	Besides, from the provided solutions is easy to calculate the respective synthetic spectra in order to evaluate their stellar and wind parameters with observations.
	
	The most important part of the work along this thesis has been performed under the m-CAK prescription (see Chapter~\ref{alfakdelta27}).
	We have developed a methodology capable to parametrise the line-acceleration calculating the force multiplier $\mathcal M(t)$ and their line-force parameters $k$, $\alpha$ and $\delta$, using as input an updated atomic database and hydrodynamic profiles from the \textsc{HydWind} code \citep{michel04}, iterating until a convergence criterion is attained.
	This prescription has the big advantage to converge in a short timescale ($\sim10$ minutes), but it includes some assumptions and approximations such as an approximated NLTE treatment for atomic populations and the assumption that line-force parameters are constant through the wind.
	Moreover, the consideration of density inhomogeneities on the wind is not formally included in the iterative calculation of the self-consistent solution and it is only introduced for the execution of the synthetic spectra with FASTWIND.
	Even with these disadvantages, m-CAK prescription has proven to be successful in the calculation of a line-acceleration capable to lead into accurate mass-loss rates, and the synthetic spectra performed by FASTWIND quickly have delivered a good fit in the neighbourhood of the "real" solution compared with observed spectra of massive stars.
	
	It has been found that self-consistent line-force parameters $k$, $\alpha$ and $\delta$ strongly depends on initial stellar parameters determined for the model.
	Dependence in temperature is known and expected because of the force multiplier (Eq.~\ref{forcemultiplier}) is implicitly dependent in temperature.
	However, $k$, $\alpha$ and $\delta$ have demonstrated to have a dependence on surface gravity too, despite the fact that there is no term related with mass for $\mathcal M(t)$.
	This new dependence comes from the iterative procedure, because this time the optical depth $t$ is properly calculated from the hydrodynamics via \textsc{HydWind} \citep[instead using a standard grid as][or a $\beta$-law]{abbott82}.
	More surprisingly, dependence on $\log g$ seems to be deeper than dependence on $T_\text{eff}$ for line-force multipliers, specially for surface gravities in the range of $\log g\sim3.4-3.6$.
	
	As a consequence, wind parameters obtained by the self-consistent m-CAK prescription have shown to be quite sensitive to the surface gravity.
	This is particularly observed for terminal velocities, whose values are not only larger than the constrained by spectral observations when $\log g\ge3.6$ but also presents error margins in the order of $\sim20\%$.
	Hence, it is difficult to perform a more detailed analysis about predictions for terminal velocities from a set of stellar parameters.
	In spite of that, it is interesting to remark the relationship between $v_\infty$ and the line-force parameter $\alpha$, where we found that self-consistent solutions satisfy Eq.~\ref{cinfinit} for low metallicities only.
	Solar metallicities presents a more spread distribution, where the linear relationship is steeper and depends again on the initial surface gravity.
	The big influence from the surface gravity (or implicitly influence from stellar mass) is a matter for future discussion and analysis, specially at the moment to evaluate evolutionary tracks at different initial masses.
	
	Concerning to mass-loss rate (the most interesting case for us), self-consistent values are higher than those observationally determined, but still below the predicted values by the Vink's formula \citep{vink01}.
	These discrepancies between the theoretical self-consistent mass-loss rate $\dot M_\text{SC}$ and the observed one $\dot M_\text{obs}$ might partially be explained by the usage of $\beta$-law on the spectral fitting, but also the consideration of inhomogeneities in the wind (i.e., clumping) plays a role: self-consistent values for $\zeta$-Puppis (HD 66811) and HD 163758 are in the order of $2$ times higher than the clumped values found by \citet{bouret12}, but $\dot M_\text{SC}$ for the unclumped stars analysed by \citet{markova18} differs only in the order of $\sim20-30\%$ with their observed $\dot M_\text{obs}$.
	However, even for the clumped cases it has been found that self-consistent mass-loss rates reproduce accurate synthetic spectra with FASTWIND adjusting clumping factor (see Figures from~\ref{comp413540} to~\ref{comp858687}).
	Therefore, it is possible to conclude that self-consistent solutions under the m-CAK procedure provide accurate mass-loss rate theoretical values, that can be used for future studies.
	Indeed, m-CAK prescription has been used as a basis to calculate new stellar and wind parameters for a set of spectra of massive stars by means of the spectral fitting with FASTWIND.
	
	Simultaneously, we have analysed the case of self-consistent solutions beyond the m-CAK prescription; that means, with a line-acceleration calculated without the assumptions implemented for the previous case such as Sobolev approximation or quasi-NLTE atomic populations.
	Besides, flux field is directly determined by solving the NLTE radiative transfer equation.
	This line-acceleration is obtained from the output of the radiative transfer code CMFGEN, whereas the new hydrodynamics are calculated with the help of the mathematical tool called Lambert $W$-function.
	Iterative combination of the execution of CMFGEN and the Lambert $W$-Function to recalculate $v(r)$ is called Lambert-procedure.
	This full NLTE prescription also provides us an accurate synthetic spectra to fit our standard star ($\zeta$-Puppis), but the self-consistent mass-loss rate obtained this time is $\sim40\%$ lower than under the m-CAK prescription, both cases using the same stellar parameters (see Table~\ref{zpuppissolutions}).
	This discrepancy may be attributed to the differences on both methods to determine the self-consistent line-acceleration, such as the treatment for atomic populations, the inclusion of clumping, the flux field used or the rotational effects.
	Concerning this last point, we have discussed the fact that differences on mass-loss rate would be only around a $\sim25\%$ if rotation had not been included for the m-CAK self-consistent solution.
	However, it is important to remark also that we are not expecting to reproduce the same self-consistent wind parameters for m-CAK and Lambert-procedure because both prescriptions are using different radiative transfer codes to calculate their synthetic spectra, and the differences between them were already outlined by \citet{massey13}.
	The main conclusion is that both self-consistent methodologies predict higher values for mass-loss rates but with a lower clumping factor than previously calculated by former studies.
	Indeed, both self-consistent values for $f_\text{cl}$ (see Table~\ref{zpuppissolutions}) are closer to the value calculated by \citet[][$f_\text{cl}=5.0$]{puls06} than to more recent extremely high values from \citet[][$f_\text{cl}=20$]{bouret12} or \citet[][$f_\text{cl}=10$]{sander17}.
	Nevertheless, it is necessary a more extended analysis in order to understand the physics about the implications of these differences on clumping.
	
%	However, influence of atomic populations is not strong enough to produce huge changes, whereas the consideration of a "clumped" line-acceleration for the m-CAK prescription should produce an even larger $g_\text{line}$ (analogous to Fig.~\ref{clumpinitial}).
%	The remaining difference on both methods that could produce a lower line-acceleration (and hence lower mass-loss rate) for the Lambert-procedure is the flux field.
%	Due to the fact that CMFGEN calculates the radiative transfer equation considering a full line-blanketing effect \citep[i.e., considering the exact impact produced for each atomic population calculated by the solution of the statistical equilibrium equations, see][]{hillier98}, resulting radiative acceleration%\footnote{See Section~\ref{lambertprocedure}, where differences between radiative acceleration $g_\text{rad}$ and line-acceleration $g_\text{line}$ are outlined.} 
%	(Eq.~\ref{glinecalc}) includes then the full influence from the line-blanketing.
%	On the other hand, \textsc{Tlusty} flux fields used on the m-CAK prescription also considers line-blanketing effect but this is more simplified and calculated on the plane-parallel approximation \citep{lanz03}, and then resulting m-CAK line-acceleration might be enhanced.
	
	In spite of the differences on the m-CAK and Lambert-procedures, the first one has the great advantage of the time and CPU saving, so we employ it to obtain theoretical values on wind parameters for subsequent studies.
	Besides the spectral fitting done over a set of HERMES spectra, we use the self-consistent mass-loss rates to perform new evolutionary tracks using the \textsc{Genec} code.
	For main-sequence stars ($\log g\ge3.9$), values for $\dot M_\text{SC}$ are below the classical ones obtained from the Vink's recipe, originating then evolutive tracks with a higher retention of the stellar mass and more luminous.
	It is interesting to notice that, even when $\dot M_{SC}>\dot M_\text{Vink}$ for $\log g\le3.5$, the opposite initial trend produced at the beginning keeps unchanged over the new resulting evolutionary track.
	Ergo, main difference on this new evolutionary tracks with the classical ones lies in the fact that self-consistent mass-loss rates are below the Vink's values for higher surface gravities.
%	Both studies are still in progress, therefore their results requires deeper discussions and analysis before present the final conclusions.
	Both studies, even when their scopes are part of the forthcoming work beyond the thesis presented here, are examples of a novel research in the field of massive stars derived from the self-consistent solutions for the stellar winds presented in this thesis.
	We aim to exploit their potential in the ongoing years.

%_____APÉNDICES_________________________________________________________________________________
%\appendix
%\chapter{Astronomical units}
%	\begin{table}[h!]
%		\centering
%		\begin{tabular}{ccl}
%			\hline\hline
%			Symbol & Name & MKS equivalence\\
%			\hline
%			kK & kilo Kelvin & 1000 K\\
%			$M_\odot$ & solar mass & $1.989\times10^{30}$ kg\\
%			$R_\odot$ & solar radius & $6.9598\times10^{8}$ m\\
%			$L_\odot$ & solar luminosity & $3.826\times10^{26}$ J s$^{-1}$\\
%			\hline
%		\end{tabular}
%	\end{table}

%_____BIBLIOGRAFÍA_______________________________________________________________________________


\begin{thebibliography}{}\addcontentsline{toc}{chapter}{Bibliography}
	\small{
	\bibitem[Abbott, 1982]{abbott82} Abbott, D. C., 1982 ApJ, 259, 282 A
	\bibitem[{{Abbott} \& {Lucy}(1985)}]{abbott85}{Abbott}, D.~C., \& {Lucy}, L.~B. 1985, ApJ, 288, 679
	\bibitem[{{Araya} {et~al.}(2014){Araya}, {Cur{\'e}}, \& {Cidale}}]{araya14}{Araya}, I., {Cur{\'e}}, M., \& {Cidale}, L.~S. 2014, ApJ, 795, 81
	\bibitem[Araya \& Cur{\'e}(2017)]{araya17} Araya, I., \& Cur{\'e}, M.\ 2017, The Lives and Death-throes of Massive Stars, 383
	\bibitem[Araya et al.(2018)]{araya18} Araya, I., Cur{\'e}, M., ud-Doula, A., et al.\ 2018, MNRAS, 477, 755	
	\bibitem[{{Asplund} {et~al.}(2009){Asplund}, {Grevesse}, {Sauval}, \& {Scott}}]{asplund09}{Asplund}, M., {Grevesse}, N., {Sauval}, A.~J., \& {Scott}, P. 2009, ARAA, 47, 481
	\bibitem[{{Bouret} {et~al.}(2005){Bouret}, {Lanz}, \& {Hillier}}]{bouret05}{Bouret}, J.-C., {Lanz}, T., \& {Hillier}, D.~J. 2005, A\&A, 438, 301
	\bibitem[Bouret et al., 2012]{bouret12} Bouret, J.-C., Hillier, D.~J., Lanz, T., \& Fullerton, A.~W.\ 2012, A\&A, 544, A67
%	\bibitem[Bouret et al., 2013]{bouret13} Bouret, J.-C., Lanz, T., Martins, F., et al.\ 2013, A\&A, 555, A1 
%	\bibitem[Bouret et al., 2015]{bouret15} Bouret, J.-C., Lanz, T., Hillier, D.~J., et al.\ 2015, MNRAS, 449, 1545
	\bibitem[Cardona et al., 2010]{cardona10} Cardona, O., Martínez-Arroyo, M., \& López-Castillo, M.~A.\ 2010, ApJ, 711, 239
	\bibitem[Carroll \& Ostlie, 1996]{carroll96} Carroll, B.~W., \& Ostlie, D.~A.\ 1996, Institute for Mathematics and Its Applications,
	\bibitem[Castor et al., 1975]{cak} Castor, J. I., Abbott, D. C. \& Klein, R. I., 1975 ApJ, 195, 157C (CAK)
	\bibitem[{Corless {et~al.}(1993)Corless, Gonnet, Hare, \& Jeffrey}]{corless93}Corless, R.~M., Gonnet, G.~H., Hare, D. E.~G., \& Jeffrey, D.~J. 1993, The Maple Technical Newsletter, 9, 12
	\bibitem[{Corless {et~al.}(1996)Corless, Gonnet, Hare, Jeffrey, \& Knuth}]{corless96}Corless, R.~M., Gonnet, G.~H., Hare, D. E.~G., Jeffrey, D.~J., \& Knuth, D.~E. 1996, Advances in Computational Mathematics, 5, 329
	\bibitem[Crowther, 2007]{crowther07} Crowther, P. A., 2007, ARAA, 45, 177.
	\bibitem[Curé, 2004]{michel04} Curé, M., 2004, ApJ, 614, 929
	\bibitem[Cur{\'e} \& Rial(2004)]{michel04b} Cur{\'e}, M., \& Rial, D.~F.\ 2004, A\&A, 428, 545
	\bibitem[Cur{\'e}, \& Rial(2007)]{michel07} Cur{\'e}, M., \& Rial, D.~F.\ 2007, Astronomische Nachrichten, 328, 513
	\bibitem[Curé et al., 2011]{michel11} Cur{\'e}, Cidale, L., \& Granada, A.\ 2011, ApJ, 737, 18 
	\bibitem[Cur{\'e} et al.(2012)]{michel12} Cur{\'e}, M., Cidale, L., \& Rial, D.~F.\ 2012, ApJ, 757, 142
	\bibitem[de Becker et al.(2010)]{debecker10} de Becker, M., Linder, N., \& Rauw, G.\ 2010, NA, 15, 76 
	\bibitem[Ekstr{\"o}m et al.(2012)]{ekstrom12} Ekstr{\"o}m, S., Georgy, C., Eggenberger, P., et al.\ 2012, A\&A, 537, A146
	\bibitem[{{Fierro-Santill{\'a}n} {et~al.}(2018){Fierro-Santill{\'a}n}, {Zsarg{\'o}}, {Klapp}, {D{\'{\i}}az-Azuara}, {Arrieta}, {Arias}, \& {Sigalotti}}]{fierro18}{Fierro-Santill{\'a}n}, C.~R., {Zsarg{\'o}}, J., {Klapp}, J., {et~al.} 2018, ApJS, 236, 38
	\bibitem[Friend \& Abbott, 1986]{friend86} Friend, D.~B., \& Abbott, D.~C.\ 1986, ApJ, 311, 701
	\bibitem[Gayley, 1995]{gayley95} Gayley, K.~G.\ 1995, ApJ, 454, 410
	\bibitem[Georgy et al.(2012)]{georgy12} Georgy, C., Ekstr{\"o}m, S., Meynet, G., et al.\ 2012, A\&A, 542, A29
	\bibitem[Georgy et al.(2013)]{georgy13} Georgy, C., Ekstr{\"o}m, S., Eggenberger, P., et al.\ 2013, A\&A 558, A103
	\bibitem[Nebot Gómez-Moran \& Oskinova(2018)]{nebot18} Nebot Gómez-Moran, A., \& Oskinova, L.~M.\ 2018, VizieR Online Data Catalog, J/A+A/620/A89
	\bibitem[Gormaz-Matamala et al.(2015)]{alex14} Gormaz-Matamala, A.~C., Herv{\'e}, A., Chen{\'e}, A.-N., et al.\ 2015, New Windows on Massive Stars, 100
	\bibitem[Gormaz-Matamala et al.(2019)]{alex19} Gormaz-Matamala, A.~C., Cur{\'e}, M., Cidale, L.~S., \& Venero, R.~O.~J.\ 2019, ApJ, 873, 131 
	\bibitem[Gray, \& Corbally(2009)]{spectralclassification} Gray, R.~O., \& Corbally, C.\ 2009, Stellar Spectral Classification by Richard O. Gray and Christopher J. Corbally. Princeton University Press
	\bibitem[Gr{\"a}fener et al.(2002)]{grafener02} Gr{\"a}fener, G., Koesterke, L., \& Hamann, W.-R.\ 2002, A\&A, 387, 244 
	\bibitem[Gr{\"a}fener, \& Hamann(2008)]{grafener08} Gr{\"a}fener, G., \& Hamann, W.-R.\ 2008, A\&A, 482, 945
	\bibitem[Groh et al.(2014)]{groh14} Groh, J.~H., Meynet, G., Ekstr{\"o}m, S., et al.\ 2014, A\&A, 564, A30
	\bibitem[Groh et al.(2019)]{groh19} Groh, J.~H., Ekstr{\"o}m, S., Georgy, C., et al.\ 2019, A\&A, 627, A24
	\bibitem[Grunhut et al.(2012)]{grunhut12} Grunhut, J.~H., Wade, G.~A., Sundqvist, J.~O., et al.\ 2012, MNRAS, 426, 2208 
	\bibitem[Grunhut et al.(2017)]{grunhut17} Grunhut, J.~H., Wade, G.~A., Neiner, C., et al.\ 2017, MNRAS, 465, 2432 
%	\bibitem[Georgy et al., 2013]{georgy13}Georgy, C., Ekstr{\"o}m, S., Eggenberger, P., et al.\ 2013, A\&A, 558, A103
	\bibitem[Hamann \& Gr{\"a}fener(2003)]{hamann03} Hamann, W.-R., \& Gr{\"a}fener, G.\ 2003, A\&A, 410, 993 
	\bibitem[Hillier(1990a)]{hillier90a} Hillier, D.~J.\ 1990, A\&A, 231, 111 
	\bibitem[{{Hillier}(1990b)}]{hillier90b}{Hillier}, D.~J. 1990, A\&A, 231, 116
	\bibitem[{{Hillier} \& {Miller}(1998)}]{hillier98}{Hillier}, D.~J., \& {Miller}, D.~L. 1998, ApJ, 496, 407
	\bibitem[{{Hillier} \& {Lanz}(2001)}]{hillier01}{Hillier}, D.~J., \& {Lanz}, T. 2001, in Astronomical Society of the Pacific Conference Series, Vol. 247, Spectroscopic Challenges of Photoionized Plasmas, ed. G.~{Ferland} \& D.~W. {Savin}, 343
	\bibitem[{{Hubeny} \& {Lanz}(1995)}]{hubeny95}{Hubeny}, I., \& {Lanz}, T. 1995, ApJ, 439, 875
	\bibitem[{{Krti{\v c}ka} \& {Kub{\'a}t}(2010)}]{kk10}{Krti{\v c}ka}, J., \& {Kub{\'a}t}, J. 2010, A\&A, 519, A50
	\bibitem[{{Krti{\v{c}}ka} {et~al.}(2015){Krti{\v{c}}ka}, {Kub{\'a}t}, \& {Krti{\v{c}}kov{\'a}}}]{krticka15}{Krti{\v{c}}ka}, J., {Kub{\'a}t}, J., \& {Krti{\v{c}}kov{\'a}}, I. 2015, A\&A, 579, A111
  	\bibitem[{{Krti{\v c}ka} \& {Kub{\'a}t}(2017)}]{kk17}{Krti{\v c}ka}, J., \& {Kub{\'a}t}, J. 2017, A\&A, 606, A31
	\bibitem [Kroupa, 2001]{kroupa01} Kroupa, P., 2001, MNRAS 322, 231K.
	\bibitem[{{Kudritzki} {et~al.}(1989){Kudritzki}, {Pauldrach}, {Puls}, \& {Abbott}}]{kppa89}{Kudritzki}, R.~P., {Pauldrach}, A., {Puls}, J., \& {Abbott}, D.~C. 1989, A\&A, 219, 205
	\bibitem[{{Kudritzki} \& {Puls}(2000)}]{kudritzki00}{Kudritzki}, R.-P., \& {Puls}, J. 2000, Annual Review of Astronomy and Astrophysics, 38, 613
	\bibitem[{{Kudritzki}(2002)}]{kudritzki02}{Kudritzki}, R.~P. 2002, ApJ, 577, 389
	\bibitem[{{Kurucz}(1979)}]{kurucz79}{Kurucz}, R.~L. 1979, ApJS, 40, 1
	\bibitem [Lamers \& Cassinelli, 1999]{stellarwind} Lamers, H. \& Cassinelli, J., \textit{Introduction to Stellar Winds}, Cambridge University Press, 1999.
	\bibitem[Lanz, 2000]{lanz00} Lanz, T., 2000, EAA Book E, 2102 L, \textit{Stellar Atmospheres}.
	\bibitem[{{Lanz} \& {Hubeny}(2003)}]{lanz03}{Lanz}, T., \& {Hubeny}, I. 2003, ApJS, 146, 417
	\bibitem[Liermann(2015)]{liermann15} Liermann, A.\ 2015, Wolf-rayet Stars: Proceedings of an International Workshop Held in Potsdam, 129
	\bibitem[Lucy \& Solomon, 1970]{lucy70} Lucy, L.~B., \& Solomon, P.~M.\ 1970, ApJ, 159, 879
	\bibitem[Maeder \& Meynet, 1987]{maeder87} Maeder, A., \& Meynet, G., 1987, A\&A, 182, 243M
	\bibitem[{{Markova} \& {Puls}(2008)}]{markova08}{Markova}, N., \& {Puls}, J. 2008, A\&A, 478, 823
	\bibitem[{{Markova} {et~al.}(2018){Markova}, {Puls}, \& {Langer}}]{markova18}{Markova}, N., {Puls}, J., \& {Langer}, N. 2018, A\&A, 613, A12
	\bibitem[{{Mazzali} \& {Lucy}(1993)}]{mazzali93}{Mazzali}, P.~A., \& {Lucy}, L.~B. 1993, A\&A, 279, 447
	\bibitem[Meynet et al., 1994]{meynet94} Meynet, G., Maeder, A., Schaller, G., Schaerer, D., \& Charbonnel, C.\ 1994, A\&AS, 103,  
	\bibitem[Mihalas et al.(1975)]{mihalas75} Mihalas, D., Kunasz, P.~B., \& Hummer, D.~G.\ 1975, ApJ, 202, 465 
	\bibitem[{{Mihalas}(1978)}]{mihalas78}{Mihalas}, D. 1978, {Stellar atmospheres /2nd edition/}
	\bibitem[{{Mokiem} {et~al.}(2005){Mokiem}, {de Koter}, {Puls}, {Herrero}, {Najarro}, \& {Villamariz}}]{mokiem05}{Mokiem}, M.~R., {de Koter}, A., {Puls}, J., {et~al.} 2005, A\&A, 441, 711
	\bibitem[{{M{\"u}ller} \& {Vink}(2008)}]{muller08}{M{\"u}ller}, P.~E. \& {Vink}, J.~S. 2008, A\&A, 492, 493
	\bibitem[{{Najarro} {et~al.}(2011){Najarro}, {Hanson}, \& {Puls}}]{najarro11}{Najarro}, F., {Hanson}, M.~M., \& {Puls}, J. 2011, A\&A, 535, A32
	\bibitem[Maeder(1983)]{maeder83} Maeder, A.\ 1983, A\&A, 120, 113
	\bibitem[Maeder(1987)]{maeder87a} Maeder, A.\ 1987, A\&A, 173, 247
	\bibitem[Maeder, \& Meynet(1987)]{maeder87b} Maeder, A., \& Meynet, G.\ 1987, A\&A, 182, 243
	\bibitem[Maeder, \& Meynet(2000)]{maeder00} Maeder, A., \& Meynet, G.\ 2000, Astronomy and Astrophysics, 361, 159
	\bibitem[Marcolino et al.(2017)]{marcolino17} Marcolino, W.~L.~F., Bouret, J.-C., Lanz, T., et al.\ 2017, MNRAS, 470, 2710
	\bibitem[Martins et al.(2005)]{martins05} Martins, F., Schaerer, D., \& Hillier, D.~J.\ 2005, A\&A, 436, 1049 
	\bibitem[Martins et al.(2015)]{martins15} Martins, F., Herv{\'e}, A., Bouret, J.-C., et al.\ 2015, A\&A, 575, A34
	\bibitem[Massey et al.(2013)]{massey13} Massey, P., Neugent, K.~F., Hillier, D.~J., et al.\ 2013, ApJ, 768, 6
%	\bibitem[Martins \& Palacios, 2016]{martins16} Martins, F., \& Palacios, A.\ 2016, arXiv:1612.03044
	\bibitem[Nobili \& Turolla(1988)]{nobili88} Nobili, L., \& Turolla, R.\ 1988, ApJ, 333, 248
	\bibitem[Noebauer \& Sim, 2015]{noebauer15} Noebauer, U. M., \& Sim, S. A. 2015, MNRAS, 453, 3120
	\bibitem[Pauldrach et al., 1986]{ppk} Pauldrach, A., Puls, J., \& Kudritzki, R.~P.\ 1986, A\&A, 164, 86 
	\bibitem[{{Pauldrach}(2003)}]{pauldrach03}{Pauldrach}, A.~W.~A. 2003, Reviews in Modern Astronomy, 16, 133
	\bibitem[{{Puls}(1987)}]{puls87}{Puls}, J. 1987, A\&A, 184, 227
	\bibitem[{{Puls} {et~al.}(1996){Puls}, {Kudritzki}, {Herrero}, {Pauldrach}, {Haser}, {Lennon}, {Gabler}, {Voels}, {Vilchez}, {Wachter}, \& {Feldmeier}}]{puls96}{Puls}, J., {Kudritzki}, R.-P., {Herrero}, A., {et~al.} 1996, A\&A, 305, 171
	\bibitem[Puls et al., 2000]{puls00} Puls, J., Springmann, U., \& Lennon, M.\ 2000, A\&AS, 141, 23
	\bibitem[{{Puls} {et~al.}(2005){Puls}, {Urbaneja}, {Venero}, {Repolust}, {Springmann}, {Jokuthy}, \& {Mokiem}}]{puls05}{Puls}, J., {Urbaneja}, M.~A., {Venero}, R., {et~al.} 2005, A\&AS, 435, 669
	\bibitem[{{Puls} {et~al.}(2006){Puls}, {Markova}, {Scuderi}, {Stanghellini}, {Taranova}, {Burnley}, \& {Howarth}}]{puls06}{Puls}, J., {Markova}, N., {Scuderi}, S., {et~al.} 2006, A\&A, 454, 625
	\bibitem [Puls et al.(2008)]{puls08} Puls, J., Vink, J.~S., \& Najarro, F.\ 2008, A\&ARv, 16, 209 
	\bibitem[Prialnik(2009)]{prialnik10} Prialnik, D.\ 2009, An Introduction to the Theory of Stellar Structure and Evolution by Dina Prialnik. Cambridge University Press
	\bibitem[Prinja et al.(1990)]{prinja90} Prinja, R.~K., Barlow, M.~J., \& Howarth, I.~D.\ 1990, ApJ, 361, 607
	\bibitem[{{Repolust} {et~al.}(2004){Repolust}, {Puls}, \& {Herrero}}]{repolust04}{Repolust}, T., {Puls}, J., \& {Herrero}, A. 2004, A\&A, 415, 349
	\bibitem[{{Sahu} \& {Blaauw}(1993)}]{sahu93}{Sahu}, M., \& {Blaauw}, A. 1993, in Astronomical Society of the Pacific Conference Series, Vol.~35, Massive Stars: Their Lives in the Interstellar Medium, ed. J.~P. {Cassinelli} \& E.~B. {Churchwell}, 278
	\bibitem [Salpeter, 1955]{salpeter55} Salpeter, E., 1955 ApJ 121, 161S
	\bibitem[{{Sander} {et~al.}(2017){Sander}, {Hamann}, {Todt}, {Hainich}, \& {Shenar}}]{sander17}{Sander}, A.~A.~C., {Hamann}, W.-R., {Todt}, H., {Hainich}, R., \& {Shenar}, T. 2017, A\&A, 603, A86
  	\bibitem[{{Santolaya-Rey} {et~al.}(1997){Santolaya-Rey}, {Puls}, \& {Herrero}}]{santolaya97}{Santolaya-Rey}, A.~E., {Puls}, J., \& {Herrero}, A. 1997, A\&A, 323, 488
	\bibitem [Schaerer \& Schmutz, 1994]{schaerer94} Schaerer, D., \& Schmutz, W.\ 1994, A\&A, 288, 231
	\bibitem[Smith(2014)]{smith14} Smith, N.\ 2014, ARAA, 52, 487
	\bibitem[Sobolev(1960)]{sobolev60}{Sobolev}, V.~V. 1960, {Moving envelopes of stars}
	\bibitem[{{Sota} {et~al.}(2011){Sota}, {Ma{\'{\i}}z Apell{\'a}niz}, {Walborn}, {Alfaro}, {Barb{\'a}}, {Morrell}, {Gamen}, \& {Arias}}]{sota11}{Sota}, A., {Ma{\'{\i}}z Apell{\'a}niz}, J., {Walborn}, N.~R., {et~al.} 2011, ApJS, 193, 24
	\bibitem[Sota et al.(2014)]{sota14} Sota, A., Ma{\'\i}z Apell{\'a}niz, J., Morrell, N.~I., et al.\ 2014, ApJS, 211, 10
	\bibitem[Springmann \& Pauldrach(1992)]{springmann92} Springmann, U.~W.~E., \& Pauldrach, A.~W.~A.\ 1992, A\&A, 262, 515 
	\bibitem[Sundqvist, \& Puls(2018)]{sundqvist18} Sundqvist, J.~O., \& Puls, J.\ 2018, A\&A, 619, A59
%	\bibitem [Tramper et al., 2014]{tramper14} Tramper, F., Sana, H., de Koter, A., Kaper, L., \& Ram{\'{\i}}rez-Agudelo, O.~H.\ 2014, AAP, 572, A36
	\bibitem[Venero et al.(2016)]{venero16} Venero, R.~O.~J., Cur{\'e}, M., Cidale, L.~S., et al.\ 2016, ApJ, 822, 28
	\bibitem[Vink et al., 1999]{vink99} Vink, J.~S., de Koter, A., \& Lamers, H.~J.~G.~L.~M.\ 1999, A\&A, 350, 181
	\bibitem[Vink et al., 2000]{vink00} Vink, J.~S., de Koter, A., \& Lamers, H.~J.~G.~L.~M.\ 2000, A\&A, 362, 295
	\bibitem[Vink et al., 2001]{vink01} Vink, J.~S., de Koter, A., \& Lamers, H.~J.~G.~L.~M.\ 2001, A\&A, 369, 574 
	}
\end{thebibliography}
\end{document}